**FACULTE POLYTECHNIQUE DE MONS**

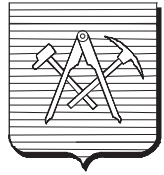

# Thèse de Doctorat

*présentée en vue de l'obtention du titre de*
**Docteur en Sciences Appliquées**
*par*

**Ir Fabien ROGISTER**

---

## Nonlinear Dynamics of Semiconductor Lasers Subject to Optical Feedback

---

**Membres du Jury**

Prof. **J. LOBRY**, Président
Dr **A. GAVRIELIDES**,
Air Force Research Laboratory (USA)
Dr **C.R. MIRASSO**,
Universitat de les Illes Balears (Espagne)
Prof. **I. VERETENNICOFF**, VUB
Dr **Th. ERNEUX**, ULB
Dr **O. DEPARIS**
Dr **P. MEGRET**, co-promoteur
Prof. **M. BLONDEL**, promoteur
Prof. **J. HANTON**, Doyen

**Octobre 2001**

Thèse préparée au Service d'Electromagnétisme et de Télécommunications

# Nonlinear Dynamics of Semiconductor Lasers Subject to Optical Feedback

## Abstract

The three goals of this PhD thesis are to improve the understanding of the mechanisms underlying the dynamical instabilities observed in semiconductor lasers subject to external optical feedback, to propose a technique to suppress these instabilities and finally to take advantage of them, for instance for secure communications or to generate microwave time-periodic oscillations. Highlights of the thesis can be summarized as follows:

1. We demonstrate numerically and experimentally an all-optical technique for stabilization of an external cavity semiconductor laser operating in the so-called low-frequency fluctuation regime. This technique, which uses a second optical feedback to stabilize a laser diode pumped close to threshold and subject to a first coherent optical feedback, has several advantages over existing methods. Most important is that it does not require any modification of the laser operating parameters or of the first optical feedback; it is moreover robust, reliable and easy to implement.
2. We report on experimental generation of high-frequency periodic oscillations in a laser diode subject to coherent optical feedback from two external cavities. This observation has motivated analytical and numerical studies of unusual Hopf frequencies in laser diodes subject to coherent optical feedback from both single and double external cavities. The latter have revealed that high-frequency periodic oscillations are associated to a beating mechanism between modes of the compound cavity formed by the laser and the external cavities. As a practical consequence, laser diodes coupled to external cavities can be viewed as promising all-optical sources of microwave oscillations.
3. The necessity of including multimode dynamics and stochastic noise sources in models for the study of low-frequency fluctuations in semiconductor lasers subject to coherent optical feedback has attracted much attention these past years. We demonstrate that a multimode model that takes spontaneous emission into account and assumes a parabolic gain profile reproduces most of the features that have been experimentally observed, in particular the coexistence of in-phase and out-of phase fast dynamics. We demonstrate furthermore that, in the frame of our model, the out-of-phase dynamics is by no means related to the intrinsic stochastic nature of spontaneous emission.
4. In agreement with experiment, the same model also predicts the simultaneous occurrence of bursts in the free modes and dropouts in the selected mode which is observed when a laser diode is subjected to a mode-selective optical feedback. We establish that the selective mode-induced LFF is associated with collisions of the system trajectory in phase-space with saddle-type antimodes preceded by a chaotic itinerancy of the system trajectory among external-cavity modes.

5.      Finally, we demonstrate numerically anticipative synchronization between a first diode subject to incoherent optical feedback and a second diode driven by the first one through incoherent optical injection. We show that this synchronization scheme can be applied to secure communications. The proposed method requires no fine tuning of the diode optical frequencies, a clear advantage over other schemes based on coherent optical feedback and injection. It is therefore attractive for experimental realization.

# De la dynamique non-linéaire des lasers à semi-conducteurs soumis à rétroaction optique

## Résumé

L'objectif de cette thèse consiste à améliorer la compréhension des mécanismes sous-jacents aux instabilités dynamiques observées dans les lasers soumis à des rétroactions optiques, à proposer une technique permettant leur suppression dans le cas d'une rétroaction optique cohérente et, finalement, à tirer profit de ces instabilités. Les principaux apports originaux de cette thèse peuvent être résumés comme suit :

1. Nous démontrons théoriquement et expérimentalement une technique permettant la stabilisation d'un laser à semi-conducteur soumis à une rétroaction optique cohérente et fonctionnant dans le régime dit des fluctuations de puissance à basses fréquences. Cette technique, qui utilise une seconde rétroaction optique pour stabiliser le laser, a plusieurs avantages. Le plus important de ceux-ci est sans doute qu'elle ne nécessite aucune modification des paramètres accessibles du laser, tels que courant d'injection et température, ou de la première rétroaction optique; la méthode est en outre robuste, fiable et de réalisation aisée.
2. Nous observons expérimentalement la génération d'oscillations périodiques de grandes fréquences par une diode laser soumise à une double rétroaction optique cohérente. Ce résultat expérimental a motivé l'étude analytique et numérique de fréquences de Hopf inhabituelles dans les cas d'une diode laser soumise à rétroactions optiques simple et double. Ces études théoriques ont montré que les oscillations de grandes fréquences observées résultent d'un mécanisme de battement entre des modes de la cavité composée du laser et des cavités externes. Une application prometteuse de ce phénomène consiste en la réalisation de sources complètement optiques à grand débit.
3. La nécessité d'utiliser des modèles prenant en compte les nombreux modes longitudinaux des lasers à semi-conducteurs et incorporant des sources de bruit pour l'étude du régime des fluctuations à basses fréquences a fait l'objet de nombreux débats, parfois passionnés, au cours des dernières années. Nous démontrons qu'un modèle qui prend en compte l'émission spontanée et utilise une approximation quadratique de la dépendance fréquentielle du gain permet de reproduire la quasi-totalité des observations expérimentales faites à ce jour, notamment la coexistence de fluctuations rapides en phase et hors-phase des modes longitudinaux du laser. Nous montrons de plus que la dynamique hors-phase des modes n'est pas liée à la nature intrinsèquement stochastique de l'émission spontanée.

4.   Dans le cas d'un laser soumis à une rétroaction optique agissant sur un seul de ses modes, le même modèle prédit la brusque activation des modes libres simultanément aux chutes de puissance du mode sélectionné par la rétroaction. Nous établissons que les fluctuations à basses fréquences sont associées aux collisions de la trajectoire du système dans l'espace des phases avec des points de selle.

5.   Nous démontrons numériquement la synchronisation par anticipation entre deux diodes lasers en régime chaotique, la première étant soumise à une rétroaction optique incohérente et la seconde à un couplage optique incohérent avec la première. Nous montrons que ce schéma de synchronisation peut être appliqué à des fins cryptographiques. La méthode que nous proposons ne nécessite aucun réglage précis des fréquences optiques des deux lasers, ce qui constitue un avantage important par rapport à d'autres systèmes cryptographiques basés sur des diodes lasers soumises à des rétroactions et à des couplages optiques cohérents. Notre schéma de communications sécurisées est donc particulièrement attrayant d'un point de vue expérimental.

# Contents

*Contents*

*Contents*

# Preface

Semiconductor lasers are not only key components in modern technology; they are also highly nonlinear systems. Both features make them a fascinating and challenging topic of research. This thesis – the first in this field at the Service d'Electromagnétisme et de Télécommunications of the Faculté Polytechnique de Mons – deals with the nonlinear dynamics of these devices when they are subjected to external delayed optical feedback.

Most of the material presented here has been published in refereed scientific journals as rapid communications or letters and presented at international conferences. My publications are listed hereafter.

My thesis and publications are the most visible concretization of my work over the past four years. I would like to mention besides two other tasks from which I also learned a great deal: firstly, I supervised the research work of a few engineering students in their last year; two of those students are now working on their own PhD thesis; secondly, during a few months, I selected and ordered a complete and performing experimental setup which will allow new relevant experiments to be carried out in the near future.

I wish to express my sincere gratitude to Professor M. Blondel, promoter of this thesis, for his warm hospitality in the department that he leads and for his firm support during these four years. I thank Dr P. Mégret, my co-promoter, for his accessibility, his judicious comments on my research work and his patient reading of this manuscript. Dr O. Deparis has also contributed to my guidance. I have especially appreciated his help when writing my first papers. O. Deparis has taught me that the work of the scientist goes beyond research and discovery and helped me to improve my communication skills. I whish to thank Dr T. Erneux (Université Libre de Bruxelles) with whom I have collaborated on different occasions. He made me aware of the power of the method of asymptotic mathematical expansion. I thank Dr A. Gavrielides for the opportunity he gave me to spend a few weeks at the Nonlinear Optics Center of the Air Force Research Laboratory in Albuquerque and for enlightening discussions. I thank also Dr D.W. Sukow and Dr D. Pieroux with whom I had the pleasure to work respectively on the experiment in Albuquerque and on the theory of laser diodes subject to incoherent optical feedback.

I have enjoyed the comradeship of my colleagues, in particular Drs R. Kiyan, A. Gusarov and A. Fotiadi, Ir O. Pottiez, Ir J. Hanoteau, Ir M. Sciamanna and Mr. O. Aubry; they have contributed in making my stay in the department very enjoyable.

Last but not least, I thank Djamila for her continuous support and her patience and my parents for ensuring the logistic.

Fabien Rogister,
October 28, 2001.

# Publications in refereed journals

1. F. Rogister, P. Mégret, O. Deparis, M. Blondel, and T. Erneux, "Suppression of low-frequency fluctuations and stabilization of a semiconductor laser subjected to optical feedback from a double cavity: theoretical results," *Opt. Lett.*, vol. 24, pp. 1218-1220, 1999.
2. F. Rogister, D.W. Sukow, A. Gavrielides, P. Mégret, O. Deparis, and M. Blondel, "Experimental demonstration of suppression of low-frequency fluctuations and stabilization of an external cavity laser diode," *Opt. Lett.*, vol. 25, pp. 808-810, 2000.
3. T. Erneux, F. Rogister, A. Gavrielides and V. Kovanis, "Bifurcation to mixed external cavity mode solutions for semiconductor lasers subject to optical feedback," *Opt. Commun.*, vol. 183, pp. 467-477, 2000.
4. F. Rogister, P. Mégret, O. Deparis, and M. Blondel, "Coexistence of in-phase and out-of-phase dynamics in a multimode external cavity laser diode operating in the low-frequency fluctuations regime," *Phys. Rev. A*, vol. 62, pp. 061803 (R) 1-4, 2000.
5. F. Rogister, A. Locquet, D. Pieroux, M. Sciamanna, O. Deparis, P. Mégret, and M. Blondel, "Secure communication scheme using laser diodes subject to incoherent optical feedback and incoherent optical injection," *Opt. Lett.*, vol. 26, pp. 1486-1488, 2001.
6. A. Locquet, F. Rogister, M. Sciamanna, P. Mégret, and M. Blondel, "Two types of synchronization of two distant unidirectionally coupled chaotic external-cavity semiconductor lasers," *Phys. Rev. E*, vol. 64, pp. 045203 – 045206 (R), 2001.
7. F. Rogister, M. Sciamanna, O. Deparis, P. Mégret, and M. Blondel, "Low-frequency fluctuation regime in a multimode semiconductor laser subject to a mode-selective optical feedback," *Phys. Rev. A*, vol. 65, pp. 015602 1-4, 2002.
8. F. Rogister, D. Pieroux, M. Sciamanna, P. Mégret, and M. Blondel, "Anticipating synchronization of two chaotic laser diodes by incoherent optical coupling and its application to secure communications", accepted for publication in *Optics Communications*.
9. M. Sciamanna, F. Rogister, O. Deparis, P. Mégret, M. Blondel, T. Erneux, "Bifurcation to polarization self-modulation in vertical-cavity surface-emitting lasers," *Opt. Lett.*, vol. 27, pp. 261-263, 2002.
10. M. Sciamanna, T. Erneux, F. Rogister, O. Deparis, P. Mégret, M. Blondel, "Bifurcation bridges between external cavity modes lead to polarization self-

# Publications in conference proceedings

# 1. Introduction

Since the seminal paper by Schawlow and Townes in 1958 [1], the LASER (Light Amplification by Stimulated Emission of Radiation) effect has found applications in a still increasing number of fields. For instance, lasers are key components in data storage systems such as compact-disk audio players and CD-ROMs, holography, remote sensing, material processing, surface treatment, fiber-optics communication systems, supermarket bar-code scanners, medical diagnostics, surgery, interferometry, laser printers…

Semiconductor lasers, also referred to as laser diodes, have several advantages over other types of lasers. Their length is very small (about 250 µm for edge-emitting laser diodes and 1 µm for vertical cavity surface emitting laser diodes instead of several tenths of cm for gas lasers). They are easy to produce in large quantities and at relatively low prices. They are also low power consume and very efficient. Due to all these technological advantages, laser diodes have become indispensable in fiber-based optical networks and optical data storage applications.

Laser diodes differ also from conventional lasers by their much more pronounced sensitivity to external perturbations owing to their high gain per unit of length and, in the case of edge-emitting laser diodes, to the relatively low reflectivity of their facets (0.3 instead of 0.99 in conventional lasers) or, in the case of vertical cavity surface emitting laser diodes, to their short internal roundtrip time. These perturbations, which are undesirable in most applications, lead to dynamical instabilities, such as chaos, that in turn can degrade severely spectral and temporal performances of the laser diodes.

The study of nonlinear dynamics, which has been prompted by the discovery of deterministic chaos by Lorenz in 1963 in a simplified model of convection rolls in the atmosphere [2], has become these last three decades one of the most active fields in modern science. Deterministic chaos denotes irregular motion that is generated by nonlinear systems whose dynamical laws uniquely determine the time evolution of a state of these systems from knowledge of their previous history [3]. At the beginning, chaos was considered as a mathematical curiosity, or as a nuisance. It is nowadays well established that many irregular and unpredictable behaviors in physical systems that were attributed in the past to random external perturbations, are in fact associated to deterministic chaos. They are the consequence of the sensitive dependence on the initial conditions that is a potential property of nonlinear systems to separate exponentially fast initially close trajectories. As a result, the exact prediction of the long-time behavior of such systems is practically impossible. Lorenz called metaphorically this limitation for predictability in chaotic systems the butterfly effect ("Does the flap of a butterfly's wings in Brazil set off a tornado in Texas?"). The impact of the conceptual implications due to the discovery that the dynamics of deterministic systems can be inherently unpredictable has often been compared to those spread when it was found that quantum mechanics only allows statistical predictions.

Due to the spread of high-speed computers and refined experimental techniques, deterministic chaos has been found in multiple systems as diverse as forced pendulums, chemical reactions, stimulated heart cells, biological models for population dynamics, fluids near the onset of turbulence, classical many-body systems, lasers and nonlinear optical devices (Refs. 3-7 and references therein). Yet more surprising, intensive studies have revealed that the mechanisms leading to the transition from regular to chaotic behaviors as a control parameter is varied are universal: completely different systems follow similar routes to chaos. It is generally accepted that all routes to chaos have not yet been discovered but three of them are quite recurrent. One of these was proposed in 1971 by Ruelle and Takens [8] and modified in 1978 with the assistance of Newhouse [9]. They showed that only three bifurcations are needed to observe chaotic motion: from steady state to periodic behavior, from periodic to two-frequencies quasiperiodic behavior, and then from quasiperiodic behavior to chaos. In 1979, Manneville and Pomeau discovered a second frequent route to chaos, the so-called intermittency route [10,11]. Beyond a critical value of a control parameter, the regular behavior of the system is interrupted by randomly distributed periods of irregular bursts; the frequency of these intermittent events increases as the control parameter departs from its critical value, leading eventually to fully developed chaotic behavior. In the period-doubling route to chaos, which was proposed by Feigenbaum [12] in 1978, the periodic behavior of a system undergoes a succession of bifurcations as a control parameter is varied. The period doubles at each bifurcation until it becomes infinite at a finite value of the parameter. The behavior of the system then becomes chaotic.

The study of nonlinear dynamics in lasers was initiated by Haken in 1975 [13] when he established the mathematical isomorphism between the Maxwell-Bloch equations, which describe the dynamics of the electric field, the mean polarization of the atoms and the population inversion, and the Lorenz equations for atmospheric convective flow [2]. Since the work of Haken, much attention has been devoted to the nonlinear dynamics of lasers in general, and laser diodes in particular.

Single-mode semiconductor lasers are generally characterized by three variables: the optical power and phase and the electron-hole pair number (or carrier number) in the active layer. In isolated laser diodes, the optical phase is a slaved dynamical variable. Since the occurrence of chaos is not possible in two-dimensional autonomous systems, isolated laser diodes cannot exhibit chaotic motion as long as the injection current is not modulated. By contrast, chaos can be observed when laser diodes are current modulated [14] (in which case they constitute two-dimensional non-autonomous systems) or optically injected from another laser [15] (in which case they constitute three dimensional autonomous systems). Subjected to delayed optical [16,17] or optoelectronic feedbacks [18], they become infinite-dimensional systems and chaos can again be observed.

Optical feedback can be of different kinds. Coherent optical feedback may be caused by unwanted reflections of the laser output on the reflective surface of a compact disk or a fiber facet in optical fiber transmission systems. In the case of incoherent optical feedback, the polarized output field emitted by the laser undergoes a 90° polarization rotation before reinjection into the laser cavity. When the laser diode is subjected to optical feedback from a phase conjugating mirror, the mirror inverts the phase of the field. In the case of opto-electronic feedback, the output of the laser is detected by a high-speed photodetector and a feedback current is added on or deducted from the bias current of the laser diode.

Although effects of coherent optical feedback on laser diodes were reported earlier, the research on such systems was actually initiated in 1980 by Lang and Kobayashi [19] who demonstrated multistability and hysteresis phenomena. Moderate amount of coherent optical feedback can considerably affect the spectral and temporal features of semiconductor lasers. Under adequate phase matching of the external cavity, it can lead to a linewidth narrowing of several orders of magnitudes. The coherent optical feedback can however be also detrimental, inducing a tremendous broadening of the spectral linewidth up to several tenths of gigahertz. This phenomenon, which was referred to as coherence collapse by Lenstra et al. in 1985 [20], has been proved to be a manifestation of chaotic dynamics [16,20-23]. Experimental and theoretical works have shown the existence of period-doubling [23] and quasiperiodic (Ruelle-Talkens-Newhouse scenario) [16] routes to chaos. Another regime that has been widely studied is the so-called low-frequency fluctuation regime. Characterized by sudden dropouts of the laser output followed by gradual, stepwise recoveries [24], it has been experimentally classified as a time-inverted type-II intermittency [25]. The deterministic nature of this regime has been shown theoretically by Sano in 1984 [26].

As concerns laser diodes with incoherent optical feedback, the idea to use these systems to generate high-frequency optical pulses was theoretically proposed by Otsuka and Chern [27] in 1991. The experimental demonstration of this prediction has been carried out recently [28]. Investigation of the equations proposed by Otsuka and Chern has also anticipated period doubling and quasiperiodic routes to chaos [17].

This thesis is devoted to the study of nonlinear dynamics in laser diodes subject to coherent and incoherent optical feedback. As we have seen, the study of nonlinear dynamics in semiconductor lasers subject to coherent and incoherent optical feedback is important from a practical point of view. On the one hand, optical feedback can lead to severe degradations of the temporal and spectral performances of laser diodes. A better understanding of the feedback-induced instabilities allows to propose techniques for their suppression or their control. On the other hand, these instabilities can also be useful, for instance for cryptographic purposes. From a fundamental-physics point of view, the dynamics of laser diodes subject to optical feedback is very challenging because it can be of a highly complicated nature: the delay time involved by the feedback indeed causes the system to have an infinite number of degrees of freedom.

Hereafter, we summarize the main results of our Ph.D. research work following the chronological sequence in which they were obtained. We shall use from now on "optical feedback" for coherent optical feedback, as most of the thesis and much of the scientific literature are indeed concerned with that sort of feedback.

## All-optical technique for stabilization of an external cavity laser diode: numerical and experimental demonstrations

As stated above, one of the instabilities that are observed in laser diodes subject to coherent optical feedback is characterized by sudden dropouts followed by gradual, stepwise recoveries of the optical intensity occurring on a time scale much larger that the period of the relaxation oscillations or the external-cavity round trip time [24]. For that reason, this regime is usually referred to as the low-frequency fluctuation (LFF) regime. External cavity laser diodes are commonly modeled with the deterministic Lang-Kobayashi (LK) equations [19] that assume single-mode operation of the laser and weak or moderate amount of external optical feedback. Relying on this deterministic model, Sano proposed that the intensity dropouts are caused by crises between local chaotic attractors and saddle-type antimodes [26]. In his interpretation, the process of the intensity recoveries is associated to a chaotic itinerancy of the system trajectory in phase space among the attractor ruins of external-cavity modes with a drift towards the maximum gain mode close to which collisions with antimodes occur.

Since optical-feedback-induced instabilities degrade the laser diode performances, it is highly desirable to control such behaviors. We have proposed, and demonstrated at first numerically, a method that uses a second external optical feedback [29] to stabilize a chaotic laser diode operating in the low-frequency fluctuation regime, paying close attention to the antimodes and the external cavity modes. The LFF can be suppressed and the laser stabilized as antimodes responsible for the crises are destroyed or shifted away from the external-cavity modes; moreover, the laser is steered to lock onto new, stable maximum gain modes that are created as the second feedback strength increases. Our method has two main advantages: stabilization may in theory be achieved regardless of the first feedback strength and it does not require modification of any laser parameter or the first optical feedback.

We had then the opportunity to demonstrate experimentally our technique at the Nonlinear Optics Center of the Air Force Research Laboratory in Albuquerque, USA. There, we confirmed the above numerical results: in good agreement with our theoretical predictions, we have shown that the suppression of LFF and the stabilization of the laser can indeed be achieved by means of a second optical feedback [30].

**High-frequency periodic oscillations in laser diodes subject to optical feedback**

In addition to this experimental demonstration of the technique we had proposed, we also observed high-frequency periodic oscillations. These have been interpreted to be the result of a beating between an external cavity mode and an antimode. In a first step, this observation motivated an analytical study on unusual Hopf frequencies in laser diodes subject to a single optical feedback [31]. A bifurcation analysis of the LK equations was then proposed and the existence of time-periodic solutions that are combinations of an external cavity mode and an antimode was demonstrated. At the same time, we found numerically that similar behaviors can be anticipated by an extension of the Lang-Kobayashi equations to the double-feedback problem and that they are of the same nature as those observed in the case of a single feedback. Laser diodes subject to optical feedback and exhibiting almost sinusoidal high-frequency oscillations are of great practical interest since they can constitute all-optical sources of microwave oscillations.

**Coexistence of in-phase and out-of-phase dynamics in a multimode external cavity laser diode operating in the low-frequency fluctuation regime**

The necessity of including multimode dynamics and stochastic noise sources in mathematical models for the study of low-frequency fluctuations has been the subject of many, sometimes passionate, discussions in the past few years.

Using a multimode extension of the Lang-Kobayashi equations that takes spontaneous emission into account and assumes a parabolic gain profile, we have investigated the low frequency fluctuation regime in a multimode laser diode subject to optical feedback [32]. In a first step, we have numerically investigated the case of global feedback and found, in agreement with experiments, that all the longitudinal modes of the laser display low-frequency fluctuations. We have shown that two qualitatively different behaviors may take place within the LFF regime in multimode lasers on a picosecond time scale, namely in-phase and out-of-phase oscillations of the longitudinal modes of the laser, depending on the operating parameters. Each of these two behaviors corresponds to a specific statistical distribution of the laser output. As a result, we have demonstrated that two statistical distributions of the picosecond laser intensity measured by two different experimental teams [33,34] were not conflicting but complementary.

We have then investigated the role of spontaneous emission on the emergence of out-of-phase dynamics. Although spontaneous emission is intrinsically a stochastic process, we have found that it does not act as a random perturbation sustaining the out-of-phase oscillations of the laser modes but as an emission source necessary to multimode operation.

## Low-frequency fluctuation regime in a multimode semiconductor laser subject to a mode-selective optical feedback

A frequency-selective optical component such as an etalon or a grating can be placed in the external cavity in order to select a single longitudinal mode and re-inject it into the laser cavity. In this way, the laser is restricted to oscillate essentially in the selected mode, the free modes being depressed most of the time. The occurrence of intensity bursts in the free modes simultaneous to dropouts in the mode selected by the feedback has recently been reported in similar experimental arrangements [35]. Several interpretations of the intensity dropouts in the selected mode and bursts in the free modes have been given. One of them is based on the reduced stability of the selected mode with regards to perturbations occurring in the free modes. More recently, relying on an adaptation of the Tang, Statz and deMars equations [36] to semiconductor lasers [37], the LFF induced by a mode-selective optical feedback has been interpreted as being associated with a heteroclinic connection between the selected-mode steady state (saddle focus) and the free-mode steady states (saddle node) [38]. In the frame of this model, chaotic itinerancy with a drift is not a possible mechanism as long as intensity bursts in the free modes are observed. A question that deserved investigation is thus the following: may the dynamical instability induced by a selective feedback still be related to the interpretation given by Sano for the conventional LFF? Since many of the techniques for controlling the LFF regime are linked to the so-called chaotic itinerancy with a drift, answering this question is indeed an important issue.

In good agreement with experiments, the multimode extension of the Lang-Kobayashi equations that we have used predicts intensity bursts in the free modes simultaneously to dropouts in the selected mode. we have found moreover that the bursts in the free modes and dropouts in the selected mode are associated with collisions of the system trajectory in phase-space with saddle-type antimodes preceded by a chaotic itinerancy of the system among external-cavity modes [39].

## Anticipative synchronization of two chaotic laser diodes by incoherent optical coupling and its application to secure communications

Synchronization between chaotic oscillators has attracted great attention this last decade for its application to secure communications [40]. Laser diodes subject to optical feedback are considered to be good candidates for chaotic cryptography. Due to the feedback delay, they can exhibit chaotic behaviors of high complexity (hyperchaos) leading to a high degree of confidentiality. Furthermore, such systems have been shown to be suitable for high encoding/decoding rates [41]. However, in cryptographic schemes based on laser diodes subject to coherent optical feedback, the synchronization that is needed for the decoding can be strongly degraded in the presence of detuning between the optical frequencies of the transmitter and the receiver because the fields emitted by both lasers interact coherently inside the receiver laser cavity. It was therefore important to investigate alternative cryptographic schemes with high degree of complexity but without sensitivity on optical frequency detuning.

In Refs. 42 and 43, we have numerically demonstrated anticipative synchronization between a first diode subjected to incoherent optical feedback and a second diode driven by the first one through incoherent optical injection. In this scheme, the feedback and injected fields act on the carrier population in the diode active layers but do not interact with the intracavity lasing fields. As a consequence, the phases of the feedback and injection fields do not intervene on the lasers dynamics. For that reason, this synchronization scheme requires no fine tuning of the diode optical frequencies. This is a clear advantage in regard to other schemes based on coherent optical feedback and injection. It is therefore attractive for experimental realization. We have shown furthermore that this synchronization scheme can be applied to chaos-embedded communication. With a view to secure communications, we have checked how difficult it is to intercept a message encoded by chaos shift keying without an adequate replica of the transmitter laser.

This thesis is organized as follows. Chapter 2 presents the basic features of solitary laser diodes and laser diodes subject to optical feedback that are needed for the understanding of our results. In Chapter 3, we present an analytical and numerical study of a regime of high-frequency periodic oscillations in laser diodes subject to a single, coherent, optical feedback. The all-optical technique for stabilization of an external cavity laser diode is numerically and experimentally demonstrated in Chapter 4. In this Chapter, we report furthermore the observation of high-frequency periodic oscillations in the double-feedback configuration. We show that this regime can be anticipated by using an extension of the Lang-Kobayashi equations and that the high-frequency oscillations are of the same nature as those reported in Chapter 3 in the case of a single-optical feedback. Chapter 5 is devoted to the low-frequency fluctuation regime in multimode laser diodes subject to global optical feedback and mode-selective optical feedback. In Chapter 6, we propose an original secure communication scheme based on anticipative synchronization between two laser diodes respectively subjected to incoherent optical feedback and incoherent optical injection. We summarize our results and propose several promising perspectives in Chapter 7. The goal of Appendixes A and B is to present some of the basic concepts underlying nonlinear dynamics and chaos. We finally describe the numerical method we have used in the resolution of the rate equations in Appendix C.

# 2. Theoretical description of laser diodes with and without optical feedback

La théorie est le general,
l'application le soldat.

<div align="right">Léonard de Vinci</div>

## 2.1. Introduction

In this chapter, we describe the theoretical approach to the study of the equilibrium and stability of laser diodes subject or not to optical feedback from an external cavity.

After recalling the underlying principles of laser diodes (Section 2.2), the equations appropriate to single mode (Section 2.3) and multimode (Section 2.4) lasers without feedback are obtained following the procedure outlined in Petermann [1]. The derivation of the Lang-Kobayashi equations for single mode operation with coherent optical feedback (Section 2.5) is inspired from the review article of van Tartwijk and Lenstra [2]; they are then extended to the description of multimode laser operation with coherent optical feedback. The Lang-Kobayashi equations allow to explain the main features of the coherence collapse regime (Section 2.6) and the low-frequency fluctuation regime (Section 2.7). Finally, the rate equations for single mode laser diodes subject to incoherent optical feedback and incoherent optical injection are obtained in Section 2.8.

## 2.2. Operating principles

The acronym "laser" stands for Light Amplification by Stimulated Emission of Radiation. A laser is an optical oscillator that comprises an externally pumped active medium and a resonant cavity. The resonant cavity provides an adequate feedback mechanism. The role of the active medium is twofold. On the one hand, it provides a spontaneous emission noise input that may initiate an oscillation process. On the other hand, it can amplify the light propagating within the cavity. For the type of lasers under study, the active medium consists of a semiconductor material that can be pumped optically or electrically. The feedback mechanism regenerates part of the light to increase the amplification and select the modes of the electromagnetic field to be amplified.

In the following, we focus on a Fabry-Perot type laser diode. Fig. 2.1 sketches a typical edge-emitting semiconductor laser with a double heterostructure. The resonator is a Fabry-Perot cavity where the light travels parallel to the surface of the wafer on which the laser is built and is partially reflected on laser facets. These facets, whose Fresnel reflectivity is about 30%, are obtained by cleaving the wafer along a crystallographic axis. The length of the cavity is typically 250 µm. A thin layer (thickness of about 0.1 µm) of a direct-band-gap semiconductor material, for instance $In_{1-x}Ga_xAs_{1-y}P_y$, is embedded between two p-type and n-type cladding layers with larger band-gaps. Under forward bias, electrons and holes move freely to this region but cannot cross over to the other side because of the potential barrier resulting from the band-gap difference (Fig. 2.2). This allows for a large concentration of electrons and holes in this region, usually referred to as active region. The double heterostructure also allows optical confinement because the cladding layers have a smaller refractive index than the active layer. The resulting waveguide confines the light and therefore reduces the internal loss. In the active region, electrons and holes can recombine through radiative or nonradiative mechanisms. Nonradiative recombinations, such as Auger recombination, trap and surface recombinations are not helpful for laser operation. Radiative recombinations can be either spontaneous or stimulated [Fig. 2.3 (a,b)]. In the case of spontaneous emission, photons are emitted in random directions and without phase relation; thus the emitted light is incoherent. By contrast, when electron-hole recombination is stimulated by an already existing photon, the emitted photon has the same wavelength, phase and direction as the incoming photon. This results in coherent amplification of the incident light.

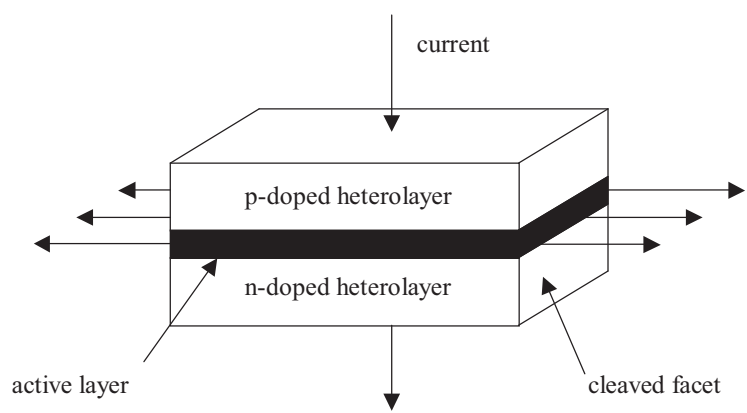

**Fig. 2.1** Sketch of a typical edge-emitting semiconductor laser with a double heterostructure.

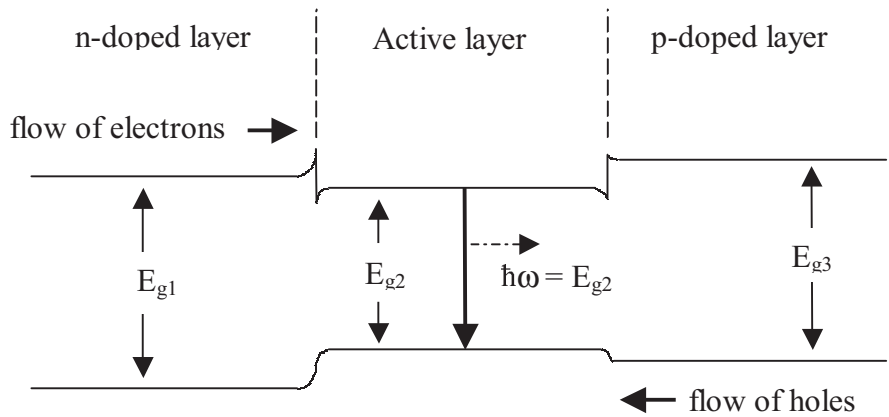

**Fig. 2.2** Energy-band diagram of a double-heterostructure laser diode under forward bias. $E_{g1}$, $E_{g2}$, $E_{g3}$ denote the energy gap of the n-doped layer, the active layer and the p-doped layer, respectively. $E_{g1}$ and $E_{g3}$ are considerably larger than $E_{g2}$.

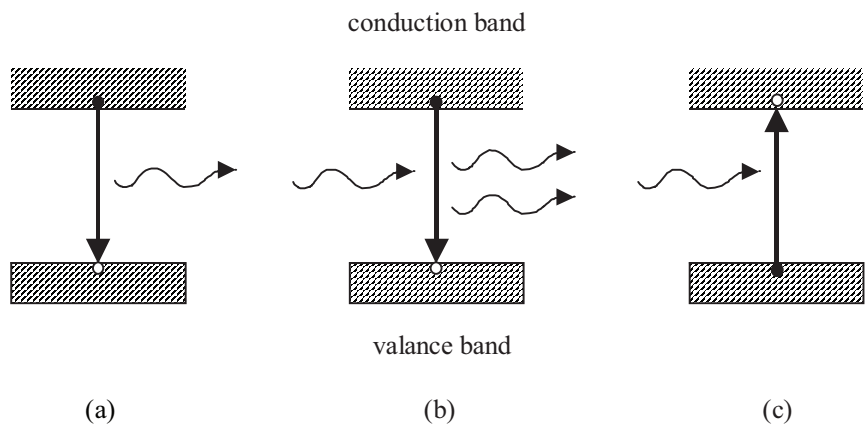

**Fig. 2.3** Schematic illustration of (a) spontaneous emission, (b) stimulated-emission and (c) absorption processes.

   Since a photon in the active region may be absorbed to generate an electron-hole pair [Fig. 2.3 (c)], stimulated emission must compete with absorption. At low values of the injection current, the number of electron-hole pairs supplied by the injection current is small; absorption dominates amplification and the net rate of stimulated emission, namely the difference between the stimulated photon rate and the absorption rate, is negative. The net rate of stimulated emission increases with the injection current. The semiconductor is said to be optically transparent when the rates of stimulated emission and photon absorption are equal. Beyond transparency, the condition known as population inversion is achieved when the net rate of stimulated emission is positive. However, depending of the value of the injection current, the net rate of stimulated emission can be positive but not sufficient to overcome the transmission losses at the laser facets, the internal losses due to photon absorption inside the cavity without generation of carrier and the loss due to light scattering. The diode starts lasing when the injection current exceeds a critical value called threshold current at which the net gain overcomes all losses. Above threshold, the number of electron-hole pairs remains almost clamped to its threshold value. The laser output increases almost linearly with the injection current as most of the electron-hole pairs that are injected in excess recombine through stimulated emission within the selected cavity modes. In the same time, the spectral width of the laser narrows considerably because of the coherent nature of the stimulated emission.

In addition to transverse confinement of both carriers and photons, it must be noted that mechanisms for lateral confinement can be incorporated in the laser. In index-guided lasers, lateral optical confinement is achieved by lateral variation of the refractive index. In gain-guided lasers, the current is injected only over a narrow region using, for instance, a strip contact that leads to a lateral variation of the optical gain that confines in turn the light to the stripe vicinity.

We have so far considered Fabry-Perot type laser diodes in which the light is partially reflected on laser facets. The feedback mechanism can however be achieved by other ways. In distributed-feedback (DFB) lasers, the feedback is not localized at the cavity facets but is distributed throughout the cavity length. This is achieved by using a periodic index perturbation integrated along the laser structure. In distributed bragg reflector (DBR) lasers, gratings are etched outside the active medium. In these last two types of lasers, gratings are used to select a single longitudinal mode of oscillation. Due to this spectral feature, those lasers are of particular interest for optical fiber communications.

## 2.3. Single-mode solitary laser diode

### 2.3.1. Rate equations

In this section, we derive rate equations describing the temporal behavior of the slowly-varying complex amplitude $E(t)$ of the electric field $\mathcal{E}(t)$ and the total number of electron-hole pairs in the laser cavity $N(t)$ of a single longitudinal mode laser diode. $E(t)$ is normalized in such a way that $|E(t)|^2$, the square of the absolute value of the complex field amplitude, corresponds to the photon number inside the laser cavity.

For simplicity, we assume that the laser structure is designed to support a single lateral and transverse mode. Considerable simplification occurs if we assume that the material response is instantaneous so that the polarization can be adiabatically eliminated from the model. This approximation is justified since the material response, which is governed by the intraband scattering processes, is relatively fast (~0.1 ps) compared to the photon lifetime (~1 ps) and the carrier recombination time (~1 ns) [3].

We model the laser diode as a Fabry-Perot resonator consisting of a dielectric medium confined between two mirrors. The light traveling both in the forward and in the backward directions is amplified due to stimulated emission; it is furthermore partially reflected and partially transmitted at each mirror (Fig. 2.4).

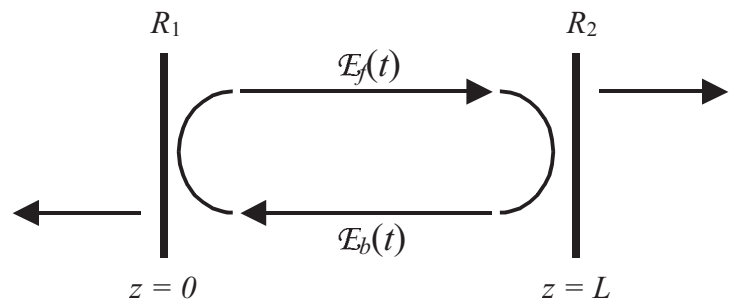

**Fig. 2.4** Schematic representation of a Fabry-Perot cavity of length *L*. $R_1$ and $R_2$ denote the reflectivities of the left and right facets, respectively.

The forward and backward propagating power, $\mathcal{P}_f(z)$ and $\mathcal{P}_b(z)$, are respectively denoted as

$$\mathcal{P}_f(z) = \mathcal{P}_{f0} \exp(gz - \alpha_s z) \tag{2.1}$$

and

$$\mathcal{P}_b(z) = \mathcal{P}_{b0} \exp[g(L-z) - \alpha_s(L-z)] \tag{2.2}$$

where *g* is the gain due to stimulated emission, $\alpha_s$ accounts for any optical loss inside the laser cavity that does not yield a generation of carriers within the active layer, and *L* is the length of the cavity.

In order to derive the lasing (i.e. the threshold) conditions, the field amplitude must be considered. The amplitudes of the forward and backward traveling complex electric fields are respectively denoted as

$$E_f(z) = E_{f0} \exp\left[-i\frac{n\omega}{c}z + \tfrac{1}{2}(gz - \alpha_s z)\right] \tag{2.3}$$

and

$$E_b(z) = E_{b0} \exp\left\{-i\frac{n\omega}{c}(L-z) + \tfrac{1}{2}[g(L-z) - \alpha_s(L-z)]\right\} \quad (2.4)$$

where *c* is the velocity of light in vacuum, *n* the real part of the refractive index and $\omega = 2\pi v$ the optical angular frequency, $v$ being the optical frequency. The difference $(g-\alpha_s)$ is linked to the imaginary part *n'* of the index of refraction by the relation

$$n' = \frac{c}{2\omega}(g - \alpha_s). \quad (2.5)$$

The amplitudes $E_f(z)$ and $E_b(z)$ are related by the relations

$$E_{f0} = r_1 E_b(0) \text{ and } E_{b0} = r_2 E_f(L) \quad (2.6)$$

where $r_1$ and $r_2$ are the reflection coefficients at the laser facets. The condition for sustained laser oscillation

$$r_1 r_2 \exp\left[-2i\frac{n\omega}{c}L + (g-\alpha_s)L\right] = 1 \quad (2.7)$$

is a consequence of Eqs. (2.3),(2.4) and (2.6). In the following, we refer to the left-hand side of the previous equation as the round trip gain:

$$G = r_1 r_2 \exp\left[-2i\frac{n\omega}{c}L + (g-\alpha_s)L\right]. \quad (2.8)$$

The modulus of Eq. (2.7) yields the gain $g = g_{th}$ at laser threshold. If $r_1$ and $r_2$ are real, one obtains

$$g_{th} = \alpha_s + \frac{1}{2L}\ln\left(\frac{1}{R_1 R_2}\right) \quad (2.9)$$

where $R_1 = r_1^2$ and $R_2 = r_2^2$. The second term of the right-hand side of Eq. (2.9) corresponds to losses at the laser facets. Eq. (2.9) states merely that at laser threshold, the gain balances exactly the losses. The phase of Eq. (2.9) yields the resonance condition

$$\omega_m = m\frac{\pi c}{nL} \quad (2.10)$$

where $m$ is an integer and $\omega_m$ denotes the angular frequency of the $m$th longitudinal mode of the Fabry-Perot cavity. The longitudinal mode spacing frequency can be found as follows. From Eq. (2.10), one find

$$d\left(\frac{nL\omega}{m\pi c}\right) = \frac{L}{\pi c}\left(\frac{n}{m}d\omega + \frac{\omega}{m}dn - \frac{n\omega}{m^2}dm\right) = 0 \qquad (2.11)$$

where the dispersion of the refractive index is taken into account. Introducing $dm = 1$ and the group refractive index

$$n_g = n + \omega \frac{dn}{d\omega}, \qquad (2.12)$$

yields, also in view of Eq. (2.10), the mode spacing frequency:

$$\Delta v = \frac{c}{2n_g L}. \qquad (2.13)$$

The roundtrip time of a mode with frequency $v = \omega_m/2\pi$ is the inverse of the mode spacing

$$\tau_{in} = \frac{2n_g L}{c} = \frac{1}{\Delta v}. \qquad (2.14)$$

The refractive index $n$ depends on the optical frequency and on the carrier density so that the resonance angular frequency at threshold of the $m$th longitudinal mode may be introduced according to

$$\omega_{m,th} = m\frac{\pi c}{n(\omega_{m,th}, N_{th})L} \qquad (2.15)$$

where $N_{th}$ is the carrier number at threshold. For clarity, we will write in the following $\omega_{th}$ instead of $\omega_{m,th}$ and $n_{th}$ instead of $n(\omega_{m,th}, N_{th})$.

We have derived so far the condition for sustained monochromatic laser oscillation. In order to study the dynamical behavior of the laser diode, the round trip gain $\mathcal{G}$ must now be understood as an operator for the time-dependent electric field acting differentially on each frequency. $g$, $\alpha_s$, $r_1$ and $r_2$ can be considered to be frequency independent at least for frequencies close to the frequency of the longitudinal mode under consideration [1]. By contrast, we must take into account the dependence of the refractive index on both the carrier number and the optical frequency. We expend $n\omega/c$ in terms of the optical angular frequency and the carrier number at threshold:

$$\begin{aligned}\frac{n\omega}{c} &\approx \frac{(n\omega)_{th}}{c} + \frac{1}{c}\left.\frac{\partial n\omega}{\partial \omega}\right|_{th}(\omega-\omega_{th}) + \frac{1}{c}\left.\frac{\partial n\omega}{\partial N}\right|_{th}(N-N_{th}) \\ &= \frac{n_{th}\omega_{th}}{c} + \frac{n_g}{c}(\omega-\omega_{th}) + \frac{\omega_{th}}{c}\left.\frac{\partial n}{\partial N}\right|_{th}(N-N_{th})\end{aligned}$$  (2.16)

Using Eq. (2.16), the round trip gain $\mathcal{G}$ can be written as

$$\mathcal{G} \equiv \mathcal{G}_1 \mathcal{G}_\omega$$ (2.17)

where

$$\mathcal{G}_1 = \exp\left[(g-\alpha_s)L + \frac{1}{2}\ln(R_1 R_2) - 2i\frac{\omega_{th}L}{c}\left.\frac{\partial n}{\partial N}\right|_{th}(N-N_{th})\right]$$ (2.18)

is frequency independent but

$$\mathcal{G}_\omega = \exp\left[-2i\frac{n_{th}\omega_{th}L}{c} - 2i\frac{n_g L}{c}(\omega-\omega_{th})\right]$$ (2.19)

is frequency dependent. Since the term $2n_{th}\omega_{th}L/c$ [Eq. (2.15)] is an integer multiple of $2\pi$ and $2n_g L/c$ is equal to the round trip time $\tau_{in}$ [Eq. (2.14)], $\mathcal{G}_\omega$ can be rewritten as

$$\mathcal{G}_\omega = \exp[-i(\omega-\omega_{th})\tau_{in}].$$ (2.20)

For monochromatic fields with frequency $\omega$, introducing $i\omega = d/dt$ yields

$$\mathcal{G}_\omega = \exp(i\omega_{th}\tau_{in})\exp\left(-\tau_{in}\frac{d}{dt}\right).$$ (2.21)

The round trip gain reads therefore

$$\mathcal{G} = \mathcal{G}_1 \exp(i\omega_{th}\tau_{in})\exp\left(-\tau_{in}\frac{d}{dt}\right).$$ (2.22)

At this point, we must introduce the relation between the gain and the carrier number. The optical gain is observed to increase almost linearly with the total number of hole-electron pairs *N* for all values of the injection current [1]. The gain can then be approximated as

$$g(N) = \left.\frac{\partial g}{\partial N}\right|_{N_0} (N - N_0) \qquad (2.23)$$

where $N_0$ is the carrier number at transparency and $\partial g/\partial N$ is the gain coefficient. $g(N)$ is also often linearized around its threshold value. In this case,

$$g(N) = g_{th} + \left.\frac{\partial g}{\partial N}\right|_{N_{th}} (N - N_{th}). \qquad (2.24)$$

The gain per second is defined as

$$\begin{aligned} G(N) &= v_g g(N) \\ &= G_N (N - N_0) \end{aligned} \qquad (2.25)$$

where we have introduced the differential gain $G_N = v_g \left.\frac{\partial g}{\partial N}\right|_{N_0} \cong v_g \left.\frac{\partial g}{\partial N}\right|_{N_{th}}$ and the group velocity $v_g = c/n_g$. Using Eq. (2.9) and defining the photon lifetime as the inverse of the total loss rate of photons

$$\tau_p = v_g \left[\alpha_s + \frac{1}{2L} \ln\left(\frac{1}{R_1 R_2}\right)\right]^{-1}, \qquad (2.26)$$

Eq. (2.25) can be rewritten as

$$G(N) = \frac{1}{\tau_p} + G_N (N - N_{th}). \qquad (2.27)$$

Application of the round trip gain to the time dependent electric field $\mathcal{E}_f$ of the forward traveling wave at $z = 0$ yields

$$\mathcal{E}_f(t) = \mathcal{G}\mathcal{E}_f(t) = \mathcal{G}_1 \exp(i\omega_{th}\tau_{in}) \exp\left(-\tau_{in}\frac{d}{dt}\right)\mathcal{E}_f(t); \qquad (2.28)$$

the operator $\exp(-\tau_{in} d/dt)$ implying a time shift of $-\tau_{in}$, we have

$$\mathcal{E}_f(t) = \mathcal{G}_1 \exp(i\omega_{th}\tau_{in})\mathcal{E}_f(t-\tau_{in}). \quad (2.29)$$

Since the optical field oscillates essentially at $\omega \approx \omega_{th}$, we introduce the slowly-varying complex amplitude $E(t)$ according to

$$\mathcal{E}_f(t) = E(t)\exp(i\omega_{th}t). \quad (2.30)$$

If Eq. (2.30) is inserted in Eq. (2.29), one obtains

$$E(t-\tau_{in}) = (1/\mathcal{G}_1)E(t). \quad (2.31)$$

The variation of $E(t)$ during one round trip being small, it can be approximated by

$$E(t) = E(t-\tau_{in}) + \tau_{in}\frac{dE(t)}{dt}. \quad (2.32)$$

Insertion of Eq. (2.32) into Eq. (2.31) yields

$$\frac{dE(t)}{dt} = \frac{1}{\tau_{in}}\left(1 - \frac{1}{\mathcal{G}_1}\right)E(t). \quad (2.33)$$

Since $\mathcal{G}_1$ is close to unity for laser operation, the expansion of $1/\mathcal{G}_1$ yields, with the help of Eq. (2.18):

$$\begin{aligned}\frac{1}{\mathcal{G}_1} &= \exp\left[-(g-\alpha_s)L - \frac{1}{2}\ln(R_1R_2) + 2i\frac{\omega_{th}L}{c}\frac{\partial n}{\partial N}\bigg|_{th}(N-N_{th})\right] \\ &\cong 1 - (g-\alpha_s)L - \frac{1}{2}\ln(R_1R_2) + 2i\frac{\omega_{th}L}{c}\frac{\partial n}{\partial N}\bigg|_{th}(N-N_{th}).\end{aligned} \quad (2.34)$$

Using Eqs. (2.9),(2.14) and (2.24), one finds after some manipulations:

$$\frac{1}{\tau_{in}}\left(1-\frac{1}{\mathcal{G}_1}\right) = \frac{c}{2n_g}\frac{\partial g}{\partial N}\bigg|_{th}(N-N_{th}) - i\frac{\omega_{th}}{n_g}\frac{\partial n}{\partial N}\bigg|_{th}(N-N_{th}). \quad (2.35)$$

Eq. (2.33) can therefore be rewritten as

$$\frac{dE(t)}{dt} = \left[ \frac{c}{2n_g} \frac{\partial g}{\partial N}\bigg|_{th} (N - N_{th}) - i \frac{\omega_{th}}{n_g} \frac{\partial n}{\partial N}\bigg|_{th} (N - N_{th}) \right] E(t) \qquad (2.36)$$

In order to complete the derivation of the rate equation for the field, we must introduce here a new parameter. The real and imaginary parts of the refractive index are linked by the Kramers-Kronig relations [4]. The real part (i.e. *n*) is noticeably more sensitive to variations of the electron-hole population than the imaginary part (i.e. *n'*), which is expressed by the inequality

$$\alpha = \frac{\delta n}{\delta n'} = -2 \frac{\omega}{c} \frac{\partial n/\partial N}{\partial g/\partial N} > 1. \qquad (2.37)$$

In laser diodes, this parameter, namely the linewidth enhancement factor, is quite large [5]. Its typical value is between 3 and 7 that leads to substantial linewidth broadening as we shall see in the following.

Using Eq. (2.27) and after insertion of Eq. (2.37) into Eq. (2.36), the rate equation for the slowly varying envelope of the electric field reads

$$\frac{dE(t)}{dt} = \frac{1 + i\alpha}{2} \left[ G(N) - \frac{1}{\tau_p} \right] E(t). \qquad (2.38)$$

This equation is normalized such that $P(t) = |E(t)|^2$ is the number of photons inside the laser cavity. Assuming that both laser facets have equal reflectivity, the power (in W) that is emitted per facet is

$$P_{OUT} = \frac{1}{2} \hbar \omega \alpha_m v_g P. \qquad (2.39)$$

Eq. (2.38) alone is not sufficient to describe the behavior of the laser. Indeed, the gain, and therefore the electric field, depends on the carrier number. Assuming charge neutrality, the general form of the rate equation describing the dynamical behavior of the electron-hole pair number *N* is

$$\frac{\partial N}{\partial t} = D \nabla^2 N + \frac{I}{e} - \frac{N}{\tau_s} - R_{st}(N, |E|^2). \qquad (2.40)$$

The first term on the right-hand side accounts for carrier diffusion, *D* being the diffusion coefficient. We assume in the following that the carrier distribution does not vary significantly within the active-region; the diffusion term in Eq. (2.40) is negligible in that case. The second term accounts for the electron-hole pairs that are injected into the active regions by means of the electrical current *I*. *e* is the magnitude of the electron charge. The third term accounts for losses of carriers by spontaneous emission and non-radiative transitions. The last term stands for the stimulated loss rate of carriers (Ref. 3 and references therein). The carrier lifetime is approximated by

$$\tau_s(N)^{-1} = A_{nr} + BN + CN^2 \tag{2.41}$$

Here, $A_{nr}$, $BN$ and $CN^2$ are related to mechanisms such as trap or surface recombination, to radiative recombination and to the Auger recombination process, respectively. Above lasing threshold, $\tau_s$ can be considered as a constant to a good approximation because the carrier number is nearly clamped to its threshold value. Finally, $R_{st}$ accounts for the loss of electron-hole pairs due to stimulated recombination. It is given by

$$R_{st}(N,|E|^2) = G(N)|E|^2. \tag{2.42}$$

The rate equation for the electron-hole pair number can therefore be rewritten as

$$\frac{dN}{dt} = \frac{I}{e} - \frac{N}{\tau_s} - G(N)|E|^2. \tag{2.43}$$

We have assumed so far that the gain increases linearly with the population inversion. Experiments show however a saturation at high intensities (Ref. 1 and references therein). Gain saturation is induced by different mechanisms. Most important are thought to be spatial hole burning and dynamic carrier heating, which we now briefly describe. The optical field intensity is not constant along the laser active layer. The electron-hole pairs recombine faster where the intensity is maximum, leading to local depletion. Spatial hole burning occurs when the carrier diffusion is not sufficiently effective to supply enough new carriers; that in turn induces gain saturation. Dynamic carrier heating occurs when electrons in the conduction band are excited to higher energy states from which they relax slowly through electron-phonon interactions. During all this process, they cannot recombine with holes in the valance band and therefore do not participate to the stimulated emission.

In order to account for saturation, the gain per second for a single-mode laser diode is often rewritten as

$$G(N,|E|^2) = G_N(N - N_0)(1 - \varepsilon|E|^2) \tag{2.44}$$

where $\varepsilon$ is the gain saturation coefficient of the laser. Taking into account gain saturation, the rate equations can be rewritten

$$\frac{dE(t)}{dt} = \frac{1+i\alpha}{2}\left[G(N,|E|^2) - \frac{1}{\tau_p}\right]E(t). \qquad (2.45)$$

$$\frac{dN}{dt} = \frac{I}{e} - \frac{N}{\tau_s} - G(N,|E|^2)|E|^2. \qquad (2.46)$$

In the literature, the saturation is sometimes considered to not affect directly the phase variation. In that case, the equation for the electric field is rewritten as

$$\frac{dE(t)}{dt} = \frac{1}{2}\left[G(N,|E|^2) - \frac{1}{\tau_p} + i\alpha G_N(N-N_{th})\right]E(t). \qquad (2.47)$$

As will be shown in the following, gain saturation can lead to an important damping of the relaxation oscillations in solitary lasers pumped far above threshold. It has also a stabilizing effect in the dynamics of semiconductor lasers subject to optical feedback [13,14]: the increase of $\varepsilon$ shifts the onset of chaos toward higher feedback levels. We have found in addition that gain saturation plays an important role in the synchronization of two laser diodes respectively subject to incoherent optical feedback and incoherent optical injection (see Chapter 6). The role of gain saturation is less relevant when the laser subject to feedback is operated close to threshold [13]. We will not consider Eq. (2.47) in this work.

## 2.3.2.    Laser noise

Eqs. (2.45) and (2.46) are entirely deterministic. To be complete, they must also take stochastic fluctuations arising from laser noise into account. Lax [6,7] and Haken [8] have proposed a quantum-mechanical description of the noise. A more intuitive explanation based on physical arguments and leading to the same result has been provided by Henry [5,9]. The role of stochastic fluctuations is included by adding Langevin noise forces to Eqs. (2.45) and (2.46), which yields

$$\frac{dE(t)}{dt} = \frac{1+i\alpha}{2}\left[G(N,|E|^2) - \frac{1}{\tau_p}\right]E(t) + F_E(t), \qquad (2.48)$$

$$\frac{dN}{dt} = \frac{I}{e} - \frac{N}{\tau_s} - G(N,|E|^2)|E|^2 + F_N(t). \qquad (2.49)$$

The term $F_E(t)$ describes the role of spontaneous emission: the amplitude and phase of the modal field are randomly perturbed by each spontaneously emitted photon that participates to the mode. $F_N(t)$ arises from the discrete nature of the carrier generation and recombination processes, i.e. shot noise. Shot noise has much less impact on the laser dynamics than spontaneous emission noise. It will therefore often be neglected in this work. Since the noise terms in Eqs. (2.48) and (2.49) result from the superposition of a large number of independent events, the corresponding probability densities tend to a normal law. Under Markovian assumption, i.e. assuming that the correlation time of the noise sources is much shorter than the relaxation times $\tau_s$ and $\tau_p$ (the system has no memory), the Langevin forces satisfy the following relations

$$<F_i(t)> = 0, \qquad (2.50)$$

$$<F_i(t)F_j(t')> = 2D_{ij}\delta(t-t'). \qquad (2.51)$$

where $<>$ denotes ensemble average. The diffusion coefficients appearing in Eq. (2.51) are given by [1,15]

$$2D_{EE^*} = R_{sp}, \qquad (2.52)$$

$$2D_{EE} = 0, \qquad (2.53)$$

$$2D_{EN} = -R_{sp}|E|, \qquad (2.54)$$

$$2D_{NN} = 2\left(R_{sp}|E|^2 + \frac{N}{\tau_s}\right). \qquad (2.55)$$

$R_{sp}$ is the rate at which photons that participate to the considered oscillating laser mode are spontaneously emitted; it is proportional to the carrier number:

$$R_{sp} = \beta N. \qquad (2.56)$$

Until now, we have considered rate equations for the slowly-varying complex field amplitude $E(t)$. It is often convenient to consider the dynamics of the photon number $P(t)$ and phase $\phi(t)$ instead of the complex electric field. Introducing the relation

$$E(t) = \sqrt{P(t)}\exp(i\phi(t)), \qquad (2.57)$$

the complex equation (2.48) is equivalent to [1,15]

$$\frac{dP(t)}{dt} = \left[G(N,P) - \frac{1}{\tau_p}\right]P(t) + R_{sp} + F_P(t) \tag{2.58}$$

and

$$\frac{d\phi(t)}{dt} = \frac{\alpha}{2}\left[G(N,P) - \frac{1}{\tau_p}\right] + F_\phi(t), \tag{2.59}$$

whereas Eq. (2.49) becomes

$$\frac{dN}{dt} = \frac{I}{e} - \frac{N}{\tau_s} - G(N,P)P + F_N(t). \tag{2.60}$$

where the nonlinear gain (2.44) is rewritten as

$$G(N,P) = G_N(N - N_0)(1 - \varepsilon P). \tag{2.61}$$

$F_P(t)$, $F_\phi(t)$ and $F_N(t)$ are real Langevin noise terms with diffusion coefficients [3]

$$2D_{PP} = 2R_{sp}P, \tag{2.62}$$

$$2D_{\phi\phi} = \frac{R_{sp}}{2P}, \tag{2.63}$$

$$2D_{NN} = 2\left(R_{sp}P + \frac{N}{\tau_s}\right), \tag{2.64}$$

$$2D_{P\phi} = 0, \tag{2.65}$$

$$2D_{PN} = -2R_{sp}P, \tag{2.66}$$

$$2D_{N\phi} = 0. \tag{2.67}$$

### 2.3.3. Steady state solutions

We study in this section the dependence of the steady state solutions of the rate equations (2.58)-(2.60) on the injection current. We neglect the stochastic terms $F_P$, $F_\phi$ and $F_N$ and assume a nonlinear gain given by Eq. (2.61).

The steady state solutions are characterized by a constant photon number $P_s$, a constant carrier number $N_s$ and a constant frequency shift $\Delta\omega_s$ with respect to the threshold frequency. By setting the derivatives of $P$ and $N$ to zero and the time derivative of $\phi$ to $\Delta\omega_s$, the steady state photon number, carrier number and frequency shift are respectively given by

$$\left(\varepsilon G_N \frac{\tau_s}{\tau_p}\right)P^3 - \left\{G_N \frac{\tau_s}{\tau_p} + \varepsilon G_N\left[\tau_s \frac{I}{e} + N_0(\tau_s \beta - 1)\right]\right\}P^2$$
$$+ \left\{G_N\left[\tau_s \frac{I}{e} + N_0(\tau_s \beta - 1)\right] - \frac{1}{\tau_p}\right\}P + \tau_s \beta \frac{I}{e} = 0, \qquad (2.68)$$

$$N_s = \frac{\frac{\tau_s}{\tau_p}P_s - \frac{I}{e}\tau_s}{\tau_s \beta - 1}, \qquad (2.69)$$

$$\Delta\omega_s = \frac{\alpha}{2}\left[G_N(N - N_0)(1 - \varepsilon P) - \frac{1}{\tau_p}\right]. \qquad (2.70)$$

In particular, for $\beta = 0$ and $\varepsilon = 0$, there are two steady state solutions given by

$$P_s = 0, \ N_s = \tau_s \frac{I}{e} \text{ and } \Delta\omega_s = \frac{\alpha}{2}G_N\tau_s\left(\frac{I}{e} - \frac{I_{th}}{e}\right), \qquad (2.71)$$

and

$$P_s = \frac{\tau_p}{e}(I - I_{th}), \ N_s = N_{th} \text{ and } \Delta\omega_s = 0, \qquad (2.72)$$

where $I_{th} = \frac{e}{\tau_s}N_{th}$ is the threshold current.

The second solution has no physical meaning below threshold. For $I < I_{th}$, the carrier number and the frequency shift of the first solution increase linearly with the injection current while there are no photons in the laser cavity. Above threshold, both solutions are possible. As will be shown in the next section, the first solution, which corresponds to a laser turned off, is unstable. By contrast, the second solution, which corresponds to a linear increase of the photon number with the injection current while the carrier number and the frequency shift are clamped to their threshold values, is stable.

Due to spontaneous emission, a real laser already emits light below threshold. Fig. 2.5 shows the light-current (L-I) and carrier number-current (N-I) curves obtained from Eqs. (2.68) and (2.69) for two non-zero values of $\beta$, namely $\beta = 10^6$ s$^{-1}$ and $\beta = 10^7$ s$^{-1}$, as well the L-I and N-I curves obtained from Eqs. (2.71) and (2.72) with $\beta = 0$. The other parameters are given in Table 2-I. The figure reveals that in presence of spontaneous emission, the threshold is no longer well defined, as it is for $\beta = 0$. It is however customary to define the threshold current $I_{th}$ and carrier number $N_{th}$ for the limiting case $\beta = 0$.

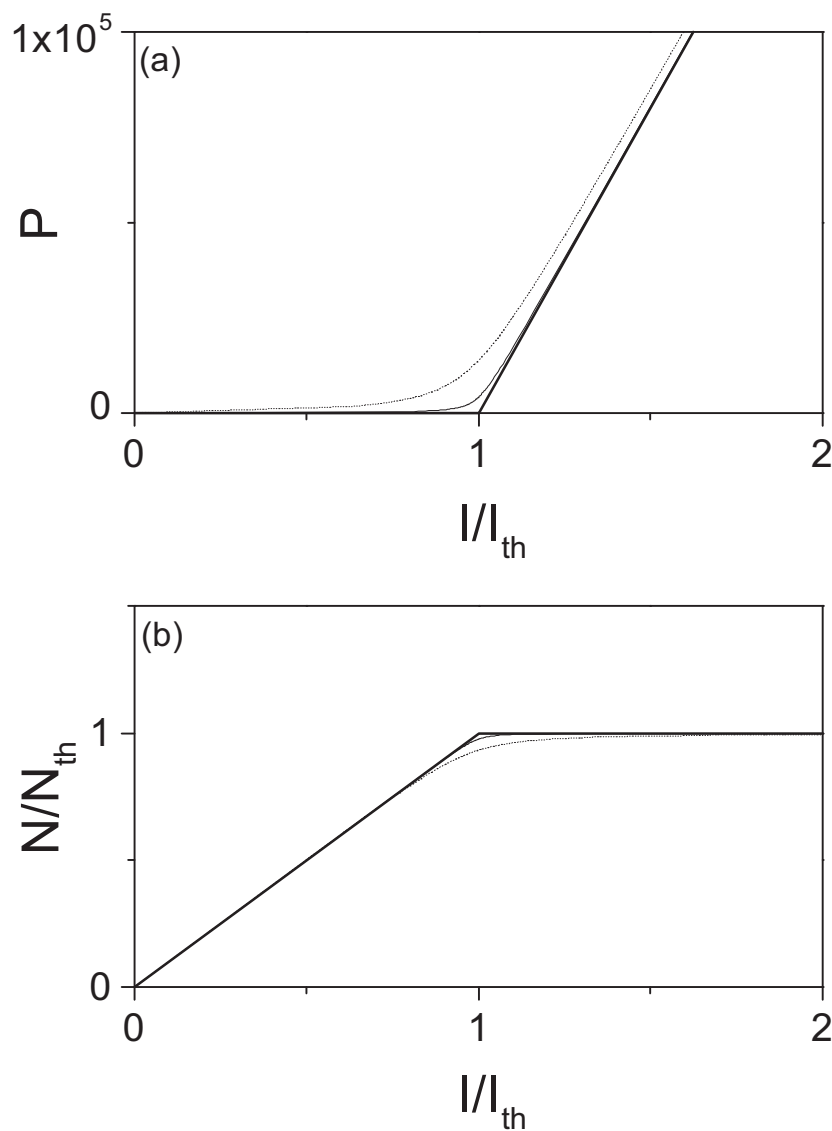

**Fig. 2.5** Light-current (a) and carrier number-current (b) curves for three different values of the spontaneous emission rate, namely $\beta = 0$ (thick line), $10^6$ (thin line) and $10^7$ (dotted line).

Fig. 2.6 shows the influence of gain saturation on the L-I and N-I curves calculated for $\varepsilon = 10^{-7}$ and $\varepsilon = 10^{-5}$. The progressive saturation of the light intensity and the increase of the carrier number are clearly visible for $\varepsilon = 10^{-5}$. For $\varepsilon = 10^{-7}$, which is a more realistic value, the effect of the saturation on the curves is almost imperceptible.

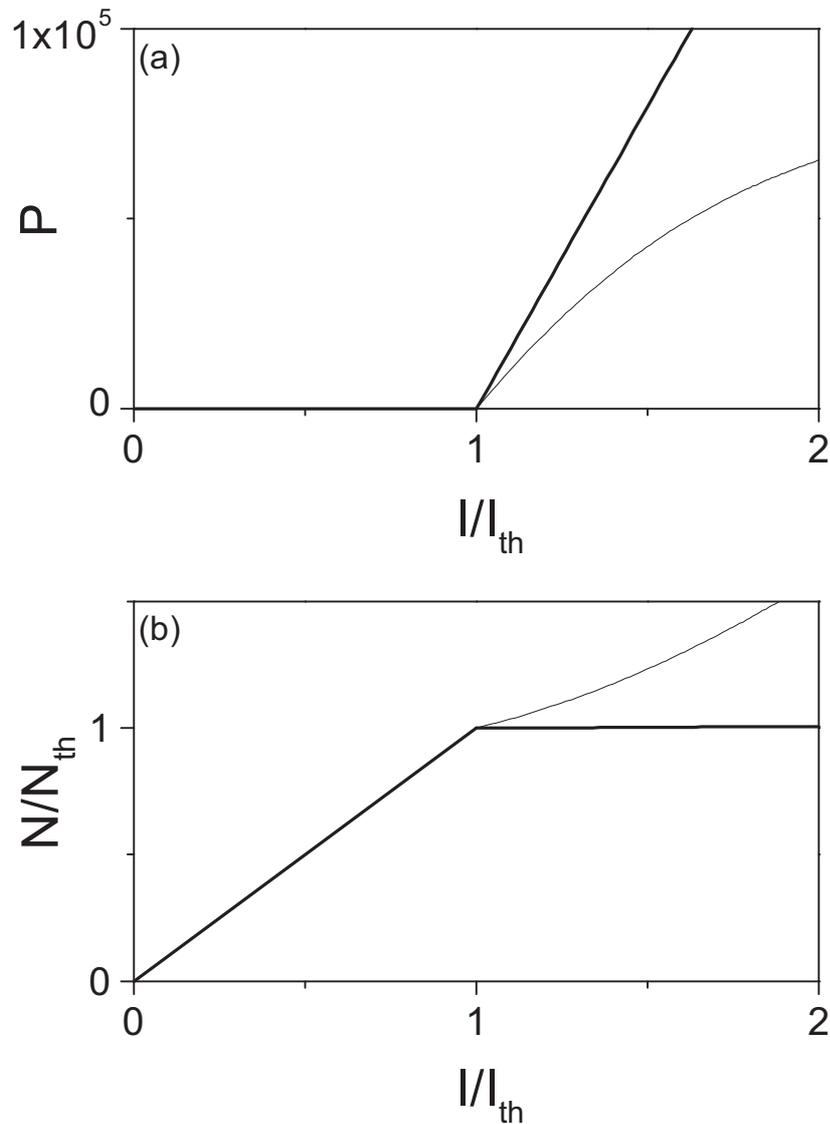

**Fig. 2.6** Influence of the gain saturation on the L-I (a) and N-I (b) curves for $\varepsilon = 10^{-7}$ (thick line) and $\varepsilon = 10^{-5}$ (thin line). $\beta = 0$.

**Table 2-I**

| Symbol | Parameter | Value |
|---|---|---|
| $\alpha$ | Linewidth enhancement factor | 4 |
| $G_N$ | Differential gain | $10^4$ s$^{-1}$ |
| $N_0$ | Carrier number at transparency | $1.1 \times 10^8$ |
| $\tau_{in}$ | Laser roundtrip time | 8 ps |
| $\tau_s$ | Carrier lifetime | 2 ns |
| $\tau_p$ | Photon lifetime | 2 ps |
| $I_{th}$ | Solitary laser threshold | 12.82 mA |

## 2.3.4. Stability of steady state solutions

The linearized rate equations for deviations $\delta P$, $\delta N$ and $\delta \phi$ around the steady-state values $P_s$, $N_s$ and $\Delta \omega_s t$ are found by inserting the time dependent photon number, carrier number and phase

$$P(t) = P_s + \delta P(t), \tag{2.73}$$

$$N(t) = N_s + \delta N(t), \tag{2.74}$$

$$\phi(t) = \Delta \omega_s t + \delta \phi(t), \tag{2.75}$$

in Eqs.(2.58)-(2.60). Neglecting quadratic terms with respect to the deviations, we obtain

$$\frac{d}{dt}\begin{pmatrix} \delta P(t) \\ \delta N(t) \\ \delta \phi(t) \end{pmatrix} = \begin{pmatrix} -\Gamma_p & G_N P_s + \beta & 0 \\ -(G_P P_s + G_s) & -\Gamma_n & 0 \\ \alpha G_P /2 & \alpha G_N & 0 \end{pmatrix} \begin{pmatrix} \delta P(t) \\ \delta N(t) \\ \delta \phi(t) \end{pmatrix} \quad (2.76)$$

with

$$G_P = \left.\frac{\partial G}{\partial P}\right|_s = -\varepsilon\, G_N (N_s - N_0), \quad (2.77)$$

$$G_N = \left.\frac{\partial G}{\partial N}\right|_s = G_N (1 - \varepsilon\, P_s), \quad (2.78)$$

$$\Gamma_P = -\left(G_s - \frac{1}{\tau_p}\right) - G_P P_s, \quad (2.79)$$

$$\Gamma_N = \frac{1}{\tau_s} + G_N P_s, \quad (2.80)$$

where

$$G_S = G(N_s, P_s). \quad (2.81)$$

The general solution of this system of homogeneous linear equations is

$$\begin{pmatrix} \delta P(t) \\ \delta N(t) \\ \delta \phi(t) \end{pmatrix} = \sum_{m=1}^{3} c_m \begin{pmatrix} \delta P_m \\ \delta N_m \\ \delta \phi_m \end{pmatrix} \exp(s_m t) \quad (2.82)$$

where $s_m$ and $(\delta P_m, \delta N_m, \delta \phi_m)$ are respectively the eigenvalues and the eigenvectors of the matrix appearing in Eq. (2.76). The constants $c_m$ are determined by considering the initial conditions. The eigenvalues $s_m$ are the roots of the characteristic equation of the system

$$s[s^2 + (\Gamma_P + \Gamma_N)s + \Gamma_P \Gamma_N + (G_P P_s + G_s)(G_N P_s + \beta)] = 0. \quad (2.83)$$

The solution $s_1 = 0$ accounts for the fact that the optical phase is undetermined. The general expression of the two other solutions is

$$s_{2,3} = -\frac{(\Gamma_P + \Gamma_N)}{2} \pm \sqrt{\frac{(\Gamma_P - \Gamma_N)^2}{4} - (G_P P_s + G_s)(G_N P_s + \beta)}. \quad (2.84)$$

For simplicity, we consider at first the stability of the steady state solution above and below threshold in the case $\beta = 0$, $\varepsilon = 0$. We will later consider their stability above threshold taking $\beta$ and $\varepsilon$ into account.

If $\beta = \varepsilon = 0$, two real eigenvalues are associated to the first steady state solution $[P_s = 0,$ Eq. (2.71)] :

$$s_2 = -\frac{1}{\tau_s}, \tag{2.85}$$

$$s_3 = \frac{G_N \tau_s}{e}(I - I_{th}). \tag{2.86}$$

Below threshold, both eigenvalues are negative and the equilibrium solution $P_s = 0$ is stable. Above threshold ($I > I_{th}$), $s_3$ is positive and the equilibrium solution $P_s = 0$ is unstable.

The eigenvalues associated to the second steady state solution [$N_s = N_{th}$, Eq.(2.72)] read if $\beta = \varepsilon = 0$:

$$s_{2,3} = -\Gamma_R \pm i\Omega_R \tag{2.87}$$

where

$$\Gamma_R = \frac{1}{2}\Gamma_N = \frac{1}{2}\left(\frac{1}{\tau_s} + G_N P_s\right) \geq 0 \tag{2.88}$$

and

$$\Omega_R = \sqrt{\omega_R^2 - \Gamma_R^2} \tag{2.89}$$

with

$$\omega_R = \sqrt{\frac{G_N P_s}{\tau_p}}. \tag{2.90}$$

$\Gamma_R$ and $\omega_R$ are respectively the damping rate and the angular frequency of relaxation oscillations. These oscillations are the result of the interplay between the photon and electron-hole pair populations in the laser active layer. $\Gamma_R$ is always positive: the relaxation oscillations are thus damped, meaning that the second steady state solution [Eq. (2.72)] is stable.

We now consider the case $\beta \neq 0$ and $\varepsilon \neq 0$ when the laser operates above threshold. The eigenvalues $s_{2,3}$ read

$$s_{2,3} = -\Gamma_R \pm i\Omega_R \quad (2.91)$$

where

$$\Gamma_R = \frac{1}{2}(\Gamma_N + \Gamma_P) = \frac{1}{2}\left(\frac{1}{\tau_s} + G_N P_s + \frac{\beta N_s}{P_s} - G_P P_s\right) \quad (2.92)$$

and

$$\Omega_R = \sqrt{\left[(G_s + G_P P_s)(G_N P_s + \beta) - \frac{(\Gamma_N - \Gamma_P)^2}{4}\right]}. \quad (2.93)$$

In the previous equation, the term $G_s G_N P_s$ dominates the other terms by several orders of magnitude; hence, to a good degree of approximation,

$$\Omega_R \cong \sqrt{G_s G_N P_s}. \quad (2.94)$$

Above threshold, one has $G_s \cong 1/\tau_p$ and $N_s \cong N_{th}$. Using Eq.(2.60), the steady state value of $P$ can thus be approximated by

$$P_s \cong \frac{I - I_{th}}{eG_s} \quad (2.95)$$

yielding

$$\Omega_R \cong \sqrt{\frac{G_N(I - I_{th})}{e}}. \quad (2.96)$$

From Eq. (2.92), it appears that the steady state solution is stable as long as the gain saturation coefficient is positive. However, when the gain increases with intensity, i.e. when $G_P$ is positive, $\Gamma_R$ can become negative. The laser then exhibits self-pulsing that has been reported in gain-guided ALGaAs lasers [3].

It follows also from Eqs. (2.92) and (2.93) that $\Gamma_R$ increases whereas $\Omega_R$ decreases almost linearly with the gain saturation coefficient $\varepsilon$, as shown in Fig. 2.7. Although the effect of the saturation on the steady state values $P_s$ and $N_s$ is generally almost imperceptible for $\varepsilon = 10^{-7}$, this parameter affects strongly the damping rate $\Gamma_R$. For instance, Fig. 2.7 shows that $\Gamma_R$ is almost five times larger when $\varepsilon = 10^{-7}$ than when $\varepsilon = 0$.

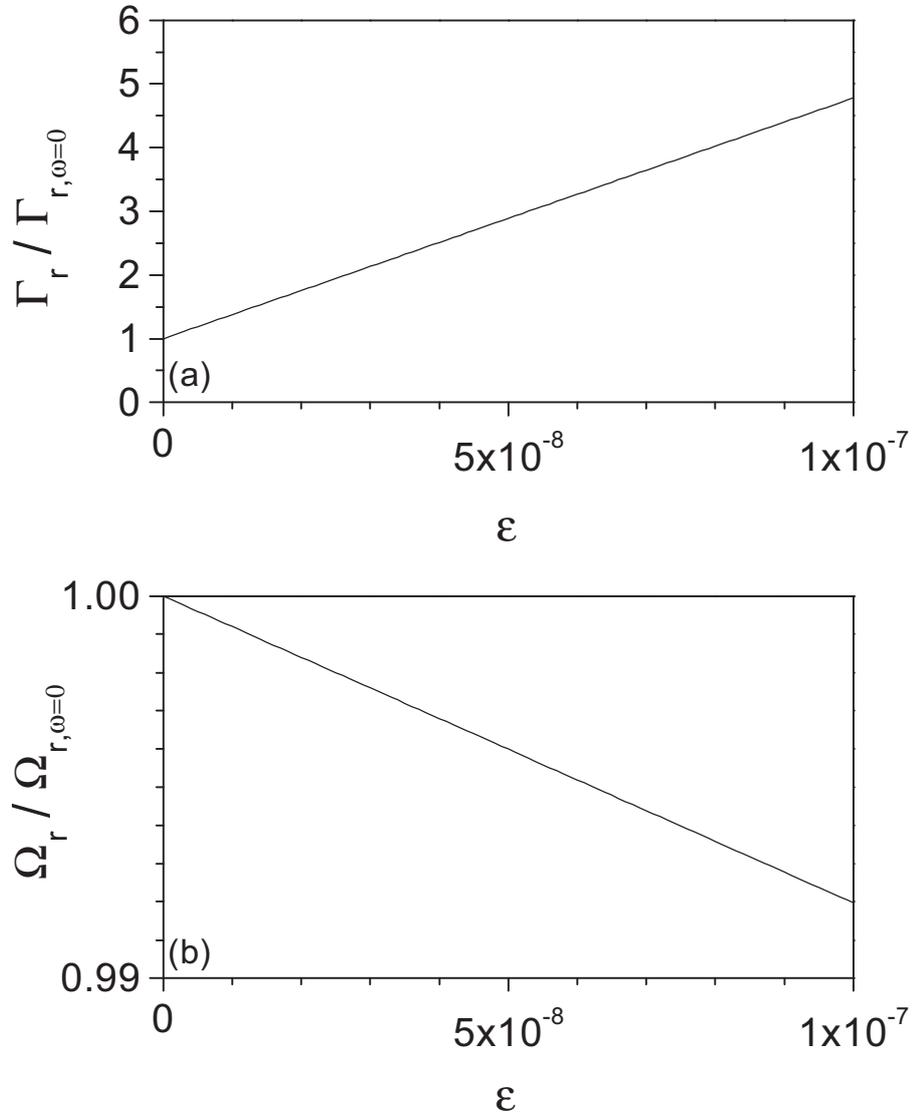

**Fig. 2.7** Variation of the damping rate $\Gamma_R$ (a) and of the angular frequency $\Omega_R$ (b) with the gain saturation coefficient $\varepsilon$ for $I = 2 \times I_{th}$ and $\beta = 10^4$ s$^{-1}$. The other parameters are given in Table 2-I.

## 2.3.5. Intensity noise spectrum and optical spectrum

When a solitary laser diode operates well above threshold, the fluctuations arising from the spontaneous emission process and the carrier-generation-recombination process are small compared to the laser field and the carrier population. The stochastic rate equations (2.58)-(2.60) can then be linearized, yielding

$$\frac{d}{dt}\begin{pmatrix}\delta P(t)\\ \delta N(t)\\ \delta\phi(t)\end{pmatrix}=\begin{pmatrix}-\Gamma_p & G_N P_s+\beta & 0\\ -(G_P P_s+G_s) & -\Gamma_n & 0\\ 0 & \alpha G_N & 0\end{pmatrix}\begin{pmatrix}\delta P(t)\\ \delta N(t)\\ \delta\phi(t)\end{pmatrix}+\begin{pmatrix}F_P(t)\\ F_N(t)\\ F_\phi(t)\end{pmatrix} \quad (2.97)$$

where $G_N$, $G_P$, $\Gamma_p$ and $\Gamma_n$ are defined by Eqs. (2.77)-(2.80). The contribution to the phase arising from the power dependence of the gain has been neglected because it is much smaller than the contribution of the carrier dependence of the gain [3]. In order to obtain the relative intensity noise (RIN) and the optical spectra, the linear system must be solved in the frequency domain. Defining the Fourier transform of a function $f(t)$ as

$$\tilde{f}(\omega)=\int_{-\infty}^{\infty}f(t)\exp(-i\omega t)dt, \quad (2.98)$$

the Fourier transform of Eqs. (2.97) yields

$$\begin{pmatrix}\widetilde{\delta P}(\omega)\\ \widetilde{\delta N}(\omega)\\ \widetilde{\delta\phi}(\omega)\end{pmatrix}=\begin{pmatrix}i\omega+\Gamma_p & -(G_N P_s+\beta) & 0\\ G_P P_s+G_s & i\omega+\Gamma_n & 0\\ 0 & -\alpha G_N & i\omega\end{pmatrix}^{-1}\begin{pmatrix}\tilde{F}_P(\omega)\\ \tilde{F}_N(\omega)\\ \tilde{F}_\phi(\omega)\end{pmatrix}. \quad (2.99)$$

After some rewriting, one finds

$$\widetilde{\delta P}(\omega)=\frac{(\Gamma_n+i\omega)\tilde{F}_P(\omega)+(G_N P_s+\beta)\tilde{F}_N(\omega)}{\Omega_R^2+\Gamma_R^2-\omega^2+2i\omega\Gamma_R}, \quad (2.100)$$

$$\widetilde{\delta N}(\omega)=\frac{-(G_P P_s+G_s)\tilde{F}_P(\omega)+(\Gamma_p+i\omega)\tilde{F}_N(\omega)}{\Omega_R^2+\Gamma_R^2-\omega^2+2i\omega\Gamma_R}, \quad (2.101)$$

$$\widetilde{\delta\phi}(\omega)=\frac{1}{i\omega}\left(\tilde{F}_\phi(\omega)+\frac{\alpha}{2}G_N\widetilde{\delta N}(\omega)\right). \quad (2.102)$$

The relative intensity noise spectrum is defined as the normalized spectral density of the intensity fluctuations:

$$RIN(\omega)=\frac{S_{\delta P}(\omega)}{<P>^2} \quad (2.103)$$

with

$$S_{\delta P}(\omega) = \int_{-\infty}^{\infty} <\delta P(t+\tau)\delta P(t)> \exp(-i\omega\tau)d\tau . \quad (2.104)$$

In the case of a stationary and ergodic random function $f(t)$, one can demonstrate [10] that

$$<\tilde{f}(\omega)\tilde{f}^*(\omega')> = S_f(\omega)\delta(\omega-\omega'). \quad (2.105)$$

Using this relation and Eq. (2.51) for the Langevin forces, we find

$$<\widetilde{\delta P}(\omega)\widetilde{\delta P}^*(\omega')> = S_{\delta P}(\omega)\delta(\omega-\omega') \quad (2.106)$$

and

$$<\tilde{F}_i(\omega)\tilde{F}_j^*(\omega')> = 2D_{ij}\delta(\omega-\omega'). \quad (2.107)$$

Combining Eqs. (2.61),(2.62),(2.64),(2.100),(2.106) and (2.107), we obtain

$$RIN(\omega) = \frac{2\beta N_s[\Gamma_n^2 + \omega^2 + (G_N P_s + \beta)^2[1+(1/\tau_s \beta P_s)] - 2\Gamma_n(G_N P_s + \beta)]}{P_s[(\Omega_R^2 + \Gamma_R^2 - \omega^2)^2 + (2\omega\Gamma_R)^2]}. \quad (2.108)$$

Eq. (2.108) reveals that the relative intensity noise is significantly enhanced in the vicinity of the relaxation oscillation frequency $\omega_R = \sqrt{\Omega_R^2 + \Gamma_R^2}$. An increase of the injection current leads to a shift of the RIN peak, since the relaxation oscillation frequency depends on the current, as well as to a decrease of the noise at a given frequency. Fig. 2.8 shows the RIN spectrum for two different values of the gain saturation coefficient, namely $\varepsilon = 10^{-7}$ and $\varepsilon = 0$. As expected, the gain saturation strongly damps the noise-induced fluctuations of the laser intensity.

The optical spectrum is defined mathematically as the Fourier transform of the autocorrelation function of the optical field:

$$S_E(\omega) = \int_{-\infty}^{\infty} <E^*(t+\tau)E(t)> \exp(-i\omega\tau)d\tau \quad (2.109)$$

where

$$E(t) = \sqrt{P_s + \delta P(t)} \exp[i(\omega_{th}t + \phi_s(t) + \delta\phi(t))]. \quad (2.110)$$

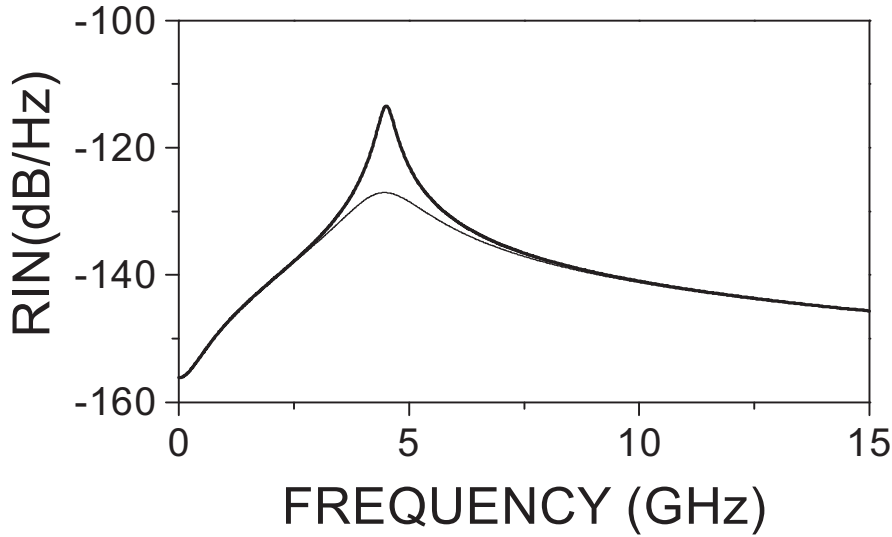

**Fig. 2.8** RIN spectrum for two different values of the gain saturation coefficient, namely $\varepsilon = 10^{-7}$ (thin line) and $\varepsilon = 0$ (thick line). $I = 2 \times I_{th}$ and $\beta = 10^4$ s$^{-1}$. The other parameters are given in Table 2-I.

For simplicity, we neglect intensity fluctuations and assume that the field fluctuates with respect to the phase only. Using Eq. (2.110), $S_E$ can be rewritten as

$$S_E(\omega) = P_s \int_{-\infty}^{\infty} <\exp(i\Delta_\tau \phi)> \exp[-i(\omega - \omega_s)\tau] d\tau \qquad (2.111)$$

where

$$\Delta_\tau \phi = \delta\phi(t-\tau) - \delta\phi(t) \text{ and } \omega_s = \omega_{th} + \Delta\omega_s. \qquad (2.112)$$

Assuming that $\Delta_\tau \phi$ has a normal distribution [11], which is the case if many photons are spontaneously emitted during the time interval $\tau$,

$$<\exp(i\Delta_\tau \phi)> = \exp\left[-\frac{1}{2}<(\Delta_\tau \phi)^2>\right]. \qquad (2.113)$$

Since the cross correlation between the frequency components $\widetilde{\delta\phi}(\omega)$ vanishes, one finds:

$$<(\Delta_\tau \phi)^2> = \frac{1}{\pi} \int_{-\infty}^{\infty} <|\widetilde{\delta\phi}(\omega)|^2> (1-\cos(\omega\tau)) d\omega. \qquad (2.114)$$

The general expression for $<|\widetilde{\delta\phi}(\omega)|^2>$, to be found from Eq. (2.102), is rather complicated. It can be simplified by noting that the term $(G_P P_s + G_s)\widetilde{F}_P(\omega)$ is dominant in Eq. (2.101) for $\widetilde{\delta N}(\omega)$. The approximate expression for $<|\widetilde{\delta\phi}(\omega)|^2>$ is

$$<|\widetilde{\delta\phi}(\omega)|^2> \cong \frac{R_{sp}}{2P_s\omega^2}\left(1 + \frac{\alpha^2\Omega_R^2}{\left((\Omega_R^2 + \Gamma_R^2 - \omega^2)^2 + (2\omega\Gamma_R)^2\right)}\right). \qquad (2.115)$$

The function $\omega^2 <|\widetilde{\delta\phi}(\omega)|^2>$, which is in fact the spectral density of the frequency noise, is relatively flat for $\omega \ll \Omega_R$, yielding

$$<(\Delta_\tau\phi)^2> = \frac{(1+\alpha^2)R_{sp}|\tau|}{2P_s} \qquad (2.116)$$

when the delay $\tau$ is not too short. Using Eqs. (2.111), (2.113) and (2.116), one finds that the optical spectrum can be approximated by a Lorentzian centered at $\omega_s$:

$$S_E(\omega) = 2P_s \frac{\Delta\omega}{(\omega - \omega_s)^2 + \Delta\omega^2} \qquad (2.117)$$

with a full width at half maximum of $\Delta\omega = \frac{(1+\alpha^2)R_{sp}}{2P_s}$. Using $\Delta\omega = 2\pi\Delta f$, the linewidth $\Delta f$ of the optical spectrum is

$$\Delta f = \frac{(1+\alpha^2)R_{sp}}{4\pi P_s}. \qquad (2.118)$$

Eq. (2.118) expresses that two mechanisms contribute to the laser linewidth. The direct contribution of the noise-induced phase fluctuations results in the term $R_{sp}/(4\pi P_s)$. The contribution of the fluctuations in the intensity is accounted by $\alpha^2 R_{sp}/(4\pi P_s)$. The latter can be explained as follows. Intensity fluctuations perturb the carrier number and consequently the refractive index. This in turn affects the phase owing to the amplitude-phase coupling. The significant broadening of the linewidth in semiconductor lasers (for which $\alpha$ is larger than 1) with respect to that expected on the basis of the Schawlow-Townes theory [12] was explained for the first time by Henry [5].

## 2.4. Multimode solitary laser diode

### 2.4.1. Rate equations

We have derived so far rate equations for single-mode laser diodes. These equations can be generalized to describe multimode operation as follows [1]:

$$\frac{dE_m}{dt} = \tfrac{1}{2}(1+i\alpha)[G_m(N) - \gamma_m]E_m(t) + F_m(t), \tag{2.119}$$

$$\frac{dN}{dt} = \frac{I}{e} - \frac{N}{\tau_s} - \sum_{i=-M}^{M} G_i(N)|E_i(t)|^2 + F_N(t), \tag{2.120}$$

$E_m(t)$ is the slowly varying complex electric field of the $m$th mode oscillating at the frequency $\omega_m$. $E_m(t)$ is normalized in such a way that $P_m(t) = |E_m(t)|^2$ is the photon number in the $m$th mode. $\gamma_m$ and $G_m(N)$ are the mode-dependent loss and gain. In conventional Fabry-Perot laser diodes, $\gamma_m$ is mode independent: $\gamma_m = \tau_p^{-1}$ for all modes. The integer $m$ varies from $-M$ to $+M$. In Eq.(2.120), the summation is over all modes participating in the stimulated emission process. $F_m(t)$ and $F_N(t)$ are Langevin noise forces accounting for spontaneous emission noise and shot noise respectively. The Langevin forces satisfy the following relations

$$<F_i(t)> = 0, \tag{2.121}$$

$$<F_i(t)F_j(t')> = 2D_{ij}\delta(t-t'). \tag{2.122}$$

The diffusion coefficients $D_{ij}$ are given by

$$2D_{E_m E_n^*} = R_{sp}\delta_{mn}, \tag{2.123}$$

$$2D_{E_m E_n} = 0, \tag{2.124}$$

$$2D_{E_m N} = -R_{sp}|E_m|, \tag{2.125}$$

$$2D_{NN} = 2\left(R_{sp}\sum_{m=-M}^{M}|E|^2 + \frac{N}{\tau_s}\right), \tag{2.126}$$

where we have assumed a mode-independent spontaneous emission rate.

Assuming that the gain profile is homogenously broadened and neglecting four wave mixing interactions between the modes as well as cross and self-saturations, we approximate the gain by a quadratic function of the form:

$$G_m(N) = G_c(N - N_0)\left[1 - \left(m\frac{\Delta\omega_L}{\Delta\omega_g}\right)^2\right]. \tag{2.127}$$

$G_c$ and $m = 0$ are the gain coefficient and the longitudinal mode number at the gain peak. $\Delta\omega_L = 2\pi/\tau_{in}$ and $\Delta\omega_g$ are the longitudinal mode spacing and the gain width of the laser material respectively.

When the laser is pumped above threshold, the photon density in laser diodes is very high owing to the small cross-section of the active region. Therefore, nonlinear optical phenomena (see Ref. 1 for a description) can occur, yielding a gain that depends also on the photon number. In this case, self and cross-saturations of the gain must be taken into account and the gain reads

$$G_m(N, |E|^2) = G_c(N - N_0)\left[1 - \left(m\frac{\Delta\omega_L}{\Delta\omega_g}\right)^2 - \psi|E_m|^2 - \sum_{n=-M}^{M}\zeta_{nm}|E_n|^2\right]. \tag{2.128}$$

$\psi$ is the mode-independent self-saturation parameter. The cross-saturation matrix is given by (see Ref. 16 and references therein)

$$\zeta_{nm} = C\frac{1 + \alpha\tau_{pol}\Delta\omega_L(m-n)}{1 + \tau_{pol}[\Delta\omega_L(m-n)]^2} \tag{2.129}$$

where $C$ is a constant and $\tau_{pol}$ is the polarization relaxation time. Although these effects can be important on the dynamics of multimode lasers, we will not consider the cross and self-saturations in this work.

As for the single-mode case, Eqs. (2.119) and (2.120) can be rewritten in terms of the photon number $P_m(t)$ and the phase $\phi_m(t)$ associated with the $m$th mode instead of the modal complex electric fields. Neglecting cross and self-saturations as well as stochastic fluctuations arising from the spontaneous emission process and shot noise, the rate equations then read [1,3]

$$\frac{dP_m(t)}{dt} = [G_m(N) - \gamma_m]P_m(t) + R, \tag{2.130}$$

$$\frac{d\phi_m(t)}{dt} = \frac{\alpha}{2}[G_m(N) - \gamma_m], \quad (2.131)$$

$$\frac{dN}{dt} = \frac{I}{e} - \frac{N}{\tau_s} - \sum_m G_m(N)P_m(t). \quad (2.132)$$

## 2.4.2. Light-current curve

The light-current curve is found by considering the steady state modal intensities and carrier number. Setting to zero the time derivatives in Eqs. (2.130) and (2.132) and taking into account that in conventional Fabry-Perot laser diodes $\gamma_m$ is mode independent (i.e. $\gamma_m = \tau_p^{-1}$, $\tau_p$ being the photon lifetime), we obtain

$$P_{ms} = \frac{R_{sp}}{\frac{1}{\tau_p} - G_m(N_s)} \quad (2.133)$$

and

$$I = e\left(\frac{N_s}{\tau_s} - \sum_{i=-M}^{M} G_i(N_s)P_{is}\right). \quad (2.134)$$

Fig. 2.9 and Fig. 2.10 show respectively the variation of the photon numbers in the seven brightest modes and the variation of the total number of photons emitted by the laser with the injection current. The figures have been obtained considering $N_s$ as a parameter. Below or near threshold, the photon number increases in all modes. Above threshold, the side modes saturate whereas the main mode, which corresponds to the maximum of the gain curve, increases continuously with the current. Moreover, the modal intensities are proportional to the spontaneous emission rate.

The mode suppression ratio (MSR) is defined as the ratio of the photon number in the main mode to the photon number in the brightest side mode:

$$\text{MSR} = \frac{P_0}{P_1} = \frac{\gamma_1 - G_1(N_s)}{\gamma_0 - G_0(N_s)}. \quad (2.135)$$

The MSR is a measure of the spectral purity of the laser diode. In index-guided Fabry-Perot lasers, the MSR is typically in the range 10-20 whereas the cavity losses are mode independent and the relative gain difference is less than 0.5%. As a comparison, for DFB lasers with mode dependent losses, a MSR of the order of 1000 can be achieved with a relative change $(\gamma_1-\gamma_0)/\gamma_0$ in the modal losses of the order of 10%.

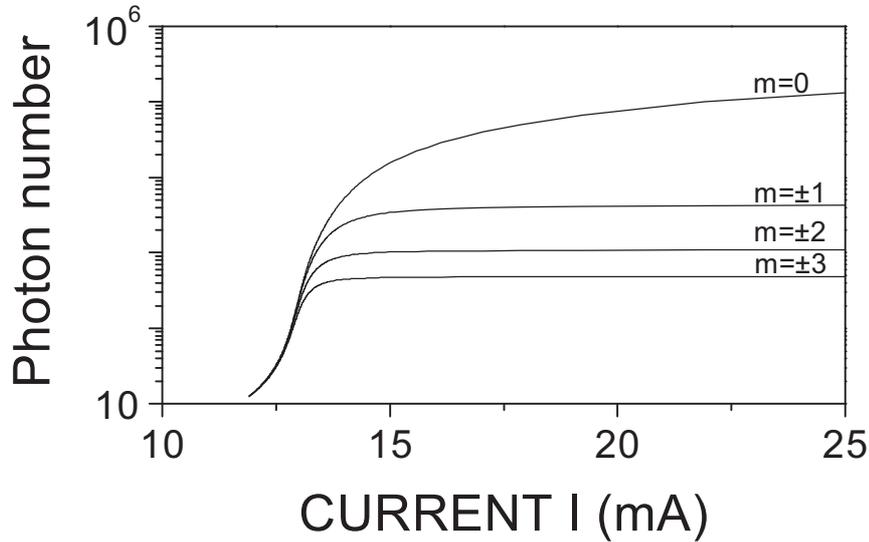

**Fig. 2.9** Variation of the photon numbers in the seven brightest modes of a laser diode with current. $\Delta\omega_g = 2\pi \times 4.7$ THz, $\beta = 10^4$ s$^{-1}$. The other parameters are given in Table 2-I.

Finally, it is important to notice that, in the frame of the previous model, the multimode nature of the laser diode is attributed to spontaneous emission as can be seen from Eq. (2.133).

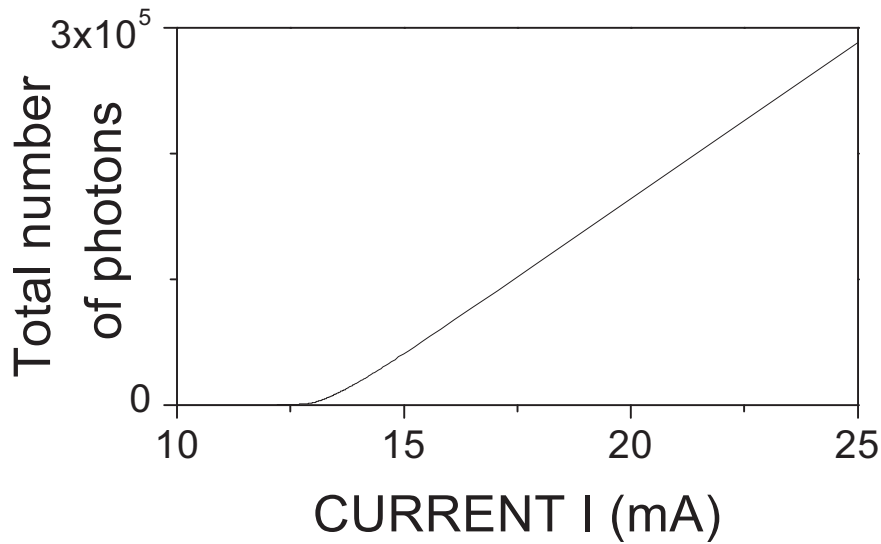

**Fig. 2.10** Variation of the total number of photons with current. $\Delta\omega_g = 2\pi \times 4.7$ THz, $\beta = 10^4$ s$^{-1}$. The other parameters are given in Table 2-I.

## 2.5. Laser diode subject to coherent optical feedback

### 2.5.1. Rate equations for a single mode laser diode subject to coherent optical feedback: the Lang-Kobayashi equations

For simplicity, we derive the field equation for a laser diode subject to a coherent optical injection in a first step and the field equation for a laser diode subject to an external, coherent, optical feedback in a second step.

We assume that the external field $\mathcal{E}_{ext}(t) = E_{ext}(t)\exp(i\omega_{ext}t)$ is incident on the laser diode left facet ($z = 0$) that has an amplitude transmission coefficient $t_1'$ for an electromagnetic field entering the laser (Fig. 2.11). The field $t_1'\mathcal{E}_{ext}(t)$ that enters the laser cavity is normalized in the same way as the laser field itself. The difference equation describing the field amplitude after one round-trip [Eq. (2.29)] has to be rewritten as

$$\mathcal{E}_f(t) = G_1 \exp(i\omega_{th}\tau_{in})\mathcal{E}_f(t - \tau_{in}) + t_1'\mathcal{E}_{ext}(t). \qquad (2.136)$$

The rate equation for the slowly-varying complex amplitude of the electric field $E(t)$ can then be derived in the same way than in section 2.3.1, still for $G_1$ close to 1 (as is the case under normal conditions). Thus [18]

$$\frac{dE(t)}{dt} = \frac{1+i\alpha}{2}\left[G_N(N-N_0) - \frac{1}{\tau_p}\right]E(t) + \frac{t_1'}{\tau_{in}}E_{ext}(t)\exp(i\Delta\omega t) + F_E(t) \qquad (2.137)$$

where $\Delta\omega = \omega_{ext} - \omega_{th}$ is the frequency detuning.

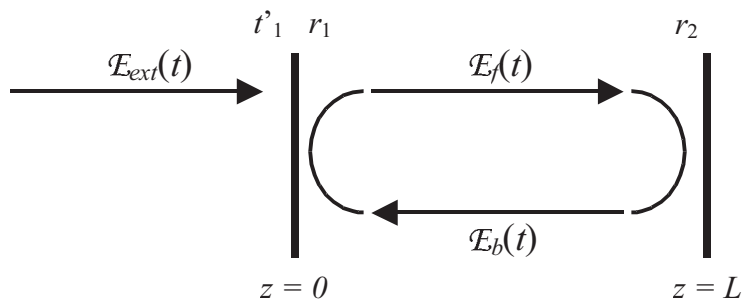

**Fig. 2.11** Schematic representation of an optically injected laser diode. $t_1'$ denotes the amplitude transmission coefficient from outside of the cavity. $r_1$ and $r_2$ are the amplitude reflection coefficients of the left and right laser facets from inside of the laser cavity.

We now consider the case of external optical feedback from an external mirror with amplitude transmission coefficient $r_3$ positioned at a distance $L_{ext}$ from the left facet. The round-trip time in the external cavity is $\tau = 2L_{ext}/c$ (Fig. 2.12).

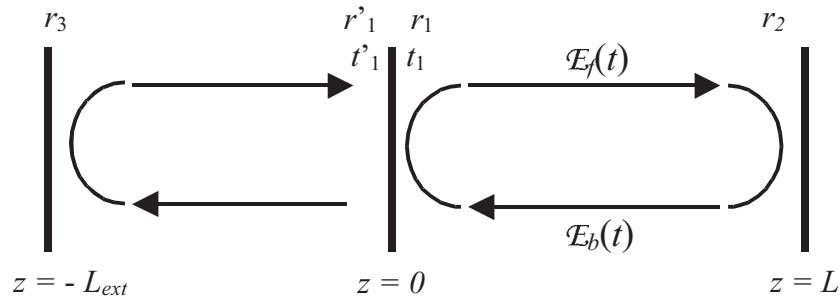

**Fig. 2.12** Schematic representation of a laser diode subject to coherent optical feedback. $r_1'$ is the amplitude reflection coefficient from outside of the cavity. $t_1$ is the amplitude transmission coefficient from inside of the cavity. $r_3$ is the amplitude reflection coefficient of the external mirror.

The field $t_1' E_{ext}(t)$ that enters the laser cavity contains the contributions from an infinite number of round-trips in the external cavity:

$$t_1' E_{ext}(t) = r_3 t_1 t_1' E_b(t-\tau) + r_1' r_3^2 t_1 t_1' E_b(t-2\tau) + \ldots + \frac{(r_1' r_3)^n}{r_1'} t_1 t_1' E_b(t-n\tau) \qquad (2.138)$$

where $r_1'$ is the amplitude reflection coefficient from outside of the cavity and $E_b(t)$ is the electric field of the backward travelling wave at z = 0 with

$$E_f(t) = r_1 E_b(t). \qquad (2.139)$$

Using the relations [19]

$$t_1 t_1' = 1 - r_1^2 \qquad (2.140)$$

and

$$r_1 = -r_1', \qquad (2.141)$$

the difference equation Eq. (2.29) can be rewritten as

$$E_f(t) = G_1 \exp(i\omega_{th}\tau_{in}) E_f(t-\tau_{in}) + \frac{r_1^2 - 1}{r_1^2} \sum_{n=1}^{\infty} (-r_1 r_3)^n E_f(t-n\tau) \qquad (2.142)$$

The rate equation for the slowly-varying complex amplitude of the electric field $E(t)$ can then be written as [2]:

$$\frac{dE(t)}{dt} = \frac{1+i\alpha}{2}\left[G_N(N-N_0) - \frac{1}{\tau_p}\right]E(t)$$
$$+ \frac{1}{\tau_{in}}\frac{r_1^2-1}{r_1^2}\sum_{n=1}^{\infty}(-r_1 r_3)^n E(t-n\tau)\exp(-in\omega_{th}\tau) + F_E(t). \quad (2.143)$$

When the intensity of the light reflected by the external mirror is weak, or when the laser left facet is antireflection coated, the feedback term in Eq. (2.143) can be approximated by taking into account just one round-trip in the external cavity, yielding

$$\frac{dE(t)}{dt} = \frac{1+i\alpha}{2}\left[G_N(N-N_0) - \frac{1}{\tau_p}\right]E(t) + \frac{\kappa}{\tau_{in}}E(t-\tau)\exp(-i\omega_{th}\tau) + F_E(t). \quad (2.144)$$

Eq. (2.144) was first derived by Lang and Kobayashi in 1980 [20] and has become the paradigm for investigating the dynamical behavior of laser diodes subject to coherent optical feedback. The so-called Lang-Kobayashi equations comprise the above equation and the equation for the carrier number, the latter being not modified by the feedback. In Eq. (2.144), the feedback parameter $\kappa$ is defined as

$$\kappa = (1-r_1^2)\frac{r_3}{r_1}. \quad (2.145)$$

$\kappa^2$ is the intensity reflected from the external cavity relative to the intensity reflected from the laser mirror. The phase $\omega_{th}\tau$ measures the sensitivity of the system on the precise position of the external mirror within half a wavelength. Since a tiny increase of the delay $\tau$ can lead to a large modification of the phase $\omega_{th}\tau$, it is convenient to consider $\tau$ and the product $\omega_{th}\tau$ as two independent parameters. The value of $\omega_{th}\tau$ is often given in the range $[-\pi,\pi[$.

Depending on the problem under consideration, the Lang-Kobayashi equations might have to be generalized to take spontaneous emission and gain saturation into account. Eq. (2.144) can then be rewritten as

$$\frac{dE(t)}{dt} = \frac{1+i\alpha}{2}\left[G(N,|E|^2) - \frac{1}{\tau_p}\right]E(t) + \frac{\kappa}{\tau_{in}}E(t-\tau)\exp(-i\omega_{th}\tau) + F_E(t) \quad (2.146)$$

or as

$$\frac{dE(t)}{dt} = \frac{1}{2}\left[G(N,|E|^2) - \frac{1}{\tau_p} + i\alpha G_N(N-N_{th})\right]E(t)$$
$$+ \frac{\kappa}{\tau_{in}}E(t-\tau)\exp(-i\omega_{th}\tau) + F_E(t) \tag{2.147}$$

## 2.5.2. Steady state solutions of the Lang-Kobayashi equations

Although this mechanism can be of great importance in some cases, for instance by postponing the destabilization of the laser diode due to optical feedback, we neglect the saturation of the gain in this section and the next one. The analytical treatment of the rate equations is indeed much more easy in that case [2]. We neglect furthermore the spontaneous emission and shot noise terms. The Lang-Kobayashi equations then read

$$\frac{dE(t)}{dt} = \frac{1+i\alpha}{2}\left[G_N(N-N_0) - \frac{1}{\tau_p}\right]E(t) + \frac{\kappa}{\tau_{in}}E(t-\tau)\exp(-i\omega_{th}\tau), \tag{2.148}$$

$$\frac{dN(t)}{dt} = \frac{I}{e} - \frac{N(t)}{\tau_s} - G_N(N-N_0)|E(t)|^2. \tag{2.149}$$

These equations are equivalent to

$$\frac{dP(t)}{dt} = \left[G_N(N-N_0) - \frac{1}{\tau_p}\right]P(t)$$
$$+ \frac{2\kappa}{\tau_{in}}\sqrt{P(t)P(t-\tau)}\cos[\phi(t)-\phi(t-\tau)+\omega_{th}\tau], \tag{2.150}$$

$$\frac{d\phi(t)}{dt} = \frac{\alpha}{2}\left[G_N(N-N_0) - \frac{1}{\tau_p}\right]$$
$$- \frac{2\kappa}{\tau_{in}}\sqrt{\frac{P(t-\tau)}{P(t)}}\sin[\phi(t)-\phi(t-\tau)+\omega_{th}\tau], \tag{2.151}$$

$$\frac{dN(t)}{dt} = \frac{I}{e} - \frac{N(t)}{\tau_s} - G_N(N-N_0)P(t).\qquad(2.152)$$

where we consider the photon number $P(t)$ and the phase $\phi(t)$ instead of the complex electric field $E(t)$.

The steady state solutions are obtained by introducing

$$P(t) = P_s,\ N(t) = N_s,\ \phi_s(t) = (\omega_s - \omega_{th})t \qquad(2.153)$$

in the rate equations (2.150)-(2.152). The steady state value of the carrier number is found from the equation for the photon number:

$$G_N(N_s - N_0) = 1/\tau_p - 2\frac{\kappa\tau}{\tau_{in}}\cos(\omega_s\tau)\qquad(2.154)$$

or equivalently, after substitution of $1/\tau_p$ by $G_N(N_{th}-N_0)$ [Eqs. (2.25) and (2.27)]:

$$N_s = N_{th} - 2\frac{\kappa}{G_N\tau_{in}}\cos(\omega_s\tau)\qquad(2.155)$$

The steady state angular frequency $\omega_s$ is obtained by substituting Eq. (2.154) into the phase equation (2.151); thus

$$\omega_s - \omega_{th} = -\frac{\kappa}{\tau_{in}}\sqrt{1+\alpha^2}\sin(\omega_s\tau + \arctan\alpha).\qquad(2.156)$$

Finally, the steady state value of the photon number is obtained by substituting Eq. (2.155) into the equation for the carriers:

$$P_s = \frac{\dfrac{I}{e} - \dfrac{N_s}{\tau_s}}{G_N(N_s - N_0)}.\qquad(2.157)$$

It is often convenient to consider the steady state value of the phase difference over one round-trip, $\Delta\phi_s$, defined as

$$\Delta\phi_s = \phi_s(t) - \phi_s(t-\tau) + \omega_{th}\tau = \omega_s\tau \qquad(2.158)$$

instead of the angular frequency $\omega_s$. The phase differences $\omega_s\tau$ can easily be found by graphical construction (Fig. 2.13) of the function

$$f(\omega\tau) \equiv \omega\tau - \omega_{th}\tau + \frac{\kappa\tau}{\tau_{in}}\sqrt{1+\alpha^2}\sin(\omega\tau + \arctan\alpha) \tag{2.159}$$

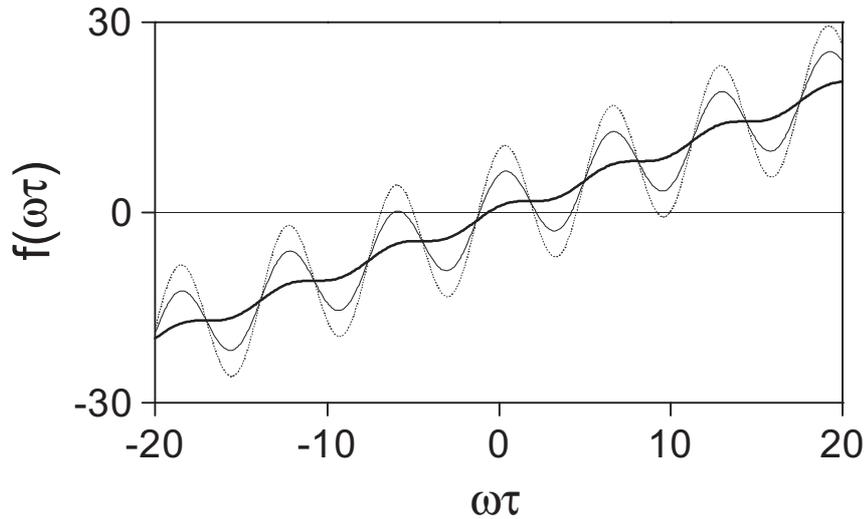

**Fig. 2.13** Function $f(\omega\tau)$ and for $\tau = 2$ ns, $\omega_{th}\tau = 0$ and $\kappa = 10^{-3}$ (thick line), $\kappa = 6 \times 10^{-3}$ (thin line) and $\kappa = 10^{-2}$ (dots).

New roots of the function $f$ are progressively created in pairs when $\kappa$ or $\tau$ are increased (the number of steady state solutions thus increases). The minimum feedback level that is needed for the creation of the first pair of additional steady state solutions is found from Eq. (2.159) in a straightforward manner; thus

$$\kappa = \frac{\tau_{in}}{\tau\sqrt{1+\alpha^2}} \tag{2.160}$$

when

$$\omega_{th}\tau = \pi - \arctan\alpha \mod 2\pi \tag{2.161}$$

It appears from Fig. 2.13 and Eq. (2.158) that the phase difference between two neighboring steady state solutions tends to $\pi$ when the feedback level or the feedback delay increase.

From Eqs. (2.155) and (2.156), one finally finds that the steady state solutions lie on an ellipse in the $(\omega_s\tau, N-N_{th})$ plane:

$$\left[(\omega_s\tau - \omega_{th}\tau) + \frac{\alpha G_N \tau}{2}(N_s - N_{th})\right]^2 + \left[\frac{G_N \tau}{2}(N_s - N_{th})\right]^2 = \left(\frac{\kappa}{\tau_{in}}\right)^2. \tag{2.162}$$

### 2.5.3. Stability of the steady state solutions of the Lang-Kobayashi equations

The linear stability analysis of the steady state solutions is performed by introducing infinitesimal perturbations $\delta P(t)$, $\delta N(t)$ and $\delta\phi(t)$ around the steady state solutions $E_s$, $N_s$ and $\phi_s$ in the rate equations; the nonlinear terms, e.g. $\delta P \delta N$ or $\delta P^2$, are neglected for the purpose. Eqs. (2.150)-(2.152) then lead to the linear and homogeneous system of equations:

$$\delta\dot{\mathbf{x}}(t) = \mathbf{A}\delta\mathbf{x}(t) - \mathbf{B}\delta\mathbf{x}(t-\tau) \tag{2.163}$$

where

$$\delta\mathbf{x}(t) = \begin{bmatrix} \delta P(t) & \delta N(t) & \delta\phi(t) \end{bmatrix}^{\mathrm{T}}, \tag{2.164}$$

$$\mathbf{A} = \begin{pmatrix} -\dfrac{\kappa}{\tau_{in}}\cos(\omega_s\tau) & G_N P_s & -2\dfrac{\kappa}{\tau_{in}}P_s\sin(\omega_s\tau) \\ -\left[\dfrac{1}{\tau_p}+G_N(N_s-N_{th})\right] & -\left(\dfrac{1}{\tau_s}+G_N P_s\right) & 0 \\ \dfrac{\kappa}{2\tau_{in}}\dfrac{\sin(\omega_s\tau)}{P_s} & \dfrac{\alpha G_N}{2} & -\dfrac{\kappa}{\tau_{in}}\cos(\omega_s\tau) \end{pmatrix}, \tag{2.165}$$

$$\mathbf{B} = \begin{pmatrix} -\dfrac{\kappa}{\tau_{in}}\cos(\omega_s\tau) & 0 & -2\dfrac{\kappa}{\tau_{in}}P_s\sin(\omega_s\tau) \\ 0 & 0 & 0 \\ \dfrac{\kappa}{2\tau_{in}}\dfrac{\sin(\omega_s\tau)}{P_s} & 0 & -\dfrac{\kappa}{\tau_{in}}\cos(\omega_s\tau) \end{pmatrix}. \tag{2.166}$$

Introducing $\delta\mathbf{x}(t) = \delta\mathbf{x}_0 e^{st}$ ($\delta\mathbf{x}_0$ and $s$ are complex) as the general solution of the system, we obtain

$$(\mathbf{A} - \mathbf{B}e^{-s\tau} - s\mathbf{I})\delta\mathbf{x}_0 = 0 \tag{2.167}$$

where

$$\mathbf{A} - \mathbf{B}e^{-s\tau} - s\mathbf{I} =$$

$$\begin{pmatrix} -\dfrac{\kappa}{\tau_{in}}\cos(\omega_s\tau).(1-e^{-s.\tau}) - s & G_N P_s & -\dfrac{2\kappa}{\tau_{in}} P_s \sin(\omega_s\tau)(1-e^{-s.\tau}) \\ -\left[\dfrac{1}{\tau_p} + G_N(N_s - N_{th})\right] & -\left(\dfrac{1}{\tau_s} + G_N P_s\right) - s & 0 \\ \dfrac{\kappa}{2\tau_{in}}\dfrac{\sin(\omega_s\tau)}{P_s}.(1-e^{-s.\tau}) & \dfrac{\alpha G_N}{2} & -\dfrac{\kappa}{\tau_{in}}\cos(\omega_s\tau).(1-e^{-s.\tau}) - s \end{pmatrix}$$

(2.168)

The characteristic determinant of the matrix $\mathbf{A} - \mathbf{B}e^{-s\tau} - s\mathbf{I}$ must vanish for non-trivial solutions $\delta x_0 \neq 0$:

$$s^3 + s^2\left[\dfrac{1}{\tau_s} + G_N P_s + \dfrac{2\kappa}{\tau_{in}}\cos(\omega_s\tau).(1-e^{-s\tau})\right]$$

$$+ s\left[\left(\dfrac{\kappa}{\tau_{in}}.(1-e^{-s\tau})\right)^2 + 2\left(\dfrac{1}{\tau_s} + G_N P_s\right) + \dfrac{\kappa}{\tau_{in}}\cos(\omega_s\tau)(1-e^{-s\tau}) + \dfrac{G_N}{\tau_p} P_s\right]$$

$$+ \left[\left(\dfrac{\kappa}{\tau_{in}}\right)^2\left(\dfrac{1}{\tau_s} + G_N P_s\right)(1-e^{-s\tau})^2 + \dfrac{\kappa}{\tau_{in}}\dfrac{G_N}{\tau_p} P_s\sqrt{1+\alpha^2}\cos(\omega_s\tau + \arctan\alpha)(1-e^{-s\tau})\right] = 0.$$

(2.169)

The stability of steady state solutions is determined by the sign of the real part of the eigenvalues of the matrix $\mathbf{A} - \mathbf{B}e^{-s\tau}$ that are the roots of the transcendental equation Eq. (2.169). Due to the presence of the exponential terms, the characteristic equation has an infinite number of roots. When all roots have negative real parts, any initial perturbation decreases with time. By contrast, when the real part of one or more roots is positive, perturbations increase until the non-linear terms in Eqs. (2.150)-(2.152) enter into play. We note that Eq. (2.169) always admits the solution $s = 0$; the latter accounts for the fact that the phase of steady state solutions is arbitrary. The solution $s = 0$ is therefore irrelevant in our study.

We present here two kinds of instabilities.

### Saddle-node instability

We have shown in section 2.5.2 that Eqs. (2.154)-(2.156) admit only one solution at weak feedback level. As the feedback level increases, new steady state solutions are created in pairs (Fig. 2.13), usually from saddle-node bifurcations[1,2] [21,25]. This suggests that half of the steady state solutions are saddle points, one root of the characteristic determinant [Eq. (2.169)] being always real and positive indicating instability referred to as saddle-node instability. These fixed points lie on the upper part of the ellipse in the phase difference-carrier number plane (Fig. 2.14). The criterion for a fixed point of the Lang-Kobayashi equations to be a saddle point is [26,27]

$$\left.\frac{\partial f(\omega\tau)}{\partial \omega\tau}\right|_{\omega_s\tau} \equiv 1 + \frac{\kappa\tau}{\tau_{in}}\sqrt{1+\alpha^2}\cos(\omega_s\tau + \arctan\alpha) \leq 0. \quad (2.170)$$

The always unstable fixed points are often called *antimodes* and correspond physically to destructive interference between the field inside the laser and the feedback field. The other fixed points correspond to constructive interference and are identified as the *external cavity modes*.

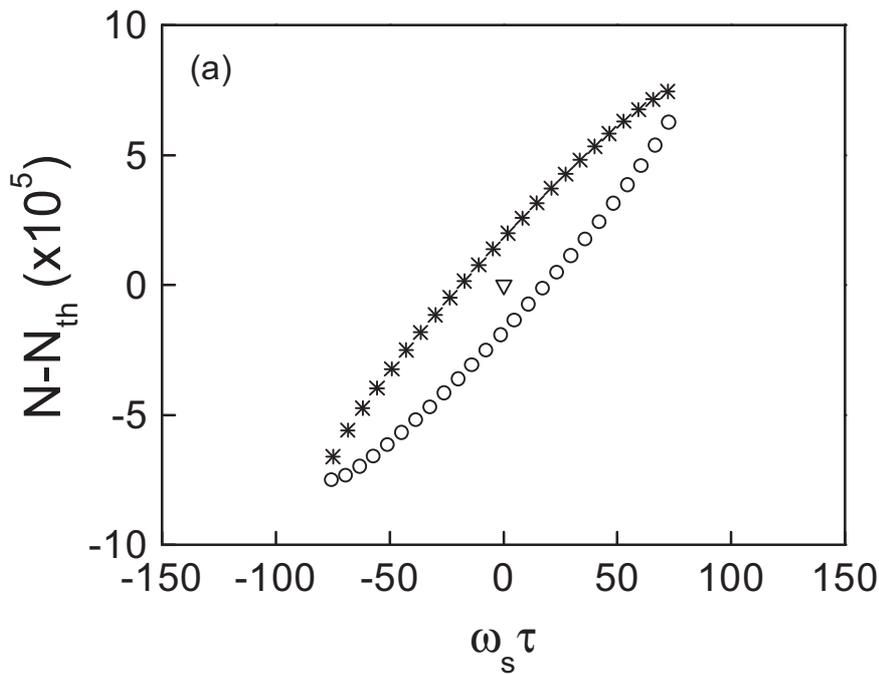

**Fig. 2.14** Steady state solutions in the $(\omega_s\tau, N-N_{th})$ plane. Circles (o) and stars (∗) represent respectively the external cavity modes and the antimodes. $I = 2 \times I_{th}$, $\tau = 5$ ns, $\kappa = 0.03$, $\omega_{th}\tau = 0$. The other parameters are given in Table 2-I.

---

[1] The reader can find more information on saddle-node bifurcations in Appendix B.
[2] Pitchfork bifurcations [21,22] occur in the particular case $\omega_{th}\tau = \pi - \arctan(\alpha) \mod 2\pi$.

## Hopf instability

A Hopf instability (or Hopf bifurcation [21,25]) occurs when the sign of the real part of two complex-conjugate eigenvalues associated to a given steady state solution becomes positive as one of the control parameters is varied. The conditions for a Hopf bifurcation are determined by substituting $s = \pm j\Omega$ in the characteristic equation [Eq. (2.169)] and separating real and imaginary parts. One obtains two equations for the feedback level $\kappa_H$ and the frequency $\Omega_H$ of oscillations at the Hopf bifurcation point. These equations are transcendental and difficult to solve numerically. Several analytical approximations of the Hopf bifurcation points have been proposed [28-30]. Asymptotic approximations based on the smallness of $\chi = \tau_p/\tau_s$, i.e. the ratio between the photon lifetime and the carrier lifetime[3], have been carried out [31-34]. Rather simple expressions of $\kappa_H$ and $\Omega_H$ are obtained by considering a suitable approximate form of Eq. (2.169) in power series of $\chi^{1/2}$; those are

$$\kappa_H = \frac{2\Gamma_R \tau_{in}}{B[1-\cos(\omega_R \tau)]} = \frac{\left(\frac{1}{\tau_s} + \omega_R^2 \tau_p\right)\tau_{in}}{B[1-\cos(\omega_R \tau)]}, \quad (2.171)$$

with

$$B = -\sqrt{1+\alpha^2}\cos(\omega_s \tau - \arctan\alpha) \quad (2.172)$$

and

$$\Omega_H = \omega_R + \frac{1}{2}\left(\frac{1}{\tau_s} + \omega_R^2 \tau_p\right)\cot\left(\frac{\omega_R \tau}{2}\right) \quad (2.173)$$

where $\omega_R$ is the angular frequency of the relaxation oscillations of the solitary laser. In the parameter ranges where these asymptotic approximations are valid, Eq. (2.173) implies that the relaxation oscillations become undamped and self-sustained oscillations appear with angular frequencies close to the relaxation oscillation frequency of the solitary laser when the feedback level passes the value $\kappa_H$ given by Eq. (2.171). The steady state solution becomes unstable and a stable periodic solution, i.e. a limit cycle, is created around this destabilized fixed point. Fig. 2.15 shows the approximate $\kappa_H$ and $\Omega_H$ as functions of $\omega_R \tau/2\pi$. Both figures clearly show that the approximations fail, $\kappa_H$ and $\Omega_H$ being unbounded, when

$$\omega_R \tau = 2n\pi \ (n = 0,1,2,\ldots), \tag{2.174}$$

that is when the relaxation oscillation frequency of the solitary laser is a multiple of the external cavity frequency or when it tends to 0, i.e. for low value of the injection current [Eq. (2.90)]. In the lastter case, the Hopf frequency may deviate considerably from $\omega_R$: when the injection current is at a value close to the solitary laser threshold, the Hopf frequency is proportional to the inverse of the external cavity round-trip time $\tau$ [34]. Hopf frequencies which differ from $\omega_R$ are also possible if the current is arbitrary as we will discuss in the next chapter.

---

[3] This ratio is a $O(10^{-3})$ quantity.

The above forms of $\kappa_H$ and $\Omega_H$ are rigorously correct when the relaxation oscillation frequency is exactly in between two external cavity resonances, i.e. when

$$\omega_R \tau = \pi + 2n\pi \ (n = 0,1,2,\ldots). \tag{2.175}$$

When the injection current and the delay are such that Eq. (2.175) is satisfied, the minimum feedback level that leads to a Hopf bifurcation is

$$\kappa_{H,\min} = \frac{\Gamma_R \tau_{in}}{\sqrt{1+\alpha^2}}. \tag{2.176}$$

Finally, Eq. (2.171) reveals that only external cavity modes for which $B > 0$ can be destabilized by increasing the feedback level. By contrast, external cavity modes that have a negative $B$ cannot be destabilized. Such modes, when they exist, are found in the vicinity of the lowest extremity of the ellipse in the phase difference-carrier number plane. In particular, the so-called *maximum gain mode* (MGM) for which

$$\omega_s \tau = 0 \bmod 2\pi \text{ when } \omega_{th}\tau = \frac{\kappa}{\tau_{in}}\tau\alpha \bmod 2\pi \tag{2.177}$$

is always stable [35]. This mode corresponds to the maximum possible suppression of carriers and to the maximum intensity. It must be noted that this definition of the maximum gain mode is given for very particular values of $\omega_{th}\tau$ that depend on the delay and the feedback level. As a consequence, it is always possible for any value of the feedback strength to adjust the phase $\omega_{th}\tau$ by careful positioning (within the wavelength range) of the external mirror so that at least one mode (the maximum gain mode) is stable. For other feedback phases, the highest intensity modes may not be stable against relaxation oscillations. For very large values of $\kappa$, $\tau$ or $\alpha$, however, there will be a range of stable modes satisfying the condition $B < 0$ [35]. Moreover, new highest gain modes are stable when they emerge from saddle-node bifurcations [36,37]. By extension, we will call in the following *maximum gain mode* the mode for which the intensity is maximum regardless of $\omega_{th}\tau$.

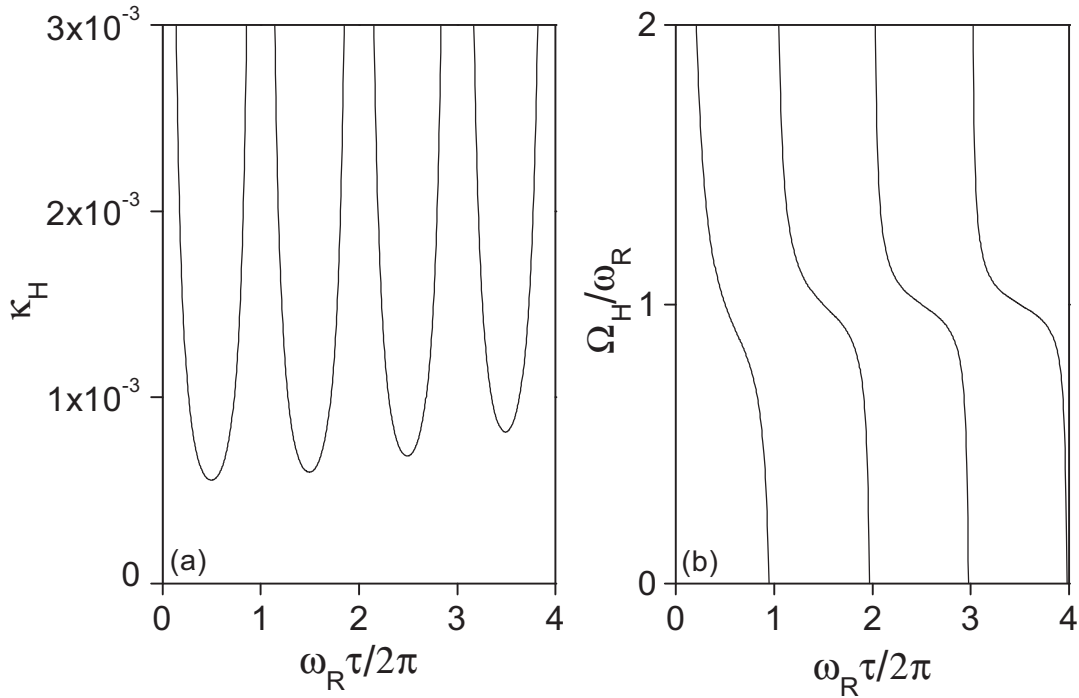

**Fig. 2.15** (a) Hopf bifurcation curves. The figure represents $\kappa_H$ as a function of the angular frequency of the relaxation oscillations of the solitary laser. (b) Hopf bifurcation frequencies. The figure represents $\Omega_H$ as function of the angular frequency of the relaxation oscillations of the solitary laser. $I = 1.3 \times I_{th}$, $\tau = 2$ ns, $\omega_{th}\tau =$ -arctan($\alpha$). The other parameters are given in Table 2-I. We consider the external cavity mode for which $\omega_s\tau =$ -arctan($\alpha$).

### 2.5.4. Potential model and minimum linewidth mode

In Section 2.5.2, we have seen that several external cavity modes coexist for feedback levels satisfying

$$\kappa \geq \frac{\tau_{in}}{\tau\sqrt{1+\alpha^2}}. \tag{2.178}$$

For a sufficiently weak feedback, the spontaneous emission noise may induce hopping between external cavity modes. When all the external cavity modes are stable, i.e. when $\kappa$ satisfies $\kappa < \kappa_{H,min}$, the dynamics of mode hopping has been demonstrated by Mørk et al. [38,39] and by Lenstra [40] to be governed by the following equation for the phase difference over one external cavity round-trip $\Delta\phi(t) = \phi(t)-\phi(t-\tau)+\omega_{th}\tau$:

$$\frac{d}{dt}\Delta\phi(t) = -\frac{1}{\tau}\frac{d}{d\Delta\phi}V(\Delta\phi) + F_{\Delta\phi}(t) \tag{2.179}$$

where

$$V(\Delta\phi) = (\Delta\phi - \omega_{th}\tau)^2 - 2\frac{\kappa\tau}{\tau_{in}}\sqrt{1+\alpha^2}\cos(\arctan\alpha + \Delta\phi) \tag{2.180}$$

and $F_{\Delta\phi}(t)$ is a Langevin noise force satisfying the relation

$$<F_{\Delta\phi}(t)F_{\Delta\phi}(t')> = \frac{R_{sp}(1+\alpha^2)}{2P_s}\delta(t-t'). \tag{2.181}$$

The derivation of Eq. (2.179) from the rate equations (2.150)-(2.152) is based on the assumption that the relaxation oscillations are strongly damped so that $P(t) = P_s$. Eq. (2.179) is analogous to the equation of motion of a particle moving with strong friction in a potential $V(\Delta\phi)$ and exposed to a fluctuating force $F_{\Delta\phi}(t)$. In this model, the external cavity modes and antimodes correspond respectively to local minima and maxima of the potential. The noise term $F_{\Delta\phi}(t)$ induces jumps between the minima of $V$. Within the limits of validity of the model, the mode that is the most stable against spontaneous emission noise, and therefore the dominant one, is the external cavity mode that corresponds to the deepest potential well. Moreover, it can be shown that the linewidth $\Delta f$ of an external cavity mode is related to the potential curvature and is given by

$$\Delta f = 4\left(\frac{d^2V(\Delta\phi)}{d(\Delta\phi)^2}\bigg|_s\right)^{-2}\frac{R_{sp}(1+\alpha^2)}{4\pi P_s} \tag{2.182}$$

with

$$\frac{d^2V(\Delta\phi)}{d(\Delta\phi)^2}\bigg|_s = 2 + 2\frac{\kappa\tau}{\tau_{in}}\sqrt{1+\alpha^2}\cos(\arctan\alpha + \Delta\phi_s) \tag{2.183}$$

and $\Delta\phi_s = \phi_s(t) - \phi_s(t-\tau) + \omega_{th}\tau = \omega_s\tau$. In the absence of feedback, Eq. (2.182) is equivalent to Eq. (2.118).

The *minimum linewidth mode* is defined as the external cavity mode for which [35]

$$\omega_s \tau = \omega_{th} \tau \qquad (2.184)$$

when the external mirror is positioned so that

$$\omega_{th} \tau = -\arctan \alpha \bmod 2\pi . \qquad (2.185)$$

This mode is characterized by an optimal reduction of the linewidth and is the most stable against spontaneous emission fluctuations. When the relation (2.185) is satisfied, the linewidth reduction can reach three orders of magnitude leading to a tremendous increase of the laser coherence. This mode has in general a large attraction basin and can be destabilized for relatively weak feedback [see Eq. (2.171)]. By extension, most of authors call "minimum linewidth mode" the mode that is the most stable against spontaneous emission noise regardless of $\omega_{th}\tau$. This mode is also the closest to the solitary laser frequency.

Fig. 2.16 displays potentials corresponding to three representative cases. In the absence of feedback [Fig. 2.16 (a)], the potential has a unique minimum and the curvature at the minimum is small. Additional potential wells appear in the presence of feedback [Fig. 2.16 (b)]. The depth of the well associated to the minimum linewidth mode is maximized when the condition (2.185) is fulfilled [Fig. 2.16 (c)].

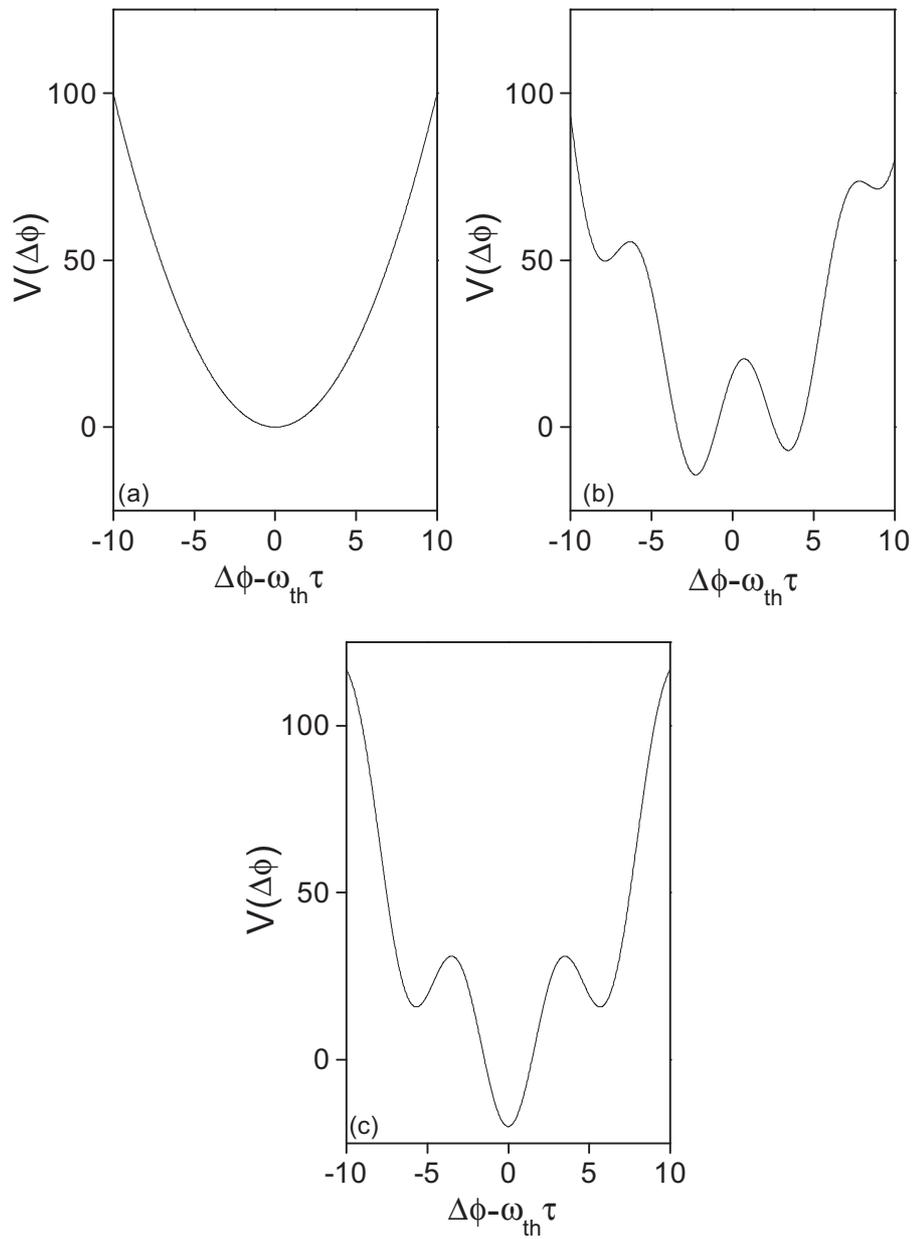

**Fig. 2.16** Potentials. (a) $\kappa = 0$. (b) $\kappa = 0.019$, $\tau = 1$ ns, $\omega_{th}\tau = 1.17$. (c) Same as in (b) but $\omega_{th}\tau = -\text{atan}(\alpha)$. $\alpha = 4$, $\tau_{in} = 8$ ps.

## 2.5.5. Rate equations for multimode semiconductor lasers subject to coherent optical feedback

We introduce coherent optical feedback in the equations for the fields of the longitudinal modes of the laser in the same way as for the single-mode problem; thus, neglecting self and cross-saturations, the equations read

$$\frac{dE_m}{dt} = \frac{1}{2}(1+i\alpha)[G_m(N)-\gamma_m]E_m(t)$$
$$+ \frac{\kappa_m}{\tau_{Lm}} E(t-\tau)\exp(-i\omega_m\tau) + F_m(t) \quad (2.186)$$

$$\frac{dN}{dt} = \frac{I}{e} - \frac{N}{\tau_s} - \sum_{i=-M}^{M} G_i(N)|E_i(t)|^2 + F_N(t), \quad (2.187)$$

where $\tau$ is the external-cavity roundtrip time, $\kappa_m$ and $\omega_m\tau$ are the feedback strength and phase of the $m$th mode, $\tau_{Lm}$ is the roundtrip time of the $m$th optical mode inside the diode cavity. The other parameters as well as the frequency dependence of the gain are defined as in the previous sections.

## 2.6. The coherence collapse regime

We have considered so far the effect of a relatively low feedback level, in which case the laser diode behavior can be described by using a potential model (section 2.5.4) and a small-signal analysis (section 2.5.3). We now discuss the extension of the theory for large feedback levels. Tkach and Chraplyvy have proposed in 1986 a phenomenological classification of five qualitatively different regimes [41] that we present hereafter. They studied in their experiment a 1.5 µm distributed feedback laser with an injection current twice larger than the threshold value. The feedback level, in terms of $\kappa^2$, that characterizes the different regimes depends strongly on the laser structure and on the other operating parameters (injection current, feedback delay and phase). As a consequence, it must be kept in mind that the numerical values of $\kappa^2$ we give here correspond to the experiment of Tkach and Chraplyvy and are only indicative for other experiments.

For very weak feedback level, below –70 dB in the experiment of Tkach et Chraplyvy, the laser operates in Regime I and can oscillate in only one external cavity mode that is stable. A careful positioning of the external mirror can lead to considerable linewidth reduction as we have seen in Section 2.5.4. When increasing slightly the feedback level, the laser diode enters regime II as additional steady state solutions are created in pairs through saddle-node bifurcations. The laser exhibits hopping induced by spontaneous emission noise between external cavity modes. The feedback level needed to enter Regime II is given by Eq. (2.178) and depends on the external cavity length. In the experiment of Tkach et Chraplyvy, $\kappa^2$ is around –70 dB. The rate of mode hopping decreases when increasing the feedback level. Provided that the latter remains sufficiently low not to excite relaxation oscillations, Regime III is reached as the laser is observed to operate in one dominant mode only, which is the mode with minimum linewidth. It should be noted however that the feedback level for which the relaxation oscillations are excited depends on the type of laser diode. For Fabry-Perot lasers, the relaxation oscillations become undamped before regime III is reached. Regime IV takes place for feedback levels in the range of –45 to –10 dB. As one reaches the condition (2.176), the relaxation oscillations become undamped (Hopf bifurcation). The behavior of the intensity, phase and carrier number associated to a particular external cavity mode is periodic; one observes a limit cycle close to the destabilized external cavity mode in phase space. The different steady state solutions loose progressively their stability. The external cavity modes with large negative frequency detuning with respect to the solitary laser frequency are the most stable [36,37]. Two well-known routes to chaos are often reported in the literature dedicated to semiconductor lasers with optical feedback. The period-doubling route to chaos, also called the Feigenbaum scenario (Refs. 22-25,42), consists of a succession of bifurcations taking place for smaller and smaller increases of the feedback level up to an accumulation point, beyond which chaos is observed. In laser diodes subject to optical feedback, this route to chaos is followed whenever the relaxation oscillation frequency is locked to a multiple of the external cavity frequency [43-44]. In absence of such locking, the laser diodes are observed to follow a quasiperiodic route to chaos [43-46]. The quasi-periodic route to chaos, also called the Ruelle-Takens-Newhouse scenario (Refs. 22-25 and 47), is characterized by a succession of three bifurcations: the attractor associated in the phase space to an external cavity mode first changes from a fixed point to a limit cycle (or equivalently a periodic orbit) characterized by a single oscillation frequency; the limit cycle then transforms into a torus characterized by two incommensurate frequencies; finally the torus transforms into an hypertorus as a third frequency appears. This hypertorus is very often unstable against perturbations and the dynamics becomes chaotic. The chaotic attractors that arise from the individual external cavity modes merge progressively, thus creating more complex chaotic attractors. The system trajectory in phase space then visits randomly the ruins of the chaotic attractors. To this complicated dynamical behavior is associated a large increase of the laser linewidth (typically from 100 MHz to 25 GHz). This behavior is therefore referred to as coherence collapse regime [49]. The latter is bounded from above by a regime of single-mode narrow linewidth operation corresponding to a feedback level higher than –10 dB. This Regime V can only be reached by applying a very low injection current or by antireflection coating the laser facet facing the external mirror. It must be pointed out that the transition to chaos from regime V to regime IV for a moderate and constant level of feedback has been suggested to follow an intermittency route as the injection current is progressively increased [50].

## 2.7. The low-frequency fluctuation regime

### 2.7.1. Historical overview

When a laser diode subject to optical feedback is pumped close to its solitary threshold, its optical power can exhibit sudden dropouts followed by gradual, stepwise recoveries occurring on a time scale much larger than the period of the relaxation oscillations or the external-cavity round trip time. For this reason, this regime is usually referred to as the low-frequency fluctuations (LFF) regime.

Already observed in 1977 by Risch and Voumard [51], the LFF regime has attracted much theoretical and experimental interest. Multiple explanations of its origin have been proposed. Henry and Kazarinov [52] introduced a model based on the concept of a potential well from which the system can escape owing to spontaneous emission noise. Although the approximate analytical results are in qualitative agreement with the dependence of the drop period with respect to the current [53], accurate statistical studies of the dropout time-interval have revealed that this model does not capture the essence of the dropout statistics [54]. Mørk et al. [55] demonstrated that a noise-driven multimode traveling wave model reproduces the intensity dropouts and interpreted the low-frequency fluctuations as resulting from a bistability between the maximum gain mode and a temporally stable (within the external-cavity round-trip time) minimum-linewidth mode that corresponds to a lower power state. Observing that turning off the noise generator does not lead to the disappearance of the LFF, they suggested that the low-frequency fluctuations may take place on a chaotic attractor. This proposition was confirmed by Sacher et al. [56] who classified low-frequency fluctuations as type-II intermittency. Recently, a scenario according to which the laser diode behaves as an excitable medium and the low-frequency fluctuations are induced by noise has been put forward [57-59]. The validity of this stochastic interpretation has however been strongly questioned [60,61].

The most generally accepted interpretation of the LFF regime was presented by Sano [62] in 1994; it relies on the Lang-Kobayashi equations [20] and assumes single-mode operation of the laser. Sano showed that the intensity dropouts are caused by crises between local chaotic attractors and saddle-type antimodes [62]. In his analysis, the process of intensity recoveries is associated to a chaotic itinerancy of the system trajectory in phase space among the attractor ruins of external cavity modes, with a drift towards the maximum gain mode close to which collisions with antimodes occur. By investigating in deeper details the Lang-Kobayashi equations, van Tartwijk et al. [63] anticipated the presence of short intensity pulses. Fischer et al. [64] confirmed this prediction by streak camera observation. Additional numerical studies based on statistical distributions of the time interval between consecutive dropouts revealed that the results predicted by the Lang-Kobayashi equations are in better agreement with experiments if the spontaneous emission noise is taken into account [65,66]. Later, the coexistence of the LFF regime with stable emission in a single high-gain external cavity mode for a large range of experimental parameters was experimentally demonstrated [67-69]. The existence of stable external cavity modes with high-gain was previously predicted by analytical study of the Lang-Kobayashi equations [35]. In Ref. 69, Hohl and Gavrielides have pointed out a route to the LFF regime for the case of a short delay (1 ns) optical feedback. They showed both experimentally and numerically that the laser undergoes a cascade of bifurcations as the feedback rate is increased, one external cavity mode becoming unstable and being replaced by the next one that is stable. The laser was found to exhibit sustained low-frequency fluctuations in the unstable regions while operating in the stable maximum gain mode in the stable ones. Very recently, evidences were given of the existence of a particular set of periodic and quasiperiodic solutions of the Lang-Kobayashi equations [70,71]. An important consequence is that the LFF regime may be regarded as a chaotic motion over a set of destabilized solutions of that type.

Despite multiple successful predictions based on the Lang-Kobayashi equations, recent experiments [72-75] revealed that multimode operation often occurs within the LFF regime contrary to a major assumption of this model. The latter observation has led to two essential issues that are the determination of the minimal model that allows to capture all the features of the LFF regime in multimode laser diode and, ultimately, to know if this regime can still be understood in terms of chaotic itinerancy and crises. These questions are addressed in Chapter 5.

A summary of the main findings of Sano [62] and van Tartwijk et al. [63] is given in the next section. They are essential for understanding Chapter 3.

### 2.7.2. Dynamics

Fig. 2.17 displays the behavior of a laser diode subject to a moderate optical feedback and pumped at its solitary threshold. The rate equations (2.148) and (2.149) are integrated numerically with feedback level, delay and phase respectively fixed at $\kappa = 0.012$, $\tau = 5$ ns and $\omega_h \tau = 0$; the injection current is $I = I_{th}$ and the other parameters are given in Table 2-I. In Fig. 2.17(a), the photon number is averaged over 2 ns to account for the finite bandwidth of the detection devices that are employed in most experimental studies. The average photon number exhibits sporadic dropouts followed by slow recoveries. As the recovery is achieved, the laser intensity is observed to fluctuate around a plateau until the next dropout. The time duration between consecutive dropouts is of the order of a few hundred nanoseconds, a time scale that is much longer than the period of the relaxation oscillations and the roundtrip time in the external cavity. In contrast to the average intensity, the instantaneous photon number [Fig. 2.17(b)] exhibits trains of short pulses, those being approximately 100 ps wide [Fig. 2.18(a)].

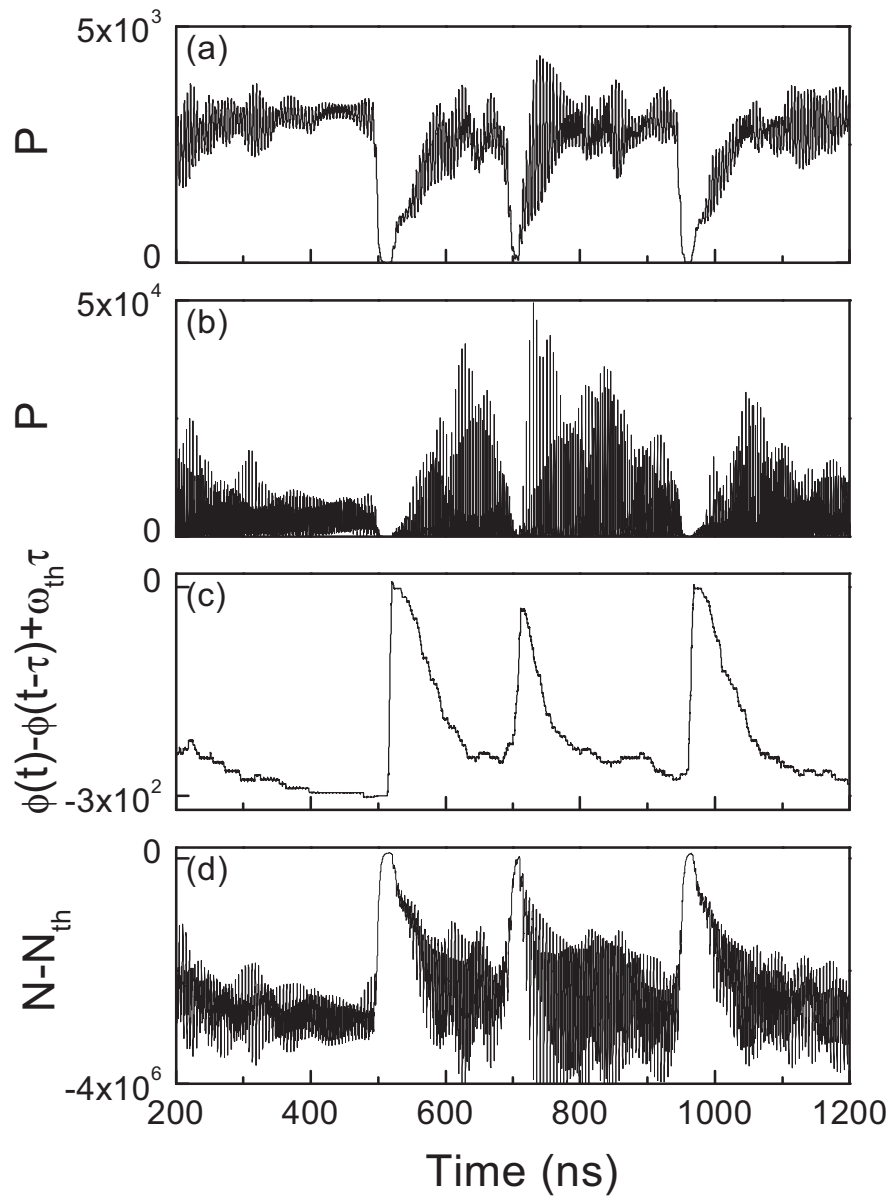

**Fig. 2.17** Time traces of the photon number averaged over 2 ns (a), instantaneous photon number (b), phase difference over one roundtrip time (c) and carrier number (d).

During the buildup of the average intensity, the phase difference function[4] $\Delta\phi(t) = \phi(t) - \phi(t-\tau) + \omega_{th}\tau$ decreases almost monotically [Fig. 2.17(c)]. The time evolution of $\Delta\phi(t)$ is characterized by a set of successive plateau values around which it fluctuates. These plateaus correspond to the successive external cavity mode frequencies multiplied by the feedback time delay. The jumps of $\Delta\phi(t)$ between different plateaus are observed when the instantaneous intensity is very low (Fig. 2.18). Fig. 2.17(d) reveals that progressive decreases and sharp increases of the carrier number accompany the intensity recoveries and dropouts, respectively.

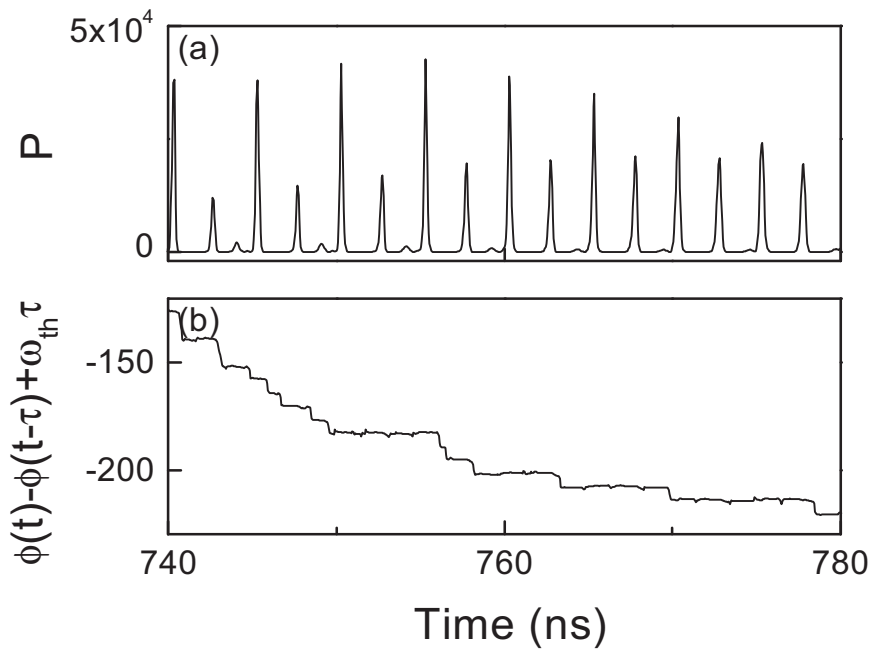

**Fig. 2.18** Time traces of the instantaneous photon number (a) and phase difference (b). These traces are blow-ups of Fig. 2.17 (b) and (c).

Following Sano [62], we can obtain more insight into the main features of the dynamics by considering the projection of the system trajectory and steady state solutions of the Lang-Kobayashi equations on the $(\Delta\phi(t), N(t)-N_{th})$ plane (Fig. 2.19). This representation reveals that the intensity buildup and, equivalently, the progressive reduction of the carrier number are associated to a succession of switching events between attractor ruins of destabilized external cavity modes. Although the direction toward the lowest frequencies is preferred, switching to higher frequencies can occur randomly in the buildup process. These inverse switching events are regarded by Sano as a sign of chaotic itinerancy[5]. The external cavity modes and antimodes with the lowest frequencies are very close to each other. A trajectory that fails to pass through the stable manifold[6] of the antimode is driven back to the closest attractor ruin. Once it succeeds to pass through the stable manifold, the trajectory is rapidly repelled along the unstable manifold of the saddle toward the threshold value of the carrier number and low intensities, initiating a dropout. The phase difference remains at first almost unchanged, owing to the feedback that compensates the variation of the carrier number. When the carrier number is high enough, the frequency rapidly increases and the laser relaxes to its solitary state. Although the instantaneous intensity has fallen, the delayed feedback is still high enough to trigger the system. The chaotic itinerancy with the drift toward the maximum gain mode, which is located at the lowest extremity of the ellipse, then restarts. Sano has demonstrated that the sudden escape of the system trajectory is associated to the collision, or merging, of a local chaotic attractor of an external cavity mode with an antimode (i.e. a saddle point). This collision, which is called *crisis* in nonlinear dynamics [21,76], changes the local chaotic attractor to an attractor ruin. Fig. 2.20 schematically illustrates this crisis.

---

[4] This function reduces to the product of the stationary angular frequency and the feedback delay $\omega_s\tau$ in case of stationary behavior.
[5] Chaotic itinerancy is deterministic random switching between attractor ruins.

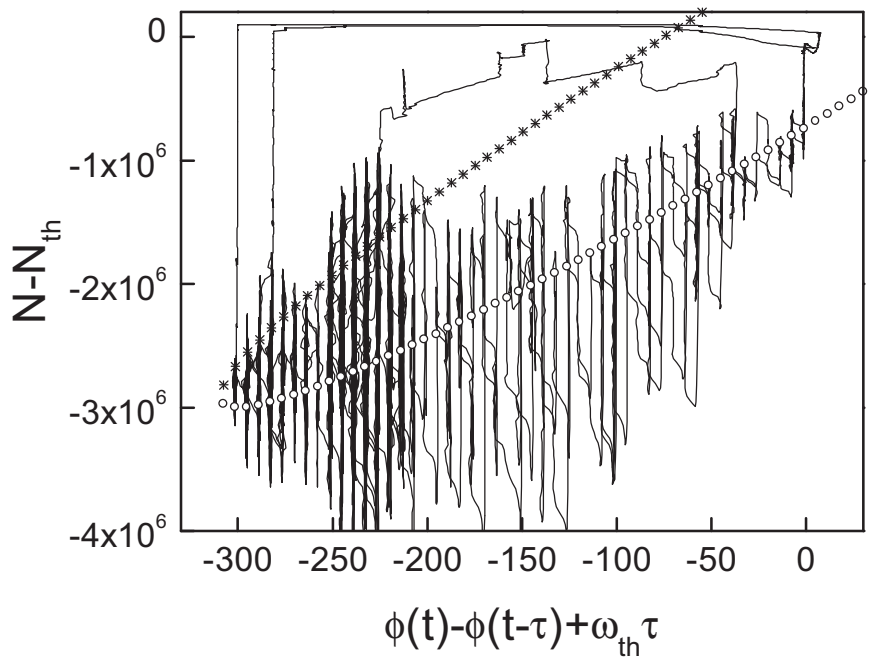

**Fig. 2.19** Projection of the system trajectory onto the $(\Delta\phi(t), N(t)-N_{th})$ plane. Circles (o) and stars (*) represent the projections of external cavity modes and antimodes, respectively.

---

[6] The reader will find the definition of stable and unstable manifolds of a saddle point in Appendix A.

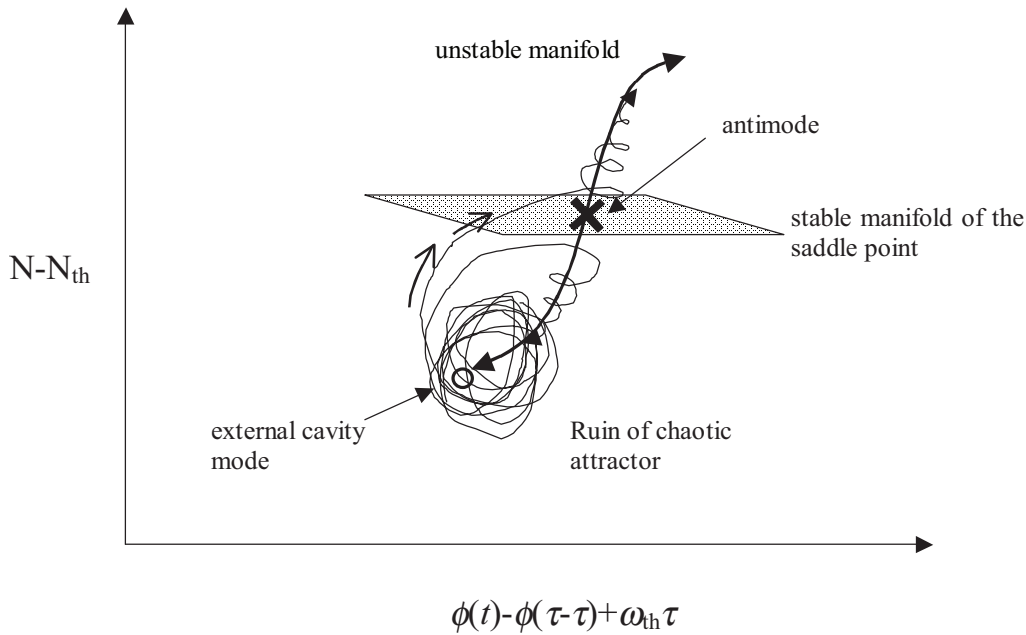

**Fig. 2.20** Schematic illustration of the antimode dynamics

In order to describe the mode-slipping in more details, we rewrite the rate equations (2.148) and (2.149) in terms of the photon number and phase difference as

$$\frac{dP(t)}{dt} = G_N[N(t) - N_{th}]P(t) + \frac{2\kappa}{\tau_{in}}\sqrt{P(t)P(t-\tau)}\cos[\Delta\phi(t)], \quad (2.188)$$

$$\frac{d\Delta\phi(t)}{dt} = \frac{\alpha\, G_N}{2}[N(t) - N(t-\tau)] - \frac{\kappa}{\tau_{in}}\sqrt{\frac{P(t-\tau)}{P(t)}}\sin[\Delta\phi(t)]$$
$$+ \frac{\kappa}{\tau_{in}}\sqrt{\frac{P(t-2\tau)}{P(t-\tau)}}\sin[\Delta\phi(t-\tau)], \quad (2.189)$$

$$\frac{dN(t)}{dt} = \frac{I}{e} - \frac{N(t)}{\tau_s} - G_N[N(t) - N_0]P(t). \quad (2.190)$$

In Ref. 63, van Tartwijk et al. observed that when the system trajectory resides on the attractor ruin of an external cavity mode, the photon and carrier numbers show strong fluctuations while the phase difference over a round-trip remains close to the value associated with the external cavity mode, albeit not being rigorously constant. As a result, the sinusoidal terms in Eq. (2.189) fluctuates rapidly. When the intensity at time $t$ or at time $t-\tau$ is low, the instantaneous variation of the phase difference becomes very large and the system can slip to a neighboring external cavity mode. Since the sinusoidal terms vary rapidly, mode-slipping can occur in either direction. Moreover, due to the last term in Eq. (2.189), a mode slipping at time $t-\tau$ favors the occurrence of a mode-slipping at time $t$. Statistically, the carrier number is smaller at time $t$ than at time $t-\tau$. As a consequence, the first term in Eq. (2.189) has, on average, an opposite effect on mode-slipping toward higher frequencies and favors a drift toward the lower frequencies. As the trajectory reaches the lower extremity of the ellipse, the difference between $N(t)$ and $N(t-\tau)$ becomes smaller and the drift reduces, raising at the same time the probability of inverse switching.

## 2.8. Rate equations for laser diodes subject to incoherent optical feedback and incoherent optical injection

In this section, we consider the case of a single-longitudinal mode laser diode subject to incoherent optical feedback. We derive rate equations that describe the dynamical behavior of this system by putting emphasis on qualitative understanding rather than rigorous theory.

In general, the lasing field may consist of a large number of longitudinal, transverse and lateral modes. In particular, the active region of a double-heterostructure semiconductor laser (which can be considered as a slab waveguide) supports two sets of transverse modes, namely the transverse electric (TE) and the transverse magnetic (TM) modes. The electric field of the TE modes and the magnetic field of the TM modes are polarized along the heterojunction plane, respectively. The selection of the TE or TM modes is based on the associated reflectivity of the laser facets and the respective transverse confinement factors [3]. Both the facet reflectivity and the transverse confinement factor (the latter represents the fraction of the mode energy available within the active layer for interaction with the charge carriers) being larger for the TE than for the TM modes, the TE modes are favored. In addition to this polarization selectivity, the laser diodes are generally designed to support only fundamental transverse and lateral modes [3].

In the case of a laser diode subject to an incoherent optical feedback (Fig. 2.21), the linearly polarized output of the laser undergoes a 90° polarization rotation through an external cavity typically formed by a Faraday rotator (FR) and a mirror. In this scheme, the polarization direction of the feedback field is orthogonal to that of the laser output. The feedback field acts on the population inversion in the laser active layer but does not interact coherently with the intracavity lasing field as it does in the case of coherent feedback. A linear polarizer (LP) can be placed between the Faraday rotator and the mirror to prevent coherent feedback induced by a second round-trip in the external cavity after reflection on the laser front facet.

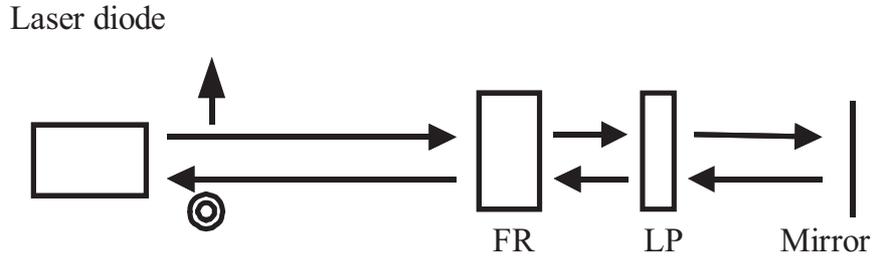

**Fig. 2.21** Schematic representation of a laser diode subject to incoherent optical feedback.

Neglecting spontaneous emission and gain saturation, the rate equations of the system can be derived in the same way as the equations for a multimode laser diode subject to coherent optical feedback and injection:

$$\frac{dE_{TE}}{dt} = \frac{1+i\alpha}{2}\left[G_{N,TE}(N-N_0) - \frac{1}{\tau_{p,TE}}\right]E_{TE}, \quad (2.191)$$

$$\frac{dE_{TM}}{dt} = \frac{1+i\alpha}{2}\left[G_{N,TM}(N-N_0) - \frac{1}{\tau_{p,TM}}\right]E_{TM}$$
$$+ \frac{\kappa_{inc}}{\tau_{in}}E_{TE}(t-\tau)\exp(i\Delta\omega t - i\omega_{TE}\tau), \quad (2.192)$$

$$\frac{dN}{dt} = \frac{I}{e} - \frac{N}{\tau_s} - G_{N,TE}(N-N_0)|E_{TE}|^2 - G_{N,TM}(N-N_0)|E_{TM}|^2 \quad (2.193)$$

where $\Delta\omega = \omega_{TE} - \omega_{TM}$ and $\kappa_{inc} = r_3 t_{TE} t'_{TM}$.

$E_{TE}(t)$ and $E_{TM}(t)$ are respectively the slowly varying complex fields of the fundamental TE and TM modes oscillating at the frequencies $\omega_{TE}$ and $\omega_{TM}$. $N(t)$ is the carrier number in the active layer. $\tau_{p,TE}$ and $\tau_{p,TM}$ are the photon lifetimes of the two polarization modes. $G_{N,TE}$ and $G_{N,TM}$ are the mode dependent gain coefficients. $\tau$ is the feedback delay. $r_3$ is the amplitude reflection coefficient of the external mirror. $t_{TE}$ and $t'_{TM}$ are the amplitude transmission coefficients from inside and outside the cavity, respectively. The other parameters are defined as in the previous sections.

Above threshold, only the TE mode lases because the gain coefficient and the photon lifetime associated with this mode are larger than those associated with the TM mode[7]. In the absence of spontaneous emission and feedback, the carrier number is clamped to its threshold value and is given by

$$N_{th} = \frac{1}{G_{N,TE}\tau_{p,TE}} + N_0.  \quad (2.194)$$

Using this relation, we can rewrite Eq. (2.192) as

$$\frac{dE_{TM}(t)}{dt} = \frac{1+i\alpha}{2}[G_{N,TM}(N-N_{th}-\beta)]E_{TM}(t) + \frac{\kappa_{inc}}{\tau_{in}}E_{TE}(t-\tau)\exp(i\Delta\omega t - i\omega_{TE}\tau) \quad (2.195)$$

where

$$\beta = \frac{1}{G_{N,TE}\tau_{TE}} - \frac{1}{G_{N,TM}\tau_{TM}}. \quad (2.196)$$

The set of equations (2.191),(2.193) and 2.195) can be simplified as follows. Firstly, the carrier number $N$ fluctuates around its threshold value and the difference $N$-$N_{th}$, which is small compared to $\beta$, can be neglected. Secondly, $E_{TM}$ can be adiabatically eliminated if we assume that the product $\beta G_{N,TM}$, which is the relaxation time of the mode $E_{TM}$, is much larger than every other characteristic time of the system. The transverse magnetic mode follows therefore the transverse electric mode:

$$E_{TM}(t) = \frac{2\kappa_{inc}}{(1+i\alpha)G_{N,TM}\tau_{in}\beta}E_{TE}(t-\tau)\exp(i\Delta\omega t - i\omega_{TE}\tau). \quad (2.197)$$

Inserting Eq. (2.197) into Eq. (2.193) yields

$$\frac{dN}{dt} = \frac{I}{e} - \frac{N}{\tau_s} - G_{N,TE}(N-N_0)\left[|E_{TE}(t)|^2 + \gamma|E_{TE}(t-\tau)|^2\right] \quad (2.198)$$

---

[7] The photon lifetime is related to the facet reflectivity by Eq. (2.26). The gain coefficient is proportional to the confinement factor. As a example, the difference of the confinement factors leads typically to a gain relative difference of 15% [3] that in turn leads to a strong polarization mode selectivity.

where

$$\gamma = \frac{1}{1+\alpha^2} \frac{4\kappa_{inc}^2}{\tau_{in}^2 \beta^2 G_{N,TE} G_{N,TM}}, \qquad (2.199)$$

is the feedback parameter. Equations (2.191) and (2.198) can also be rewritten in terms of the photon number $P(t)$ and phase $\phi(t)$ associated with the TE mode in lieu of the complex electric field; thus

$$\frac{dP}{dt} = \left\{ G_{N,TE}(N-N_0) - \frac{1}{\tau_{p,TE}} \right\} P, \qquad (2.200)$$

$$\frac{d\phi}{dt} = \frac{\alpha}{2}\left\{ G_{N,TE}(N-N_0) - \frac{1}{\tau_{p,TE}} \right\}, \qquad (2.201)$$

$$\frac{dN}{dt} = \frac{I}{e} - \frac{N}{\tau_s} - G_{N,TE}(N-N_0)\left[ P(t) + \gamma P(t-\tau) \right]. \qquad (2.202)$$

Since the equations for the photon and carrier numbers are independent of the phase, we will not consider the dynamics of this last variable in the following.

In the case of incoherent injection, the rate equations for the photon and carrier numbers of the injected laser are similarly derived in a straightforward way; they read:

$$\frac{dP}{dt} = \left\{ G_{N,TE}(N-N_0) - \frac{1}{\tau_{p,TE}} \right\} P, \qquad (2.203)$$

$$\frac{dN}{dt} = \frac{I}{e} - \frac{N}{\tau_s} - G_{N,TE}(N-N_0)\left[ P(t) + \sigma P_{ext}(t) \right] \qquad (2.204)$$

where

$$\sigma = \frac{1}{1+\alpha^2} \frac{4 t_{TM}^{'2}}{\tau_{in}^2 \beta^2 G_{N,TE}}, \qquad (2.205)$$

and $P_{ext}$ is the photon number injected from outside into the laser cavity. These equations are valid provided that the optical frequency of the injected field is compatible with the gain width of the injected laser material.

Finally, gain saturation and spontaneous emission can be taken into account in the same way as in the previous sections.

Equations (2.200) and (2.202) were proposed for the first time by Otsuka and Chern in 1991 who anticipated the feasibility of using laser diodes subject to incoherent optical feedback to generate gigabit-per-second optical pulses without external modulation or saturable absorber [77]. Recently, this proposal has been carried out experimentally [78]. Stability of the steady state solutions of Eqs. (2.200),(2.202) and the pulse regime have been investigated analytically by Pieroux et al. [79,80]. Further numerical investigations of Eqs. (2.200) and (2.202) have anticipated the coexistence of sustained periodic relaxation oscillations and regenerative periodic spikes leading to chaos [81], as well as two different transition routes to chaos, namely period doubling and quasiperiodicity [82]. The model was also extended to study the complex dynamics of an array of mutually coupled lasers [83].

# 3. Branches of periodic solutions connecting pairs of steady-state solutions in laser diodes subject to a single optical feedback

Pas un prestidigitateur n'égale la nature :
elle opère sous nos yeux, en pleine lumière,
et cependant il n'y a pas moyen de pénétrer
ses trucs.

<div align="right">Remy de Gourmont</div>

## 3.1.  Introduction

As written in the previous chapter, a semiconductor laser subject to weak to moderate optical feedback from an external reflector like the front facet of an optical fiber or an optical disk exhibits pulsating intensities that reduce the device performance in many applications. The optical feedback generates a large number of additional steady-state solutions that are created by pairs through saddle-node bifurcations. Of each pair of solutions, one is a saddle point and is referred as antimode. The other solution is called external cavity mode. External cavity modes are destabilized through Hopf bifurcations as the feedback strength is increased. In general, the Hopf bifurcation frequency is close to the relaxation oscillation frequency of the solitary laser [1-5]. There are cases, however, where this frequency may deviate considerably from the relaxation frequency, in particular when the laser is pumped close to its solitary threshold. In this case, the Hopf frequency is proportional to the inverse of the external cavity round-trip time [6]. But Hopf frequencies that differ from the relaxation oscillation frequency of the solitary laser are also possible regardless of the injection current as we will show in the present chapter.

Our renewed interest for unusual Hopf frequencies was motivated by our recent experimental study [7,8] of a laser pumped close to its solitary threshold and subject to two optical feedbacks. This experiment, presented in Chapter 4, revealed high frequency oscillations that seem to result from a beating between an external cavity mode and an antimode. The idea that such regimes could be possible was first investigated by Tager and Elenkrig [9] and by Tager and Petermann [10] for a laser subject to a single optical feedback from a short (typically less than 5 mm) external cavity. Their analysis was based on steady state solutions of the Lang and Kobayashi (LK) equations [11], approximations of the Hopf bifurcation points, and numerical simulations. In a paper [12] published in *Optics Communications* and to which we have collaborated, a bifurcation analysis of the LK equations was proposed and the existence of time-periodic solutions that are combinations of an external cavity mode and an antimode was demonstrated. Such solutions are referred to as mixed mode solutions hereafter. The total intensity was found to be modulated at a high frequency equal to the difference between the frequency of the antimode and the frequency of the external cavity mode. It was also demonstrated that mixed solutions are only possible near particular values of the feedback parameter. A very promising application of this phenomenon, already pointed out by Tager and Petermann in [10], is the all-optical generation of almost sinusoidal high-frequency oscillations with frequencies higher than 30 GHz. Those are much higher than the typical modulation bandwidth of the type of lasers under study [13].

Among the results presented in Ref. 12, of particular interest is the fact that each branch of mixed mode solutions emerges and terminates at two Hopf bifurcation points located on two distinct branches of steady state solutions: the first of these steady state solutions is an external cavity mode, the second corresponds to an antimode. This result then motivated the reconsideration of the Hopf bifurcation conditions for the steady state solutions [1]. The analysis that was carried in Ref. 12 also predicts a secondary bifurcation to quasiperiodic oscillations that was investigated numerically.

We present in this chapter the main results that were reported in Ref. 12. The chapter is organized as follows. In Section 2, we introduce the dimensionless form of the Lang and Kobayashi equations and of their steady state solutions. In Section 3, we show that a time-periodic solution that is a combination of an external cavity mode and an antimode is not an exact solution of the LK equations but is, in fact, the leading order approximation of an asymptotic solution valid when the ratio of the carrier lifetime to the photon lifetime is large. The derivation of the higher-order approximation is not presented here, but the reader can find it in Ref. 12. The main results predicted by these analytical developments are presented in Section 4 and compared with the numerical resolution of the LK equations. Our conclusions are summarized in Section 5.

## 3.2.    Dimensionless rate equations and their steady state solutions

The Lang and Kobayashi equations [11] are two equations for the slowly varying complex electric field $E(t)$ and the number $N(t)$ of electron-hole pairs in the active region of the laser diode. They read (see Chapter 2)

$$\frac{dE(t)}{dt} = \frac{1+i\alpha}{2}\left[G_N(N-N_0) - \frac{1}{\tau_p}\right]E(t) + \frac{\kappa}{\tau_{in}}E(t-\tau)\exp(-i\omega_{th}\tau), \tag{3.1}$$

$$\frac{dN(t)}{dt} = \frac{I}{e} - \frac{N(t)}{\tau_s} - G_N(N-N_0)|E(t)|^2 \tag{3.2}$$

where $\alpha$ is the linewidth enhancement factor and $G_N$ the gain coefficient. $N_{th}$ and $N_0$ are respectively the threshold and transparency values of $N$. $\tau$ is the round-trip time in the external cavity and $\tau_{in}$ the round-trip time inside the laser cavity. $\kappa$ and $\omega_{th}\tau$ are respectively the feedback level and the phase shift after one roundtrip in the external cavity. $I$ is the injection current, $e$ the magnitude of the electron charge. $\tau_s$ and $\tau_p$ are the lifetime of the electron-hole pairs and the lifetime of photons, respectively. In this chapter, the spontaneous emission is not taken into account.

Typical values of the photon and the carrier lifetimes are $\tau_p$ = 2 ps and $\tau_s$ = 2 ns. The ratio of the photon lifetime to the carrier lifetime is an $O(10^{-3})$ small quantity, which suggests to seek approximations in powers of $\chi = \tau_p/\tau_s$. To this end, it is necessary to introduce dimensionless variables $E'$ and $N'$ defined by

$$E' = \sqrt{\frac{\tau_s G_N}{2}}E, \tag{3.3}$$

$$N' = \left(\frac{\tau_p G_N}{2}\right)(N-N_{th}), \tag{3.4}$$

a time $s$ that is measured in units of the photon lifetime

$$s = \frac{t}{\tau_p}, \tag{3.5}$$

the dimensionless excess pump current

$$P = \frac{\tau_p G_N N_{th}}{2}\left(\frac{I}{I_{th}} - 1\right) \tag{3.6}$$

with

$$I_{th} = \frac{N_{th}e}{\tau_s}, \tag{3.7}$$

the dimensionless feedback rate

$$\eta = \frac{\kappa}{\tau_{in}}\tau_p, \tag{3.8}$$

the dimensionless delay of the feedback

$$\theta = \frac{\tau}{\tau_p}, \tag{3.9}$$

the ratio of the carrier lifetime to the photon lifetime

$$T = \frac{\tau_s}{\tau_p}, \tag{3.10}$$

and the angular frequency of the solitary laser normalized by $\tau_p^{-1}$:

$$\Omega = \omega_{th}\tau_p. \tag{3.11}$$

Substituting Eqs. (3.3)-(3.11) in Eqs. (3.1) and (3.2) and dropping the primes for conciseness leads to the dimensionless form of the Lang and Kobayashi equations [14]:

$$\frac{dE}{ds} = (1+i\alpha)NE + \eta E(s-\theta)\exp(-i\Omega\theta), \tag{3.12}$$

$$T\frac{dN}{ds} = P - N - (1+2N)|E|^2. \tag{3.13}$$

The steady state solutions of Eqs. (3.12)-(3.13) are of the form

$$E = A_s \exp[i(\Delta - \Omega)s] \text{ and } N = N_s \tag{3.14}$$

where $\Delta$ is a stationary angular frequency and the corresponding $A_s$ and $N_s$ are constants:

$$\Delta = \Omega - \eta[\alpha\cos(\Delta\theta) + \sin(\Delta\theta)], \tag{3.15}$$

$$A_s^2 = \frac{P - N_s}{1 + 2N_s} \geq 0, \qquad (3.16)$$

$$N_s = -\eta \cos(\Delta\theta). \qquad (3.17)$$

Although the single frequency solution satisfies Eqs. (3.12)-(3.13), a linear combination of two steady state solutions of the form

$$E(s) = A_1 \exp[i(\Delta_1 - \Omega)s] + A_2 \exp[i(\Delta_2 - \Omega)s] \qquad (3.18)$$

is not an exact solution of Eqs. (3.12)-(3.13). We will show in the next section that Eq. (3.18) is however the leading order approximation of an asymptotic solution valid when the ratio of the carrier lifetime to the photon lifetime (i.e. *T*) is large.

## 3.3. Mixed mode solutions

We seek a solution of the form

$$E(s) = E_0(s) + \chi E_1(s) + \dots, \qquad (3.19)$$

$$N(s) = N_0(s) + \chi N_1(s) + \dots \qquad (3.20)$$

where

$$\chi \equiv 1/T \qquad (3.21)$$

is the small parameter. In the limit $\chi \to 0$ ($T \to \infty$), $E_0(s)$ and $N_0(s)$ are to be obtained from the leading order equations

$$\frac{dE_0}{ds} = (1 + i\alpha)N_0 E_0 + \eta E_0(s - \theta)\exp(-i\Omega\theta), \qquad (3.22)$$

$$\frac{dN_0}{ds} = 0 \Rightarrow N_0 = \text{cst}. \qquad (3.23)$$

Substituting Eq. (3.18) into Eq. (3.22) yields

$$i(\Delta_1 - \Omega)A_1 \exp[i(\Delta_1 - \Omega)s] + i(\Delta_2 - \Omega)A_2 \exp[i(\Delta_2 - \Omega)s]$$
$$= (1 + i\alpha)N_0 \left( A_1 \exp[i(\Delta_1 - \Omega)s] + A_2 \exp[i(\Delta_2 - \Omega)s] \right) \quad (3.24)$$
$$+ \eta \exp(-i\Omega\theta)\left( A_1 \exp[i(\Delta_1 - \Omega)(s - \theta)] + A_2 \exp[i(\Delta_2 - \Omega)(s - \theta)] \right).$$

Multiplying now the left and right hand sides of Eq. (3.24) by $\left( A_{1,2} \exp[i(\Delta_{1,2} - \Omega)s] \right)^*$ and then integrating from -∞ to ∞, we obtain four equations relating the leading order approximations of the excess carrier number $N_0$, the feedback rate $\eta$, and the two frequencies $\Delta_1$ and $\Delta_2$:

$$N_0 = -\eta \cos(\Delta_1 \theta) = -\eta \cos(\Delta_2 \theta), \quad (3.25)$$

$$\Delta_1 = \Omega + \alpha N_0 - \eta \sin(\Delta_1 \theta), \quad (3.26)$$

$$\Delta_2 = \Omega + \alpha N_0 - \eta \sin(\Delta_2 \theta). \quad (3.27)$$

Comparison of Eqs. (3.25)-(3.27) with Eqs. (3.15) and (3.17) shows that the excess carrier number $N_0$ and the frequencies $\Delta_1$ and $\Delta_2$ that are associated with the mixed mode solution are in fact those of two individual single-frequency steady state solutions of the Lang-Kobayashi equations [Eqs.(3.12) and (3.13)] for a particular value of the feedback rate.

Eq. (3.25) requires that either

$$\Delta_2 \theta = \Delta_1 \theta + 2n\pi \quad (3.28)$$

or

$$\Delta_2 \theta = -\Delta_1 \theta + 2n\pi \quad (3.29)$$

where

$$n = 0, \pm 1, \pm 2, \ldots \quad (3.30)$$

An additional requirement on the difference $\Delta_2 - \Delta_1$ is found by substracting Eq. (3.26) from Eq. (3.27). Thus

$$\Delta_2 - \Delta_1 = -\eta [\sin(\Delta_2 \theta) - \sin(\Delta_1 \theta)]. \quad (3.31)$$

This condition can be satisfied only with Eq. (3.29). Substituting Eq. (3.29) into Eq. (3.31) yields the expression of the feedback rate $\eta$ as a function of $\Delta_1$:

$$\eta = \frac{n\pi - \Delta_1 \theta}{\theta \sin(\Delta_1 \theta)}. \tag{3.32}$$

We denote this critical feedback rate as $\eta_0$ in the following. $N_0$ and $\eta_0$ can be eliminated in Eq. (3.26) by using Eqs. (3.25) and (3.32). $\Delta_1$ is then obtained by solving the following transcendental equation

$$\Delta_1 \theta = \Omega \theta - (n\pi - \Delta_1 \theta)[\alpha \cot(\Delta_1 \theta) + 1]. \tag{3.33}$$

Knowing $\Delta_1$, $\Delta_2$ and $\eta_0$ are then determined from Eqs. (3.29) and (3.32), respectively. $N_0$, which must satisfy the positive intensity condition

$$\frac{P - N_0}{1 + 2N_0} \geq 0, \tag{3.34}$$

is finally found by using Eqs. (3.25) and (3.32):

$$N_0 = \frac{(\Delta_1 \theta - n\pi)\cot(\Delta_1 \theta)}{\theta}. \tag{3.35}$$

Several branches of steady state solutions are shown in Fig. 3.1. We consider parameter values that are close to those used by Tager and Petermann [10], with the exception of the gain saturation coefficient that is taken to be zero. The values of the dimensionless parameters are: $T = 1710$, $\alpha = 4$, $P = 1.155$, $\theta = 18$, $\Omega\theta = -\arctan(\alpha)$. At $\eta = \eta_0 \cong 0.104$, two steady state solutions, namely one external cavity mode and one saddle-type antimode[1], admit the same intensity [squares in Fig. 3.1(a)]. According to the above considerations, this feedback rate corresponds to the existence of an approximate solution of the LK equations that is a linear combination of the two coexisting steady state solutions: the external cavity mode whose frequency $\Delta_1$ satisfies the equality $\Delta_1 \theta = \Omega\theta$ [Eq. (3.33) with $n = -1$], and the antimode whose frequency $\Delta_2$ is determined from Eq. (3.29). This periodic solution is of the form (3.18). The corresponding intensity of the laser output is given by

$$|E(s)|^2 = |A_1|^2 + |A_2|^2 + 2|A_1||A_2|\cos(\omega s + \phi) \tag{3.36}$$

where $\phi$ is a constant phase and $\omega$ a frequency defined as

$$\omega = |\Delta_1 - \Delta_2|. \tag{3.37}$$

---

[1] The external cavity mode (antimode) satisfies the condition $1-\eta(\alpha \sin \Delta - \cos \Delta) > 0$ ($<0$) (cf. Chapter 2).

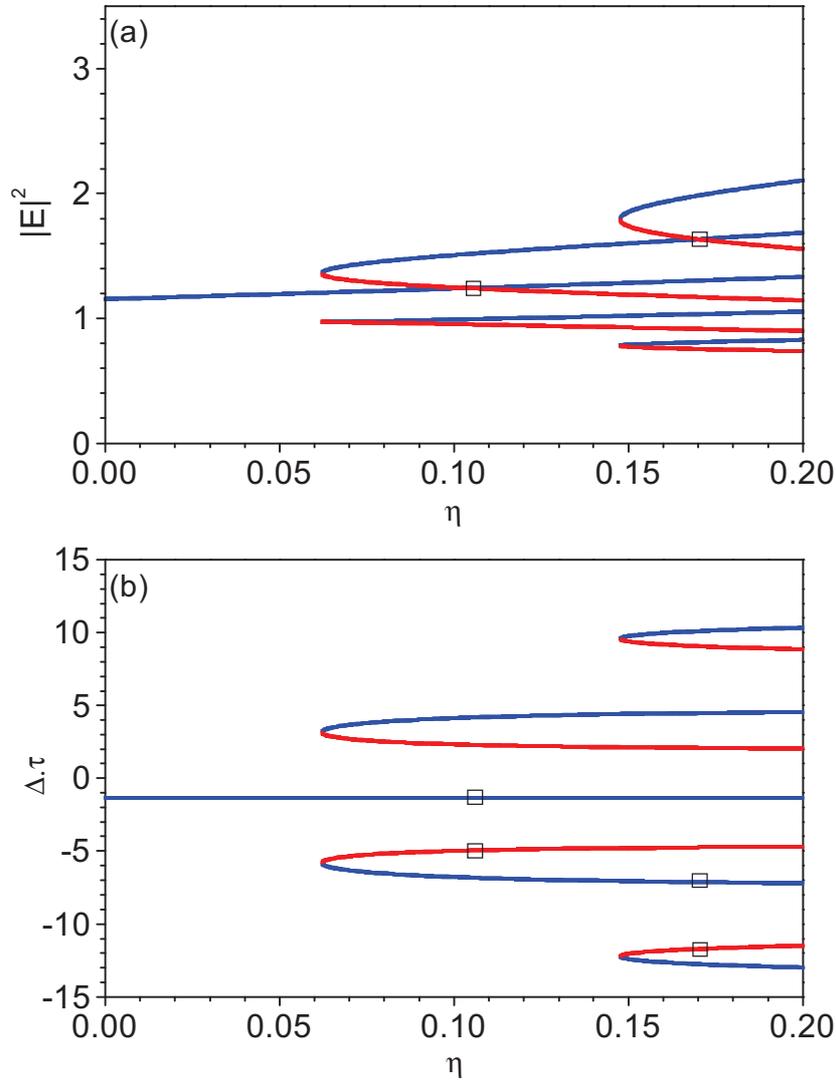

**Fig. 3.1** Intensities (a) and frequencies (b) of different branches of steady state solutions as functions of the feedback rate. Blue and red lines correspond to external cavity modes and antimodes, respectively. An external cavity mode and an antimode exhibit the same intensity at the critical feedback rates $\eta = 0.104$ and $\eta = 0.17$ (squares). For those values of $\eta$, mixed mode solutions are possible if $\chi = 0$. For $\eta = 0.104$, the mixed solution admits the two frequencies $\Delta_1\theta = -1.33$ and $\Delta_2\theta = -4.96$. The values of the dimensionless parameters are $T=1710$, $\alpha = 4$, $P = 1.155$, $\tau = 18$, $\Omega\theta = -\arctan\alpha$.

We have shown in Chapter 2 that the frequency separation between two adjacent steady state solutions is close to $\pi/\theta$ when the feedback rate is not too small. This suggests that the intensity will oscillate with, approximately, the frequency

$$f \cong 1/2\tau. \tag{3.38}$$

It is important to note that the LK equations admit multiple approximate solutions of the form (3.18). For the parameter values chosen here, this kind of solution is also observed, for instance, at $\eta = \eta_0 \cong 0.17$ with $n = -3$ (Fig. 3.1).

## 3.4. Branches of time-periodic solutions connecting pairs of steady-state solutions

In order to determine the amplitudes $A_1$ and $A_2$ appearing in the leading order solution (3.36), we need to investigate the higher order problem. For this purpose, the feedback rate $\eta$ is expanded as

$$\eta = \eta_0 + \chi \eta_1 + \ldots \tag{3.39}$$

where $\eta_0$ is given by Eq. (3.32). A two-time solution of the form

$$E = E_0(s,\upsilon) + \chi E_1(s,\upsilon) + \ldots, \tag{3.40}$$

$$N = N_0 + \chi N_1(s,\upsilon) + \ldots \tag{3.41}$$

is sought, where

$$\upsilon = \chi s \tag{3.42}$$

is a slow time that is introduced to take into account the corrections of the two frequencies $\Delta_{1,2}$ as $\chi \neq 0$. The subsequent analysis is long and tedious. We merely summarize the main results in the present section.

It is possible to show that the total intensity exhibits harmonic oscillations given by

$$|E|^2 \cong C_1^2 + C_2^2 + 2C_1 C_2 \cos(\omega s + \phi) \tag{3.43}$$

where $\omega$ is defined by Eq. (3.37), $\phi$ is a constant phase and $C_1$ and $C_2$ are functions of $\eta_1$. It is also possible to show that these branches of time-periodic solutions emerge and terminate in each case at two distinct Hopf bifurcation points. The first Hopf bifurcation point corresponds to $C_2 = 0$ and is located on a branch of an external cavity mode. The second Hopf bifurcation point is located on a branch of a saddle-type antimode and corresponds to $C_1 = 0$. The branches of time-periodic solutions overlap the points where two steady state solutions exhibit the same intensity.

Two branches of mixed mode solutions are represented in Fig. 3.2. For both branches of time-periodic solutions, the first Hopf bifurcation is from a stable external cavity mode. As a consequence, the time-periodic solutions are stable in the vicinity of the first Hopf bifurcation. However, a change of the stability of these solutions must occur between the first and the second Hopf bifurcation points since the latter is located on an unstable, saddle-type, antimode.

Fig. 3.3 presents a bifurcation diagram that we have calculated by solving numerically the Lang-Kobayashi equations [Eqs. (3.12) and (3.13)]. The figure shows the extrema of the intensity $|E|^2$ as a function of $\eta$. For $\eta \cong 0.0962$, this external cavity mode is destabilized through a Hopf bifurcation and a stable periodic branch emerges. This periodic branch compares quantitatively very well with the diagram based on the analytical approximation (Fig. 3.2). A secondary bifurcation to quasiperiodic oscillations, which are characterized by two incommensurable frequencies, occurs for $\eta \cong 0.1002$. Above $\eta \cong 0.1013$, the laser locks onto a new stable external cavity mode. This new branch appears for $\eta \cong 0.062$ and coexists with the previous branches. We observe in Fig. 3.3(a) that the mixed solution scenario repeats as the feedback rate is further increased. Figures 3.4 and 3.5 show periodic and quasiperiodic behaviors of the laser for $\eta = 0.1$ and $\eta = 0.1003$, respectively. Using $\tau_p = 1.11$ ps for the photon lifetime, the frequency of the periodic oscillations [Fig. 3.4] is found to be $f \cong 28.45$ GHz. This frequency is very close to the approximate frequency calculated analytically from Eq. (3.37) at the point $\eta_0 = 0.104$, i.e. 28.90 GHz. When the laser output is quasiperiodic, the frequency of the rapid oscillations is $f \cong 27.73$ GHz [Fig. 3.5 (b)] whereas the frequency of the slow envelope is about $f \cong 4.33$ GHz. The latter is close to the relaxation oscillation frequency of the solitary laser $f_R = 5.26$ GHz, which is calculated by using the dimensionless form of Eq. (2.88), i.e.

$$\omega_R = \sqrt{\frac{2P}{T}}, \tag{3.44}$$

and the value of the photon lifetime ($\tau_p = 1.11$ ps).

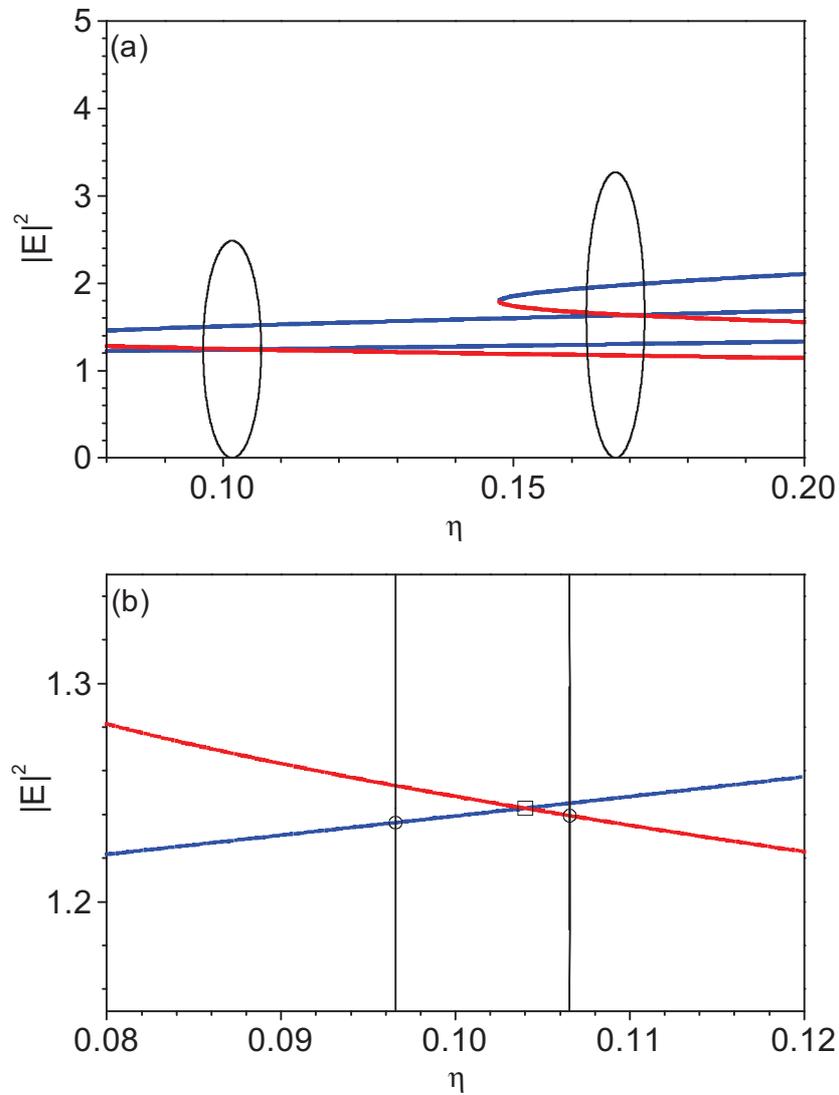

**Fig. 3.2** Bifurcation diagram of the steady state and mixed mode solutions. (a) and (b) show the maximum and the minimum of the intensity $|E|^2$. (b) is a zoom view around $\eta = 0.104$. $|E|^2$ is constant for the single frequency external cavity modes (blue lines) and antimodes (red lines) but time-periodic for the mixed mode solutions (black lines). The latter emerge and terminate at two successive Hopf bifurcation points that are located on an external cavity mode and an antimode, respectively [circles in (b)]. The square is the degenerate point shown in Fig. 3.1. Same values of the parameters as in Fig. 3.1.

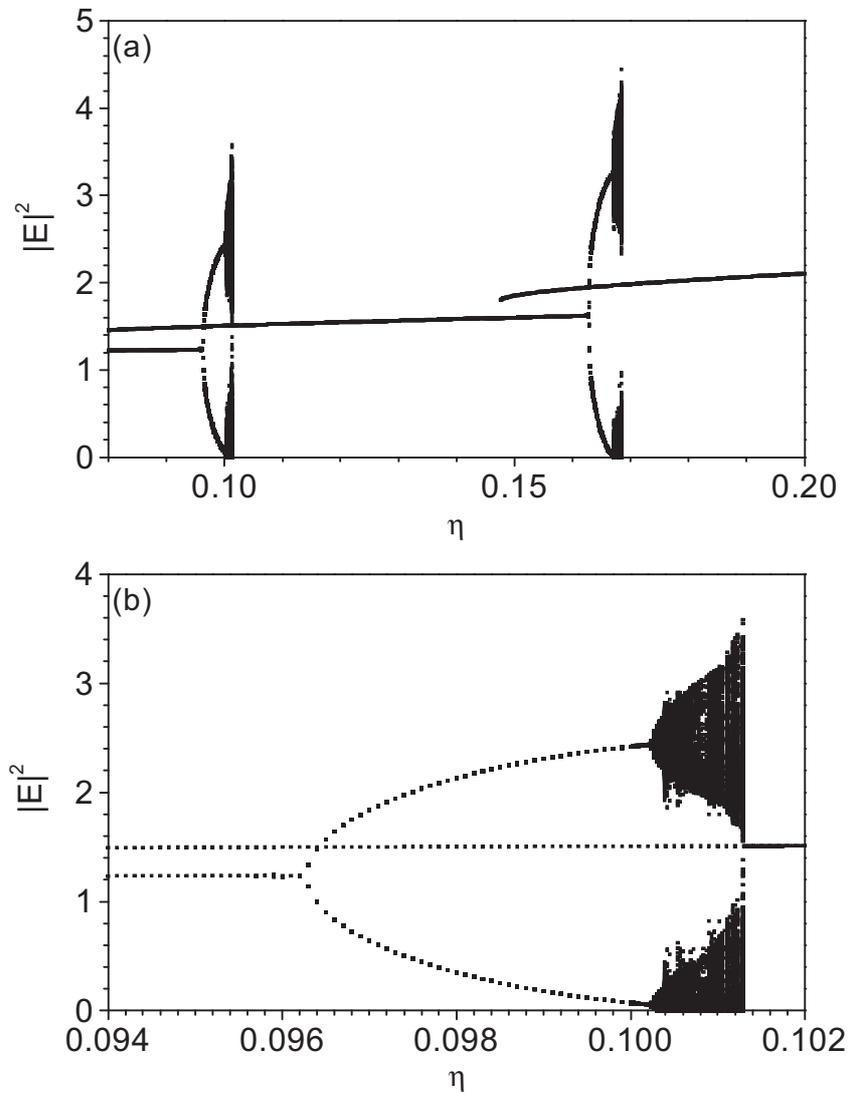

**Fig. 3.3** Bifurcation diagram obtained by solving numerically the Lang-Kobayashi equations [Eqs. (3.12) and (3.13)]. (a) shows the extrema of the intensity $|E|^2$ as a function of $\eta$; (b) is an expanded version showing the mixed mode solution undergoing a quasiperiodic bifurcation. The values of the parameters are as in Fig. 3.1.

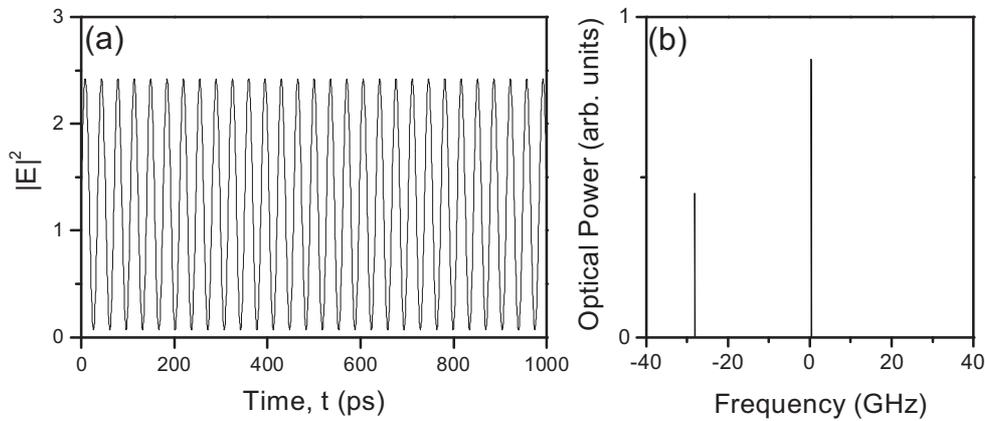

**Fig. 3.4** Periodic behavior of the laser for $\eta = 0.1$. (a) shows the temporal behavior of the dimensionless intensity, (b) the optical spectrum. The frequency of the periodic oscillation is $f \cong 28.45$ GHz. The values of the parameters are as in Fig. 3.1. In the optical spectrum, the horizontal axis has its origin at the threshold frequency of the solitary laser diode.

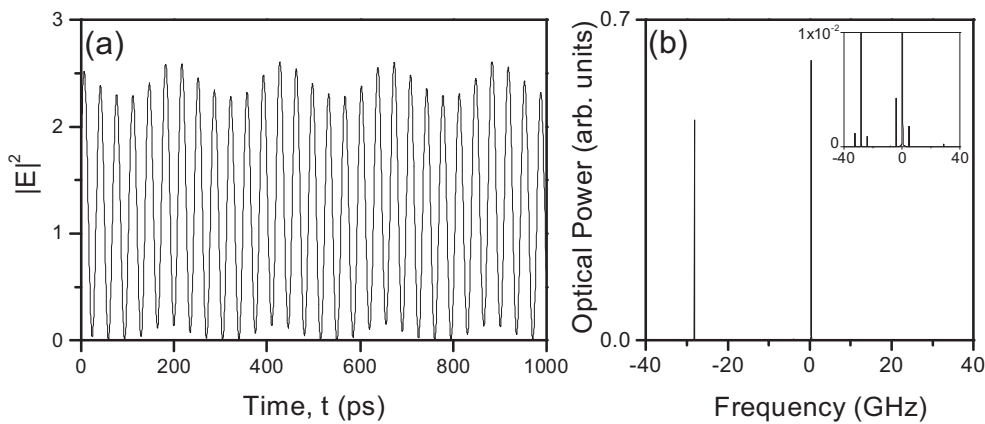

**Fig. 3.5** Quasiperiodic behavior of the laser intensity for $\eta = 0.1003$. (a) shows the temporal behavior of the dimensionless intensity, (b) the optical spectrum. The frequency of the fast oscillation is $f \cong 27.73$ GHz whereas that of the slow modulation is about $f \cong 4.33$ GHz. The latter is close to the relaxation oscillation frequency of the solitary laser ($f_R \cong 5.26$ GHz). The inset clearly shows additional lines. Those correspond to the apparition of the slow modulation. The values of the parameters are as in Fig. 3.1. In the optical spectrum, the horizontal axis has its origin at the threshold frequency of the solitary laser diode.

## 3.5. Conclusion

We have shown that a linear combination of two steady-state solutions is not an exact solution of the Lang-Kobayashi equations but is the leading approximation of an asymptotic solution valid when the ratio of the carrier lifetime to the photon lifetime is large. In addition, we have shown that branches of time-periodic solutions connect pairs of steady state solutions (one external cavity mode with one antimode, respectively). The frequency of the periodic solutions is found to be close to the inverse of twice the external cavity roundtrip time. As already pointed out by Tager and Petermann [10], the all-optical generation of almost sinusoidal high-frequency oscillations is a very promising application of laser diodes subject to optical feedback from short external cavities.

The present work has been motivated by our experimental observation of unusual high frequency periodic oscillations in a laser diode subject to two optical feedbacks [7,8]. The results that have been presented here will enlighten part of the experimental observations that we report in the next chapter.

Most of the matter of this chapter is extracted from a paper published in *Optics Communications* and to which we have collaborated [12].

# 4. All-optical technique for stabilization of an external cavity laser diode operating in the low-frequency fluctuation regime and observation of high-frequency periodic oscillations

You won't ever get there unless you first
understand where you really want to go.

## 4.1. Introduction

When subjected to external optical feedback, semiconductor lasers exhibit a large variety of dynamic behaviors such as coherence collapse [1] and low frequency fluctuations (LFF) [2] that lead to severe degradation of their temporal characteristics and increase their typical optical linewidths from 100 MHz to several tens of gigahertz. The LFF regime is typically observed when laser diodes are pumped near threshold and subjected to moderate optical reinjection from a distant reflector. The dominant features of this instability are sudden dropouts of the optical power, followed by gradual, stepwise recoveries that occur at aperiodic intervals, leading to a dramatic increase in low-frequency noise in the radiofrequency (rf) spectrum.

From an application point of view it is interesting to investigate practical methods of suppressing or controlling chaos and LFF. Among several methods (Refs. 3-11 and references therein), a dynamic targeting technique was proposed by Wieland *et al.* [3], who showed numerically that a single-mode semiconductor laser that suffers from coherence collapse when subjected to optical feedback can be steered into the maximum gain mode (MGM) by an adequate adjustment of the feedback phase as the feedback rate is varied. Because the maximum gain mode never undergoes a Hopf bifurcation [see Chapter 2 or Ref.12], the laser operates in a stable regime. Recently Hohl and Gavrielides [4] applied the dynamic targeting technique to the experimental control of a chaotic semiconductor laser biased near threshold. Although Wieland *et al.* proposed adjusting the feedback phase by accurate positioning of the reflection source (generally a mirror) within one half of an optical wavelength [3], they achieved the adjustment of the feedback phase by slightly varying the pump current [4].

One major difficulty of the targeting technique is that, because of spontaneous emission noise and the smallness of the attraction basin, the laser can be kicked out of the maximum gain mode at the beginning of the procedure that was proposed in Ref. 3 and jump to the nearest external cavity mode (ECM) that is often, but not always, stable. Furthermore, in both studies it was assumed that either the positioning of the mirror [3] or the laser pump current [4] can be adjusted as the feedback rate is varied, whereas in practical cases these parameters cannot be modified easily. It is therefore interesting to investigate alternative methods of chaos and LFF suppression that do not require modifications of the laser or feedback parameters. A method using a second optical feedback was thus proposed by Liu and Ohtsubo [11] for stabilizing a chaotic laser diode pumped far above threshold (up to twice the threshold current of the solitary laser). We have investigated analytically and numerically this parameter regime in Ref. 13 and have shown that a second optical feedback can indeed increase the stability of a laser diode subject to a first feedback.

In our first *Optics Letters* [14], we investigated the stabilizing effects of a second optical feedback specifically in the case where the laser is pumped close to threshold paying a particular attention to the bifurcation diagram of the steady state solutions. We showed numerically for what is believed to be the first time that a second optical feedback can suppress the LFF regime in a laser diode biased near threshold and subjected to a first optical feedback. In addition, we showed that one can steer a laser biased near threshold to lock into the stable maximum gain mode without any modifications of the laser or first feedback parameters. By means of numerical simulations, we have also anticipated that the second external mirror positioning does not need to be accurate; our technique is therefore easier from an experimental point of view than the dynamic targeting. We predicted moreover that the technique should work regardless the first feedback rate if the linewidth enhancement factor is not too large.

Our theoretical work was motivated by two intuitive arguments arising from the following observations. According to Sano [see Chapter 2 or Ref.15], the dropouts of the optical power in the LFF regime are caused by crises, i.e., collisions of the system trajectory in phase space with saddle-type antimodes. Each crisis is preceded by chaotic itinerancy of the system trajectory among the attractor ruins of external cavity modes, with a drift toward the maximum gain mode. For moderate feedback delays (of the order of 1 ns) and rates, a few antimodes are responsible for the crises. If correct, this interpretation suggests that one might be able to suppress LFF by shifting these antimodes away from the other external cavity modes or, better yet, by inducing them to disappear. Without modifying any parameter of the first optical feedback or the laser diode, one can achieve this by means of a second optical feedback. Moreover, in the configuration of a single optical feedback with a moderate delay (1 ns), Hohl and Gavrielides [4,16] had observed that the laser undergoes a cascade of bifurcations as the feedback rate is increased, exhibiting successively stable and unstable behaviors, such as chaos and LFF. In the stable regions of the bifurcation cascade, the laser was found to lock into the stable maximum gain mode. Our second intuition was then that such stable regions might also be observed in a double-cavity configuration when increasing the second feedback rate, and this regardless of the rate of the first feedback. Consequently laser stabilization should always be achieved in several ranges of the second feedback rate. Numerical simulations not only confirmed those intuitions but also revealed a wealth of dynamical behaviors far beyond our expectations[1].

We confirmed our theoretical predictions in a second *Optics Letters* [18], where the first experimental realization of our stabilization technique is described. We found that the method is robust, reliable and easy to implement.

Of particular interest, the experiment revealed furthermore high frequency periodic oscillations that seemed to result from the beating between two modes of the compound cavity. This observation motivated, in a first step, the theoretical study on unusual Hopf frequencies in laser diodes subject to a single optical feedback (see Chapter 3 or Ref. [19]). Simultaneously, we found numerically that similar behaviors can be anticipated by the extension of the Lang-Kobayashi equations to the double-feedback problem [20].

The present chapter is organized as follows. In Section 2, we present the extension of the Lang-Kobayashi equations to the double-feedback problem and review our theoretical results concerning the suppression of the low-frequency fluctuation regime and the stabilization of the laser diode. In Section 3, we describe the experiment confirming our theoretical predictions. In Section 4, we report on the experimental observation of high-frequency periodic oscillations and show numerically that these are associated to branches of time-periodic solutions connecting pairs of steady state solutions.

## 4.2. Stabilization of a laser diode operating in the low-frequency fluctuation regime: theoretical analysis

### 4.2.1. Model and steady state solutions

The Lang-Kobayashi equations [21] can be extended to the problem of a laser diode subject to optical feedback from a double cavity (see Fig. 4.1) by including a second delay term in the rate equation for the electric field [11,17]. Using the same normalization as in Chapter 3, the modified Lang-Kobayashi equations are

$$\frac{dE}{ds} = (1+i\alpha)NE + \eta_1 E(s-\theta_1)\exp(-i\Omega\theta_1) + \eta_2 E(s-\theta_2)\exp(-i\Omega\theta_2),\quad (4.1)$$

---
[1] This large variety of dynamical behaviors was already conjectured by Fischer et al in Ref. 17.

$$T\frac{dN}{ds} = P - N - (1+2N)|E|^2. \quad (4.2)$$

The dimensionless time $s$ is measured in units of the photon lifetime; $E(s) = A(s)\exp[i\phi(s)]$ and $N(s)$ are the normalized slowly varying complex electric field and the normalized excess carrier number. $\eta_1$ and $\eta_2$ are the normalized feedback rates of the first and second external cavities and $\theta_1$ and $\theta_2$ the ratios of the round-trip time to the photon lifetime for those cavities. $\alpha$ is the linewidth enhancement factor and $\Omega$ the dimensionless angular frequency of the solitary laser. $P$ is the dimensionless pumping current above solitary laser threshold and $T$ the ratio of the carrier lifetime to the photon lifetime.

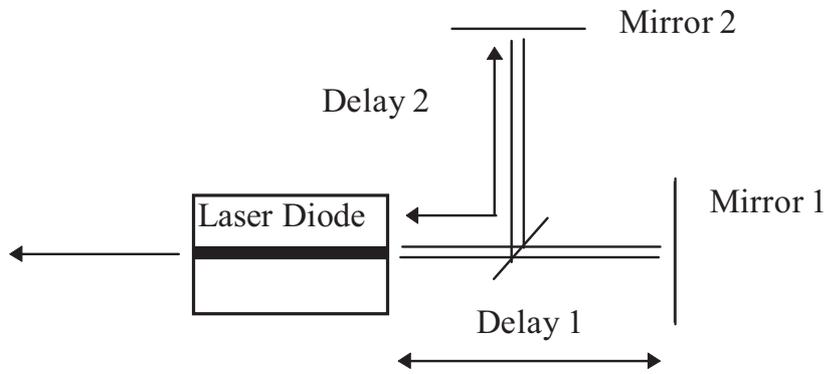

**Fig. 4.1** Schematic configuration of a laser diode subject to optical feedback from a double cavity.

The steady state solutions of Eqs. (4.1) and (4.2) can be written in the form

$$E = A_s \exp[i(\Delta - \Omega)s] \text{ and } N = N_s \quad (4.3)$$

where the stationary angular frequency $\Delta$, the amplitude $A_s$ and the normalized carrier number $N_s$ satisfy the equations

$$\Delta = \Omega - \eta_1[\alpha\cos(\Delta\theta_1) + \sin(\Delta\theta_1)] - \eta_2[\alpha\cos(\Delta\theta_2) + \sin(\Delta\theta_2)], \quad (4.4)$$

$$A_s^2 = \frac{P - N_s}{1 + 2N_s} \geq 0 \quad (4.5)$$

and

$$N_s = -\eta_1\cos(\Delta\theta_1) - \eta_2\cos(\Delta\theta_2) \quad (4.6)$$

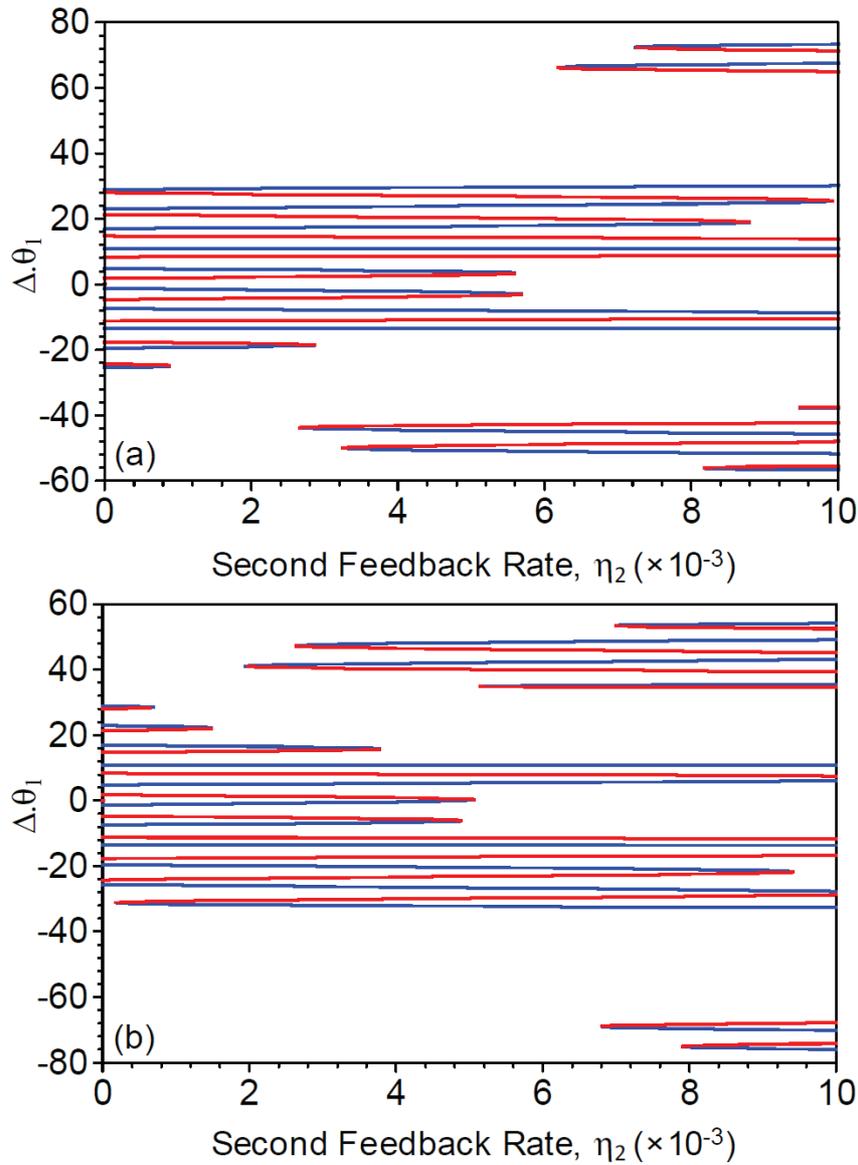

**Fig. 4.2** Stationary angular frequencies $\Delta$ as a function of $\eta_2$. In (a), the values of the parameters are $\alpha = 4$, $\eta_1 = 5 \times 10^{-3}$, $\theta_1 = 1500$, $\theta_2 = 200$ and $\Omega\theta_1 = 0.483$. In (b) $\theta_2 = 200 + \lambda/(2c\tau_p)$ with $\lambda/(2c\tau_p) = 1.3 \times 10^{-3}$, the other parameters being identical to those in (a).

Fig. 4.2, obtained from Eq. (4.4), shows typical evolutions of the product $\Delta\theta_1$ of the stationary angular frequencies ($\Delta$) and the first feedback delay ($\theta_1$) when $\eta_2$ is varied. Fig. 4.2(a) is calculated for $\alpha = 4$, $\eta_1 = 5 \times 10^{-3}$, $\theta_1 = 1500$ and $\theta_2 = 200$, assuming that the laser works at the wavelength $\lambda = 780$ nm and that the photon lifetime is $\tau_p = 1$ ps. Fig. 4.2(b) is calculated for the same parameter values, except $\theta_2$: here $\theta_2 = 200 + \lambda/(2c\tau_p)$ with $\lambda/(2c\tau_p) = 1.3 \times 10^{-3}$ ($c$ being the velocity of light in vacuum). Fig. 4.2(b) thus corresponds to a displacement of the second mirror by a quarter of wavelength with respect to the first case.

New steady state solutions are created in pairs by a saddle-node bifurcation; in each pair, one solution is an external cavity mode and the other is an antimode, the latter being always unstable. Similarly to the single-feedback case (see Ref. 22 or Chapter 2), antimodes also satisfy the condition $d\Omega/d\Delta<0$ in the double-feedback case. But, contrary to the single-feedback case, pairs of steady state solutions can disappear when the feedback rate $\eta_2$ of the second cavity increases for a given feedback rate $\eta_1$ of the first cavity. Moreover, comparison between Fig. 4.2(a) and Fig. 4.2(b) reveals that a slight displacement of one of the two cavities (the second cavity in this case) modifies strongly the ECM-antimode pattern. For instance, the two pairs of steady state solutions with the lowest frequencies disappear at first in Fig. 4.2(a) when increasing $\eta_2$. By contrast, in Fig. 4.2(b) the pairs of steady state solutions with the highest frequencies are suppressed at first.

### 4.2.2. Numerical results

Since a systematic scan of all possible values of laser and feedback parameters is out of reach, we restricted most of our investigations to typical values of the linewidth enhancement factor and of the ratio of the carrier lifetime to the photon lifetime, namely $\alpha = 4$ and $T = 1000$. Moreover, we assumed laser pumping close to threshold and considered external cavity lengths that do not exceed a few tenths of centimeters. Within these restrictions, we observed numerically that low-frequency fluctuations can be suppressed and the laser stabilized regardless of the first feedback rate and phase as well as of the second feedback phase and delay.

In the next two sections, we consider two sets of external cavity roundtrip times corresponding to two sets of external cavity lengths, namely $L_1 = 15$ cm, $L_2 = 3$ cm and $L_1 = 21$ cm, $L_2 = 19$ cm. The first set corresponds to the configuration that we studied in Ref. 14. The external cavity lengths considered in the second case are those encountered in our experimental work [18]. We point out that, in both cases, the dimensionless roundtrip time in the second cavity is chosen in such a way that the second feedback leads to the destruction of pairs of the lowest frequencies external cavity modes and antimodes.

### $L_1 = 15$ cm and $L_2 = 3$ cm

Fig. 4.3(a) is obtained from Eq. (4.4) and shows the evolution of the product of the stationary angular frequencies and the first feedback delay, $\Delta\theta_1$, with respect to $\eta_2$. In Fig. 4.3(b) we present the corresponding bifurcation diagram of the phase-difference function $\phi(t)-\phi(t-\theta_1)+\Omega\theta_1$ for $\eta_1 = 4.6\times10^{-3}$, with $\eta_2$ as the bifurcation parameter. The choice of the phase-difference function for the bifurcation diagram is convenient, since it reduces to $\Delta\theta_1$ for stationary behaviors and can be directly compared with Fig. 4.3(a). The bifurcation diagram of the phase difference function is calculated by solving numerically Eqs. (4.1) and (4.2) for

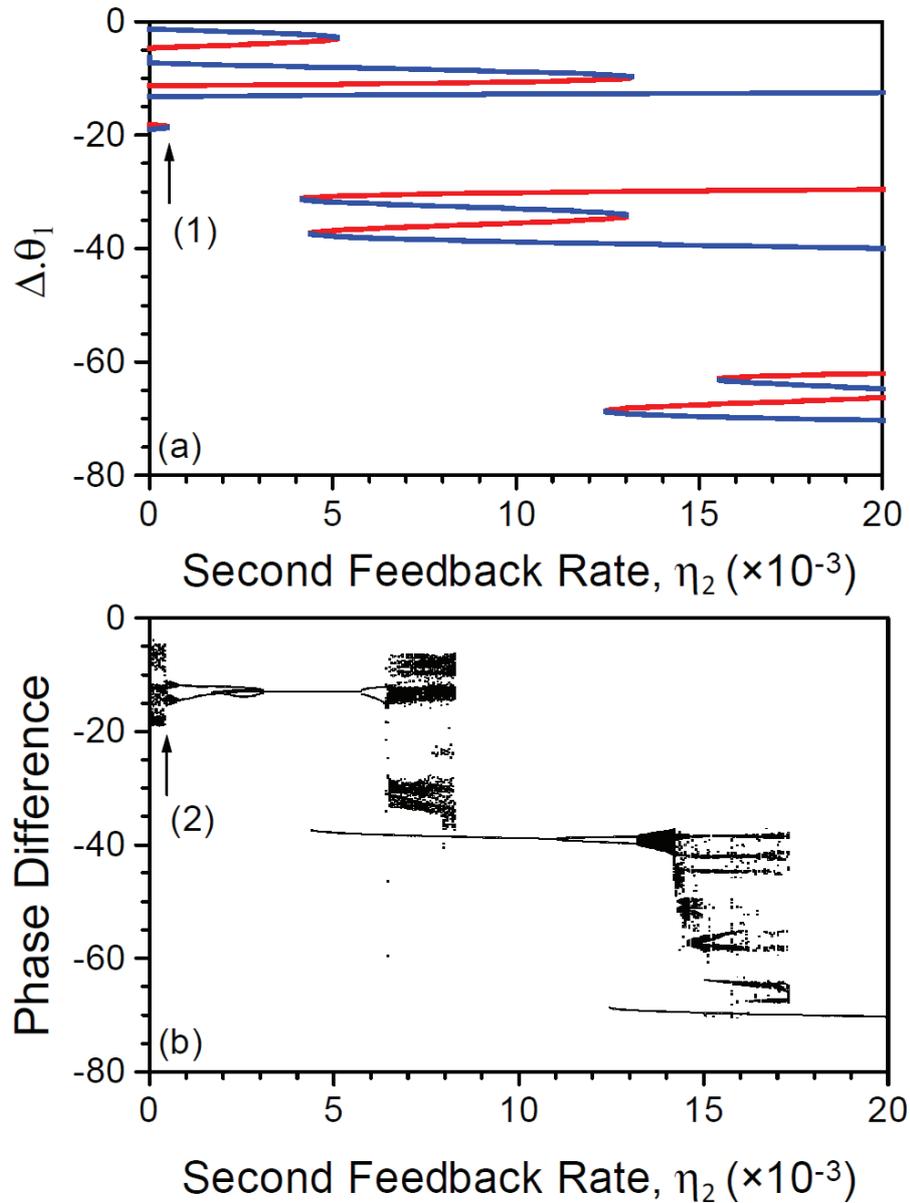

**Fig. 4.3** (a) Stationary angular frequencies Δ as function of $\eta_2$. Red curves correspond to antimodes; blue curves to external cavity modes. (b) Bifurcation diagram of the phase-difference function $\phi(t)-\phi(t-\theta_1)+\Omega\theta_1$. The second feedback rate $\eta_2$ is the bifurcation variable. Arrow 1 indicates the disappearance of the antimode that is responsible for the crisis, and arrow 2 indicates the corresponding LFF suppression.

$P = 0.001$, $\theta_1 = 1000$, $\Omega\theta_1 = -1.45$, $\theta_2 = 200$ and $\Omega\theta_2 = 0.8$. The first feedback parameters chosen here are identical to those used in Ref. 4. In Fig. 4.3(a), the first and second feedback delays were set to $\theta_1 = 999.9957999858645$ and $\theta_2 = 200 + 0.1 \times \lambda/(2c\tau_p)$, respectively[2]. For the value of $\eta_1$ chosen here, LFF is observed in the single-feedback configuration (i.e. $\eta_2 = 0$) [4]. Fig. 4.4(a) shows chaotic itinerancy of the phase trajectory among successive external cavity modes, followed by collision of the trajectory with the antimode that corresponds to the stable MGM. The trajectory is then repelled to higher values of the excess carrier number $N$, where the chaotic itinerancy towards lower values of the phase-difference function starts again.

---
[2] These values of $\theta_1$ and $\theta_2$ should in principle be used also for the calculation leading to Fig. 4.3(b); the result of the numerical integration of Eqs. (4.1) and (4.2) is however essentially not affected by small variations of $\theta_1$ and

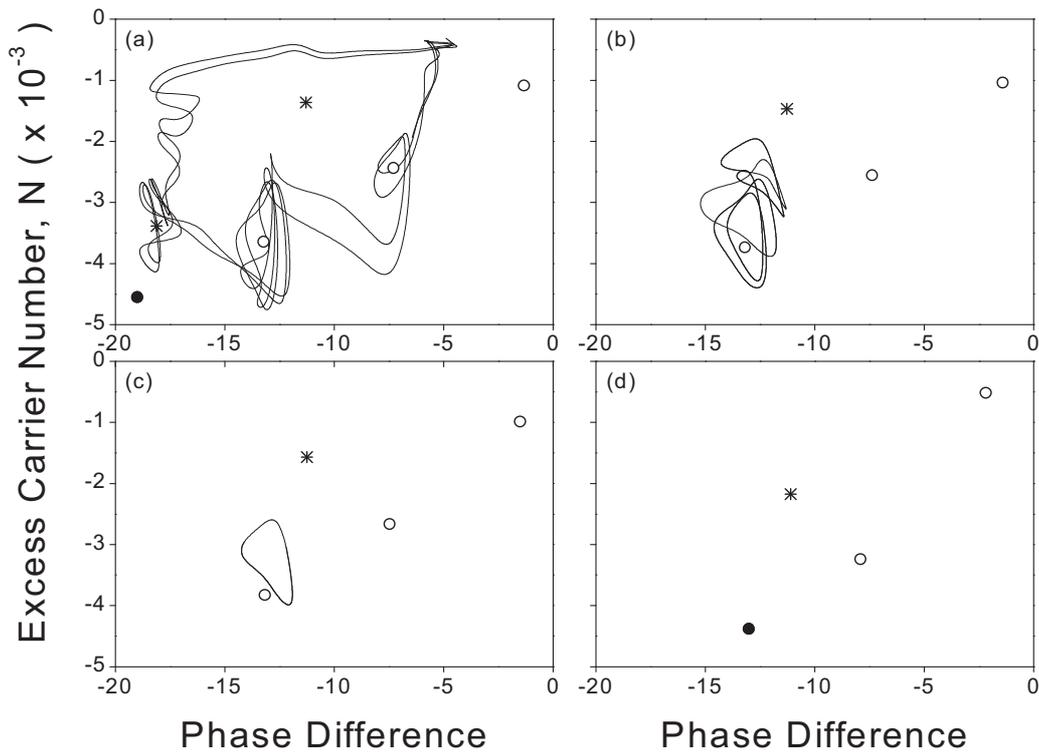

**Fig. 4.4** Phase trajectories observed in the space $[\phi(t)-\phi(t-\theta_1)+\Omega\theta_1, N(t)]$. Asterisks, antimodes; open circles unstable external cavity modes; filled circles, stable external cavity modes. (a) LFF for $\eta_2 = 0$. (b) Quasiperiodic behavior with frequency locking for $\eta_2 = 0.5 \times 10^{-3}$. (c) Limit cycle corresponding to periodic behavior for $\eta_2 = 1 \times 10^{-3}$. (d) Stationary behavior for $\eta_2 = 4 \times 10^{-3}$: the laser locks into a stable external cavity mode (filled circle). The first feedback rate is $\eta_1 = 4.6 \times 10^{-3}$ in all cases.

    We observe that several pairs of external cavity modes and antimodes are created or destructed as the second feedback strength is increased. The maximum gain mode and its antimode collide and disappear for $\eta_2 = 0.45 \times 10^{-3}$ [arrow 1 in Fig. 4.3(a)]. As this pair of modes disappears, LFF, which is observed from $\eta_2 = 0$ to $0.45 \times 10^{-3}$, suddenly stops and chaotic behavior is observed [arrow 2 in Fig. 4.3(b)] because the phase trajectory can no

---

$\theta_2$. The value of $\theta_1 = 999.9957999858645$ is used on the ground of mathematical consistency: for $\lambda = 780$ nm, we find $\Omega\theta_1 \cong -1.45$ mod $2\pi$.

longer collide with the antimode and stays close to the nearest mode. This chaotic behavior is observed in a very small range of $\eta_2$ and, as $\eta_2$ increases further, quasiperiodic behaviors appear [from $\eta_2 = 0.5 \times 10^{-3}$ to $0.8 \times 10^{-3}$ in Fig. 4.3(b)] with frequency locking near $\eta_2 = 0.5 \times 10^{-3}$ [Fig. 4.4(b)]. The quasiperiodic regime is followed by periodic behaviors up to $\eta_2 = 3 \times 10^{-3}$. Fig. 4.4(c) shows the limit cycle that corresponds to the periodic behavior observed for $\eta_2 = 1 \times 10^{-3}$. Fig. 4.3(a) and Fig. 4.3(b) show that, from $\eta_2 = 3 \times 10^{-3}$ to $5.7 \times 10^{-3}$, the laser locks into a stable external cavity mode whatever the initial conditions are [see, for instance, Fig. 4.4(d)]. For further increases of $\eta_2$ the laser undergoes a cascade of bifurcations composed of successive intervals within which it exhibits unstable behaviors, such as periodic, quasiperiodic, and chaotic behaviors and LFF (e.g., from $\eta_2 = 5.7 \times 10^{-3}$ to $8.3 \times 10^{-3}$) and stable behavior when it locks into a new stable MGM (e.g., from $\eta_2 = 8.3 \times 10^{-3}$ to $10.9 \times 10^{-3}$). Fig. 4.3(b) reveals that this scenario, with regions of laser stabilization interspersed with regions of low-frequency fluctuations, repeats as the feedback rate $\eta_2$ is further increased. In contrast to the first stabilization, the stabilizations of the laser in newly created maximum gain modes are characterized by large shifts to lower frequencies. This sequence of events is similar to the sequence that has been reported in the case of a laser diode subject to a single-feedback in Refs. 4,16,24 and 25. Similarly to the single-feedback problem, the locking of the system trajectory onto a newly created maximum gain mode is associated to the connection of the basins of attraction of this MGM and of the LFF attractor.

In the cases where the second feedback does not lead to the destruction of the antimodes responsible for the crises in the single-feedback configuration, stabilization of the laser through locking into a new maximum gain mode occurs nevertheless. As a consequence, LFF suppression and laser stabilization are achieved whatever the second feedback phase may be, and therefore accurate positioning of the second mirror is not required. We performed numerical calculations for larger first feedback rates, as large as $\eta_1 = 5 \times 10^{-2}$, and always observed laser stabilization. In all cases, the laser diode undergoes a cascade of bifurcations as the second feedback rate is increased, with successive regions in which it exhibits chaos or LFF and regions in which it exhibits stable behavior.

## $L_1 = 21$ cm and $L_2 = 19$ cm

In this section, the bifurcation diagram of the phase difference function [Fig. 4.5(b)] is calculated by solving numerically Eqs. (4.1) and (4.2) for $\eta_1 = 7.2 \times 10^{-3}$, $\theta_1 = 1400$, $\Omega\theta_1 = -2.90$, $\theta_2 = 1267$, $\Omega\theta_2 = 1.20$ and $P = 0$, with $\eta_2$ as the bifurcation parameter. The bifurcation diagram of the steady state solutions [Fig. 4.5(a)] is calculated in the same way as in the previous section, using $\theta_2 = 1267 + 0.3 \times \lambda/(2c\tau_p)$.

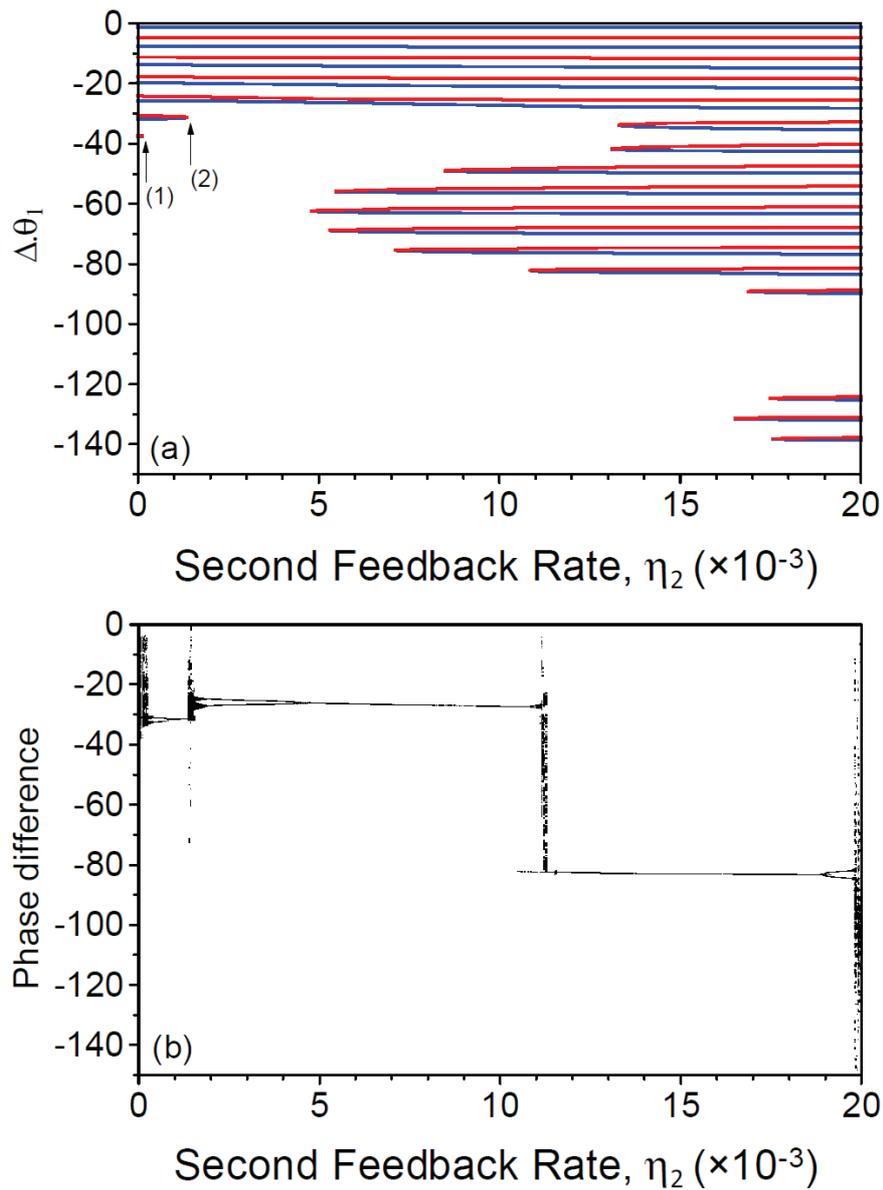

**Fig. 4.5** (a) Stationary angular frequencies $\Delta$ as function of $\eta_2$. Red curves correspond to antimodes; blue curves to external cavity modes. (b) Bifurcation diagram of the phase-difference function $\phi(t)-\phi(t-\theta_1)+\Omega\theta_1$. The second feedback rate $\eta_2$ is the bifurcation variable. Arrows 1 and 2 indicate two saddle-node bifurcations.

When the first cavity is acting alone, i.e. $\eta_2 = 0$, the trajectory displays chaotic itinerancy among five attractor ruins of external cavity modes and collisions with three different antimodes [Fig. 4.6(a)]. For $\eta_2 = 1.4 \times 10^{-4}$, the MGM and its antimode collide and disappear through a saddle-node bifurcation [arrow 1 in Fig. 4.5(a)]. The LFF continues however, as crises with remaining antimodes still occur [Fig. 4.6(b)]. As $\eta_2$ increases further, the ruin of the chaotic attractor associated to the lowest frequency external cavity mode decreases progressively in size until $\eta_2 = 2.5 \times 10^{-4}$. For this value of the second feedback rate, the attractor does no longer collide with the saddle-point. Chaotic, quasiperiodic and periodic behaviors are successively observed from $\eta_2 = 2.5 \times 10^{-4}$ to $8.4 \times 10^{-4}$. As an example, a stable chaotic attractor and a limit cycle are shown in Fig. 4.6(c) and (d) for $\eta_2 = 2.6 \times 10^{-4}$ and $4.8 \times 10^{-4}$, respectively. At $\eta_2 = 8.4 \times 10^{-4}$, the lowest frequency ECM becomes stable through a Hopf bifurcation. From $\eta_2 = 8.4 \times 10^{-4}$ to $1.37 \times 10^{-3}$, all trajectories are attracted by this stable MGM, regardless of the initial conditions. The maximum gain mode and its antimode collide at $\eta_2 = 1.37 \times 10^{-3}$ [arrow 2 in Fig. 4.5(a)]. Fig. 4.6(d) shows the steady state solutions just before the saddle-node bifurcation. As this pair of steady state solutions disappears, an atypical LFF is observed[3]: the trajectory in phase space exhibits a chaotic itinerancy but no collision with an antimode! It is however repelled toward higher values of the excess carrier number along the direction of the unstable manifold of the antimode that has disappeared at $\eta_2 = 1.37 \times 10^{-3}$ [Fig. 4.6(f)]. Additional increase of $\eta_2$ leads successively to chaotic, quasiperiodic, periodic and finally at $\eta_2 = 4.6 \times 10^{-3}$ to stationary behaviors. For $\eta_2 > 4.6 \times 10^{-3}$, the rest of the bifurcation diagram reveals a scenario similar to that described in the previous section, i.e. a succession of regions within which the laser is locked into a stable maximum gain mode and regions where the laser exhibits complex behaviors such as chaos and low-frequency fluctuations.

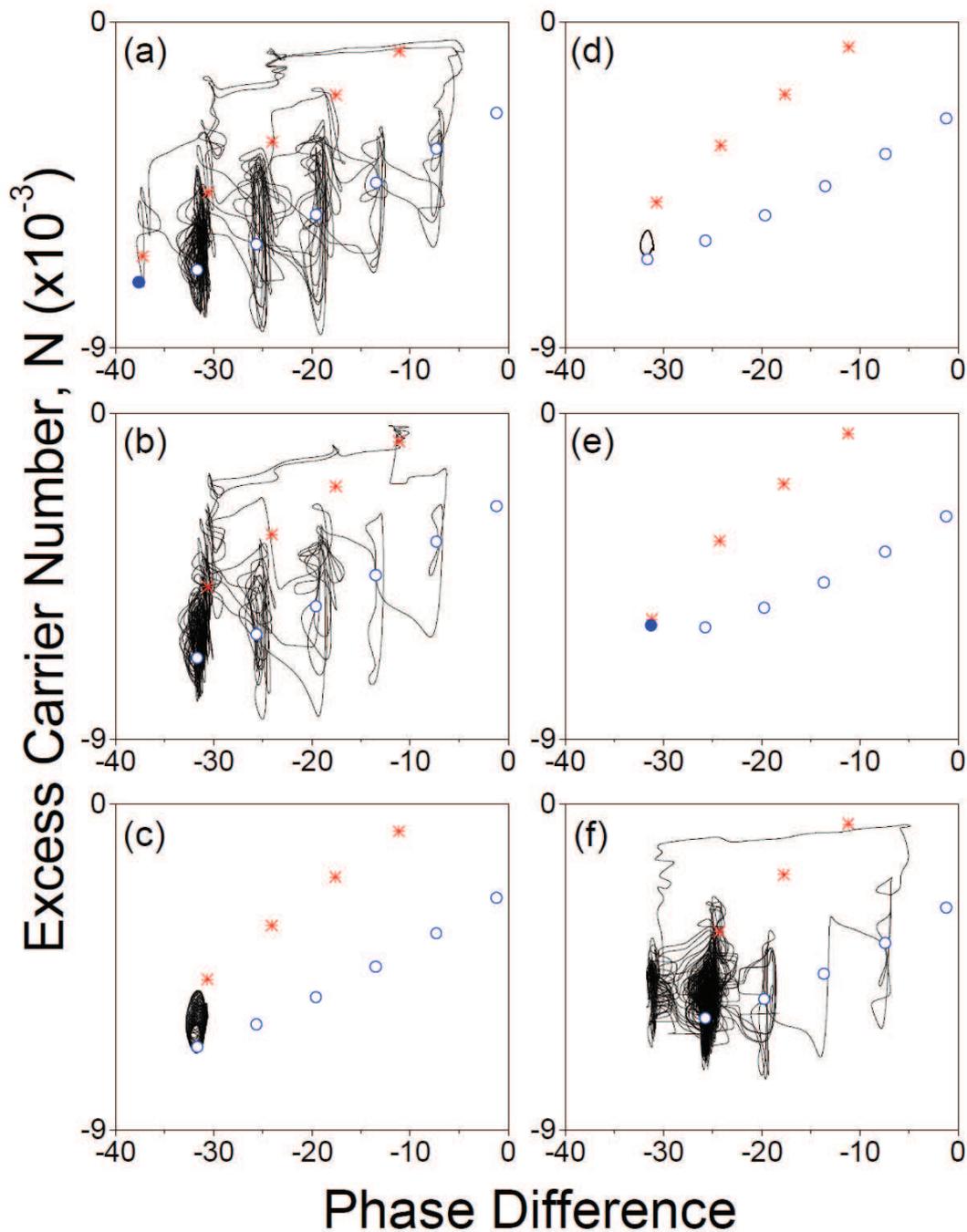

**Fig. 4.6** Phase trajectories observed in the space $[\phi(t)-\phi(t-\theta_1)+\Omega\theta_1, N(t)]$. Asterisks show the antimodes; open circles the unstable external cavity modes and filled circles the stable external cavity modes. (a) and (b): LFF for $\eta_2 = 0$ and $\eta_2 = 1.6 \times 10^{-4}$, respectively. (c) Chaotic behavior for $\eta_2 = 2.6 \times 10^{-4}$. (d) Limit cycle corresponding to a periodic behavior for $\eta_2 = 4.8 \times 10^{-3}$. (e) Stationary behavior for $\eta_2 = 1.3 \times 10^{-3}$: the laser locks into the MGM. (f) Atypical LFF for $\eta_2 = 1.4 \times 10^{-3}$. The first feedback rate is $\eta_1 = 7.2 \times 10^{-3}$ in all cases.

## 4.3. Experiment

### 4.3.1. Experimental setup

The experimental setup is illustrated in Fig. 4.7. We use an SDL-5301 laser diode operating at the wavelength $\lambda$=780 nm. The temperature-stabilized laser is biased at a pump current $I = 25.0$ mA, just below solitary threshold $I_{th} = 25.2$ mA. The laser beam is collimated by an antireflection coated lens (CL) and directed to a holographic grating (GR). The zeroth order diffracted by the grating is used to monitor the system's behavior. The first order of diffraction goes into the double cavity, which is formed by a nonpolarizing beamsplitter (NPBS) and two 99% reflectivity mirrors. The double cavity system (contained within the dashed box) is shielded from parasitic feedback from the detection branch by an optical isolator (ISO). External cavity optical lengths are $L_1 = 21$ cm and $L_2 = 19$ cm, respectively. The grating has 1200 lines per mm and is oriented at 45° with respect to the incident beam. Provided that the two mirrors are properly tilted, the grating narrows the cavity bandwidth to 50 GHz, restricting the laser diode to oscillate in only one of its longitudinal modes[4]. The laser output is monitored using a scanning Fabry-Perot interferometer (OSA; Newport SR-240C, free spectral range of 2000 GHz, finesse > 17000) and a fast ac-coupled photodiode (PD; Hamamatsu C4258, 8 GHz bandwidth). The photodiode signal is amplified and connected to a rf spectrum analyzer (RF; Hewlett-Packard HP 8596E) and a 500 MHz bandwidth digitizer (DSO; Tektronix RTD 720). The limited bandwidth of the digitizer allows the detection of the slow envelope of the optical power dynamics and, in particular, the power dropouts that characterize the LFF regime. The feedback strengths of each external cavity are controlled independently by means of a polarizing beamsplitter (PBS) and two polarizers (POL). They are characterized by measuring the effective threshold current $I_{eff}$ from which the fractional threshold reduction $\Delta I = (I_{th} - I_{eff})/I_{th}$ that is induced by each cavity acting alone is calculated. This method is convenient because the fractional threshold reductions are proportional to the feedback rates[5].

---

[3] This behavior is reported here for what is believed to be the first time.

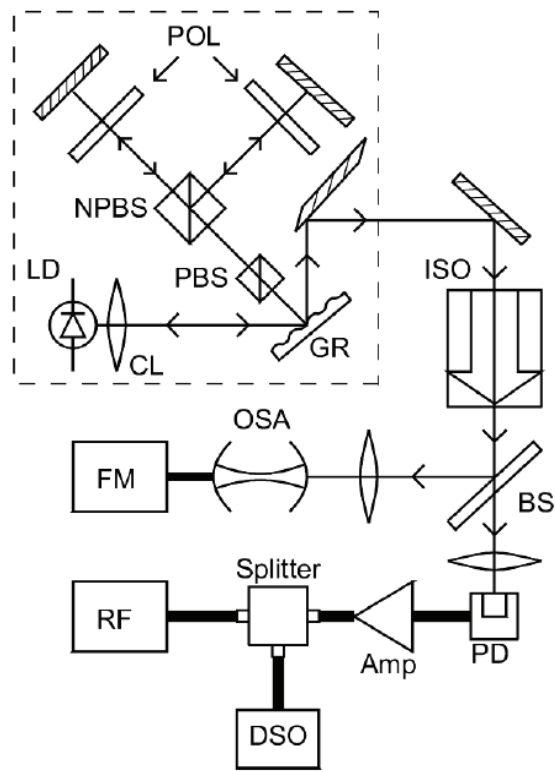

**Fig. 4.7** Experimental setup. See text for definitions.

---

[4] The two mirrors are adequately tilted in such a way that the same longitudinal mode of the laser is selected and fed back into the laser when each cavity is acting alone. The selection of the same longitudinal mode is checked by using the Fabry-Perot interferometer.

[5] According to Eqs. (4.5) and (4.6), the dimensionless pumping current must satisfy the inequality $P \geq -\kappa \cos(\Delta\theta)$ when a feedback is acting alone. At the effective lasing threshold, the maximum gain mode is the only external cavity mode that can lase since it is the one that benefits maximally from the feedback. It is possible to show that $\kappa \cos(\Delta\theta) \cong 1$ for this ECM when the feedback rate is not too low, which yields $\eta \sim \Delta I$.

## 4.3.2. Experimental suppression of the LFF regime and stabilization of the laser

Fig. 4.8 shows the experimentally observed optical spectra as a function of $\Delta I_2$, the threshold reduction that is due to the second cavity only. The horizontal axis has its origin at the frequency of the first mode that lases as the strength of the first feedback increases from zero when there is no second feedback. In the figure, the strength of the first feedback is fixed so that it corresponds to a threshold reduction $\Delta I_1 = 7.1\%$, such that the laser displays LFF in seven external cavity modes when it is not subjected to a second feedback [$\Delta I_2 = 0$, trace(a)]. Increasing slightly the strength of the second feedback [$\Delta I_2 = 0.44\%$, trace (b)] results in stabilization of the sixth external cavity mode of the *first* cavity. The limited resolution of the

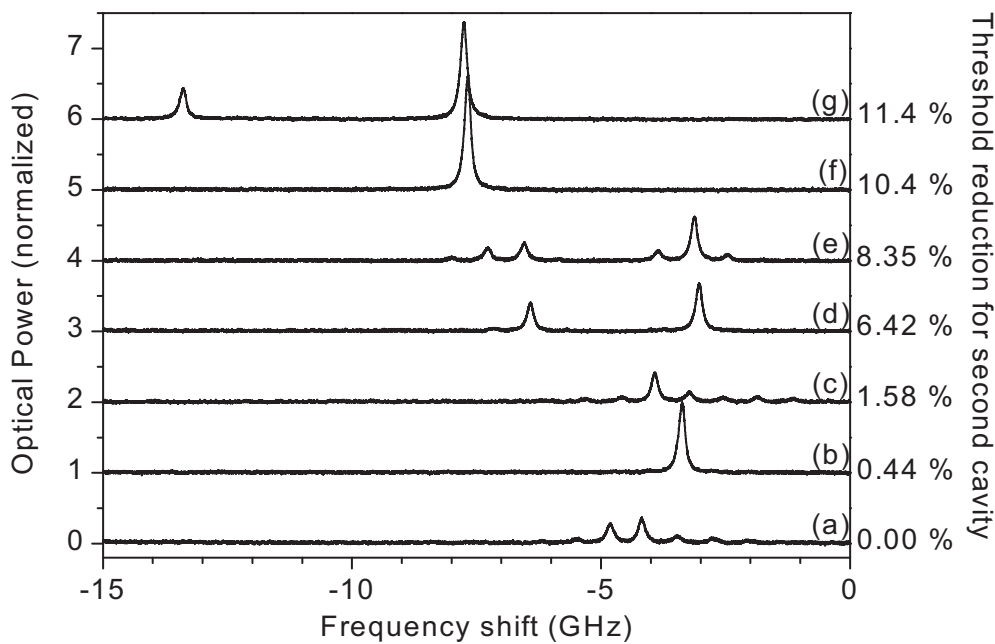

**Fig. 4.8** Experimental optical spectra measured in double-cavity configuration as a function of the threshold reduction $\Delta I_2$ owing to the second feedback strength. (a) LFF is observed in the absence of the second feedback. Increase of the second feedback strength leads to stabilization [(b) and (f)] interspersed with unstable regions [(c)-(e) and (g)]. The optical spectra have been normalized with respect to the maximum of trace (b).

Fabry-Perot interferometer, however, does not allow us to measure the linewidth of this stable mode. For $\Delta I_2 = 1.58\%$ [trace (c)], the system loses stability and once again displays LFF. For a threshold reduction of $\Delta I_2 = 6.42\%$ [trace (d)] the optical spectrum exhibits two peaks and the rf spectrum reveals that they are associated with periodic oscillations. The nature of these periodic oscillations is discussed in Section 4.4. Increasing further the second feedback strength [trace (e)], we find that the system exhibits complex dynamics until $\Delta I_2 = 10.4\%$ [trace (f)] at which the laser again becomes stable. We observe in this case a large frequency shift to lower frequencies. By comparison with the numerically calculated bifurcation diagrams, we interpret this as a locking of the system onto a newly created, stable maximum gain mode that is well separated in frequency from the original seven external cavity modes of the first cavity. Further increases of the second feedback strength lead to a continuation of this pattern, with stabilization of increasingly distant modes interspersed with regions of complex behavior, in agreement with the theoretical predictions. As can be seen, every feature observed in the numerical bifurcation cascades (Fig. 4.3 and Fig. 4.5) is also observed experimentally, despite the differences in cavity lengths or first feedback strengths.

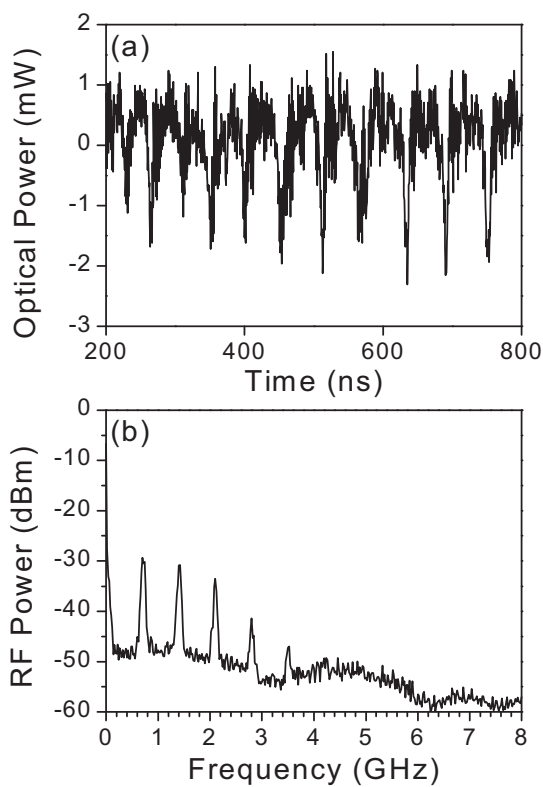
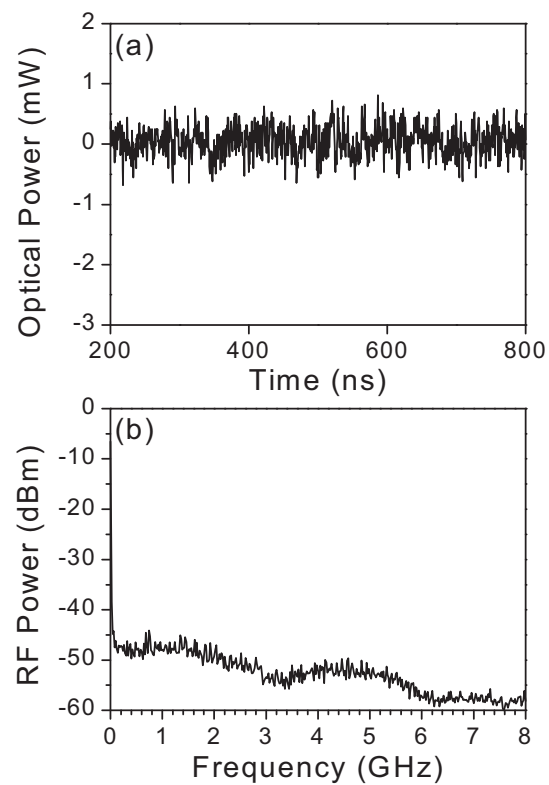

**Fig. 4.9** LFF is observed in the absence of a second optical feedback ($\Delta I_2 = 0$). (a) Fluctuations of the optical power around its average. (b) rf spectrum.

**Fig. 4.10** Stabilization of the laser by the second optical feedback ($\Delta I_2 = 0.44$ %). (a) Fluctuations of the optical power around its average. (b) rf spectrum

Time series and rf spectra that correspond to the first two steps in the bifurcation cascade are shown in Figs. 4.9 and 4.10. When the laser is subjected to the first optical feedback only, the laser output exhibits clearly strong dropouts [Fig. 4.9(a)] and the rf spectrum presents peaks at multiples of the first external cavity frequency [Fig. 4.9(b)]. When the laser is stabilized by the second feedback, the intensity shows only small fluctuations that are most likely caused by instrument noise [Fig. 4.10 (a)] and the peaks are no longer visible in the rf spectrum [Fig. 4.10 (b)]. We have observed that the control of the method is robust, limited only by the mechanical stability of the system. In agreement with numerical simulations, the method works for every first feedback strength that is accessible in the experiment. It works also without accurate positioning of the second mirror.

Fig. 4.11 illustrates the fact that it is the joint action of the two optical feedbacks that leads to stabilization of the laser. In Fig. 4.11, optical and rf spectra are recorded when the laser is subjected to feedback from the first cavity alone [traces (a) and (b)], from the second cavity alone [traces (c) and (d)], and from both cavities acting in concert [traces (e) and (f)]. While the laser suffers from LFF when subjected to feedback from the first or the second cavity alone (as is revealed by the presence of broad sidebands in the optical spectra and peaks at multiples of the external cavities frequencies in the rf spectra), it stabilizes when it is subjected to feedback from both cavities acting together; in this case, the optical spectrum exhibits a single narrow peak. By contrast, there is no peak in the rf spectrum.

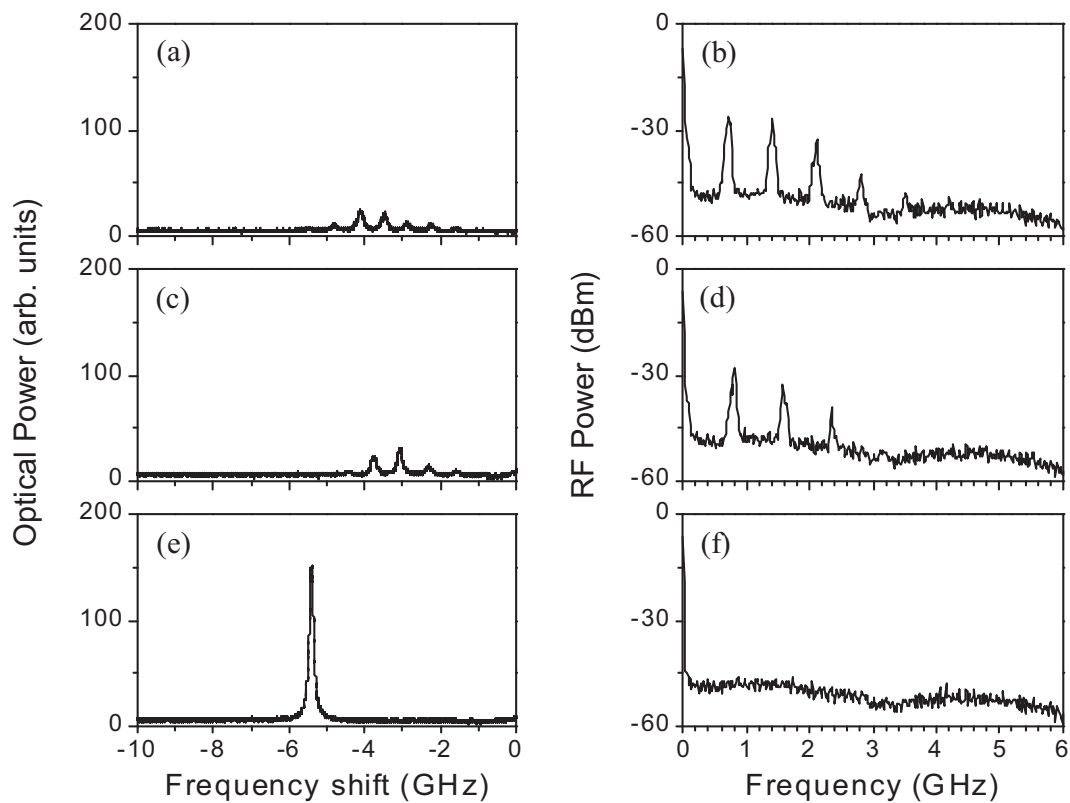

**Fig. 4.11** Experimental optical and rf spectra recorded when the laser is subjected to feedback from the first cavity alone [(a) and (b)], from the second cavity alone [(c) and (d)] and from both cavities [(e) and (f)].

## 4.4. High-frequency oscillations

The optical spectrum displays two peaks for the threshold reductions $\Delta I_2 = 6.42\%$ and $\Delta I_2 = 11.4\%$ [Fig. 4.8, traces (d) and (g)]. The corresponding rf spectra reveal that those are associated with periodic oscillations. In particular, the rf spectrum exhibits a single sharp peak at 4.86 GHz for $\Delta I_2 = 11.4\%$ (Fig. 4.12). Such a high frequency is unusual in a laser diode subject to optical feedback and pumped close to threshold; in this case, oscillations at the external cavity frequency [23] are typically expected. The rf and optical spectra suggest that the above periodic oscillations are the result of a beating between modes of the compound cavity as the frequency of 4.86 GHz corresponds to the difference between the frequencies of the two peaks in Fig. 4.8(g), within the uncertainties of the optical spectrum analyzer[6].

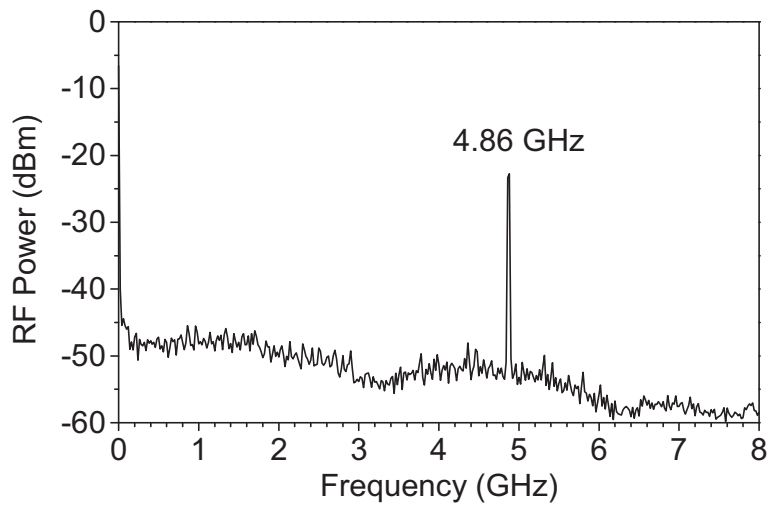

**Fig. 4.12** Experimental rf spectrum recorded when the laser is subjected to feedback from both cavities. This rf spectrum corresponds to trace (g) in Fig. 4.8.

The aim of this section is twofold. Firstly, we show that the Lang-Kobayashi equations extended to the case of the double-feedback [Eqs. (4.1) and (4.2)] admit high frequency periodic solutions. Secondly, we show that, similarly to what has been found for the single-feedback configuration (cf. Ref. 19 and Chapter 3), these periodic solutions are associated to branches connecting pairs of steady state solutions.

High-frequency periodic solutions of Eqs. (4.1) and (4.2) can be found for feedback delays compatible with the cavity lengths used in the experiment. However, the external cavity modes and antimodes are so numerous in this case that the analysis is very complex. In order to simplify the discussion, we consider here shorter cavities whose lengths are $L_1 = 15$ cm and $L_2 = 12$ cm, respectively.

---

[6] Although the supercavity SR-240C of Newport allows a high degree of finesse, it suffers from a frequency drift: determination of absolute frequencies is thus imprecise.

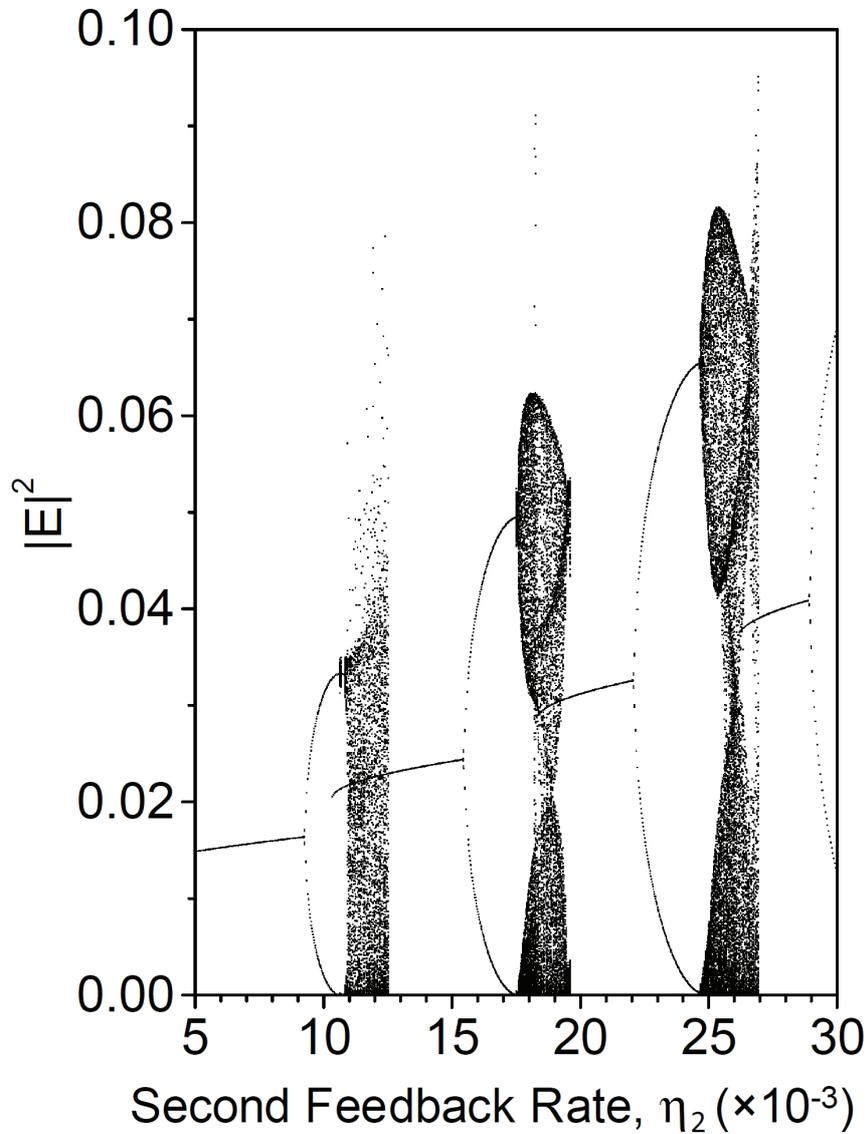

**Fig. 4.13** Bifurcation diagram obtained by solving numerically the Lang-Kobayashi equations [Eqs. (4.1) and (4.2)]. The figure shows the extrema of the intensity $|E|^2$ as a function of $\eta_2$. The diagram is obtained by progressively increasing or decreasing the bifurcation parameter.

Fig. 4.13 displays the bifurcation diagram of the normalized intensity with $\eta_2$ as the bifurcation parameter. This bifurcation diagram is obtained by numerical integration of Eqs. (4.1) and (4.2) with $\eta_1 = 0.01$, $\theta_1 = 1000$, $\theta_2 = 200$, $\Omega\theta_1 = -1.45$, $\Omega\theta_2 = 0.48$ and $P = 0.001$. Fig. 4.14, which is obtained from Eqs. (4.4)-(4.6), shows the intensities and the product of the stationary angular frequencies and the first feedback delay, $\Delta\theta_1$, for different branches of steady state solutions as functions of the second feedback rate $\eta_2$. For the value of $\eta_1$ chosen here, the laser is locked into a stable maximum gain mode (MGM 1 in Fig. 4.14) from $\eta_2 = 5 \times 10^{-3}$ to $\eta_2 = 9.3 \times 10^{-3}$. At $\eta_2 = 9.3 \times 10^{-3}$, this external cavity mode undergoes a Hopf bifurcation which is clearly visible in Fig. 4.13. The laser intensity is then periodically modulated with a frequency $f \cong 3.7$ GHz until $\eta_2 = 10.8 \times 10^{-3}$. Quasiperiodic and chaotic solutions are observed for a small increase of $\eta_2$. Low frequency fluctuations are then observed until $\eta_2 = 12.5 \times 10^{-3}$. At $\eta_2 = 12.5 \times 10^{-3}$, the laser locks into a newly created maximum gain mode (MGM 2 in Fig. 4.14), which is stable. This locking is characterized by a large shift toward lower frequencies, as revealed by Fig. 4.14(b) where we observe that the frequency difference between the MGM 1 and the MGM 2 is much larger than the frequency separation between an ECM and the corresponding antimode[7]. The bifurcation diagram (Fig. 4.13) shows that, for further increases of $\eta_2$, this external cavity mode undergoes a Hopf bifurcation at $\eta_2 = 15.5 \times 10^{-3}$ followed by a sharp bifurcation to quasiperiodic oscillations at $\eta_2 \cong 17.5 \times 10^{-3}$. A new crisis is observed for a barely larger value of the second feedback rate. The bifurcation cascade continues with large shifts toward lower frequencies as the laser locks into new stable maximum gain mode (MGM 3 and MGM 4 in Fig. 4.14).

Fig. 4.15 presents optical spectra that are computed for six increasing values of $\eta_2$ for which the behavior of the laser output is either stationary or periodic. Stationary behaviors are characterized by the presence of a single peak in the optical spectrum [Fig. 4.15 (a), (c) and (e)]. Similarly to what has been observed in our experiment, a second peak appears when the laser intensity is periodically modulated [Fig. 4.15 (b), (d) and (f)]. The frequency separation between two coexisting peaks, or equivalently the modulation frequency, is about 3.72 in Fig. 4.15(b) and (d), and 4.68 GHz in Fig. 4.15 (f).

By following the same approach as in the previous chapter, it is straightforward to demonstrate that mixed modes solutions of the form

$$E(s) \cong A_1 \exp[i(\Delta_1 - \Omega)s] + A_2 \exp[i(\Delta_2 - \Omega)s] \qquad (4.7)$$

are exact solutions of Eqs. (4.1) and (4.2) for $T \to \infty$; $\Delta_1$ and $\Delta_2$ are the frequencies of two different steady state solutions that exhibit the same intensity. In order to turn the difficulty of generalizing the higher order analysis that was followed in the single-feedback case, we use here a numerical continuation method appropriate for delay-differential equations [26]. This method allows the detection of Hopf bifurcation points, to follow stable and unstable branches of periodic solutions and to determine the modification of their stability properties. In the case under study, the direct numerical resolution of Eqs. (4.1) and (4.2) does not allow us to follow the periodic branches, the latter becoming unstable as soon as quasiperiodic behaviors are observed. Therefore, we use here the continuation method for the continuation of these periodic branches.

---

[7] The frequency separation between the MGM 1 and the MGM 2 is about 3.77 GHz for $\eta_2 = 10^{-2}$. The separation between the MGM 1 and the corresponding antimode is about 0.47 GHz.

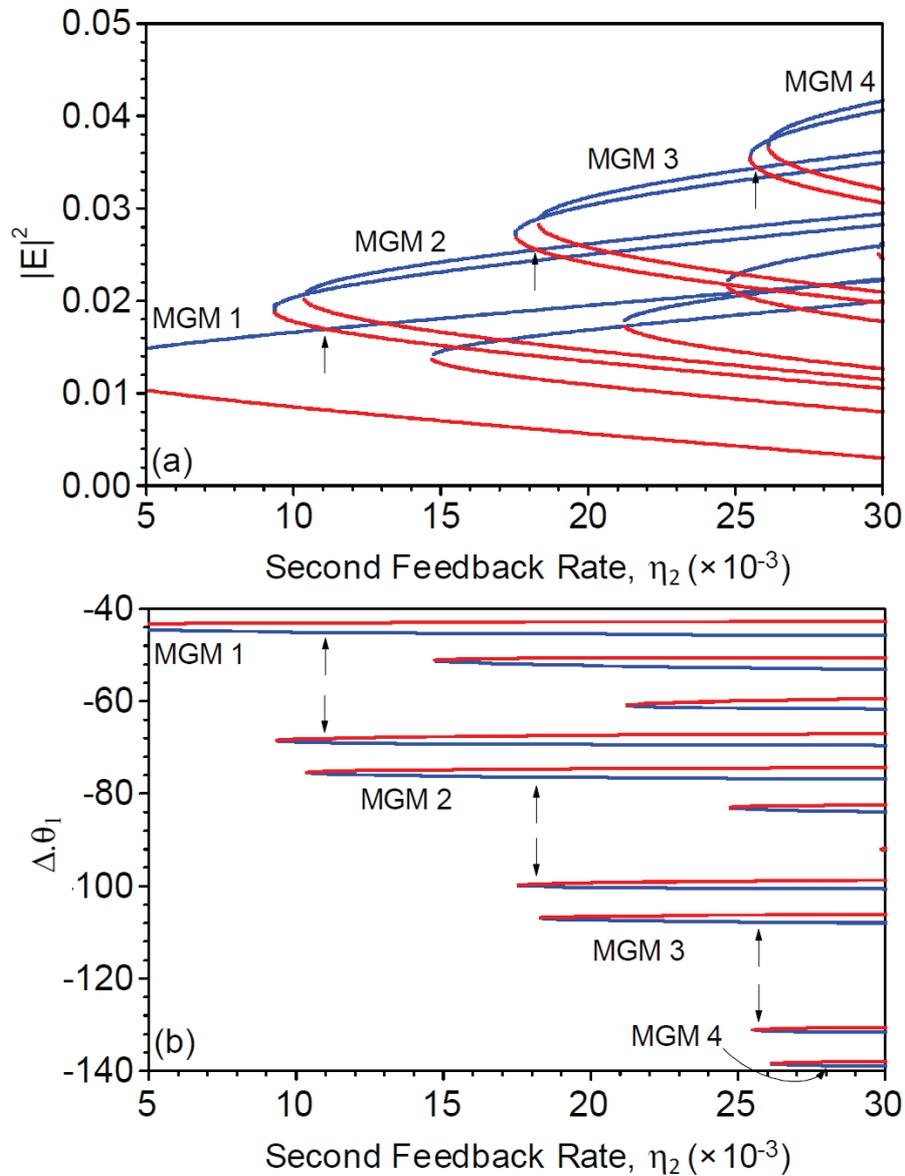

**Fig. 4.14** Intensities (a) and frequencies (b) of different branches of steady state solutions as functions of the feedback rate. Blue and red lines correspond to external cavity modes and antimodes, respectively. If $T \to \infty$, mixed mode solutions of the form (4.7) are possible for the values of $\eta_2$ at which an external cavity mode and an antimode exhibit the same intensity. Arrows indicate some of these points.

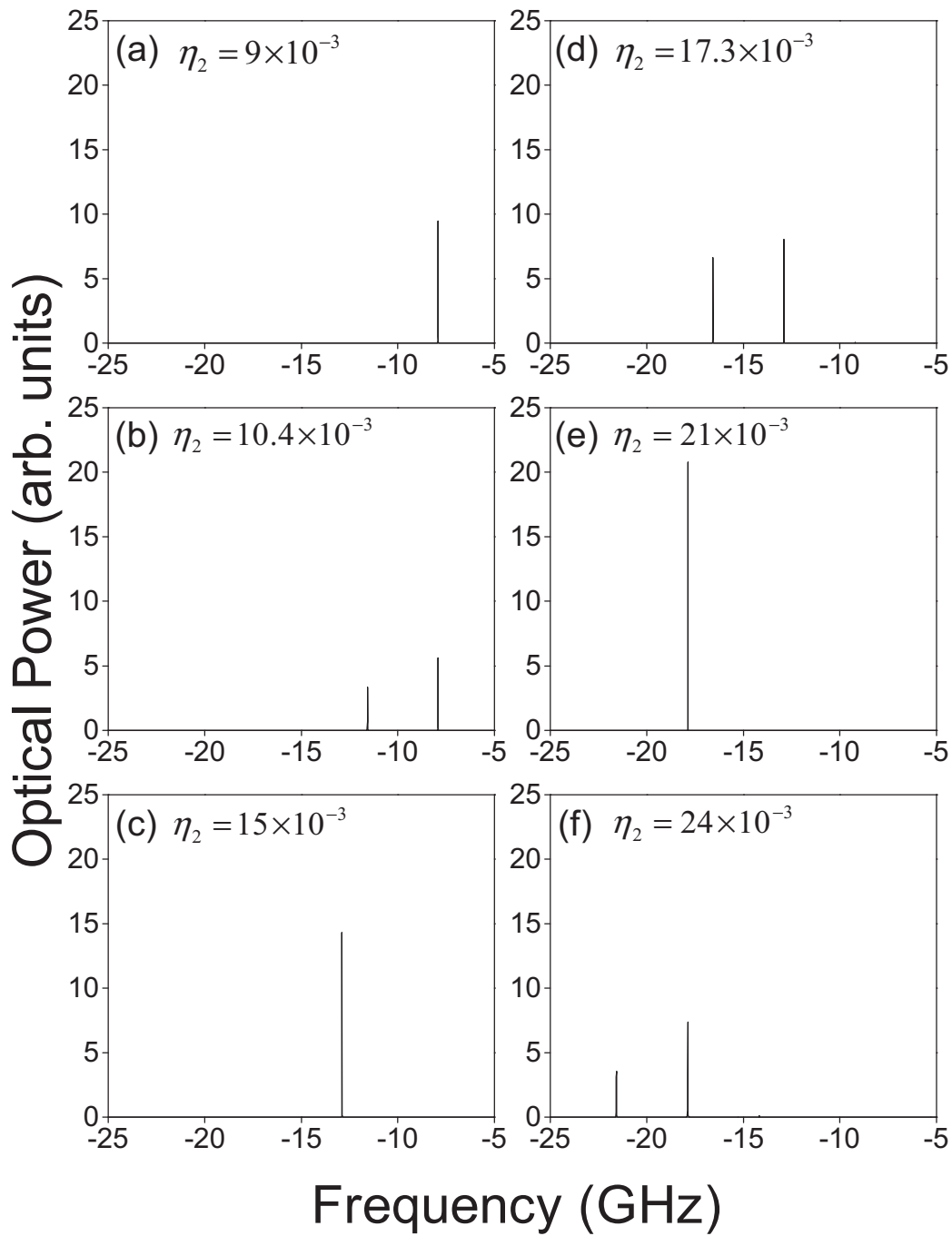

**Fig. 4.15** Numerically computed optical spectra. Spectra (a), (c) and (e) correspond to stationary behaviors. Spectra (b), (d) and (f) correspond to periodic behaviors. The corresponding modulation frequencies are 3.72 GHz, 3.72 GHz and 4.68 GHz, respectively. In each optical spectrum, the horizontal axis has its origin at the threshold frequency of the solitary laser diode.

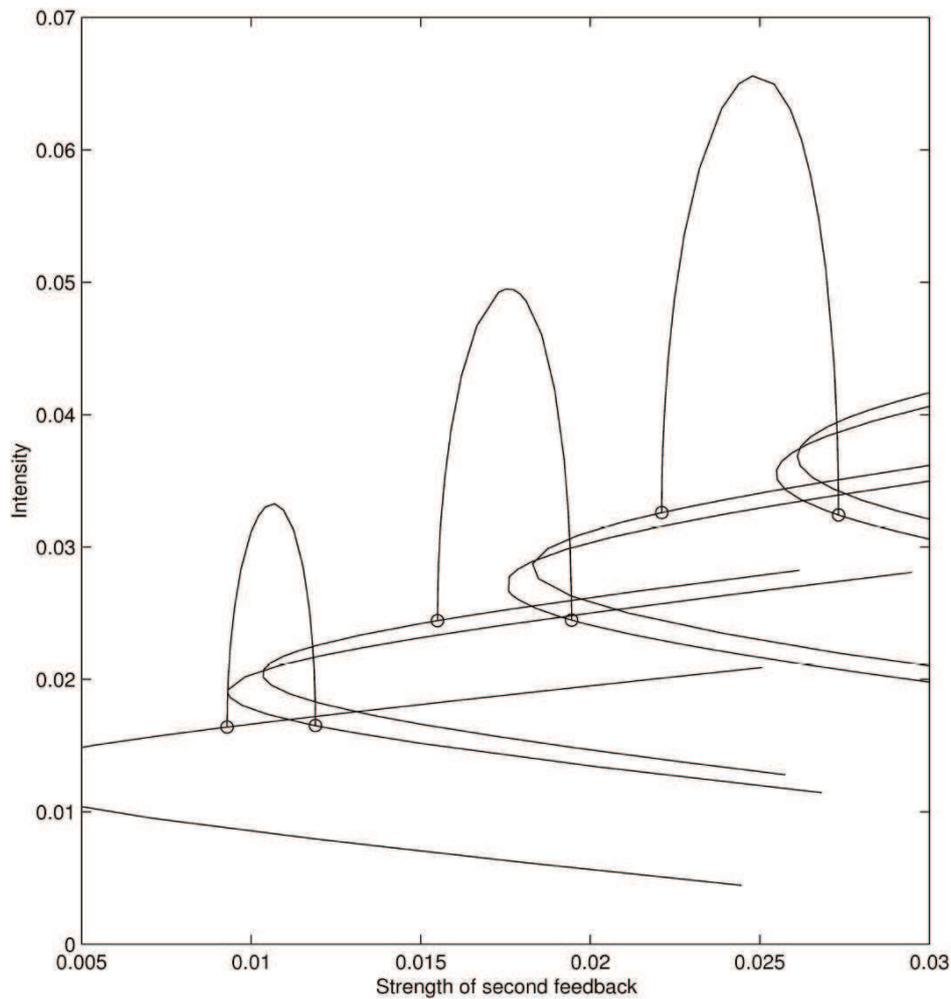

**Fig. 4.16** Bifurcation diagram obtained by using a continuation method. The figure shows the maxima of the intensity $|E|^2$ as a function of $\eta_2$. Branches of periodic solutions that are presented in this diagram connect branches of maximum gain mode with branches of antimodes. The branches of periodic solutions emerge and terminate at two successive Hopf bifurcation points (circles). This figure has been computed by Ir. B. Haegeman (Departement of Computer Science, K.U. Leuven).

Fig. 4.16 shows the bifurcation diagram calculated by using this continuation method. We observe that, as for the single-feedback case, branches of periodic solutions connect external cavity modes and antimodes. Similarly to the single-feedback case (see Chapter 3), all branches of time-periodic solutions emerge and terminate at two distinct Hopf bifurcation points. The first one is located on a stable maximum gain mode and the second on an antimode. The branches of periodic solutions overlap the points where the corresponding external cavity mode and antimode admit the same intensity (these points are indicated by arrows in Fig. 4.14). At these points, mixed mode solutions of the form (4.7) are possible if $T \rightarrow \infty$, and the differences $|\Delta_1 - \Delta_2|$ are approximations of the modulation frequencies of the laser intensity.

## 4.5. Conclusion and perspectives

We have demonstrated numerically and experimentally that a laser diode subject to a single external optical feedback and operating in the low-frequency fluctuation regime can be stabilized by means of a second optical feedback. Unlike many existing methods of control or of stabilization, our technique does not require modification of the laser or first optical feedback parameters. In good agreement with our theoretical study, the experiment reveals that the proposed stabilization scheme works for every accessible first feedback strength and without accurate control of the second cavity length. The method is therefore of practical interest; one may expect that, in many cases, it could be advantageously used instead of expensive optical isolators.

In this chapter, we have considered external cavities whose lengths do not exceed a few tenths of centimeters. In the case of a longer first external cavity, preliminary numerical simulations show that the length of the second cavity must be adequately chosen. A short second cavity (1 cm or less) seems to be propitious for laser stabilization.

In Ref. 27, we have investigated the sensitivity of our stabilization scheme to different values of $\alpha$. For $\alpha = 4$, we have observed that stabilization occurs regardless of the strength of the first feedback and in a broad range of the two feedback delays. However for larger values of $\alpha$ ($\alpha = 5$-$6$), delay and phase of the second feedback must be carefully chosen to observe stabilization. A first interpretation of this limitation of the technique is the following. Similarly to the single optical feedback problem [5,25], the overall number of external cavity modes and antimodes increases as $\alpha$ increases while the distance between them in phase space, in particular near the maximum gain mode, decreases. As a consequence the collision of attractor ruins of external cavity modes with antimodes and therefore LFF crises are more probable. A more exhaustive study of the influence of $\alpha$ is an interesting future research topic.

Our stabilization technique is fundamentally related to the trajectory drift in phase space toward low frequency external cavity modes. This mechanism occurs only when the laser is pumped close to its threshold. Liu and Ohtsubu [11] have shown that a laser pumped far above threshold and subject to a first optical feedback can also be stabilized by using a second optical feedback. In this case, however, the physical mechanisms underlying the laser stabilization are still unknown. The search of these mechanisms is an interesting issue that we shall explore in the future.

The experiment has also revealed time-periodic oscillations with frequencies much larger than those that might a priori be expected when the laser is biased close to threshold. We have demonstrated that periodic oscillations with similar frequencies are found numerically by using the extension of the Lang-Kobayashi equations to the double-feedback configuration. By using a numerical continuation method appropriate for delay-differential equations, branches of periodic oscillations were found to connect pairs of external cavity modes and antimodes. High frequency periodic solutions of the same nature have been found in laser diodes subject to a single-feedback [see Chapter 3 or Ref.19] and, more recently, in vertical-cavity surface-emitting lasers subject to a $\pi/2$ polarization rotating optical feedback [28]. These findings contribute to the idea that the existence of periodic branches connecting pairs of steady state solutions are generic features for high frequency pulsating intensities in laser diodes subject to optical feedback.

The all-optical generation of high-frequency periodic oscillations is of practical interest. The dependence of the frequency on the cavity lengths must however still be examined.

# 5. Low-frequency fluctuation regime in a multimode laser diode subject to coherent optical feedback

It is an old maxim of mine that when you have
excluded the impossible, whatever remains,
however improbable, must be the truth.

Sherlock Holmes, *The Adventure of the Beryl Coronet.*
Sir Arthur Conan Doyle

## 5.1. Introduction

As stated in Chapter 2, the purely deterministic interpretation of the origin of the LFF regime that has been brought forward by Sano [1] in 1994 relies on the Lang-Kobayashi equations [2]. These equations, which are commonly used to model external-cavity laser diodes, assume single-mode operation of the laser diode and weak to moderate external optical feedback. Sano demonstrated that, in the frame of this model, the intensity dropouts are caused by crises between local chaotic attractors and saddle-type antimodes [1]. In his interpretation, the process of intensity recovery is associated with a chaotic itinerancy of the system trajectory in phase space among the attractor ruins of external cavity modes, with a drift towards the maximum gain mode close to which collisions with antimodes occur. Additional numerical investigations of the Lang-Kobayashi equations have anticipated the presence of irregular intensity pulses [3] that have been experimentally confirmed using streak cameras [4]. Moreover, experimental studies have demonstrated the coexistence of the LFF regime with stable emission in a single high-gain external cavity mode for a large range of experimental parameters [5-7]. The existence of stable external cavity modes with high-gain was previously predicted by an analytical study of the Lang-Kobayashi equations [8]. Further studies based on statistical distributions of the time interval between consecutive dropouts have revealed that the numerical results are in better agreement with experiments if the spontaneous emission noise is taken into account in the Lang-Kobayashi equations [9,10].

Although the single-mode Lang-Kobayashi equations are able to anticipate the main features of multiple experiments, recent observations have revealed that multimode operation often occurs within the LFF regime when the Fabry-Perot type laser diodes are not restricted to oscillate in a unique longitudinal mode by a grating or an intracavity etalon [11-13]. Moreover, two statistical studies with similar experimental setups have shown that two qualitatively different behaviors on the picosecond time scale may take place within the LFF regime in multimode laser diodes: Sukow et al. [14] report no qualitative difference between statistics of single-mode and multimode lasers; by contrast, the statistical distributions found by Huyet et al. [15] suggest fast fluctuations in total intensity but no pulses. Up to now, both behaviors have not been observed in a unique experimental setup. Both have however been recently predicted by two different models. The first one, which has been introduced by Viktorov and Mandel [16], is based on the Tang, Statz, and deMars equations [17] adapted to semiconductor lasers [18]. Here, the complex modal field amplitudes are coupled to the modal moments of the carrier number that are commonly assumed not to contribute to the laser diode dynamics. In the frame of this model, the dynamical behavior of a laser with N longitudinal modes is described by N equations for the complex modal fields amplitudes and N equations for the modal moments of the carrier number. Extension to laser diodes with optical feedback predicts that the longitudinal modes of the laser can oscillate in synchronization or in antiphase [16,19]. In Ref. 20, we have investigated numerically the dynamics of a laser diode with optical feedback in the LFF regime by using a multimode extension of the Lang-Kobayashi equations in which the spontaneous emission noise is taken into account and the modal gain is assumed to decrease quadratically from its peak value. We have been able to predict the two qualitatively different behaviors that have been reported in Refs. 14 and 15. In certain parameter ranges, individual modes of the laser are found to emit pulses in a synchronous way with the consequence that the total intensity exhibits trains of pulses as well. In this case, the calculated probability density distribution of the total intensity is similar to the experimental distributions measured by Sukow et al. [14]. However, parameter ranges can also be found within which the longitudinal modes oscillate out-of-phase most of the time. In this case, the total intensity presents fast fluctuations about its average value, the amplitude of these fluctuations being much smaller than those in the individual modes. In agreement with experimental results presented in [15], the associated probability distribution of the total intensity peaks near its average value and falls off at low and high intensities. The presentation of these results is one of the two main purposes of the present chapter.

     In this introduction, we have so far mentioned experimental and theoretical works devoted to multimode laser diodes with global optical feedback, i.e. feedback that affects all longitudinal modes of the laser. However, in many experiments on the LFF regime, a frequency-selective optical component such as an etalon [5] or a grating [21] is placed in the external cavity in order to fulfill the single-mode assumption of the Lang-Kobayashi equations. This device can be adequately tuned so that only one longitudinal mode is selected and re-injected into the laser cavity. The modes that are not subject to the optical feedback are

referred to as *free modes*. In this way, the laser is restricted to oscillate essentially in the *selected mode* [5,21], the free modes being depressed most of the time. However, in similar experimental arrangements, intensity bursts in the free modes have recently been observed simultaneously with the dropouts in the mode selected by the feedback [12,22]. A first heuristic interpretation of the intensity dropouts in the selected mode and bursts in the free modes has been given by Giudici et al. [22]. This interpretation is based on the reduced stability of the selected mode with regards to perturbations occurring in the free modes. More recently, relying on an adaptation of the Tang, Statz and deMars equations [17] to a semiconductor laser [18], the selective mode-induced LFF has been interpreted by Viktorov and Mandel [23] as being associated with a heteroclinic connection between a saddle node and an unstable focus. Moreover, in the frame of this model, chaotic itinerancy with a drift is not a possible mechanism as long as intensity bursts in the free modes are observed [23]. A question that deserves investigation is then the following: may the dynamical instability induced by a selective feedback still be related to the interpretation given by Sano for the conventional LFF? Since many of the techniques for controlling the LFF regime [21,24-28] are linked to the so-called chaotic itinerancy with a drift, answering to this question is indeed an important issue. We have therefore investigated this problem in Ref. 29 by using our multimode extension of the Lang-Kobayashi equations. We present our results in this chapter. We demonstrate numerically that the model indeed predicts the occurrence of intensity bursts in the free modes simultaneously to dropouts in the selected mode. We interpret the intensity bursts in the free modes as being associated with the sudden increases of the carrier population that take place as the selected mode intensity drops. We finally show that LFF induced by the mode-selective feedback may still be associated with collisions of the system trajectory with saddle-type antimodes and chaotic itinerancies with a drift.

This chapter is organized as follows. Section 2 is devoted to the study of the LFF regime in a multimode laser diode subject to global optical feedback: we first present our multimode extension of the Lang-Kobayashi equations; we then study the slow and fast dynamics of the modal intensities of the laser diode and we show, in particular, that the model predicts the coexistence of in-phase and out-of-phase dynamics that have been reported experimentally; we finally discuss the influence of spontaneous emission on our results. In Section 3, we investigate numerically the low-frequency fluctuation regime in a laser diode subject to a mode-selective optical feedback by using our multimode model. Section 4 concludes the chapter.

## 5.2. Multimode laser diode subject to global optical feedback

### 5.2.1. Rate equations

We have already presented in Chapter 2 the multimode extension of the Lang-Kobayashi equations we use here. Assuming a parabolic gain profile and taking the spontaneous emission into account, these equations read

$$\frac{dE_m}{dt} = \frac{1}{2}(1+i\alpha)\left(G_m(N) - \gamma_m\right)E_m + \frac{\kappa}{\tau_{Lm}} E_m(t-\tau) \times \exp(-i\omega_m \tau) + F_m(t), \quad (5.1)$$

$$\frac{dN}{dt} = \frac{I}{e} - \frac{N}{\tau_s} - \sum_{j=-M}^{M} G_j(N)|E_j|^2, \quad (5.2)$$

where

$$G_m(N) = G_c(N - N_0)\left[1 - \left(m\frac{\Delta\omega_L}{\Delta\omega_g}\right)^2\right]. \quad (5.3)$$

with $m = -M \ldots M$.

$E_m(t)$ is the slowly varying complex electric field of the $m$th mode oscillating at the frequency $\omega_m$. $E_m(t)$ is normalized so that $P_m(t) = |E_m(t)|^2$ is the photon number in the $m$th mode. $\alpha$ is the linewidth enhancement factor. $G_m$ and $\gamma_m$ are the mode-dependent gain coefficient and cavity loss, respectively. $\kappa$ is the feedback level and $\tau$ is the round-trip time in the external cavity. $\omega_m \tau$ is the phase shift of the $m$th mode after one roundtrip in the external cavity and $\tau_{Lm}$ the round-trip time of the $m$th optical mode inside the diode cavity. $F_m(t)$ is a Langevin noise force that accounts for spontaneous emission noise. $N(t)$ is the number of electron-hole pairs inside the active region and $\tau_s$ is their life time. $N_0$ is the transparency value of $N$. $I$ is the injection current and $I_{th}$ the threshold current of the solitary laser. $e$ is the magnitude of the electron charge. $G_c$ and $m = 0$ are the gain coefficient and the longitudinal mode number at the gain peak. $\Delta\omega_L$ and $\Delta\omega_g$ are the longitudinal mode spacing and the gain width of the laser material, respectively. In our calculations, we consider seven active optical modes (i.e. $M = 3$)[1] and assume $\gamma_m$ and $\tau_{Lm}$ to be mode-independent. In this approximation, the mode spacing is given by $\Delta\omega_L = 2\pi / \tau_{Lm}$. We use typical values for the laser diode parameters: $\alpha = 4$, $\gamma_m = 5 \times 10^{11}$ s$^{-1}$, $\tau_s = 2$ ns, $G_c = 1 \times 10^4$ s$^{-1}$, $m_c = 0$, $N_0 = 1.1 \times 10^8$, $\Delta\omega_g = 2\pi \times 4.7$ THz. The noise level is determined through $<F_m^*(t) F_n(t')> = R_{sp} \delta_{mn} \delta(t-t')$, where the spontaneous emission rate $R_{sp} = 1.1 \times 10^{12}$ s$^{-1}$ is assumed to be mode-independent. The phase shifts of the individual modes are given by $\omega_m \tau = \omega_c \tau + m\Delta\omega_L \tau$ where $\omega_c \tau$ is the phase shift of the central mode. In the two next sections, we choose $\omega_c \tau = 0$ mod $2\pi$. We choose $\tau_{Lm} = 8.3$ ps so that the phase shift is different for every mode.

### 5.2.2. Slow dynamics of the modal intensities

To model the limited bandwidth of the photodetectors commonly used in actual experiments, the time traces of the modal intensities and of the total output of the laser obtained in this section have been averaged over 2 ns. Numerical results in good agreement with those presented here have been reported very recently in Ref. 31, using a similar model.

The LFF regime can be observed in a large range of feedback parameters when the laser is pumped close to its solitary threshold. When the laser diode operates in the LFF regime, the three longitudinal modes -1, 0 and 1 are the brightest and mode 0, which is located at the maximum of the gain curve, is generally the dominant one. By contrast, the side-modes –3, -2, 2 and 3 are depressed most of the time. Fig. 5.1 displays the dynamical behavior of the seven longitudinal modes. The feedback parameters are $\kappa = 0.1$ and $\tau = 5$ ns; the injection current is $I = 1.006 \times I_{th}$. Low-frequency fluctuations are observed in all modes. The intensities of the three brightest ones drop simultaneously. Just after a dropout, the intensity of each mode starts to rise. But, as a result of competition between the laser modes, the mode with the largest gain continues to grow almost steadily [trace (d)], draining most of the electron-hole pairs, while the photon number in the other modes saturates and begins to decrease until the next dropout event [traces (a)-(c) and (e)-(g)]. For the injection current chosen here, the time interval between two consecutive dropouts is long enough to allow modes -3, -2, 2 and 3 to reach the spontaneous-emission level. This side-mode extinction has been experimentally observed very recently [32].

In order to compare our numerical results with the experimental observations reported in [30], we present in Fig. 5.2 the temporal evolution of the intensity in the main mode [trace (a)], the intensity shared by all modes except the main mode [trace (b)] and the total intensity [trace (c)]. The first time trace shows that the dropouts in the total intensity are followed by gradual recoveries until a plateau is reached. The number of photons shared by all modes except the dominant one, increases rapidly in a first step but, as a consequence of the steady increase of the main mode, starts to decrease as soon as the total output of the laser reaches a plateau. These numerical results are in qualitative agreement with experimental traces displayed in Fig. 5.3.

---

[1] The results remain the same when more modes are considered. Indeed, additional side-modes are depressed most of the time.

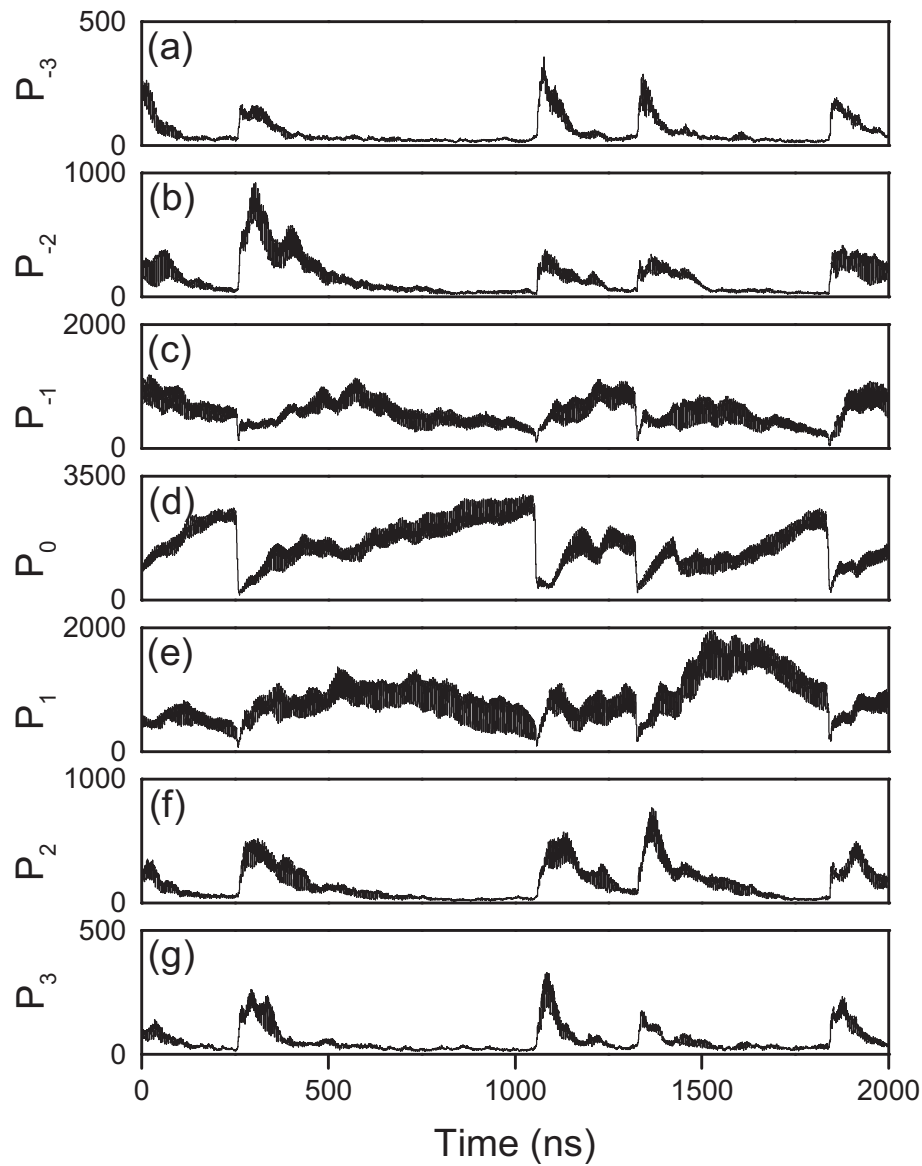

**Fig. 5.1** Intensity time traces of the 7 longitudinal modes. The time traces have been averaged over 2 ns to model the limited bandwidth of the usual photodetectors used in actual experiments. $\kappa = 0.1$, $\tau = 5$ ns and $I = 1.006 \times I_{th}$.

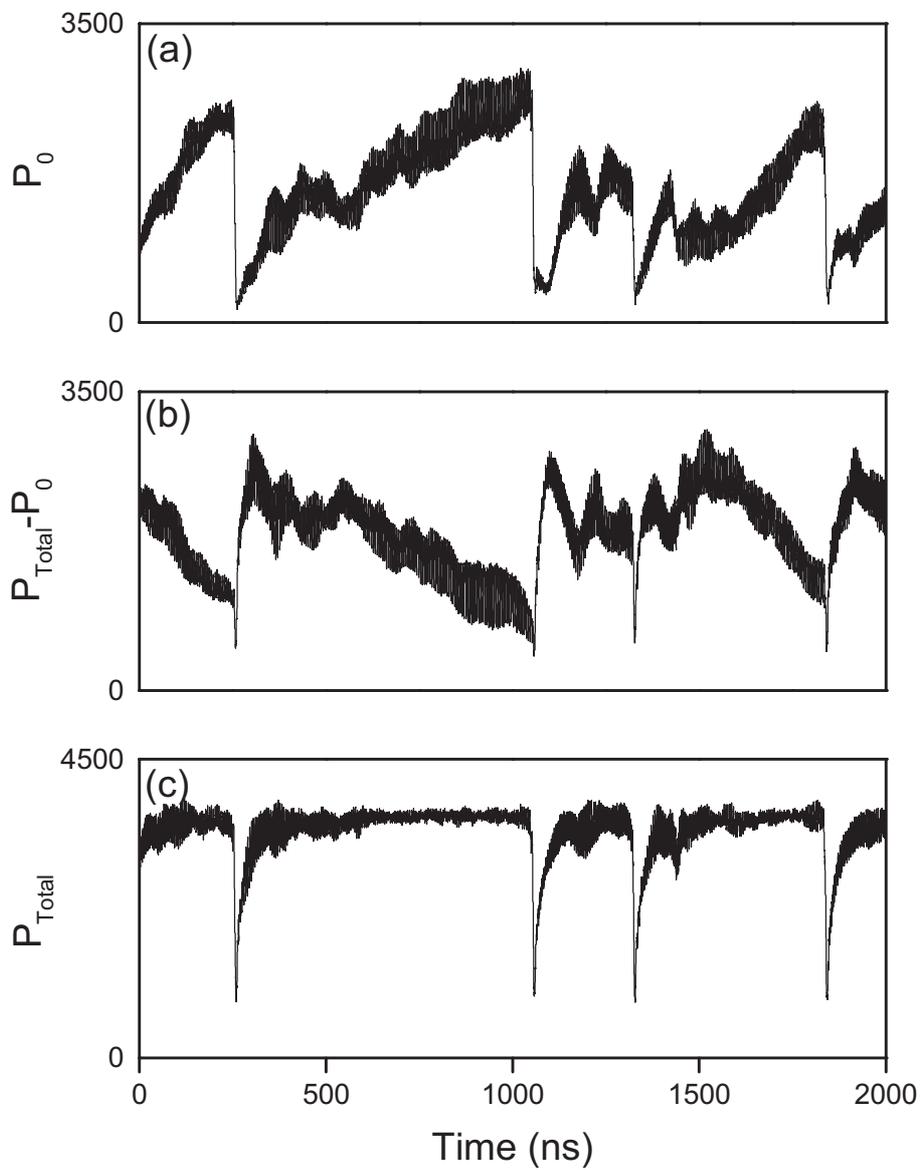

**Fig. 5.2** Time traces of the intensity of the main mode [trace (a)], the intensity shared between all modes except the dominant one [trace (b)], and the total output of the laser [trace (c)]. Traces (b) and (c) have been calculated from the traces presented in Fig. 5.1.

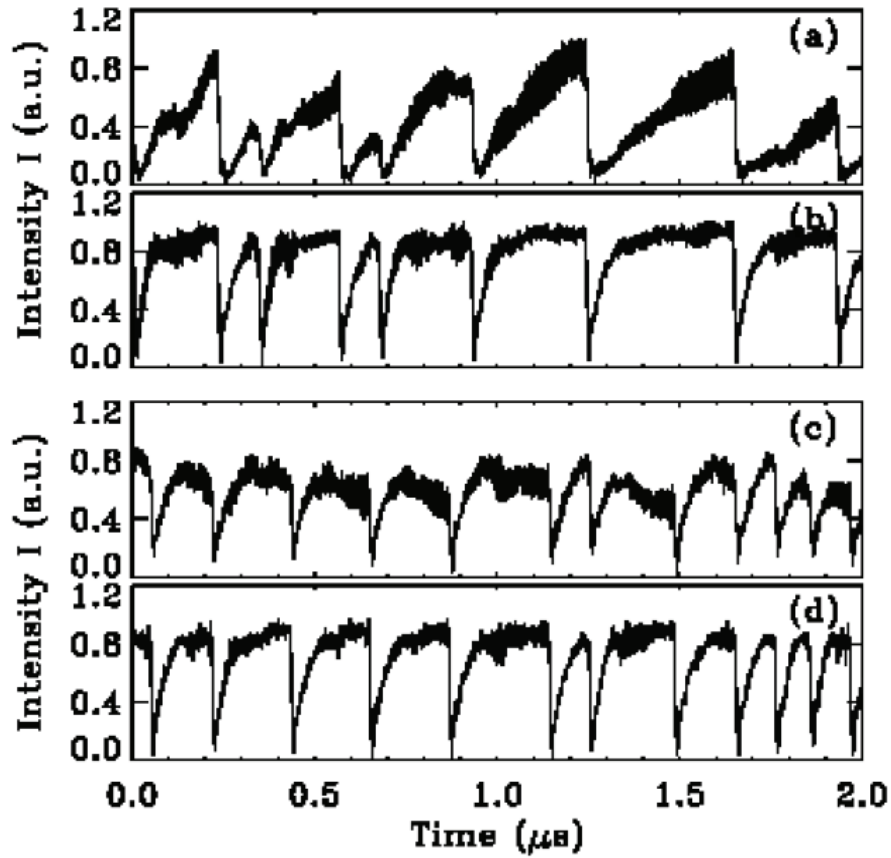

**Fig. 5.3** Experimental recording of low-frequency fluctuations in a multimode laser diode. Traces (a) and (b) show simultaneous recordings of the dominant mode behavior and the total output of the laser. Traces (c) and (d) show simultaneous recordings of the intensity shared by all modes except the dominant mode and the total intensity emitted by the laser. The experiment was realized with a Toshiba TOLD 9140 20 mW laser diode pumped at $I = 1.02 \times I_{th}$ with $I_{th} = 39.6$ mA. The feedback induced threshold reduction was of 9.68% and the external cavity length was $L_{ext} = 82$ cm corresponding to an external cavity roundtrip time $\tau \approx 5.5$ ns. Figure and data are extracted from Ref. 30.

## 5.2.3. Coexistence of in-phase and out-of-phase dynamics in the low-frequency fluctuation regime

In order to show that two different behaviors on the picosecond time scale can be found, the following will be focused on the dynamics of the total and modal intensities of the laser and their corresponding statistics for two different values of the injection current, $I = 1.12 \times I_{th}$ and $I = 1.08 \times I_{th}$, all other parameters being kept constant. The feedback strength and delay are respectively $\kappa = 0.32$ and $\tau = 3$ ns in the present section.

Fig. 5.4(a) displays the average time trace of the total intensity for $I = 1.12 \times I_{th}$. The average intensity increases steadily with time until it suddenly drops to a minimum value. After a dropout, the average intensity gradually recovers only to drop again after a random time. On the picosecond time scale, the intensities of the individual longitudinal modes

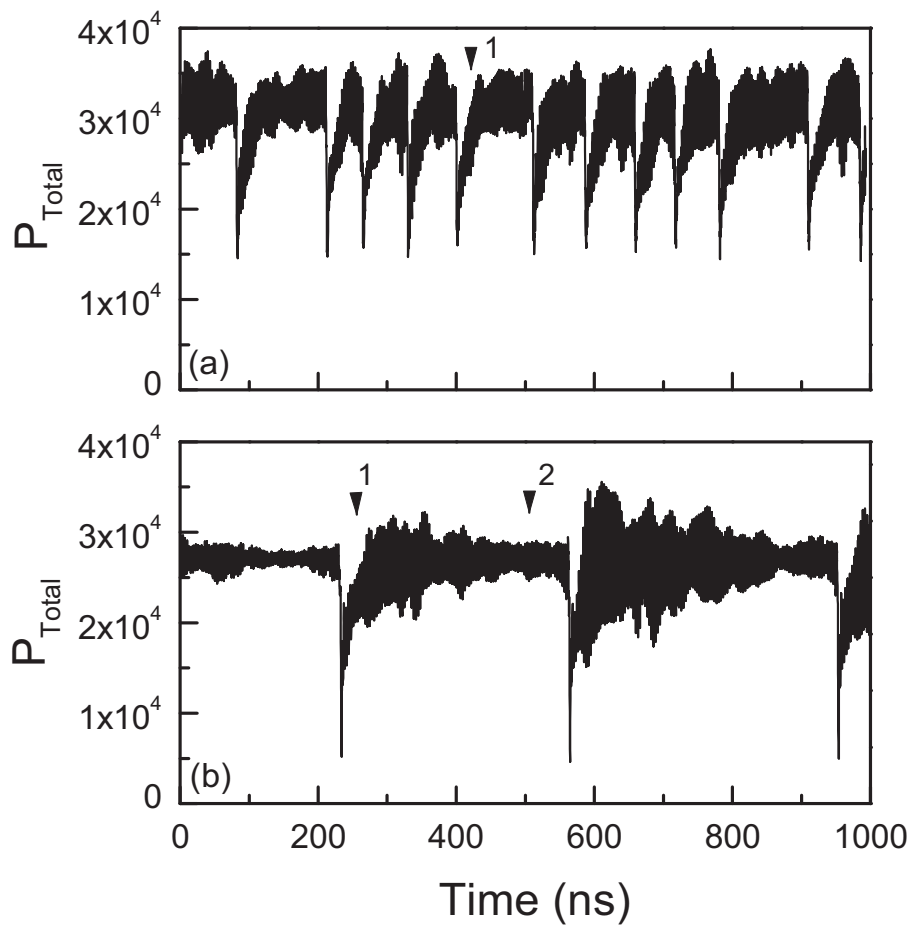

**Fig. 5.4** Time traces of the laser total intensity for two different injection currents: (a) $I = 1.12 \times I_{th}$; (b) $I = 1.08 \times I_{th}$. The time traces have been averaged over 2 ns.

exhibit fast intensity pulses [Fig. 5.5(a)-(c) where only the three brightest modes are shown]. Furthermore, these modes oscillate in phase: they emit pulses with different amplitudes but synchronously. Pulses with high intensity amplitude in one mode coincide most of the time with pulses with low intensity in the other modes. As a consequence, the total intensity exhibits a train of pulses as well [Fig. 5.5(d)], but with more regular amplitude, showing no qualitative difference with respect to a single mode laser. The modal synchronization is observed during the entirety of the recovery process, immediately after as well as before a dropout. In order to compare our numerical results to the experiments, we have calculated the

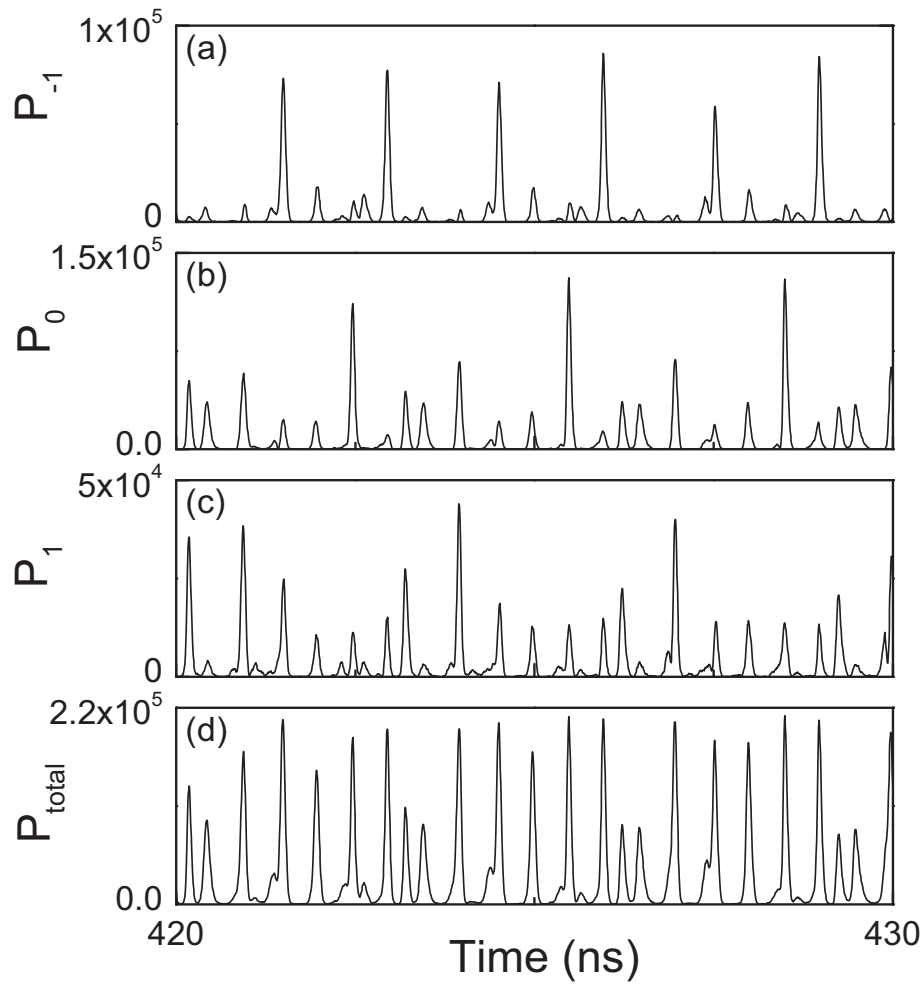

**Fig. 5.5** (a)-(c) Time traces of the unaveraged photon numbers in the three brightest modes; (d) time trace of the total number of photons emitted by the laser. The time segment of the traces corresponds to arrow 1 in Fig. 5.4(a). $I = 1.12 \times I_{th}$.

probability density distributions for the total intensity and for the individual longitudinal modes. Fig. 5.6 shows that the distribution corresponding to the total intensity, similarly to the distributions of the individual modes, is maximum at very low intensity and decreases monotonically with increasing intensity. These distributions are characteristic of pulsating behaviors with intensity peaks several times the mean intensity. The statistical distribution for the total intensity is in good agreement with the measurements that are reported in Ref. 14 and that have been achieved by two different teams with two different experimental setups. Part of these results is shown in Fig. 5.7.

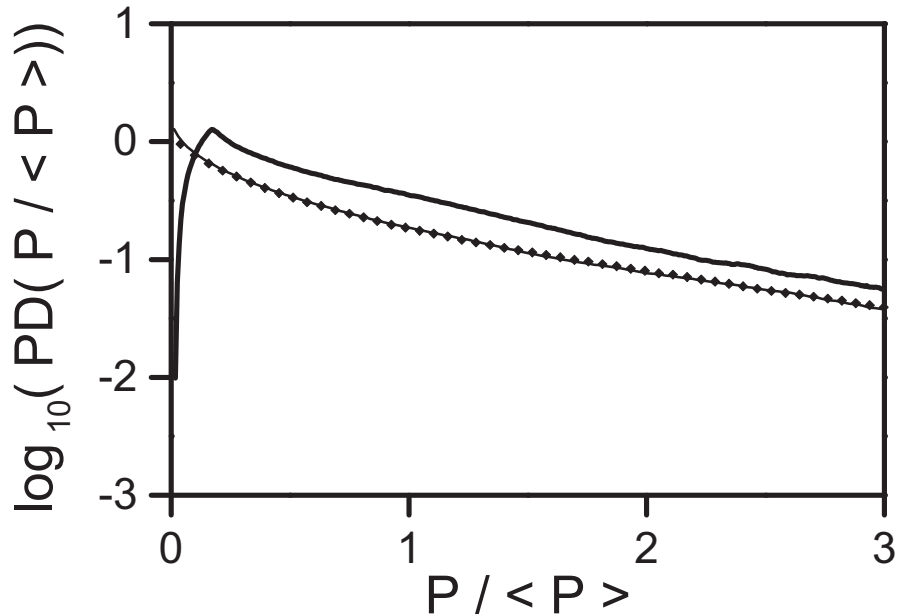

**Fig. 5.6** Probability density distributions of the total laser intensity (thick line), of the center mode (diamonds) and of mode 3 (thin line) intensities. The distributions have been calculated from 5 time series of 20-µs length. $I = 1.12 \times I_{th}$.

For $I = 1.08 \times I_{th}$, the mean time interval between two successive dropouts is considerably longer and the total intensity recoveries are not interrupted by dropouts [Fig. 5.4(b)]. In contrast to the previous case, the average total intensity oscillates around a constant value during long time intervals with respect to the duration of the recovery process. During all the recovery process, the individual modes are pulsing in a synchronous way, as in the previous case, and trains of pulses are also observed in the total intensity (Fig. 5.8). However, as the recovery process ends and the average intensity saturates, the pulses broaden. The modal synchronization is progressively lost and the individual modes are finally observed to oscillate out of phase [Fig. 5.9(a)-(c)] with repetition of an almost similar pattern each round-trip time. An exchange of energy between the individual modes then occurs, leading to a total intensity which is rarely small and exhibits fast fluctuations with smaller amplitudes than those of the individual modes about its mean value [Fig. 5.9(d)]; but no more pulses are observed until the next dropout. Consequently, the probability distribution of the total intensity peaks near its average value and falls off at low and high intensities (Fig. 5.10). By contrast, the statistical distributions for the individual modes peak at low intensity and exhibit tails, as those present high intensity pulses. These results agree very well with the probability distribution experimentally measured by Huyet et al. [15] and presented in Fig. 5.11.

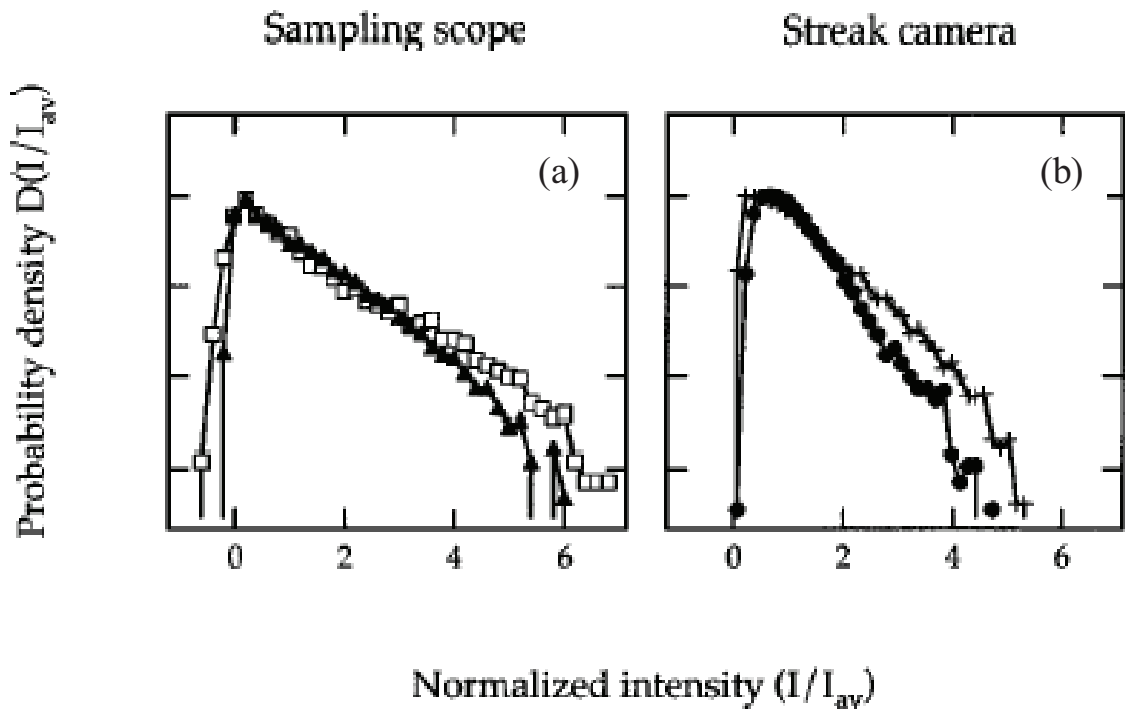

**Fig. 5.7** (a) Probability density distributions experimentally measured with a Sharp LT015MDT laser diode pumped at $I = 1.10 \times I_{th}$ (open squares) and $I = 1.50 \times I_{th}$ (triangles) with $I_{th} = 41.3$ mA. The feedback level is not specified. The external cavity length was $L_{ext} = 22$ cm corresponding to an external cavity roundtrip time $\tau = 1.46$ ns. The laser intensity was detected by a 25 GHz bandwidth photodetector connected to a 20 GHz digital sampling oscilloscope. (b) Probability density distributions experimentally measured with a Hitachi HLP1400 laser diode pumped at $I = 1.01 \times I_{th}$ (circles) and $I = 1.05 \times I_{th}$ (crosses) with $I_{th} = 57.2$ mA. The feedback level is not specified. The external cavity roundtrip time was $\tau = 3$ ns. The laser intensity was detected by a single-shot streak camera with a bandwidth from dc to more than 50 GHz. This figure is a reproduction of Fig. 4(d) and (f) presented in Ref.14.

The calculation of a complete map of dynamical behaviors in the feedback-current plane, similar to the map determined experimentally in Ref. 5 for a single mode laser, is out of reach with our numerical means for two reasons. Firstly, the smallness of the integration step (typically 1 ps) compared to the time span over which the simulations must be carried (several tenths of μs) leads to a prohibitive calculation time. Secondly, the transition between the LFF and the coherence collapse regimes, on the one hand, and the transition between in-phase and out-of-phase dynamics, on the other hand, are not sharply defined; an automatic search of the regions of parameters space corresponding to the different behavior is thus difficult, if not impossible.

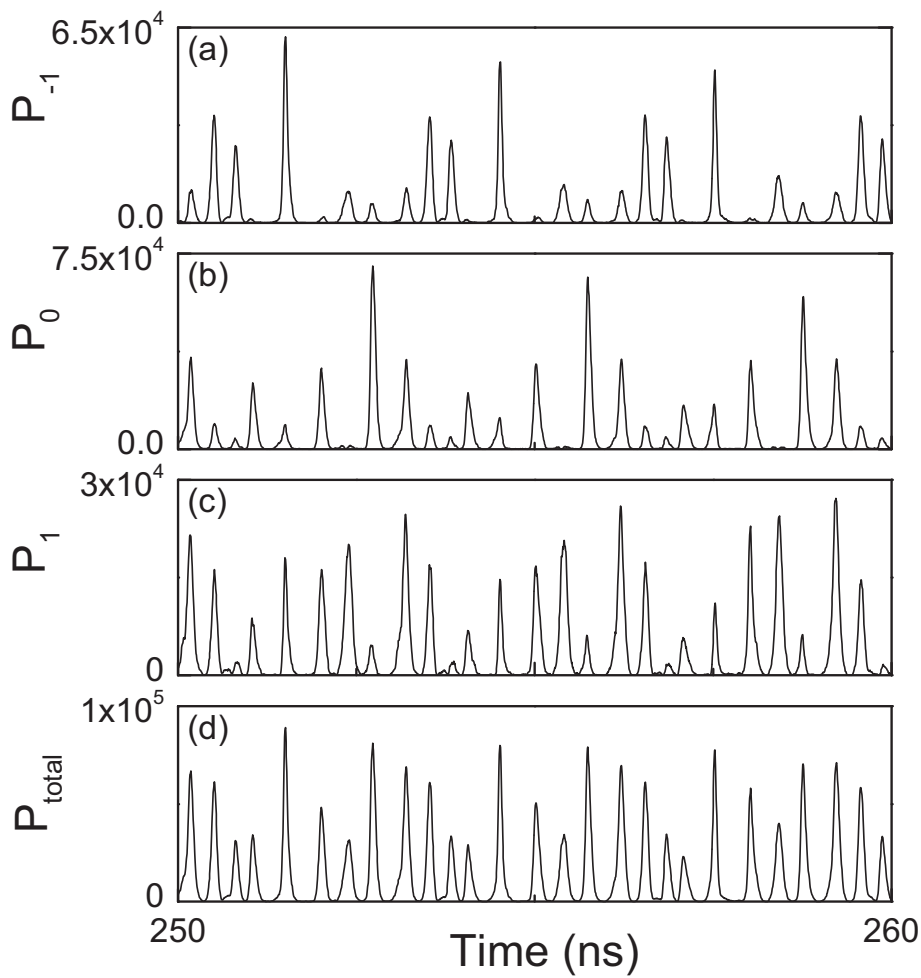

**Fig. 5.8** Same as Fig. 5.5, but with $I = 1.08 \times I_{th}$. The time segment of the traces corresponds to arrow 1 in Fig. 5.4(b).

However, extensive numerical simulations reveal that the two different kinds of behaviors we have pointed out can be found in large ranges of operating parameters, i.e., laser injection current, feedback level and external cavity length. By keeping constant the feedback parameters and decreasing continuously the injection current $I$, which is the most easily tunable parameters in real experiments, we find that the mean time interval between two consecutive intensity dropouts increases, in good agreement with experiments involving single and multimode laser diodes [6,33]. The transition between the two kinds of behaviors, that is in-phase dynamics alone and the coexistence of in-phase and out-of-phase dynamics, is observed in a small range of injection currents. For the feedback parameters chosen in this section, the transition takes place around $I = 1.10 \times I_{th}$. At higher currents, the first behavior we have depicted, i.e. a total intensity exhibiting pulses, always occurs. At lower currents, the average intensity can recover completely after a dropout and reach a constant level around which it oscillates during a long time until the next dropout. During these long time intervals, the longitudinal modes oscillate out-of-phase.

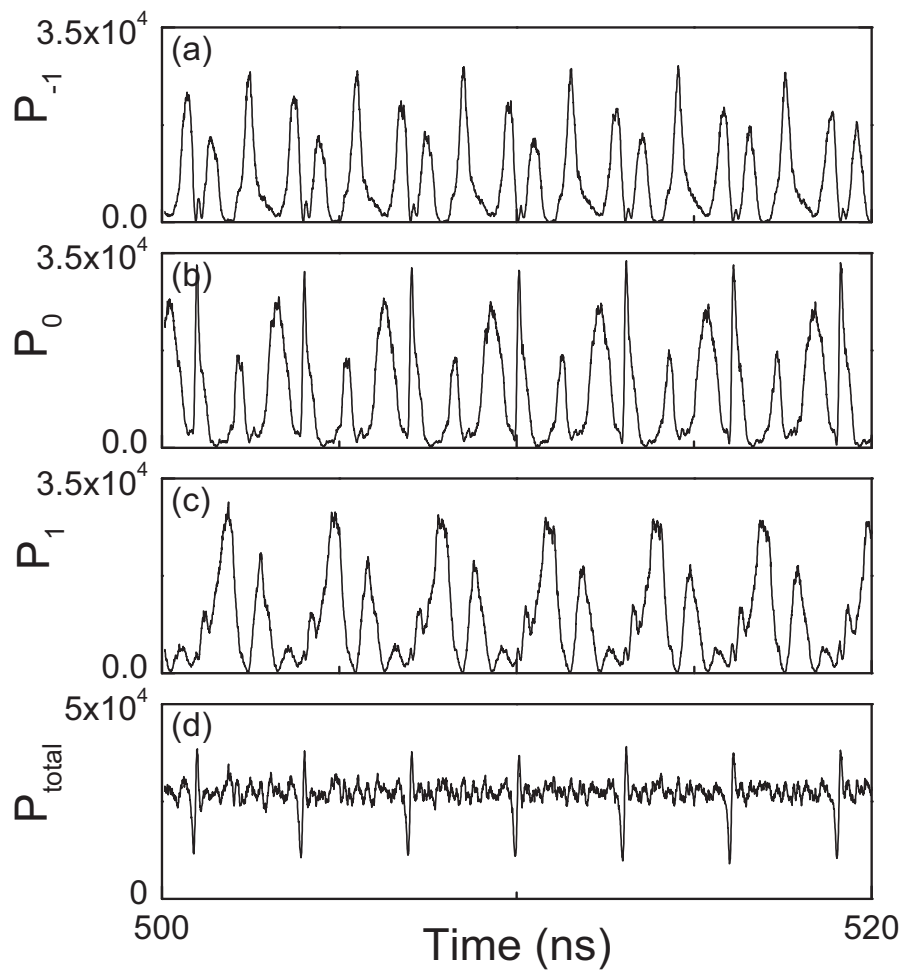

**Fig. 5.9** Same as Fig. 5.5, but with $I = 1.08 \times I_{th}$. The time segment of the traces corresponds to arrow 2 in Fig. 5.4(b).

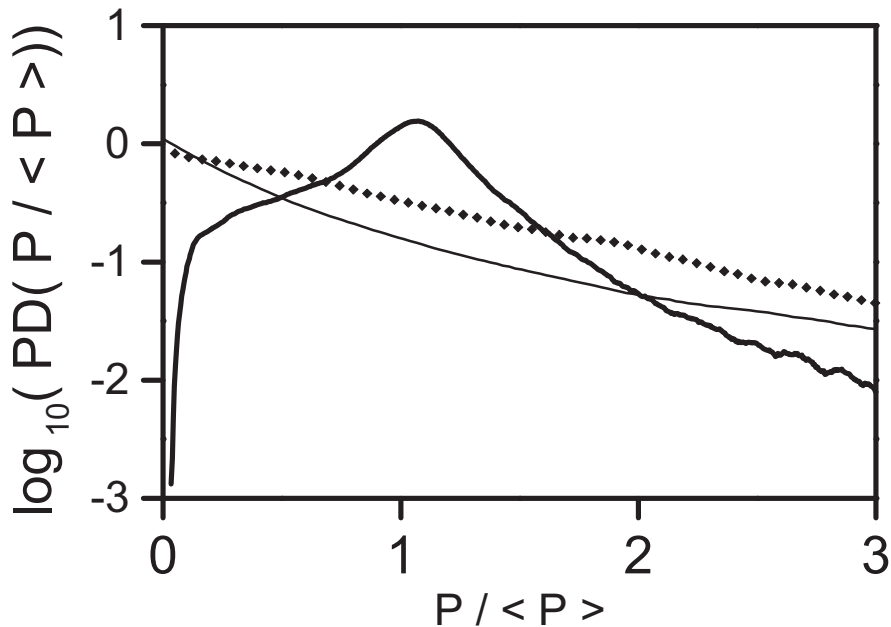

**Fig. 5.10** Probability density distributions of the total laser intensity (thick line), of the center mode (diamonds) and of mode 3 (thin line) intensities. The distributions have been calculated from 5 time series of 20-µs length. $I = 1.08 \times I_{th}$.

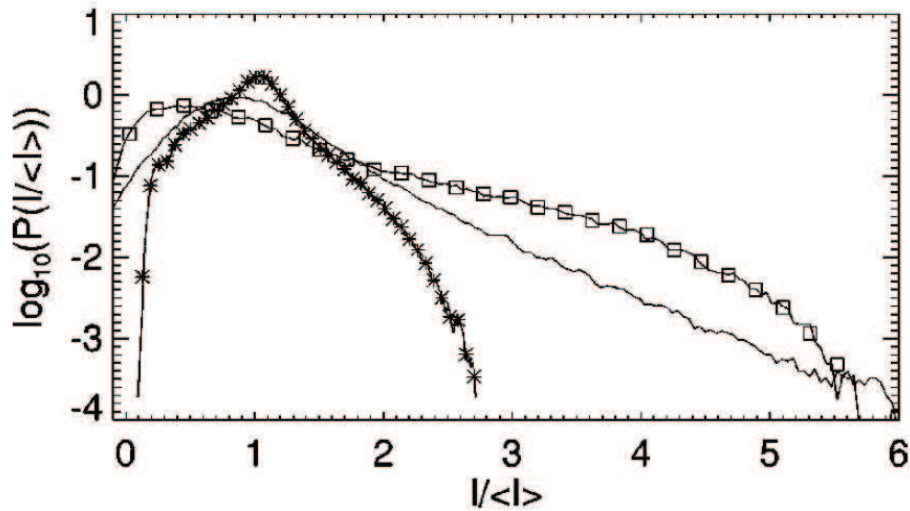

**Fig. 5.11** Measured probability density distributions of a Hitachi HLP1400 laser diode with a 0-15 GHz bandwidth. Stars: statistical distribution of the total intensity in case of multimode operation. Solid line: statistical distribution of the intensity in one longitudinal mode in case of multimode operation. Square: statistical distribution of the laser intensity when the laser is restricted by an additional etalon to oscillate in a unique longitudinal mode. The injection current and the feedback level are not specified. The external cavity length was $L_{ext} = 50$ cm corresponding to an external cavity roundtrip time $\tau = 3.3$ ns. This figure is a reproduction of Fig. 2 presented in Ref.15.

## 5.2.4. Role of the spontaneous emission

Spontaneous emission plays a crucial role in the multimode dynamics within the LFF regime and in the appearance of the out-of-phase dynamics. Indeed, the range of injection current and feedback parameters within which out-of-phase dynamics can be observed becomes smaller and smaller as the mean spontaneous emission rate is reduced. In the limiting case $R_{sp} = 0$, the laser is found to share the total intensity among the three central modes only and does not operate in the surrounding modes. The recovery of the total average intensity is always interrupted by dropouts before it can reach a plateau and the three remaining modes oscillate always in-phase. Although it is initiated by spontaneous emission, which is a stochastic process, the repetition of an almost similar pattern each round-trip (Fig. 5.9) suggests that the out-of-phase dynamics is ruled by deterministic mechanisms in the frame of the model under study. Hence, an important question is to know whether the occurrence of the out-of-phase dynamics is inherently related to the stochastic nature of the spontaneous emission. In the affirmative, the out-of-phase behavior might be interpreted as a transient that is sustained by the noise acting as a perturbation.

In order to answer that question, we reformulate our model equations in terms of the photon number $P_m(t)$ and electric field phase $\phi_m(t)$ associated to each longitudinal mode, and the carrier number $N(t)$. If the relation $E_m(t) = \sqrt{P_m(t)} \exp(i\phi_m(t))$ is inserted into Eqs. (5.1)-(5.3), one obtains

$$\frac{dP_m(t)}{dt} = [G_m(N) - \gamma_m]P_m(t) + \frac{2\kappa}{\tau_{Lm}}\sqrt{P_m(t)P_m(t-\tau)} \\ \times \cos[\phi_m(t) - \phi_m(t-\tau) + \omega_m \tau] + R_{sp},$$  (5.4)

$$\frac{d\phi_m(t)}{dt} = \frac{\alpha}{2}[G_m(N) - \gamma_m] - \frac{\kappa}{\tau_{Lm}}\sqrt{\frac{P_m(t-\tau)}{P_m(t)}} \\ \times \sin[\phi_m(t) - \phi_m(t-\tau) + \omega_m \tau],$$  (5.5)

$$\frac{dN}{dt} = \frac{I}{e} - \frac{N}{\tau_s} - \sum_{j=-M}^{M} G_j(N)P_j$$  (5.6)

where the modal gain $G_j(N)$ is given by Eq. (5.3). In these equations, spontaneous emission is taken into account by means of the term $R_{sp}$ that is the mean spontaneous emission rate. By contrast, stochastic fluctuations arising from spontaneous emission processes are neglected.

Figures 5.12-5.14 are obtained for the same feedback parameters, injection current and spontaneous emission rate as Figs. 5.8-5.10, i.e. $\kappa = 0.32$, $\tau = 3$ ns, $I = 1.08 \times I_{th}$ and $R_{sp} = 1.1 \times 10^{12}$ s$^{-1}$. Fig. 5.12 shows that, after each dropout, the total intensity recovers and reaches a plateau around which it fluctuates until the next dropout. Similarly to what has been reported in Section 5.2.3, pulses are observed as well in the individual modes as in the total output of the laser during the recovery process (Fig. 5.13). Once the average total output of the laser has reached a plateau, the modal intensities exhibit out-of-phase fluctuations that lead to much smaller fluctuations in the total intensity (Fig. 5.14). Since both sets of equations (5.1)-(5.2) and (5.4)-(5.6) predict out-of-phase dynamics, we conclude that the latter is not intrinsically related to the stochastic nature of the spontaneous emission.

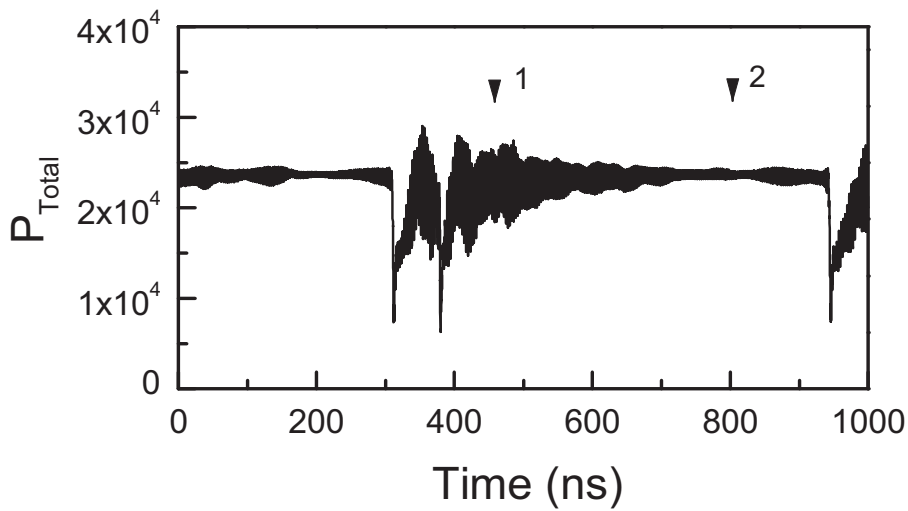

**Fig. 5.12** Time trace of the laser total intensity calculated from Eqs. (5.4)-(5.6) for $I = 1.08 \times I_{th}$. The time trace has been averaged over 2 ns.

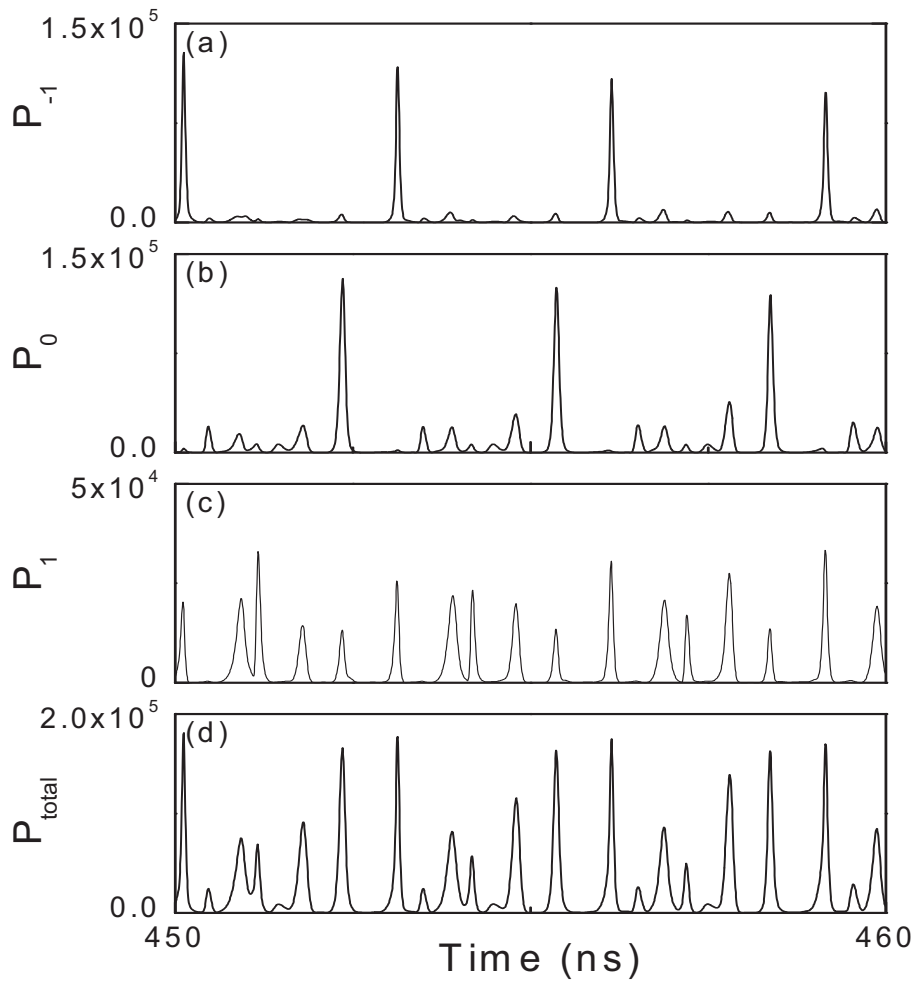

**Fig. 5.13** In-phase dynamics anticipated by Eqs. (5.4)-(5.6) for $I = 1.08 \times I_{th}$. (a)-(c) Time traces of the unaveraged photon numbers in the three brightest modes; (d) time trace of the total number of photons emitted by the laser. The time segment of the traces corresponds to arrow 1 in Fig. 5.12.

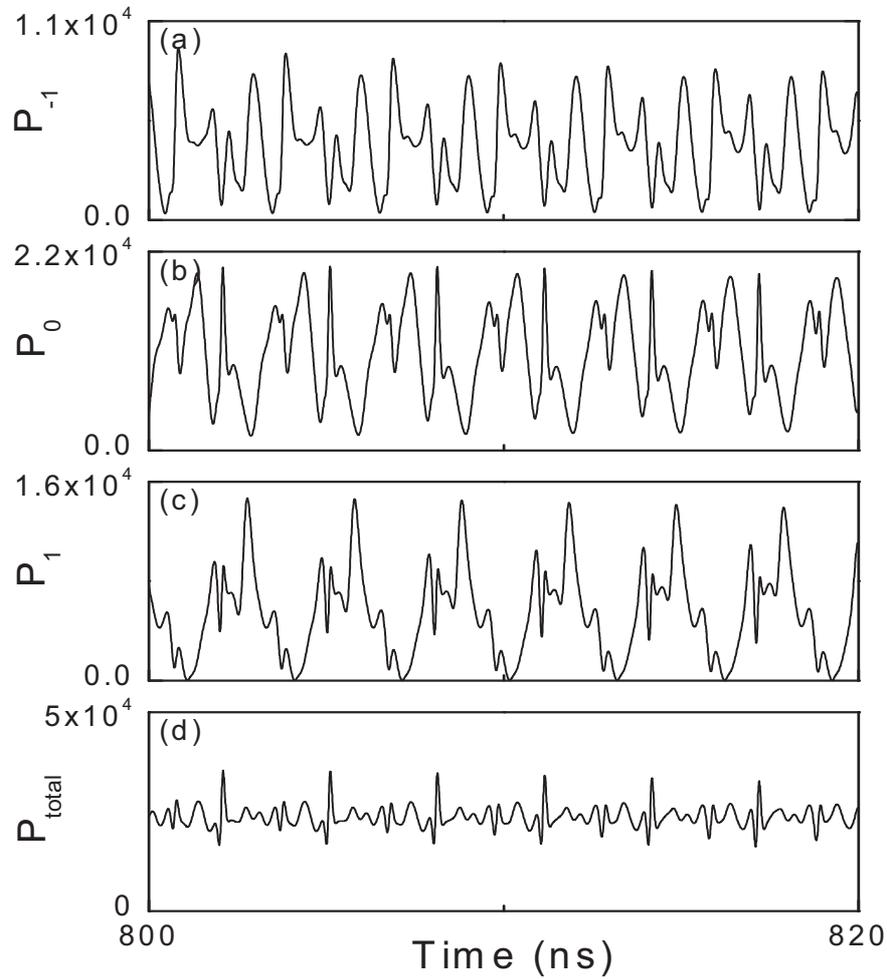

**Fig. 5.14** Out-of-phase dynamics anticipated by Eqs. (5.4)-(5.6) for $I = 1.08 \times I_{th}$. (a)-(c) Time traces of the unaveraged photon numbers in the three brightest modes; (d) time trace of the total number of photons emitted by the laser. The time segment of the traces corresponds to arrow 2 in Fig. 5.12.

## 5.3. Multimode laser diode subject to a mode-selective optical feedback

### 5.3.1. Rate equations

We study the dynamical behavior of a multimode laser diode subject to a mode-selective optical feedback by using equations in all points similar to Eqs. (5.4)-(5.6), the feedback level excepted. These equations, expressed in terms of the photon number, the electric field phase in each longitudinal mode and the carrier number, respectively $P_m$, $\phi_m$ and $N$ where $m$ is the mode number, are

$$\frac{dP_m}{dt} = (G_m(N) - \gamma_m)P_m + \frac{2\kappa_m}{\tau_{Lm}}\sqrt{P_m(t)P_m(t-\tau)} \times \cos(\phi_m(t) - \phi_m(t-\tau) + \omega_m\tau) + R_{sp},  \quad (5.7)$$

$$\frac{d\phi_m}{dt} = \frac{\alpha}{2}(G_m(N) - \gamma_m) - \frac{\kappa_m}{\tau_{Lm}}\sqrt{\frac{P_m(t-\tau)}{P_m(t)}} \times \sin(\phi_m(t) - \phi_m(t-\tau) + \omega_m\tau),  \quad (5.8)$$

$$\frac{dN}{dt} = \frac{I}{e} - \frac{N}{\tau_S} - \sum_{j=-M}^{M} G_j(N)P_j,  \quad (5.9)$$

where

$$G_m(N) = G_c(N - N_0)\left[1 - \left(m\frac{\Delta\omega_L}{\Delta\omega_g}\right)^2\right]  \quad (5.10)$$

with $m = -M \ldots M$.

In Eqs. (5.7)-(5.8), $\kappa_m \equiv \kappa\delta_{mn}$ where $n$ is the index of the mode that is selected and re-injected into the laser cavity. The other parameters are defined in Section 5.2.1. We again consider seven active optical modes and assume $\gamma_m$ to be mode-independent. The parabolic gain profile is centered on the mode 0. In this study, we choose the longitudinal mode -1 as that being selected and fed back into the laser (i.e. $n = -1$). The dynamical behavior of the laser remains qualitatively the same regardless of the mode being selected. Spontaneous emission is taken into account by the mean spontaneous emission rate term $R_{sp}$. Stochastic fluctuations arising from the spontaneous emission process are however neglected. Except for the round-trip time inside the diode cavity that is $\tau_{Lm} = 8$ ps in this section, the values of the other laser parameters are chosen identical to those of Section 5.2.1.

### 5.3.2. Free mode activation

The LFF regime can be observed for a large range of feedback parameters when the injection current is close to its threshold value. Here, we choose $I = 1.0125 \times I_{th}$, $\kappa = 0.12$, $\tau = 3$ ns and the feedback phase in the selected mode $\omega_1 \tau = 0$. Fig. 5.15(a) displays the total intensity averaged over 2 ns to model the limited bandwidth of detectors that are usually employed in experiments. The average total intensity increases steadily with time until it reaches a plateau and finally drops. After the dropout, the intensity gradually recovers until it drops again. On a

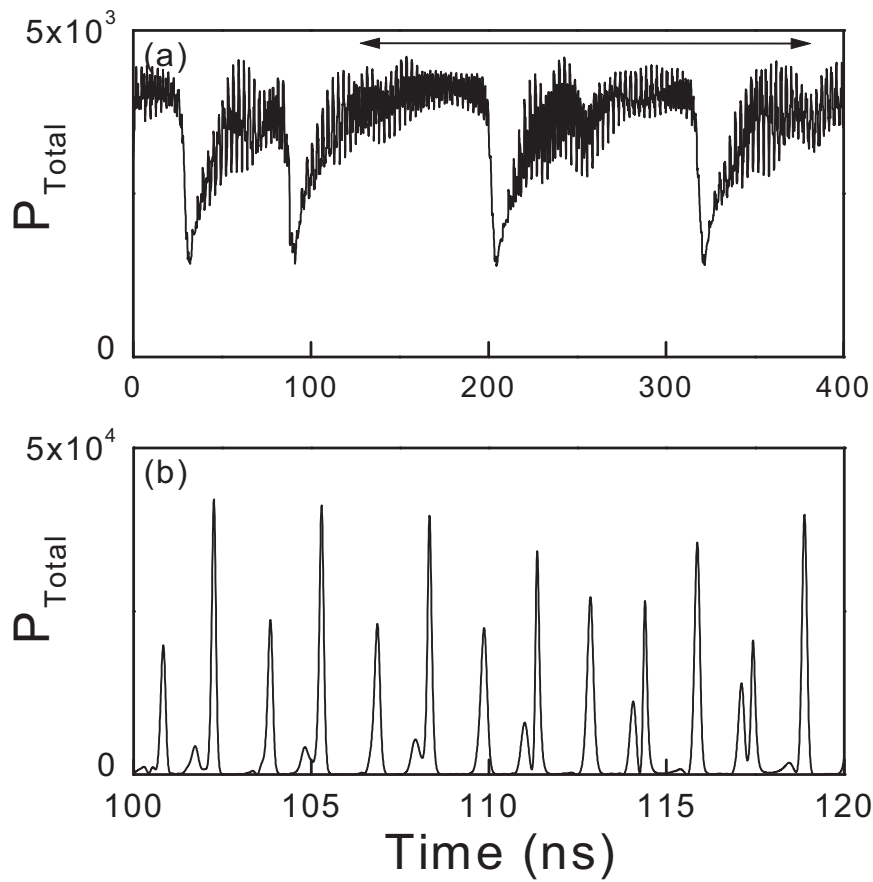

**Fig. 5.15** (a) Time trace of the laser total intensity. The trace has been averaged over 2 ns. (b) Time trace of the unaveraged total number of photons emitted by the laser.

picosecond time scale, the total intensity exhibit trains of fast pulses [Fig. 5.15(b)] which are approximately 150 ps wide and distant of about 1.5 ns. Between two consecutive pulses, the laser intensity is nearly zero. Fig. 5.16 presents the evolution of the modal intensities and the carrier number underlying the total intensity behavior shown in Fig. 5.15(a). Traces (a)-(c) show the temporal evolution of the averaged intensities in the selected mode (m = -1), in the central mode (m = 0) and in a side mode (m = 1), respectively. Trace (d) shows the behavior of the unaveraged carrier number. Similarly to the total intensity, sudden dropouts followed by long recoveries are observed in the selected mode [Fig. 5.16 (a)]. In good agreement with experiments [12,22], the free modes that are depressed most of the time present bursts simultaneous to the dropouts in the selected mode [Fig. 5.16 (b-c)]. The simultaneity of dropouts and bursts is easily checked by comparing the time trace of the total intensity [Fig. 5.15 (a)] with the trace of the selected mode [Fig. 5.16 (a)]: the dropouts in the selected mode are more pronounced than those of the total intensity. The behaviors of the free modes and of the total laser output that we report here compare well with those experimentally recorded by Giudici et al. [22] (those are reported in Fig. 5.17), and by Huyet et al. [12].

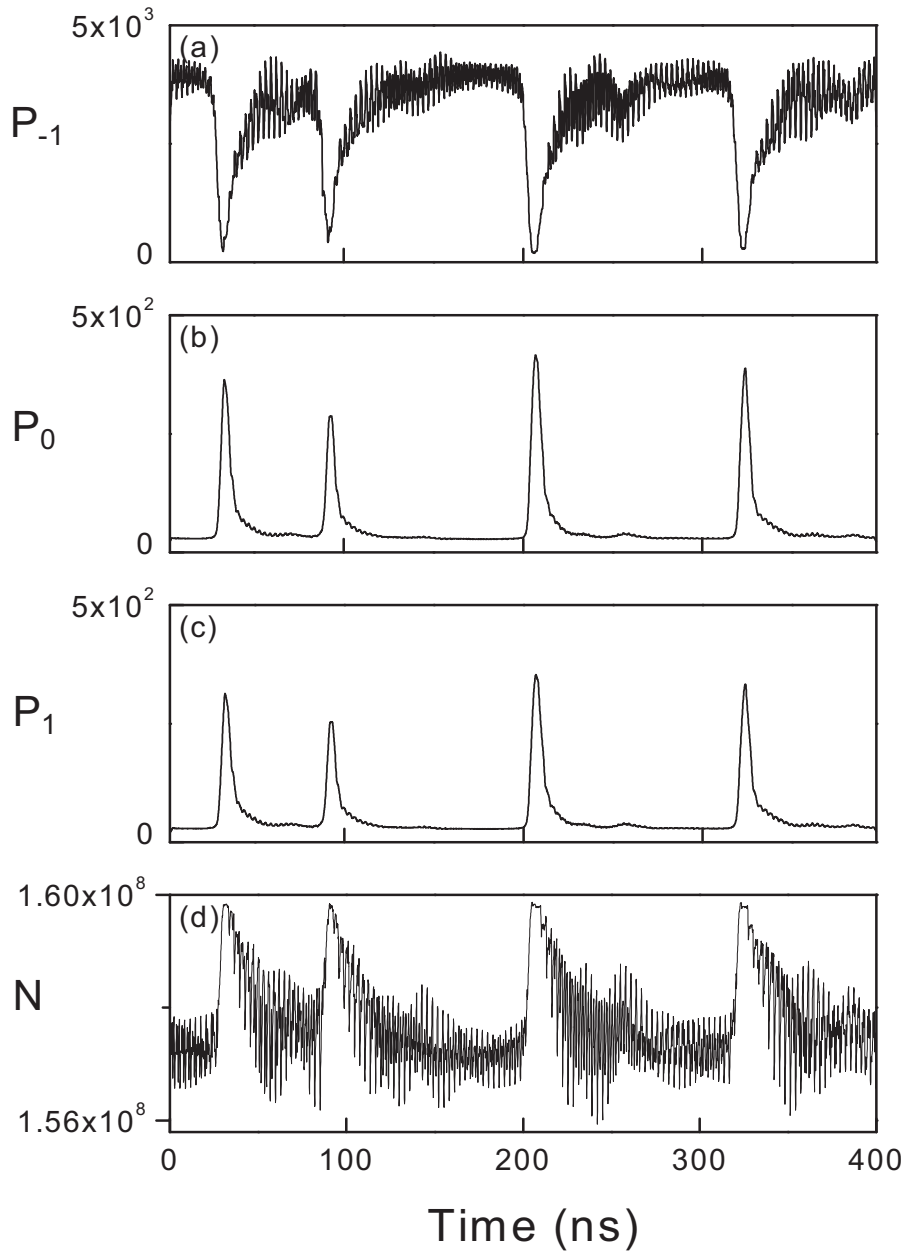

**Fig. 5.16** Time traces of the intensity of mode -1 [selected mode, trace (a)], modes 0 and 1 [free modes, traces (b)-(c)], and time trace of the carrier number [traces (d)]. Traces (a)-(c) have been averaged over 2 ns.

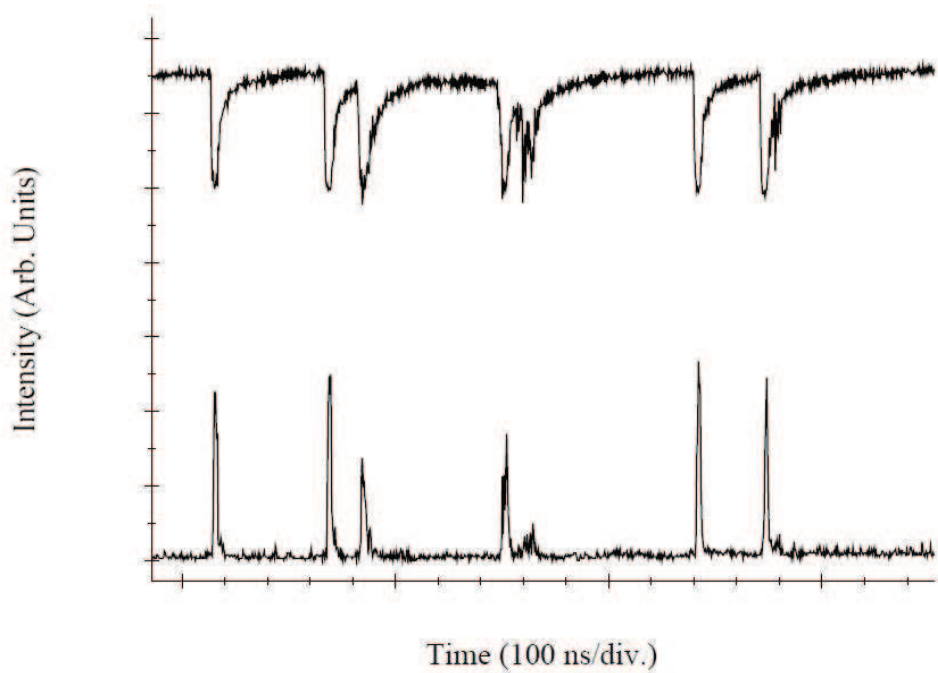

**Fig. 5.17** Simultaneous recordings of the total intensity (upper trace) and intensity of a free mode (lower trace) in a multimode laser diode subject to a frequency selective optical feedback and operating in the LFF regime. The experiment was realized with a Hitachi HLP1400 partially AR coated on the facet exposed to feedback. An holographic grating was placed in the external cavity and adequately tuned to select and reinject a single longitudinal mode in the laser cavity. The feedback delay was estimated to be $\tau = 3.25$ ns. Feedback level and injection current are not specified. Figure and data are extracted from Ref. 22.

The occurrence of bursts in the free modes simultaneously to dropouts in the selected mode may be interpreted as follows. The selected mode plays the central role. The sudden increases of the population inversion [Fig. 5.16(d)] that are associated to the intensity dropouts in the selected mode lead the free modes, triggered by spontaneous emission, to lase during a short time. The bursts last until the carrier number begins to fall. It is important to notice at this point that spontaneous emission is necessary to observe the bursts. If the mean spontaneous emission rate is set to zero, i.e. $R_{sp} = 0$, the free modes vanish and the laser is single-mode ($P_{i \neq -1} = 0$).

Steady state solutions of the rate equations are zeros of the system of nonlinear transcendental algebraic equations obtained by substituting $P_m(t) = P_{sm}$, $\phi_m(t) = (\omega_{sm} - \omega_m)t$ and $N(t) = N_s$ in Eqs. (5.7)-(5.9). They are found numerically by using a Newton-Raphson zero-finding algorithm. Similarly to the case of the single-mode Lang-Kobayashi equations [2],

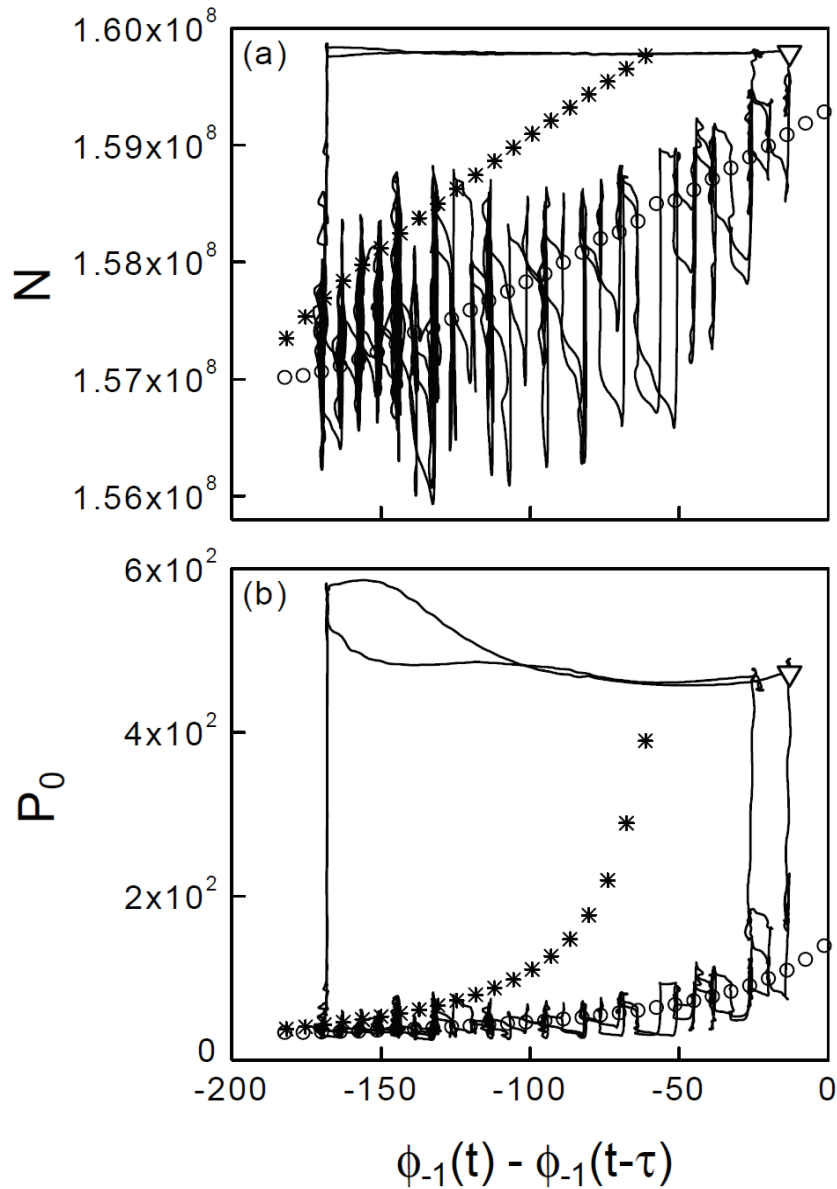

**Fig. 5.18** (a)-(b) Projection of the system trajectory onto the ($\phi_{-1}(t)-\phi_{-1}(t-\tau), N(t)$) and ($\phi_{-1}(t)-\phi_{-1}(t-\tau), P_0(t)$) planes. Circles (o) and stars represent respectively the projections of external cavity modes and antimodes. The sign $\nabla$ identifies the solitary laser steady state. The trajectory is plotted for two consecutive dropouts and corresponds to the time interval indicated by the arrow in Fig. 5.15(a).

steady state solutions of Eqs. (5.7)-(5.9) are created by pairs through saddle-node bifurcations as the feedback level increases (not shown). Of each pair of steady state solutions, one is a saddle point and is referred to as antimode. We refer to the other solution as external-cavity mode. Fig. 5.18 shows the projection of the system trajectory, the steady state solutions of Eqs. (5.7)-(5.9) as well as the steady state solution that corresponds to the solitary laser (i.e. in absence of feedback) onto the ($\phi_{-1}(t)-\phi_{-1}(t-\tau),N(t)$) and ($\phi_{-1}(t)-\phi_{-1}(t-\tau),P_0(t)$) planes. The figure reveals that the system trajectory displays a chaotic itinerancy among the external-cavity modes with a drift towards the maximum gain mode close to which collisions with saddle-type antimodes occur. Just after the collision, the trajectory is repelled towards higher values of both the carrier number [Fig. 5.18(a)] and the free modes intensities [Fig. 5.18(b)] while the intensity in the selected mode drops (not shown). The laser then relaxes to its solitary state. The delayed-feedback on the selected mode triggers the system and the chaotic itinerancy then restarts.

## 5.4. Conclusion and perspectives

Using a multimode extension of the Lang-Kobayashi equations that takes spontaneous emission into account and assumes a parabolic gain profile, we have investigated the low frequency fluctuation regime in a multimode laser diode subject to optical feedback.

The case of global feedback has been considered at first. We have shown here that all longitudinal modes of the laser display low-frequency fluctuations. The mode with the highest gain grows steadily in intensity between two consecutive dropouts while the intensity shared by the other modes saturates shortly after a dropout event in the total output of the laser and then starts to decrease until the next dropout. These results are in good qualitative agreement with a recent experimental study on the slow dynamics of a multimode laser diode subject to a global optical feedback [30].

On a picosecond time scale, we have shown that the longitudinal modes of the laser can oscillate in-phase or out-of-phase, depending on the operating parameters. Each of these two behaviors corresponds to a specific statistical distribution of the laser output. As a result, we have demonstrated that the statistical distributions experimentally measured by Sukow et al. [14], on the one hand, and by Huyet et al. [15], on the other hand, are not conflicting but complementary. Our numerical study has been published as a *rapid communication* in *Physical Review A* [20].

We have also investigated the role of spontaneous emission on the emergence of the out-of-phase dynamics. Although spontaneous emission is intrinsically a stochastic process, we have found that it does not act as a random perturbation sustaining the out-of-phase oscillations of the laser modes but rather as an emission source necessary to multimode operation.

As a second application, we have studied the case of a multimode laser diode subject to a mode-selective optical feedback. In good agreement with recent experiments [12,22], our multimode extension of the Lang-Kobayashi equations does predict intensity bursts in the free modes simultaneous to dropouts in the selected mode. Bursts in the free modes and dropouts in the selected mode are found to be associated with collisions of the system trajectory in phase-space with saddle-type antimodes preceded by a chaotic itinerancy of the system among external-cavity modes. These results will soon be published in *Physical Review A* [29].

In Section 5.2.2, we have observed that the modal dynamics predicted by our multimode model are almost symmetric with respect to the main mode. A very recent experiment reveals however that an asymmetry can occur in the modal dynamics. We have found that a more adequate description of the gain curve yields a physical explanation of this asymmetry. A paper dedicated to these experimental and theoretical results is in preparation in collaboration with the teams in Barcelona and Palma [32].

Finally, we point out that a challenging issue still remains unanswered: can low frequency fluctuations in multimode laser diodes subject to global optical feedback be interpreted in terms of chaotic itinerancy and crises? In order to answer this intriguing question, we are currently investigating the set of deterministic equations that are used in section 5.2.4 and their steady state solutions.

# 6. Anticipative synchronization of two chaotic laser diodes by incoherent optical coupling and its application to secure communications

Detailed studies of the real world impel us, albeit reluctantly, to take account of the fact that the rate of change of physical systems depends not only on their present state, but also on their past history.

R. Bellman and K.L. Cooke,
*Differential-difference equations*,
Academic Press, 1963

## 6.1. Introduction

Synchronization of two chaotic oscillators coupled in a master-slave configuration and its application to secure communications have attracted considerable interest during the past decade [1-12]. In this type of communication, the chaotic output of the transmitter oscillator is used to carry a message in an encoded way. The message can then be decoded at the receiver oscillator provided that the two oscillators synchronize. The encoding can be achieved in several ways. In chaotic masking [4,6,7], the message is added to the chaotic output of the transmitter oscillator [Fig. 6.1 (a)]. In chaos modulation [9], the chaotic output of the transmitter is modulated by the message [Fig. 6.1 (b)]. In both cases, the message decoding is based on the fact that the receiver will synchronize on the chaotic part of the received signal rather than on the full signal (output of the transmitter and message). The message can then be recovered by comparing the output of the receiver and the received signal. In chaos shift keying [5-8,12], the output of the chaotic transmitter itself directly carries the message. The transmitter switches between different chaotic orbits that correspond respectively to the bits "0" or bits "1" as one of the control parameters is modulated by the bit stream [Fig. 6.1 (c)]. Message decoding is then achieved by measuring the synchronization error between the outputs of the two oscillators.

      Most of cryptosystems that have been proposed are implemented by electronic circuits [3-12]. Synchronization and chaotic cryptography have also been demonstrated with

electrooptical circuits [13] and different types of lasers among which, for instance, solid-state lasers [14], fiber ring lasers [15-17] and laser diodes [18-33].

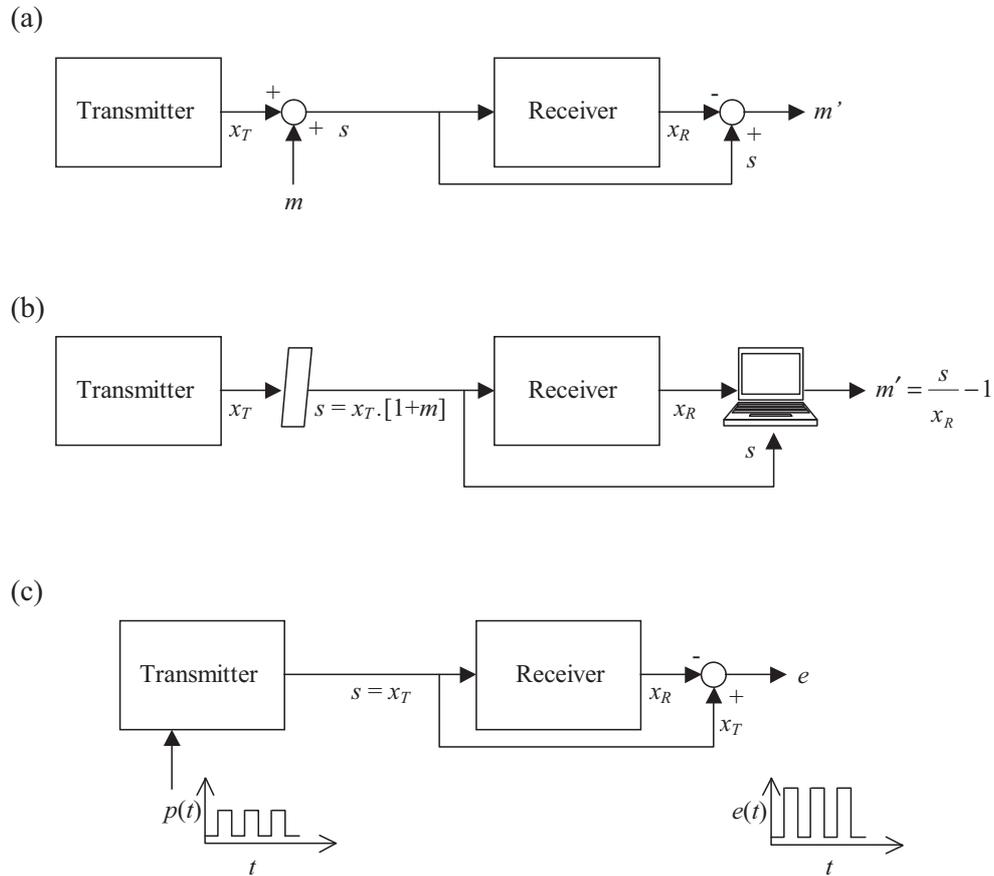

**Fig. 6.1** Schemes of secure communications based on chaotic masking (a), chaos modulation (b) and chaos shift keying (c). $x_T$ and $x_R$ are the outputs of the transmitters and the receivers, respectively. *m* is the message to be transmitted. *s* is the signal that carries the message. *m'* is the recovered message at the receiver. In the case of the communication scheme based on chaos shift keying, *p* is the control parameter at the transmitter that is modulated by the message and *e* is the synchronization error.

In recent years, much attention has been devoted to the synchronization of chaotic semiconductor lasers since these lasers are the most popular optical sources in high-speed optical communication systems. The cryptosystems involving semiconductor lasers exploit different means to drive them to chaos, for instance: conventional – coherent – optical feedback [18,20,23,25,28,29] and optical injection [19,31]. Other schemes involve semiconductor lasers with a nonlinear optical component in an optoelectronic feedback loop to generate the chaotic wave form [21,22]. Schemes implementing laser diodes with optical feedback are of particular interest. On the one hand, the all-optical feedback does not reduce the bandwidth. On the other hand, the time-delayed feedback generates a high-dimensional chaos [24] that in turn leads potentially to a high security level: indeed, the nonlinear dynamics techniques that have been proposed to unmask the messages encoded in chaos [35-38] are much less efficient in the case of high-dimensional chaos. Several experiments have demonstrated the feasibility of synchronization of two laser diodes and of message encoding/decoding in the case where the chaos was induced by coherent optical feedback [26-30,32]. However, numerical studies on such schemes implementing single-mode laser diodes subject to coherent optical feedback have revealed that the synchronization performance depends on the detuning between the free-running frequencies of the transmitter and the receiver lasers. According to Ref. 33, negative detuning by a few hundred megahertz of the receiver frequency relative to the transmitter frequency (which can easily occur under standard conditions of operation) can lead to a large degradation of the synchronization. From a practical point of view, it is therefore interesting to investigate alternative synchronization and cryptographic schemes that would not require fine tuning of the optical frequencies.

      In this chapter, we demonstrate numerically anticipative synchronization [24,39] between two unidirectionaly coupled chaotic laser diodes. The transmitter is subject to incoherent optical feedback and the receiver is coupled to the transmitter via incoherent optical injection. In this scheme, the feedback and injected fields act on the population inversion in the laser active layers but do not interact coherently with the intracavity lasing fields. As a consequence, the phases of the feedback and injection fields do not intervene on the lasers dynamics. For that reason, the synchronization scheme we propose does not require fine tuning of the optical frequencies of both lasers as it does for other schemes based on chaotic laser diodes subject to coherent optical feedback and injection. It is therefore attractive for experimental realization. We investigate the robustness of anticipative synchronization with respect to spontaneous emission noise and parameter mismatches. We demonstrate furthermore that this new synchronization scheme can be applied to secure communications: a message can be encrypted at the transmitter by chaos shift keying and, at the end of the transmission link, extracted when anticipative synchronization between both lasers is achieved. We check that the message cannot be simply obtained by low-pass filtering the transmitted signal. Finally, the sensitivity of the synchronization to large mismatches of parameters makes the replication of the scheme by eavesdropper difficult.

      This chapter is organized as follows. In Section 2, we describe the synchronization scheme we propose. In Section 3, we present the equations we use to modelize the dynamics of the transmitter and receiver lasers. In the two next sections, we give the necessary condition for anticipative synchronization and we show that the synchronization is robust to noise and small mismatches of parameters. In Section 6, we demonstrate that our scheme allows the encoding/decoding of a 250 Mbit/s message and that the communication is secure at least with respect to naïve interception.

The main results presented in this chapter have been published in *Optics Letters* [40]. The whole material of this chapter is submitted to *Optics Communications* [41].

## 6.2. Synchronization scheme

In the scheme we propose (Fig. 6.2), the linearly polarized output field of the transmitter laser first undergoes a 90° polarization rotation through an external cavity formed by a Faraday rotator (FR) and a mirror. It is then split by a non-polarizing beam splitter (BS) into two parts: one is fed back into the transmitter laser and the other is injected into the receiver laser. Polarization directions of feedback and injection fields are orthogonal to those of transmitter and receiver output fields, respectively. In other words, the transmitter laser is subjected to incoherent optical feedback while the receiver laser is subjected to incoherent optical injection. An optical isolator (ISO) shields the transmitter from parasitic reflections from the receiver. Tunable attenuators (not shown in Fig. 6.2) adjust the respective strengths of the feedback and the injection. If necessary, a linear polarizer (LP) may be placed between the Faraday rotator and the mirror to prevent coherent feedback induced by a second round-trip in the external cavity after reflection on the transmitter laser front facet.

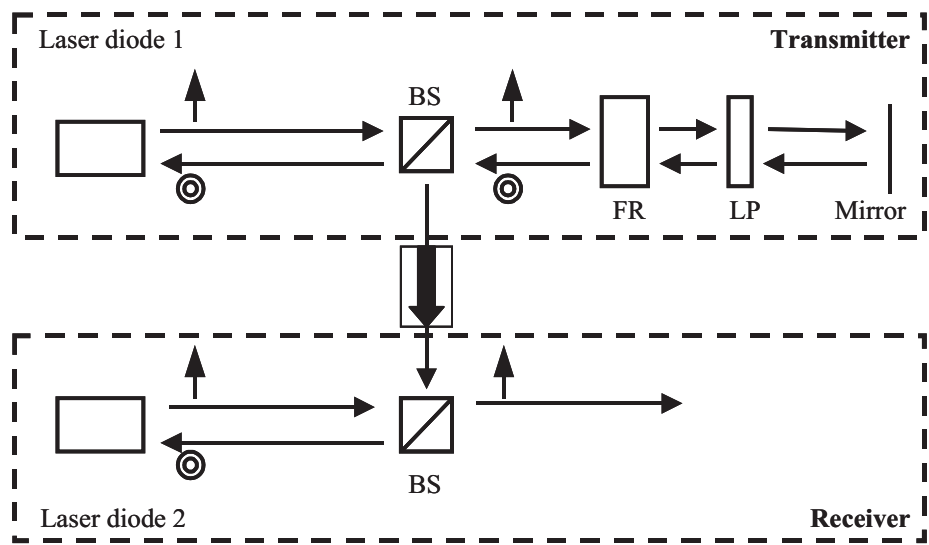

**Fig. 6.2** Schematic representation of the synchronization scheme. See text for definitions.

## 6.3. Model

The model used to describe the dynamics of the system is an extension of models proposed by Otsuka and Chern for respectively semiconductor lasers subject to incoherent optical feedback[1] [42] and semiconductor lasers with mutual incoherent coupling [43]:

$$\frac{dP_1(t)}{dt} = \left(G_1 - \frac{1}{\tau_{p1}}\right)P_1(t) + \beta_1 N_1(t) + F_1(t), \tag{6.1}$$

$$\frac{dN_1(t)}{dt} = \frac{I_1}{e} - \frac{N_1(t)}{\tau_{s1}} - G_1[P_1(t) + \gamma P_1(t-\tau)], \tag{6.2}$$

and

$$\frac{dP_2(t)}{dt} = \left(G_2 - \frac{1}{\tau_{p2}}\right)P_2(t) + \beta_2 N_2(t) + F_2(t), \tag{6.3}$$

$$\frac{dN_2(t)}{dt} = \frac{I_2}{e} - \frac{N_2(t)}{\tau_{s2}} - G_2[P_2(t) + \sigma P_1(t-\tau_c)] \tag{6.4}$$

where $G_j = G_{Nj}[1 - \varepsilon_j P_j][N_j - N_{0j}]$ with $j = 1$ for the transmitter and $j = 2$ for the receiver. In these equations, $P_j$ and $N_j$ are the photon number and the electron-hole pair number in the active region of laser $j$. $N_{0j}$ is the value of $N_j$ at transparency. $\tau_{pj}$, $\tau_{sj}$, $I_j$, $G_{Nj}$ and $\varepsilon_j$ are respectively the photon lifetime, the carrier lifetime, the injection current, the gain coefficient and the gain saturation coefficient of laser $j$. $e$ is the electronic charge. $F_j$ is a Langevin noise force that accounts for stochastic fluctuations arising from spontaneous emission process. The Langevin forces satisfy the relations $<F_j(t)F_j(t')> = 2N_jP_j\beta_j\,\delta(t-t')\delta_{ij}$, where $\beta_j$ is the spontaneous emission rate. The operating parameters $\gamma$, $\tau$ and $\sigma$ are respectively the strength and the delay of the feedback at the transmitter, and the coupling strength at the receiver. The duration taken by the light emitted by the transmitter to reach the receiver is $\tau_c$. We use typical values for the internal parameters of the transmitter laser: $\tau_{p1} = 2$ ps, $\tau_{s1} = 2$ ns, $G_{N1} = 1 \times 10^4$ s$^{-1}$, $N_{01} = 1.1 \times 10^8$, $\beta_1 = 5 \times 10^3$ s$^{-1}$ and $\varepsilon_1 = 7.5 \times 10^{-8}$. In a first step, the parameters at the receiver are chosen identical to those of the transmitter. Afterwards, we will consider slight differences between the corresponding parameters.

---

[1] A derivation of this model is presented in Chapter 2.

## 6.4. Anticipative synchronization

In absence of the stochastic terms $F_j$ and for identical internal and operating parameters, the exact synchronous solution,

$$P_2(t) = P_1(t - \Delta t), \qquad (6.5)$$

$$N_2(t) = N_1(t - \Delta t), \qquad (6.6)$$

where $\Delta t = \tau_c - \tau$ is the synchronization lag, exists only if the coupling strength at the receiver matches exactly the feedback strength at the transmitter, i.e. $\sigma = \gamma$. It should be noted that this condition is necessary but not sufficient to observe synchronization between the two lasers since the solution (6.5)-(6.6) can be stable or unstable. The latter implies that the state of the receiver at time $t$ can synchronize to the state of the transmitter at time $t - \tau_c + \tau$. In other words, the receiver anticipates the signal that will be injected at time $t + \tau$. The anticipation time is $\tau$, the feedback delay at the transmitter. Anticipative synchronization has been demonstrated recently to result from the interaction between delayed feedback and dissipation and to be a rather universal phenomenon in nonlinear dynamical systems with unidirectional coupling [39]. It has also been predicted in coupled laser diodes subject to delayed coherent optical feedback [24,33,34,44].

We first consider the synchronization of two identical lasers when the stochastic terms $F_{1,2}(t)$ are neglected in Eqs. (6.1) and (6.3). The feedback strength and delay at the transmitter are $\gamma = 0.41$ and $\tau = 9$ ns respectively. The injection current of this laser is $I_1 = 1.8 \times I_{th1}$ where $I_{th1}$ is the threshold value of $I_1$. In a first step, the receiver is shielded from the transmitter [thus, the coupling strength $\sigma = 0$ in Eq. (6.4)]. The output of the transmitter laser subject to incoherent optical feedback is chaotic [Fig. 6.3(a)], whereas that of the uncoupled receiver is steady [Fig. 6.3(b)]. In a second step, at time $t = 20$ ns, the second laser is coupled to the transmitter and the signal allowed to enter the receiver with a coupling strength that matches exactly the feedback strength at the transmitter, i.e. $\sigma = \gamma$. The receiver is then driven into chaos and synchronizes perfectly to the transmitter after a short transient; the normalized synchronization error, which is defined as

$$\Delta P(t) = \frac{P_1(t - \Delta t) - P_2(t)}{P_0}, \qquad (6.7)$$

where $P_0$ is the mean value of the receiver output in absence of optical injection, reaches 0 [Fig. 6.3(c)]. This perfect synchronization can also be shown on the synchronization diagram where the output of the receiver at time *t* is plotted versus the output of the transmitter a time *t* - Δ*t* [Fig. 6.3(d)]). After the transient, all the points lie on the 45° diagonal line.

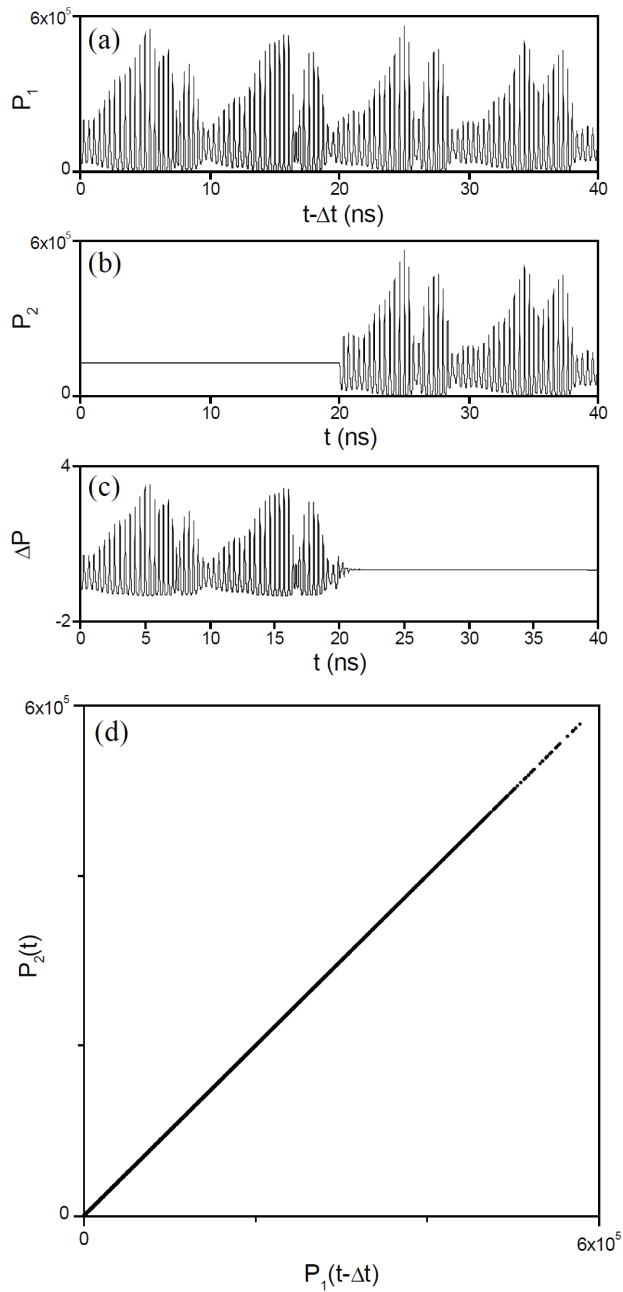

**Fig. 6.3** Simulation of synchronization. Laser parameters and internal parameters are identical. Stochastic terms $F_j(t)$ are neglected in Eqs. (6.1) and (6.3). (a) Output of the transmitter. (b) Output of the receiver. (c) Normalized synchronization error. The transmitter output is shifted by Δ*t*. (d) Synchronization diagram of the receiver output $P_2(t)$ versus the transmitter output $P_1(t-\Delta t)$. This diagram has been calculated from a simulation over 1-μs length. The transient has been discarded.

In Fig. 6.4, the stochastic terms $F_{1,2}(t)$ are taken into account. Although the output of the receiver laser [Fig. 6.4(b)] is still similar to that of the transmitter [Fig. 6.4(a)], the synchronization error is no longer null [Fig. 6.4(c)]: in the synchronization diagram [Fig. 6.4(d)], the points no longer lie exactly on the 45° diagonal line. Since they would lie on this straight line in the ideal case, the quality of the synchronization can be usefully characterized by the linear correlation coefficient r: the better the synchronization, the closer r to unity. In the present case, r = 0.993 indicating a good level of synchronization. Anticipative synchronization between the two lasers is robust with respect to stochastic fluctuations induced by spontaneous emission noise.

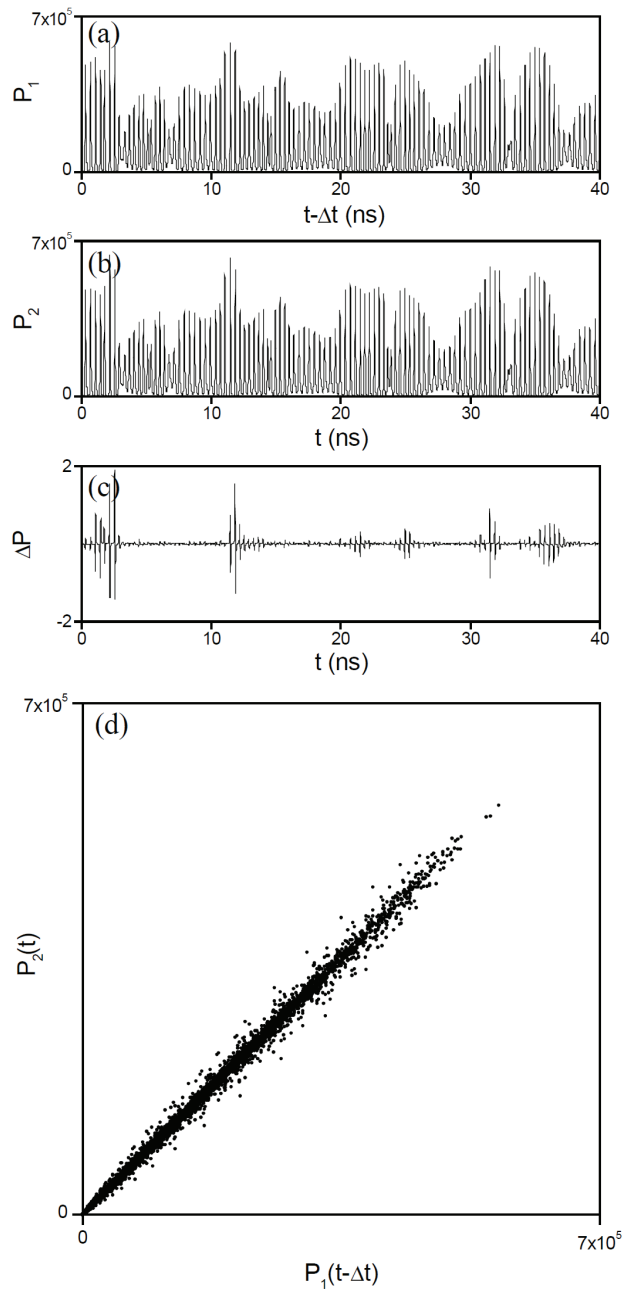

**Fig. 6.4** Same as Fig. 6.3 but stochastic terms $F_j(t)$ are taken into account.

## 6.5. Robustness of the synchronization to parameter mismatches

For cryptographic purposes, synchronization between the transmitter and receiver must only be possible for very similar internal parameters of both lasers as well as for almost identical operating parameters. Sensitivity of the synchronization to mismatches of the parameters should lead to a high level of security due to the difficulty to replicate the transmitter. However, under real world conditions, internal parameters of the two laser diodes never match exactly even if they are produced on the same wafer. Moreover, the operating parameters cannot be perfectly controlled. In practical cases, synchronization must therefore be possible also for small parameter mismatches.

Fig. 6.5 (a) shows the dependence of the correlation coefficient r between the outputs of the transmitter and receiver lasers on the relative mismatch $\delta$ between the injection currents, on the one hand, and between the coupling strength at the receiver and the feedback strength at the transmitter, on the other hand. Fig. 6.5 (b) shows the dependence of r on the mismatch between carrier and photon losses, carrier numbers at transparency and gain coefficients. The dependence of the correlation coefficient r on the relative mismatch between the gain saturation coefficients and spontaneous emission rates at the transmitter and receiver is presented in Fig. 6.5 (c). The effect of the Langevin noise forces has been taken into account in these two figures. In each case, the correlation coefficient decreases as the mismatch increases but it remains above 0.9 if the discrepancies between the transmitter and the receiver parameters are smaller than 1%. The synchronization is therefore robust to small mismatches between corresponding parameters in both systems. The injection current is the most critical parameter, followed by the carrier and the photon lifetimes. By contrast, mismatches of several percents on the gain saturation coefficient and the spontaneous emission rate do not affect the synchronization. Worth noting is that a 1% mismatch between the gain coefficients corresponds to a frequency detuning of several hundreds GHz if the frequency dependence of the gain is taken into account [45].

In order to illustrate how large parameter mismatches can degrade the synchronization, we show the time evolution of both lasers, the synchronization error and the synchronization diagram in Fig. 6.6 for $I_2 = 1.1 \times I_1$ (i.e. a 10% mismatch). This figure shows, on the one hand, that the two lasers depart from each other and that, on the other hand, the points in the synchronization diagram are strongly dispersed: the correlation coefficient is r = 0.23.

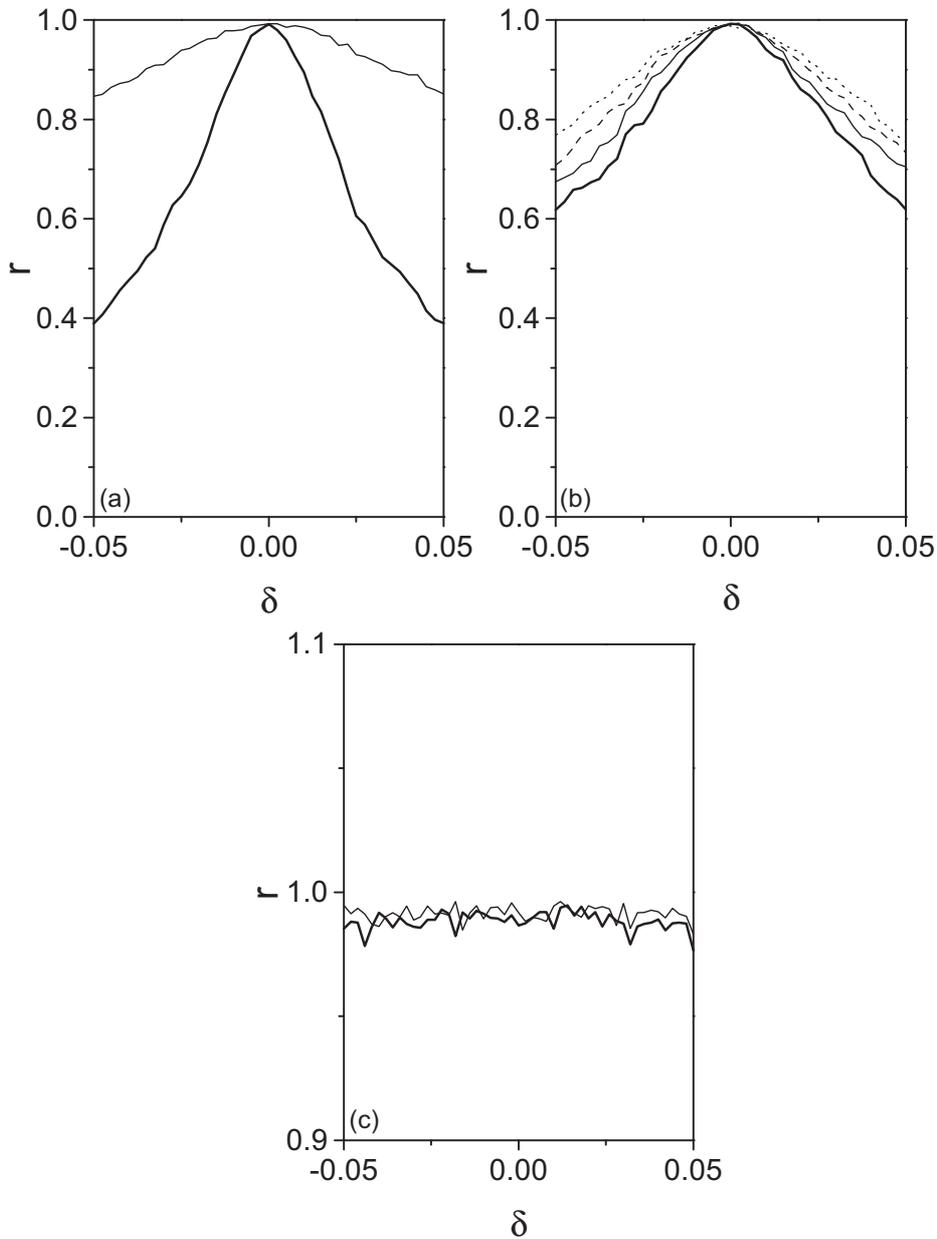

**Fig. 6.5** (a) Correlation coefficient versus relative mismatches $\delta$ of injection current $I$ (thick line) and relative mismatches between the feedback strength $\gamma$ and the coupling strength $\sigma$ (thin line), respectively. (b) Correlation coefficient versus the relative mismatches $\delta$ of carrier losses $1/\tau_s$ (thick line), photon losses $1/\tau_p$ (thin line), carrier number at transparency $N_0$ (dashed line) and gain coefficient (dotted line). (c) Correlation coefficient versus the relative mismatches $\delta$ of gain saturation coefficient (thick line) and spontaneous emission rate (thin line). The correlation coefficients have been calculated from five time series of 1-µs long.

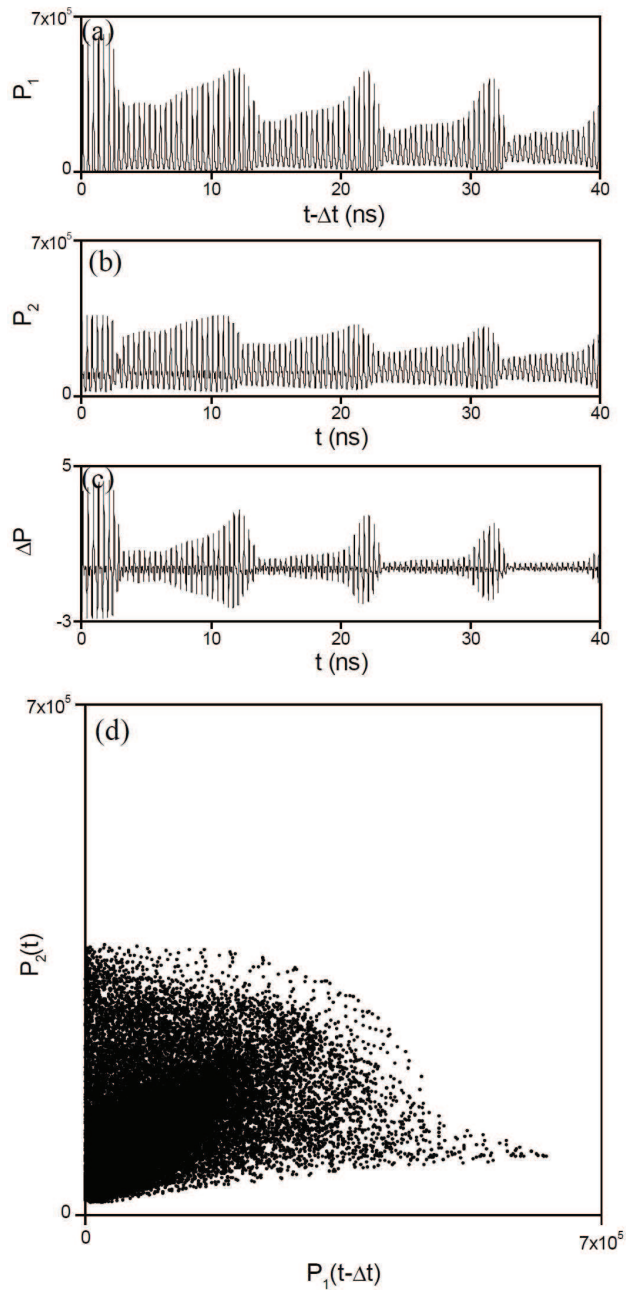

**Fig. 6.6** Same as Fig. 6.4, but with $I_2 = 1.1 \times I_1$.

We point out that the gain saturation coefficient $\varepsilon$ strongly affects the dependence of the synchronization quality on both parameter mismatches and noise level. The correlation coefficient versus mismatches of the injection currents [Fig. 6.7 (a)] and carrier losses [Fig. 6.7 (b)] is displayed for three different values of $\varepsilon$, namely $8 \times 10^{-8}$, $4 \times 10^{-8}$ and $2 \times 10^{-8}$. In the case of 1% mismatch on the carrier losses, the values of the correlation coefficient are respectively 0.95, 0.73 and 0.48. In the case of 1% mismatch on the injection current, the values of the correlation coefficient are respectively 0.87, 0.62 and 0.43. Worth noting is that the adverse role of noise increases as the gain saturation coefficient decreases: in absence of parameter mismatches (i.e. $\delta = 0$), the value of the correlation coefficient are respectively 0.99, 0.80 and 0.53 when the gain saturation coefficient is $\varepsilon = 8 \times 10^{-8}$, $4 \times 10^{-8}$ and $2 \times 10^{-8}$.

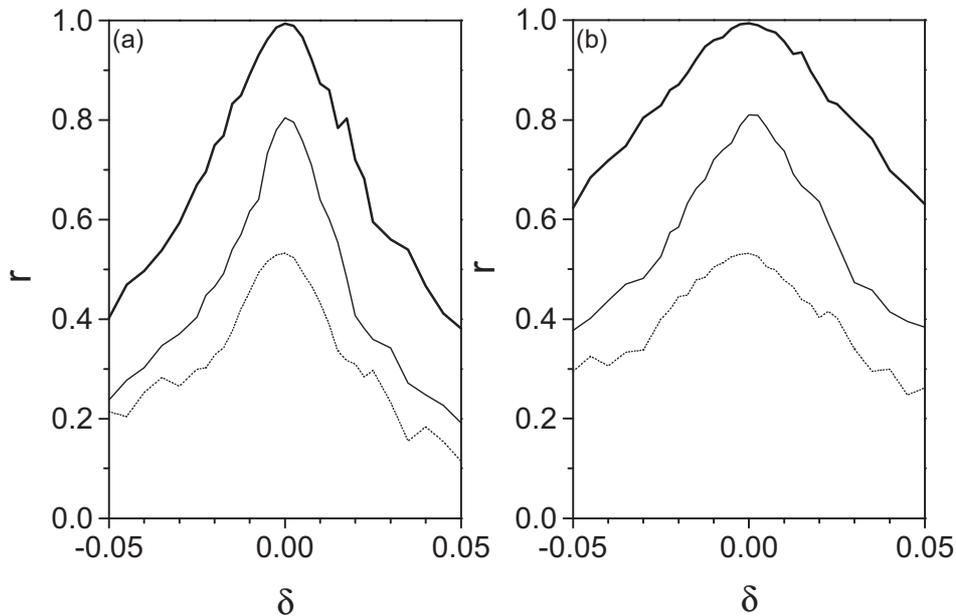

**Fig. 6.7** Correlation coefficient versus the relative mismatches $\delta$ of injection current (a) and carrier losses (b) for three different values of $\varepsilon$, namely $8 \times 10^{-8}$ (thick line), $4 \times 10^{-8}$ (thin line) and $2 \times 10^{-8}$ (dotted line). The correlation coefficients have been calculated from five time series of 1-µs length.

The role of $\varepsilon$ on synchronization robustness can be understood as follows. As shown in Chapter 2, the linear damping rate increases drastically with the gain saturation coefficient $\varepsilon$. This increase leads in turn to a more robust synchronization with respect to parameter mismatches. Indeed, as Voss [39] has demonstrated, anticipative synchronization is the result of the interplay between delayed feedback and dissipation; moreover, the larger the damping resulting from dissipation, the more robust the synchronization to perturbations and parameter mismatches [39].

## 6.6. Cryptography implementing chaos shift keying

In this study, the message encoding is achieved by chaos shift keying (Fig. 6.8). The bit stream modulates the injection current at the transmitter, i.e. bits "0" and "1" correspond to two different values of the injection current $I_1$. Here we choose to use $\iota_0 = 1.8 \times I_{th,1}$ and $\iota_1 = 1.003 \times \iota_0$ respectively. At the receiver, a replica of the transmitter laser is used. The injection current $I_2$ at the receiver is set to $\iota_0$. Ideally, i.e. in the absence of parameter mismatch and noise, message decoding is achieved by measuring the synchronization error. However, when spontaneous emission noise is taken into account or in presence of parameter mismatches, even small, the synchronization error fluctuates strongly and needs low-pass filtering in order to recover the message. We have tried different filters and found that a fourth-order Butterworth filter with a cut-off frequency of 1.3 B, where B is the bit rate, is a convenient choice. Bits "0" are then detected when the filtered synchronization error is close to zero. By contrast, bits "1" are detected when the synchronization error is large due to the mismatch between the injection currents. Fig. 6.9 shows a 250 Mbit/s message transmission for lasers chosen identical and with stochastic terms $F_{1,2}(t)$ taken into account.

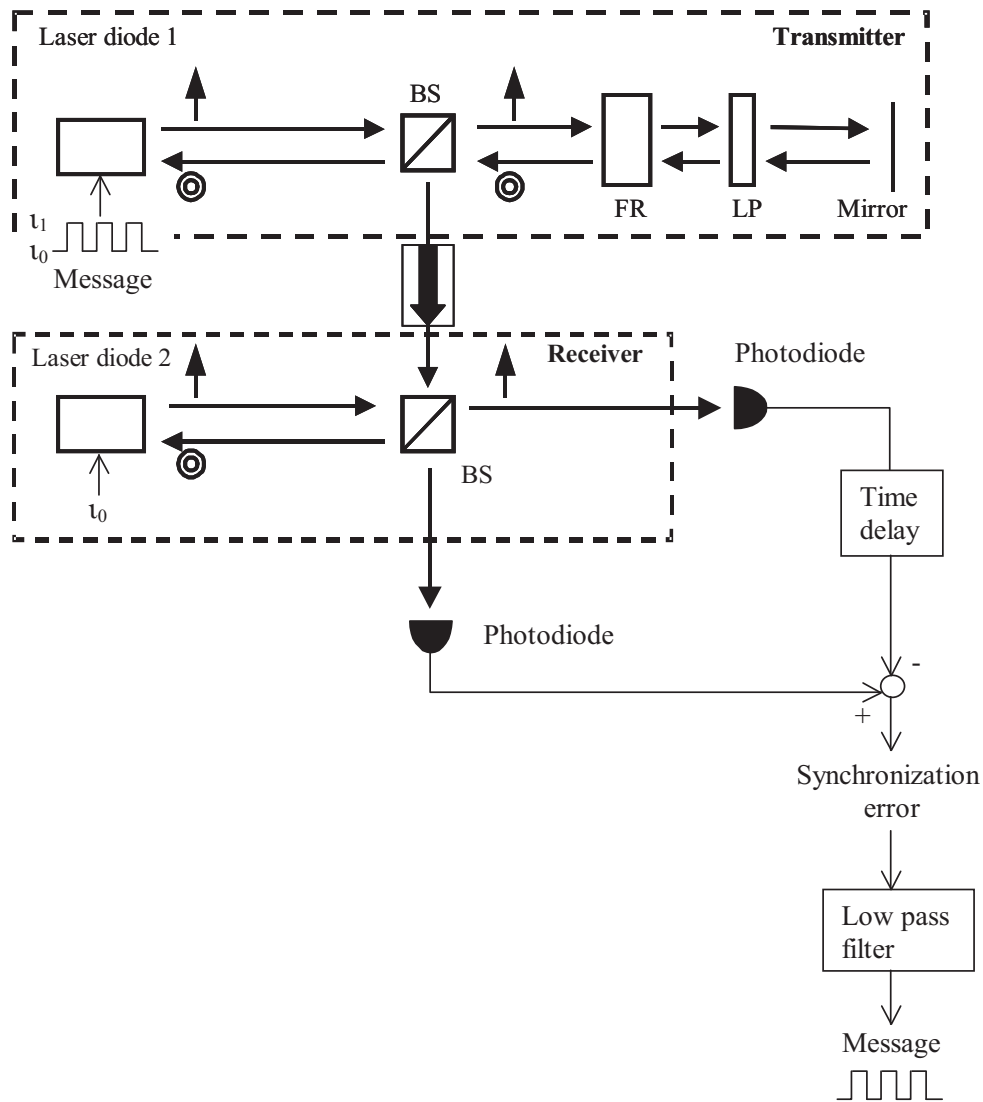

**Fig. 6.8** Schematic representation of our secure communication scheme. The message encoding is achieved by chaos shift keying: the bit stream modulates the current that is injected in the transmitter laser.

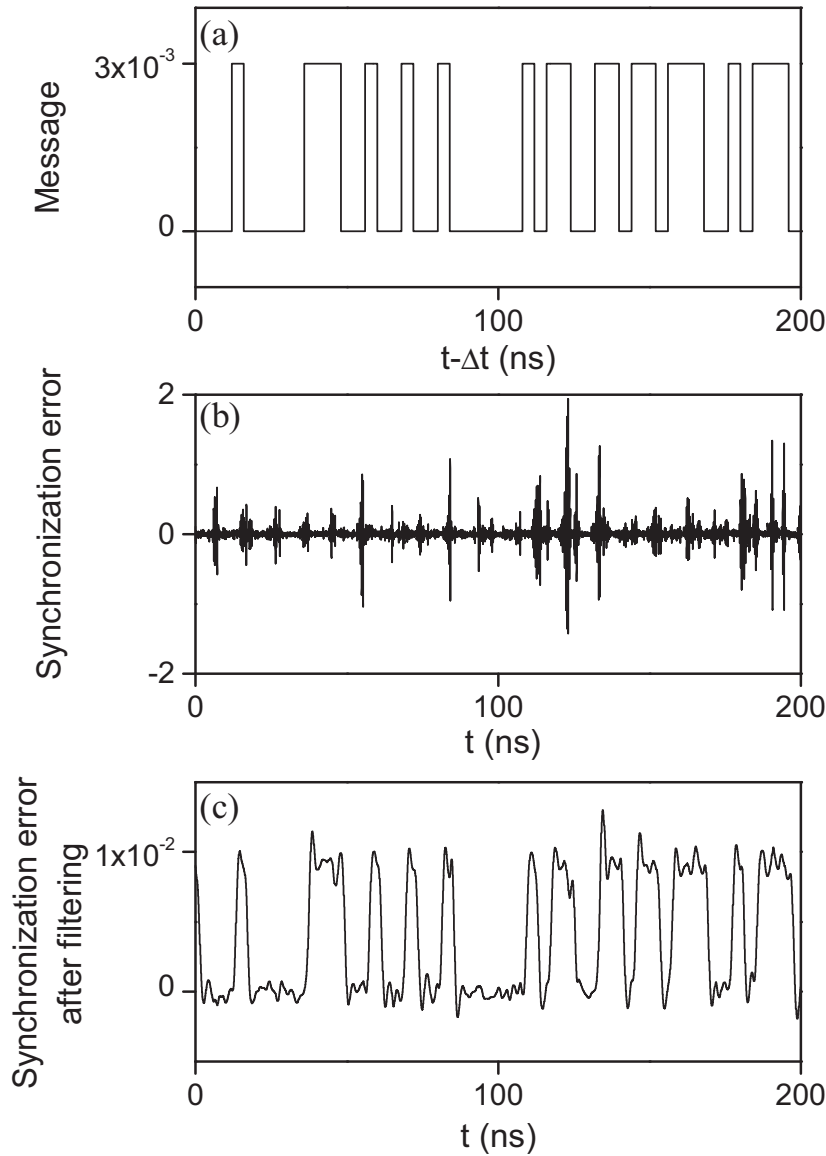

**Fig. 6.9** (a) Encoded message at a bit rate of 250 Mbit/s. (b) Synchronization error without filtering. (c) Synchronization error after filtering.

The robustness of the synchronization to small mismatches between homologous internal and operating parameters makes the secure communication implementation practically feasible. Indeed, two lasers are never identical even if they are produced on the same wafer. Moreover, the operating parameters cannot be perfectly controlled. The robustness is illustrated in Fig. 6.10 where we assume a 0.5% positive mismatch on the carrier and the photon losses: the message can still be recovered after filtering the synchronization error. The filtered synchronization error is shifted to lower values with respect to the case where the parameters match exactly. This result can be understood as follows: an increase of both photon and carrier losses leads to an increase of the threshold and therefore to a decrease of the average power.

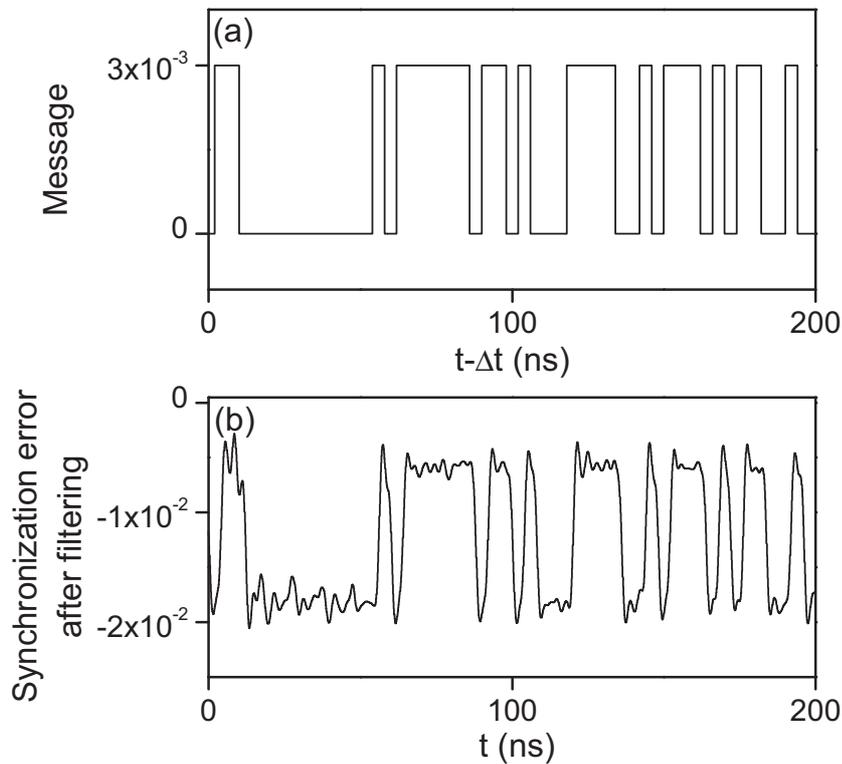

**Fig. 6.10** (a) Encoded message at a bit rate of 250 Mbit/s. (b) Synchronization error after filtering in the case of 0.5% mismatches on the carrier losses $1/\tau_s$ and the photon losses $1/\tau_p$.

The relative intensity noise (RIN) spectrum of the transmitter output under injection current modulation by the message is displayed in Fig. 6.11 (a-b) along with a reference RIN spectrum in absence of modulation (Fig. 6.11 (c-d)). Although a slight difference can be observed at low frequencies, both spectra look very similar so that the extraction of the message should be quite difficult. The encoded bits [Fig. 6.12 (a)] are also undetectable by direct observation in the time domain of the output of the transmitter [Fig. 6.12 (b)]. In order to check that the message can no longer be found by low-pass filtering the transmitted signal, Butterworth and Chebyshev filters have been used without success. As an example, Fig. 6.12 (c) depicts the transmitter output after smoothing with a fourth-order low-pass Butterworth filter with 1.3 B cut-off frequency. The message cannot be recognized.

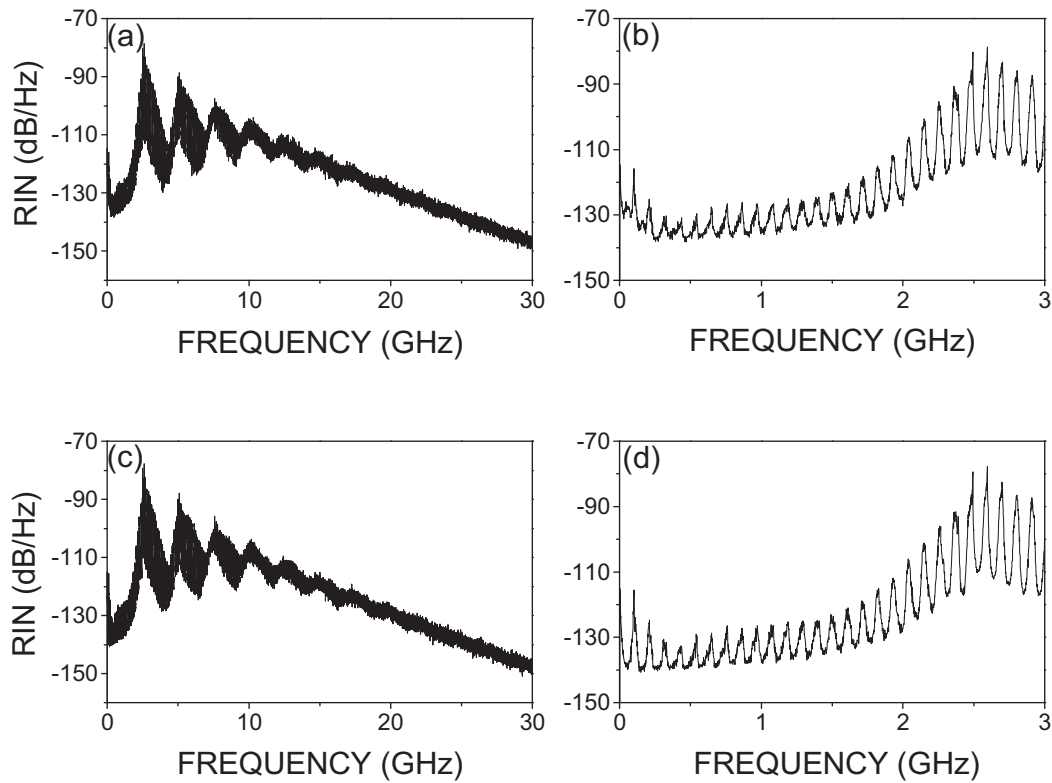

**Fig. 6.11** (a-b) RIN spectrum of the transmitter output when the injection current is modulated by a message. (c-d) Reference RIN spectrum of the transmitter output when the injection current is not modulated. The spectra have been averaged over 10 time series of 2 µs length.

Assuming that the transmitter parameters are unknown, replication of the system by an eavesdropper would be extremely difficult owing to a combination of two reasons. Firstly, parameter mismatches of only a few percents (typically 5%) lead to such severe degradations of the synchronization quality that recovery of the message is not possible. Secondly, from a laser chip to another, critical parameters such as carrier and photon lifetimes vary considerably (1-3 ns and 1-2 ps, respectively [45]). In order to illustrate the difficulty of intercepting the message that is encoded in the chaotic transmitted signal without an adequate replica of the transmitter laser, we assume in Fig. 6.13 a 5% positive mismatch on the carrier losses and the photon losses. We smooth the synchronization error with the same filter as above. Fig. 6.13(b) shows that the message is not recovered. In Fig. 6.14 we assume a 5% positive mismatch on the injection current; in this case neither, the message is not recovered. We have also investigated the influence of negative parameter mismatches on the message recovery and found qualitatively similar results.

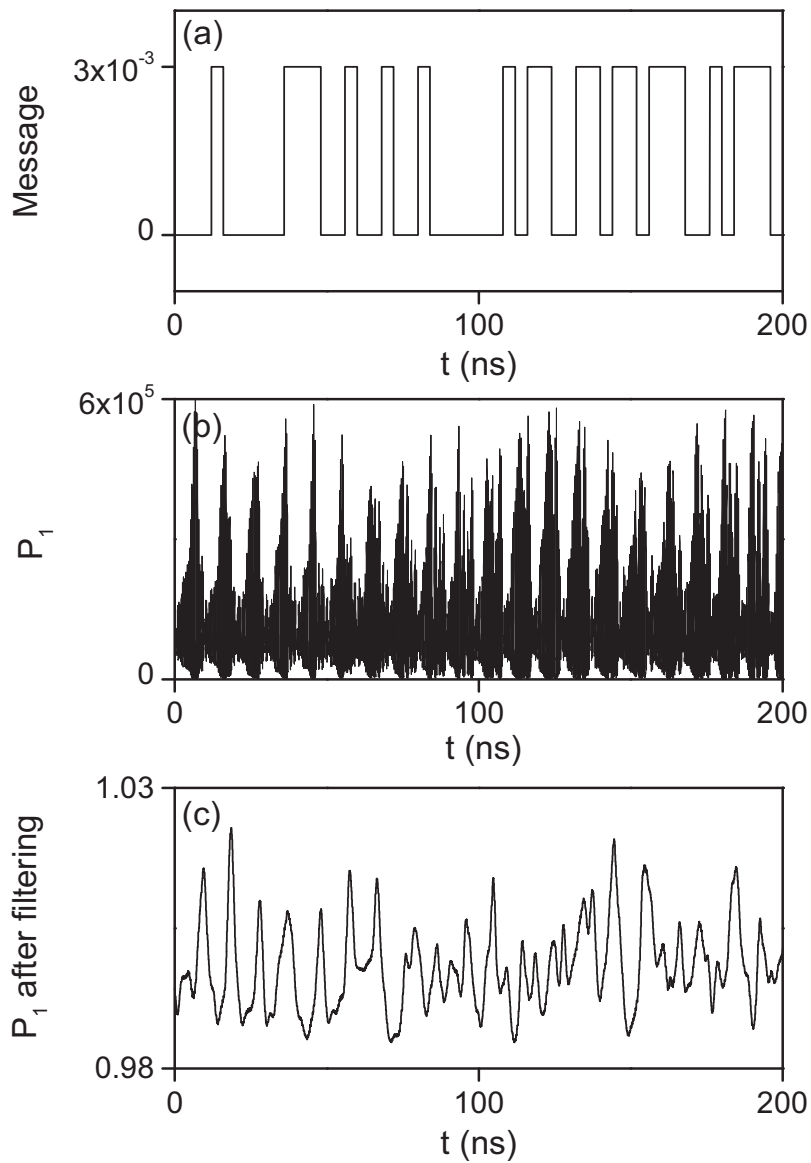

**Fig. 6.12** (a) Encoded message at a bit rate of 250 Mbit/s. (b) Output of the transmitter without filtering. (c) Output of the transmitter after filtering.

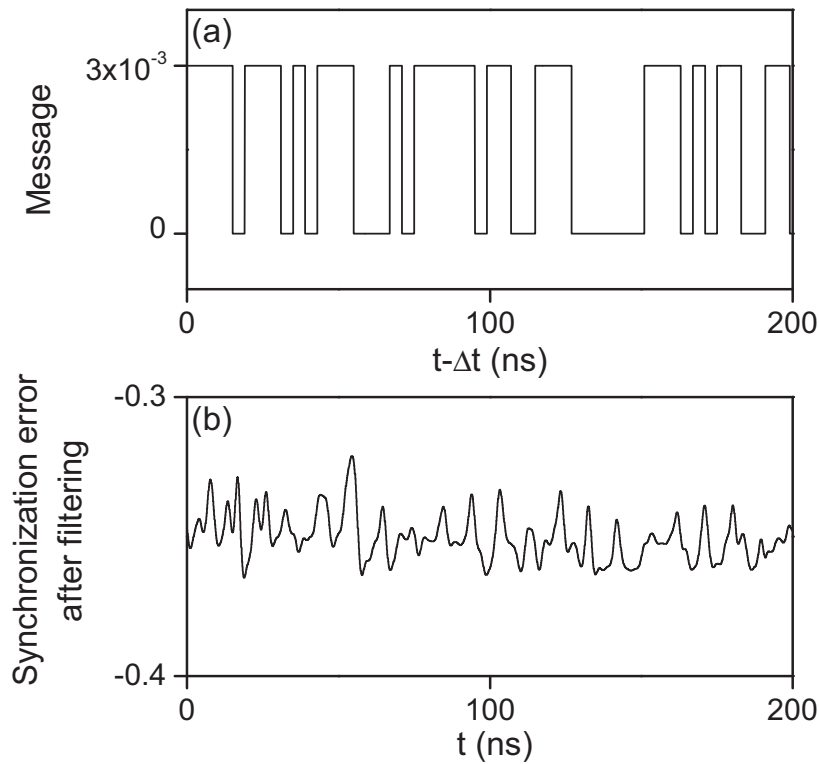

**Fig. 6.13** (a) Encoded message at a bit rate of 250 Mbit/s. (b) Synchronization error after filtering in the case of 5% mismatches on the carrier losses $1/\tau_s$ and the photon losses $1/\tau_p$.

We have finally investigated the effects of the bit rate and the depth of the current modulation at the transmitter on the quality of the message restored at the receiver. In agreement with [19,20,23], the period of modulation cannot be smaller than the transient time that the receiver laser takes to synchronize on the transmitter. The quality of the decoded message degrades as the bit rate increases until the message can no longer be restored. Simulations show that a message with a bit rate larger than 1 Gbit/s cannot be decoded at the receiver. Furthermore, the modulation depth must be small enough. Indeed, symbols can be detected by observation of the chaotic signal coming from the transmitter if the depth of modulation is larger than a few percents.

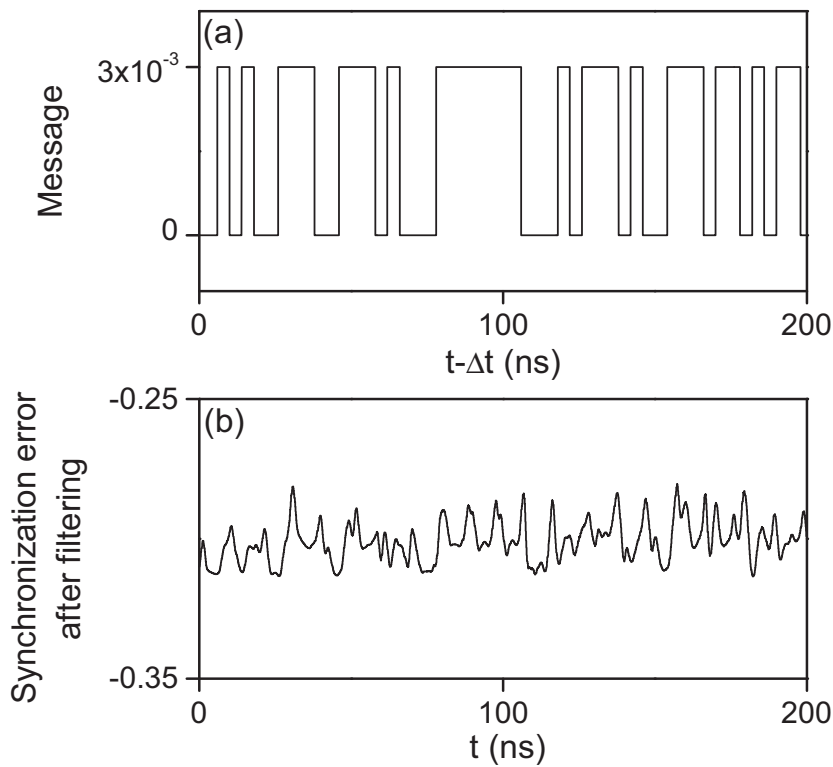

**Fig. 6.14** (a) Encoded message at a bit rate of 250 Mbit/s. (b) Synchronization error after filtering in the case of 5% mismatches on the injection current.

## 6.7. Conclusion and perspectives

We have proposed a novel synchronization scheme involving laser diodes subject to incoherent optical feedback and injection. This scheme is remarkable in that it requires no fine tuning of the laser optical frequencies, unlike other schemes based on laser diodes subject to coherent optical feedback. This is due to the absence of interaction between, on the one hand, the intracavity fields and, on the other hand, the injected and fed back fields; the latter interact only with the carrier population. Our synchronization scheme is therefore attractive for experimental and operational investigations. We have reported on anticipative synchronization between the two lasers provided that the laser parameters and the operating parameters are adequately chosen. We have shown that anticipative synchronization is robust with respect to spontaneous emission noise and small parameter mismatches. We have also observed that the gain saturation drastically affects the robustness of the synchronization. Furthermore, by implementing the former synchronization scheme, we have numerically demonstrated message encoding/decoding by chaos shift keying. The difficulty to intercept the message without an adequate replica of the transmitter laser has also been checked.

A few issues of practical importance, including the effect of the transmission channel and the bit-error rate of secure communications, are not addressed in this chapter. They will be the subjects of future investigations. Another important issue is to determine the dimension of the chaos induced by the incoherent optical feedback. Indeed, due to the delayed nature of this feedback, one can expect a high dimensional chaos that should strongly complicate eavesdropping via reconstruction of the embedding phase space [21,35-38].

# 7. Conclusions

## 7.1. Summary of results

Laser diodes are key components in many applications such as data storage and fiber-optics communication systems. These devices are particularly sensitive to external optical feedback that can induce a large variety of dynamical instabilities. Since those can degrade severely the temporal and spectral performances of laser diodes, it is of outmost importance to understand the underlying mechanisms in order to propose control techniques or, even better, to take advantage of them, for instance in secure communications. Contributing to these goals is the challenge of this thesis

For completeness, we have first derived, in Chapter 2, the basic equations that describe the dynamics of laser diodes subject to coherent and incoherent optical feedbacks. These equations have been analyzed from the point of view of stationary solutions for the particular case of single-mode laser diodes subject to coherent optical feedback (Lang-Kobayashi equations). Linearization of the basic equations in the vicinity of the stationary conditions has then led to the characteristic equation from which the stability of the stationary solutions can be discussed. In the same chapter, the basic notions concerning the low-frequency fluctuation (LFF) regime and its deterministic interpretation by Sano [Phys. Rev. A **50**, 2719 (1994)] are reviewed, given the importance of LFF studies in this thesis.

Inspired by Sano's interpretation of the LFF, we have demonstrated, as a first application of the equations derived in Chapter 2, that a laser diode subject to a first coherent optical feedback and operating in the low-frequency fluctuation regime can be stabilized by means of an adequate second optical feedback. The second feedback modifies the external cavity mode-antimode pattern in such a way that pairs of those can be created or destroyed, depending on the strength of this feedback. Owing to the drift of the system trajectory in phase space, the laser can lock into newly created, stable, maximum gain modes. Moreover, the destruction of those antimodes that are responsible for crises can also lead to LFF suppression.

The above numerical predictions have been confirmed experimentally during a stay at the Nonlinear Optics Center of the Air Force Research Laboratory in Albuquerque, USA. There, an appropriate experimental setup was designed and overall agreement between theory and experiment obtained; in particular, our stabilization scheme was confirmed; it proved to be robust, reliable and easy to implement. These numerical and experimental results are presented in Chapter 4.

In addition to the demonstration of our stabilization technique, the experiment in Albuquerque revealed high-frequency periodic oscillations with frequencies much higher than those expected when the laser is pumped close to threshold. These periodic oscillations appeared to result from a beating process. This supposition has motivated an analytical study of unusual Hopf frequencies in laser diodes subject to a *single* coherent optical feedback. The latter revealed that the Lang-Kobayashi equations indeed admit time-periodic solutions that are combinations of an external cavity mode and an antimode. Simultaneously, we found numerically that high-frequency periodic oscillations similar to those observed in the experiment could be anticipated from the Lang-Kobayashi equations extended to the problem of a feedback from a *double* cavity. Furthermore, a numerical continuation method appropriate to delay-differential equations allowed us to demonstrate that these oscillations are of the same nature as those predicted analytically for the single-feedback configuration. Hence, our observations can be viewed as the first experimental evidence of a beating process between an external cavity mode and an antimode. They demonstrate that external cavity laser diodes can constitute all-optical sources of microwave oscillations, a conclusion of great practical interest. Those results are described in Chapter 3 (single optical feedback) and Chapter 4 (double cavity), respectively.

The necessity of including multimode dynamics and stochastic noise sources in models for the study of low-frequency fluctuations has attracted much attention these past few years. We have demonstrated that a multimode extension of the Lang-Kobayashi equations that takes spontaneous emission into account and assumes a parabolic gain profile reproduces most of the features that have been experimentally observed up to now. In particular, we have shown that two qualitatively different behaviors may take place within LFF in multimode lasers, namely in-phase and out-of-phase oscillations of the longitudinal modes. The corresponding statistical distributions are in good agreement with two recent experiments. We have furthermore demonstrated that spontaneous emission acts mainly as a source necessary to sustain multimode operation; the out-of-phase dynamics is thus by no means related to its intrinsic stochastic nature. These results are presented in Chapter 5.

In the same chapter, we have considered the LFF regime in multimode laser diodes subject to mode-selective optical feedback. In good agreement with experiment, the above model predicts also the simultaneous occurrence of bursts in the free modes and dropouts in the selected mode. We have furthermore established that the selective mode-induced LFF is associated with collisions of the system trajectory in phase-space with saddle-type antimodes preceded by a chaotic itinerancy of the system trajectory among external-cavity modes. This result is of importance, not only from a fundamental point of view, but also from a practical one. Indeed, most of the techniques that have been proposed for controlling LFF rely on the single-mode Lang-Kobayashi equations and are linked to the so-called chaotic itinerancy with a drift. A corollary of our results is that similar stabilization techniques apply to multimode lasers restricted to oscillate in a single mode.

Finally, in Chapter 6, we have numerically demonstrated anticipative synchronization between a first diode subject to incoherent optical feedback and a second diode driven by the first through incoherent optical injection. This synchronization scheme requires no fine tuning of the diode optical frequencies. This is a clear advantage over other schemes based on coherent optical feedback and injection. It is therefore attractive for experimental realization. We have also shown that this synchronization scheme can be applied to chaos-embedded communication and have checked the difficulty to intercept a message encoded by chaos shift keying without an adequate replica of the transmitter laser.

## 7.2. Perspectives

We have shown in Chapter 4 that a second, adequate optical feedback can suppress LFF in a laser diode pumped close to threshold and subject to a first, given, coherent optical feedback. Laser stabilization is in this case intrinsically related to the drift of the system trajectory in phase space toward low frequency external cavity modes. According to Liu and Ohtsubo [IEEE J. Quantum Electron. **33**, 1163 (1997)], a second feedback can also stabilize a laser diode pumped far above threshold. For this parameter regime, the drift of the trajectory cannot however lead to stabilization. Worth investigating are therefore the mechanisms that are responsible for laser stabilization when the latter is pumped far above threshold.

In Chapter 5, which is dedicated to multimode dynamics within the LFF regime, we have shown that a particular extension of the Lang-Kobayashi equations that takes spontaneous emission into account and assumes a parabolic gain profile allows to anticipate most of the experimental behaviors reported in the literature. However, a recent experiment realized at the Universitat Politècnica de Catalunya (Barcelona) has revealed that an asymmetric behavior of the individual modes with respect to the main longitudinal mode can occur. We have recently found that a more accurate description of the modal gain can explain this asymmetry. A paper dedicated to these experimental and theoretical results is in preparation in collaboration with the research teams at the universities of Barcelona and Palma de Mallorca.

In the same Chapter, we have shown that low-frequency fluctuations in multimode laser diodes subject to a mode-selective optical feedback can still be interpreted in terms of chaotic itinerancy with a drift toward the maximum gain mode and of crises with saddle-type antimodes. A future important issue will be to investigate the mechanisms that are responsible for LFF in multimode laser diodes subject to global optical feedback. Vertical-cavity surface-emitting lasers (VCSELs) being characterized by two orthogonal polarization modes, one might argue that studying the LFF observed in those devices can be a useful intermediate step before attacking the much more complex case of edge-emitting lasers operating in multiple longitudinal modes. Research on LFF in VCSELs is presently underway at the Faculté Polytechnique de Mons.

Some issues of practical importance for the secure communication scheme that we have proposed in Chapter 6 have not been investigated in this thesis and will be the subject of future investigations. These include the effect of the transmission channel, the bit-error rate of secure communications and the dimension of the chaos induced by incoherent optical feedback. A high dimensional chaos, as may be expected owing to the delayed nature of the feedback, should strongly complicate eavesdropping via reconstruction of the embedding phase space.

Finally, during the last year of this thesis we have ordered and received a complete and performing experimental setup that will allow to pursue experimental and theoretical studies in parallel.